\documentstyle[11pt,aaspp4,paper,psfig,rotating]{article}

\begin{document}

\title {\bf Radio-loud Active Galaxies in the Northern ROSAT All-Sky Survey\\
III: New Spectroscopic Identifications from the RGB BL~Lac Survey}

\author{S.\ A.\ Laurent-Muehleisen\altaffilmark{1} \\
(slauren@igpp.llnl.gov)}
\affil{University of California-Davis and\\
The Institute for Geophysics and Planetary Physics,\\
Lawrence Livermore National Laboratory\\
7000 East Ave., Livermore, CA 94550, USA}
\authoraddr{L-413\\LLNL/IGPP\\7000 East Ave.\\Livermore, CA 94550}
\altaffiltext{1}{Visiting Astronomer, Kitt Peak National Observatory, National
Optical Astronomy Observatories, which is operated by the Association of 
Universities for Research in Astronomy, Inc.\ (AURA) under cooperative
agreement with the National Science Foundation.}

\author{R.\ I.\ Kollgaard \\
(rik@fnal.gov)}
\affil{Fermi National Accelerator Laboratory, Batavia, IL 60510, USA}

\author{R.\ Ciardullo, E.\ D.\ Feigelson \\
(rbc@astro.psu.edu, edf@astro.psu.edu)}
\affil{Dept.\ of Astro.\ \& Astrophys., The Pennsylvania State University,\\
University Park, PA 16802, USA}

\author{W.\ Brinkmann and J.\ Siebert\\
(wpb@rzg.mpg.de, jos@mpe.mpg.de)}
\affil{Max-Planck-Institut f\"ur extraterrestrische Physik,\\
Giessenbachstrasse, D-85740, Garching, Germany}

\begin{abstract}

We present new spectroscopic identifications for 169 objects in the RASS-Green
Bank (RGB) catalog of radio$-$ and X-ray$-$emitting AGN.  The data presented
here significantly increase the fraction of bright RGB objects with
classifications.  Specifically, we report and discuss the classification of 66
radio-loud quasars, 53 BL~Lacs, 33 Broad Line Radio Galaxies, 5 Narrow Line
Radio Galaxies, 1 Seyfert~I galaxy and 11 galaxies or galaxies in clusters.
Over 78\% of the identifications we present here are the first published
classifications for these sources.  The observations we report were undertaken
as part of our targeted search program to identify a new, large unbiased
sample of BL~Lac Objects and we therefore discuss the BL~Lac sample
extensively.  Unlike many previous surveys, we impose no selection criteria
based on optical morphology, color or broadband spectral energy distribution.
Our classifications are based solely on a carefully defined set of
self-consistent spectroscopic classification criteria.  These criteria are
then carefully evaluated and particular attention is paid to issues involving
the classification and description of BL~Lacs.  The criteria yielded BL~Lac
classifications for 53 RGB objects, 38 of which were are newly discovered
BL~Lacs.  We show these RGB BL~Lacs exhibit transitional properties between
normal galaxies and BL~Lacs discovered in previous radio and X-ray surveys.
We briefly discuss the broadband flux distributions of these new RGB BL~Lacs
and the range of \ion{Ca}{2}~H \& K break contrasts (Br$_{4000}$) they
exhibit.  We show that there is no clear separation in Br$_{4000}$ between
BL~Lacs and galaxies detected in the RGB survey, with the distribution of
break strengths varying smoothly between 0\% and 50\%.  We also present and
use a simple method based on the break strength to estimate the contribution
of both the host galaxy and AGN to the 4000\,\AA\ flux.  We also show that the
newly discovered RGB BL~Lacs reside in a ``zone of avoidance'' in the $\log
({\rm S}_{\rm x}/{\rm S}_{\rm r})$ vs.\ $\log ({\rm S}_{\rm o}/{\rm S}_{\rm
r})$ diagram.  This has important implications for BL~Lac search strategies
since it shows that RASS BL~Lac samples will be severely incomplete if
candidates are chosen only from among those objects with the highest S$_{\rm
x}$/S$_{\rm r}$ flux ratios.

\end{abstract}

\keywords{galaxies: active --- BL Lacertae objects: general --- quasars:
general --- radio continuum: galaxies --- surveys --- X-ray: general}

\begin{center}
{\em Submitted for publication in the Astrophysical Journal Supplement Series}
\end{center}

\clearpage

\section {Introduction}

The identification of X-ray and radio sources has proven to be an effective
means of creating samples of quasars, Seyfert and radio galaxies, and
BL~Lacertae Objects (BL~Lacs).  The ROSAT All-Sky Survey (RASS) is an
unusually large flux-limited survey which contains thousands of previously
unidentified AGN.  The original catalog created from it consists of over
60,000 soft X-ray (0.08$-$2.4\,keV) sources with a flux limit that varies with
ecliptic latitude from 1 to 5$\times 10^{-13}$\,ergs\,s$^{-1}$cm$^{-2}$
\cite[]{voges}.  Our goal in the study of this catalog is to create a new,
large unbiased sample of BL~Lacs.  

BL~Lacs are relatively rare objects with only $\sim$250 known prior to the
launch of the ROSAT satellite \cite[]{PGcat}.  Although several samples exist,
study of the BL~Lac phenomena has concentrated on two main samples: the
``1\,Jy'' radio$-$selected \cite[]{stickel} and the Einstein Extended Medium
Sensitivity Survey \cite[EMSS;][]{gioia,stocke,maccacaro} X-ray$-$selected
samples.  Study of these samples has shown that while radio$-$ and
X-ray$-$selected objects (RBLs and XBLs, respectively) generally share many
observational characteristics, the two subclasses are statistically different
in a variety of ways \cite[]{stocke85,morris,perlman,lmheao,heao,jannuzi}.
These results are also valid when more astrophysical classifications are made
based on the frequency at which the synchrotron spectrum peaks.  In this
taxonomy, XBLs are essentially equivalent to ``High frequency-peaked BL~Lacs''
(HBLs) and RBLs correspond well with ``Low frequency-peaked BL~Lacs''
\cite[LBLs;][]{padovanigiommi,rita}.  Few objects with intermediate properties
are known \cite[see, e.g.,][]{rita} and it is these transitional objects which
likely hold important clues about the intrinsic nature of BL~Lacs.

Creation of a large, unbiased sample of BL~Lacs of any particular subclass is,
however, a difficult problem because BL~Lacs are so rare.  They are therefore
far outnumbered by other classes of objects in large area radio or X-ray
surveys.  Their identification in a catalog as large as the RASS therefore
requires efficient selection criteria.  Techniques based on the characteristic
multiwavelength spectral energy distribution (SED) of BL~Lacs discovered in
EMSS \cite[]{gioia,stocke,maccacaro} have been used successfully in
conjunction with the Einstein Slew Survey \cite[]{slew} and the Hamburg Quasar
Survey \cite[]{NA}.  The most common variant of the SED-based search criteria
requires prior knowledge of the radio flux to identify candidates
\cite[]{schachter}.  Although the requirement for detectable radio flux could
theoretically introduce a bias, many studies have shown that radio silent
BL~Lacs either do not exist or are extremely rare
\cite[e.g.,][]{NA,tapia,impeybrand,borracorriveau,radioquiet,jannuzisearch}.
Therefore, in order to increase our BL~Lac search efficiency, we have chosen
to study the correlation of the RASS with the Green~Bank 6\,cm radio survey of
the northern sky \cite[]{gnc,gnc2}.  We then distinguish our dual radio and
X-ray$-$selected BL~Lacs from other emission-line AGN based on their optical
spectra (see \S\ref{sect:spec_class}), imposing no other selection criterion
except the presence of a bright optical counterpart.   We refer to the
resulting BL~Lac sample as the RASS-Green Bank or RGB BL~Lac sample.

Sections \ref{candidate_selection} and \ref{spectroscopy} review the creation
and followup VLA\footnote{The NRAO is operated by Associated Universities,
Inc., under a cooperative agreement with the National Science Foundation.}
observations of the RGB catalog.  Section \ref{sect:spec_class} presents the
new spectroscopic data and classifications.  Individual sources are discussed
in \S\ref{sect:indiv}.  The final section presents the broadband
characteristics of all the newly classified objects.  A more complete
discussion of the RGB BL~Lac sample appears in a companion paper
\cite[]{rgbbllacs}.  In what follows, we assume H$_{\rm
o}$=100~km\,s$^{-1}$\,Mpc$^{-1}$, q$_{\rm o}$$=$0.5 and define spectral
indices, $\alpha$, such that ${\rm S}_{\nu} \propto \nu^{-\alpha}$.

\section{Selection of Candidate Objects}\label{candidate_selection}
The RGB catalog was constructed by correlating the RASS with a new
$\sim$150,000 source radio catalog created from the 1987 Green~Bank (GB)
survey maps \cite[]{gnc,gnc2}.  Details of the radio source detection and
extraction procedure can be found in \cite{neumann}.  The radio catalog
consists of $\ge$3$\sigma$ confidence sources and has a flux density limit of
$\sim$15\,mJy in the declination range from $30^{\circ}-75^{\circ}$ and
increases to $\sim$24\,mJy at low declinations.

The cross-correlation of the RASS and the 3$\sigma$ GB survey yielded 2,127
matches with angular separations $<$100$\arcsec$.  The complete source lists
are given in \cite{rgb}.  Because the positional accuracy of both the RASS and
GB surveys is insufficient for unambiguous identification of optical
counterparts, 5\,GHz VLA observations were made of all 2,127 fields
\cite[]{rgb} and organized into two catalogs.  The first consists of 1,861
sources for which new subarcsecond positions and core radio flux densities
were obtained; the second consists of 436 sources for which only low
resolution data ($\sim$8$\arcsec$ positional accuracy) were obtained.  We
refer to this second catalog as the ``low resolution VLA catalog''.  All
sources whose radio/X-ray position difference was less than 40$\arcsec$ were
compared with Automatic Plate Measuring (APM) scans of the high Galactic
latitude ($>$25$^{\circ}$) POSS\,I photographic plates \cite[]{apm}.  Optical
counterparts within 3$\arcsec$ of RGB sources were identified and both the O
(blue) and E (red) magnitudes measured \cite[]{thesis,siebert}.  A looser
criterion of 5$\arcsec$ was used for sources in the low resolution VLA
catalog.  Approximately 80\% of the high Galactic latitude RGB sources have
optical counterparts brighter than the POSS\,I plate limit of
O$\sim$21.5\,mag.

Given 1,567 RGB objects which satisfy our radio/X-ray position criterion of
40$\arcsec$ and a probability of $\sim$5\% that a random optical counterpart
will reside within 3$\arcsec$ of an RGB radio source \cite[]{siebert}, we
expect 78 spurious optical identifications in the RGB catalog.  However, our
spectroscopic observations are limited to objects brighter than O$=$18.5\,mag.
An analysis of random high Galactic latitude objects present on the POSS\,I
plates shows that $\sim$18\% of these are brighter than 18.5\,mag
\cite[]{thesis}.  We therefore expect $\sim$14 spurious radio/optical
associations among the optically bright RGB objects.  Including the
classifications presented in this paper, 80\% of RGB sources meeting all the
above criteria have been spectroscopically identified.  Therefore, only 11 of
the bright RGB objects with spectroscopic classifications (0.9\%) are likely
to be false coincidences.

Optical spectra were obtained for many of the RGB sources which lacked
spectroscopic classifications (\S\ref{sect:spec_class}).  Unlike other similar
surveys, we select our candidate BL~Lacs without imposing any criteria other
than a limiting optical magnitude of O$\le$18.5\,mag, required by our use of
2-m class telescopes.  In particular, candidates were not selected based on
their optical morphologly, color, or broadband SEDs.  While this greatly
increases the number of candidate objects which require spectroscopic
classifications, it avoids any presumption that either radio$-$ or
X-ray$-$selected BL~Lacs are restricted to particular regions of the
$\alpha_{\rm ro}$ vs.\ $\alpha_{\rm ox}$ diagram where previous studies have
found either RBLs or XBLs often reside \cite[e.g.,][Figure
10]{stocke85,siebert}.  One of the principal results of this survey, discussed
below and in \cite{rgbbllacs}, is that many of the characteristics of the RGB
BL~Lacs such as their broadband SEDs and distribution of \ion{Ca}{2} break
contrasts, differ significantly from those exhibited by previously compiled
samples.

\section{Spectroscopic Observations}\label{spectroscopy}
We obtained low dispersion optical spectra of 171 objects in the RGB catalog
over the course of six observing runs from February 1994 through September
1995 with two facilities: the McDonald 2.7-m telescope using the Large
Cassegrain Spectrometer with the TI1 CCD and the Kitt Peak National
Observatory's 2.1-m telescope using GoldCam and the F3KC Ford CCD (Table
\ref{tab:runs}).  While most of our observations were of previously unknown
bright (O$<$18.5\,mag) objects, some previously known BL~Lacs and fainter
unclassified objects were also observed as time and observing conditions
allowed.  Spectra were taken through a 2$\arcsec$ slit, resulting in a
resolution of 10\,\AA\ for the McDonald and 14\,\AA\ for the Kitt Peak
spectra.  Table \ref{tab:runs} gives the details of each run including date,
wavelength coverage and resolution.  The slit was not rotated during the
night, but kept at a constant orientation.  The effect of this should be
small, however, since significant effort was made to observe objects as they
crossed the meridian.

\placetable{tab:runs}
\begin{table}
\dummytable \label{tab:runs}
\end{table}

The goal of these observations is to distinguish BL~Lacs from other kinds of
radio$-$ and X-ray$-$emitting objects.  We therefore require spectra with 
sufficiently high S/N ratios to unambiguously detect emission lines
characteristic of quasars, Seyfert and radio galaxies.  Simulated spectra
consisting of various line widths and strengths plus a Poisson noise
contribution show that a broad emission line of W$_{\lambda}$$\simeq$5\,\AA\
can be detected at a $\sim$4\,$\sigma$ confidence level in a 10\,\AA\
resolution spectrum when the S/N$\approx$30.  We therefore obtained spectra
with S/N$\gtrsim$30 on all objects which we classify here as BL~Lacs.  Our
observing strategy consisted of taking an initial exposure (nominally 20$-$30
minutes) after which the raw spectrum was examined.  If no emission lines or
galaxy aborption features were evident, the source was observed repeatedly
until a S/N ratio of $\approx$30 was reached.  For the brighter sources which
had reached a S/N ratio greater than 30 in one exposure, we obtained a second
exposure if the initial raw spectrum lacked any strong features in order to be
able to unambiguously distinguish very narrow lines from cosmic rays (see
Table \ref{tab:obs_times}).

\placetable{tab:obs_times}
\begin{table}
\dummytable \label{tab:obs_times}
\end{table}

The raw two-dimensional data were reduced following standard procedures using
the IRAF (V2.10.4) analysis package.  Wavelength calibration was carried out
using Helium-Neon-Argon lamps taken at the beginning and end of the night.
Comparison of our measured redshifts with published values for previously
identified sources shows our redshifts are accurate to 0.001 for z$<$1.0 and
0.005 for z$>$1.0.  The latter limit is more uncertain because the high
redshift objects generally exhibit only broad emission lines.

Final calibrated spectra (including line identifications) for the 171 sources
are shown in Figures 1.1$-$1.174.  Many of the spectra were taken under
nonphotometric conditions so that the flux density scale given in the figures
should only be used as a rough guide to the brightness of the sources.  The
vast majority of the objects we observed (132 objects or 78\%) have no
previous reported classification in the literature and we report redshifts
for 9 additional objects which previously lacked this information.  We also
looked at a small number of previously known objects.  The majority these
objects were previously known BL~Lacs without known redshifts.

In Table \ref{tab:obs_times} we give our observing log which consists of the
(1) RGB Source Name; (2) Designation for the run on which the data were
obtained (see Table \ref{tab:runs}); (3) Total Exposure Time (4) Number of
Exposures per Source; (5) Spectroscopic Classification (see \S
\ref{sect:spec_class}); (6) Redshift and (7) Notes.  Individual spectra with
notes too extensive to list in Table \ref{tab:obs_times} are discussed in
Section \ref{sect:indiv}.

\section{Results and Spectroscopic Classification} \label{sect:spec_class}
Although both optical and broadband multifrequency colors can yield
approximate classifications for extragalactic objects
\cite[e.g.,][]{stocke,mcmahon,schachter}, unambiguous classification can only
be achieved via spectroscopic observations.  This is particularly true for
BL~Lacs which exhibit a wide range of optical colors and broadband spectral
energy distributions, but whose main distinguishing characteristic is their
featureless optical continuum \cite[see][and references
therein]{weedman,BLcolor,vista,unified}.  Below we present the results of
spectroscopic observations of a large sample of bright, previously
unidentified objects in the RGB catalog and our criteria used to classify
these objects.  Figure \ref{fig:flow_chart} summarizes these criteria and
Table \ref{tab:obs_times} (column 5) gives the results for the individual
sources.  

\subsection{Emission-line Objects} \label{sect:agn}
Most classes of AGN exhibit strong emission lines.  Objects with strong broad
permitted lines are classified as Type~I AGN while those with only narrow
lines are designated as Type~II.  Distinctions are also generally made between
radio-loud and -quiet objects and between objects with high and low optical
luminosities (e.g., quasars and Seyfert galaxies, respectively).  While recent
studies show transitional objects do exist and that these divisions are
somewhat arbitrary, they do correlate with important differences between the
various classes \cite[e.g.,][and references therein]{loudquiet,qsoseyfert}.
We therefore classify objects with these divisions in mind.  Specifically, we
classify all emission-line AGN as follows:
\begin{itemize}
\item {Following \cite{KSS}, we define radio-loud objects as those having
$\log ({\rm S}_{\rm r}/{\rm S}_{\rm opt})$$>$1.0 where the radio flux density
is the core emission measured at 5\,GHz and the optical flux density is
measured from the POSS O plates.  This parameter is then used to distinguish
radio-quiet vs.\ radio-loud quasars and also Seyfert vs.\ radio galaxies.  
Alternately, radio luminosity can be used to define radio-loudness.  These two
methods produce nearly the same classification for the objects presented here,
provided the limiting radio power is chosen to be $\log {\rm P}_{\rm
r}$$=$23.4\,W\,Hz$^{-1}$, equivalent to the definition used by \cite{UW89} and
\cite{MRS}, after converting to our choice of cosmology.  The eight RGB
objects whose classification is dependent on the differences in these two
definitions are denoted in Table \ref{tab:obs_times}.}

\item{The distinction between Type I and Type II low power AGN (the
radio-quiet Seyfert~I and II galaxies and the radio-loud Broad and Narrow
Line Radio Galaxies) is based on the width of optical emission lines.  Type\,I
AGN exhibit both broad permitted and narrow forbidden lines.  Type\,II AGN
exhibit only narrow lines.  We choose as a dividing point the presence of
lines with FWHM velocity of 1000\,km\,s$^{-1}$ \cite[]{Osterbrock}.  This
criterion conveniently matches our resolution which roughly corresponds to a
FWHM of 1000\,km\,s$^{-1}$.}

\item{The division between quasars and low power AGN (Seyfert and radio
galaxies) is traditionally based on the Johnson B magnitude (4400\,\AA\
effective wavelength) and lies between $-23.0$$<$M$_{\rm B}$$<$$-21.0$\,mag
\cite[]{Osterbrock}.  The uncertainties in the POSS O magnitudes are such that
we consider them to be equivalent to B magnitudes and we choose M$_{\rm
O}$$=$$-$22.0\,mag as the faintest absolute magnitude of RGB quasars.  Using
this definition, we find no ``narrow-line'' quasars in our sample, since our
brightest narrow-line object (RGB\,J0207+295A) has M$_{\rm
O}$$=$$-$20.9\,mag.}
\end{itemize}

The classification of objects is then determined as follows (see Figure
\ref{fig:flow_chart}): Objects with $\log ({\rm S_r}$/S$_{\rm opt}$)$\ge$1.0
are classified as radio-loud, and those objects which also have M$_{\rm
O}$$\le$$-$22.0\,mag are classified as radio-loud quasars.  Radio-loud objects
with M$_{\rm O}$$>$$-$22.0\,mag are classified as radio galaxies, with the
width of the lines distinguishing between Broad Line Radio Galaxies (BLRGs)
and Narrow Line Radio Galaxies (NLRGs).

For radio-quiet objects ($\log$S$_{\rm r}$/S$_{\rm opt}$$<$1.0), the optically
luminous (M$_{\rm O}$$\le$$-$22.0\,mag) objects would have been classified as
radio-quiet quasars, but we found no such objects in our sample.  The
optically less luminous sources are either Seyfert~I or II galaxies, depending
on the line widths.  We find only one such object of this type
(RGB\,J1518+407) which, because it exhibits a broad
(FWHM$\approx$1900~km\,s$^{-1}$) H$\alpha$ line, we classify as a Seyfert~I
galaxy (also see \S\ref{sect:indiv}).

We note that the ratio of quasars to low power emission-line AGN (1.7:1) is
similar to that reported for other radio-/X-ray$-$selected samples which have
flux limits similar to the RGB survey \cite[]{morris,2sigma}.  The nearly
complete absence of radio-quiet objects is also not surprising, given the
radio flux limit of the RGB catalog ($\sim$20~mJy).

\subsection{Galaxies and Clusters} \label{sect:galaxies}
The classification of galaxies has been made solely on the basis of the
optical spectra presented in the figures.  If the object lacked broad AGN-like
emission lines and showed no indications of an underlying nonthermal powerlaw
component (see \S\ref{sect:bllacs}), we classify the object as a galaxy (see
Figure \ref{fig:flow_chart}).  However, these RGB galaxies exhibit both radio
and X-ray emission far in excess of that expected from normal elliptical
galaxies.  It is possible the X-ray emission arises not from the radio source
itself, but from a surrounding diffuse X-ray cluster.  Unfortunately,
available optical images are not deep enough to determine whether or not a
cluster is present nor does our X-ray data have sufficient resolution to
evaluate this hypothesis.  However, because previously known clusters are
associated with four of these galaxies, we classify them as ``Clusters'' in
Tables \ref{tab:obs_times} and \ref{tab:lum}.

The other eight objects with seemingly normal elliptical galaxy spectra 
(albeit with occasional weak emission lines; see Figures 1.8, 1.14 \& 1.165),
but very high X-ray and radio luminosties, may be as yet undetected clusters
or they may be optically weak, but radio and X-ray luminous AGN.  These RGB
sources have $\log~{\rm P}_{\rm r}$$>$23.0\,W~Hz$^{-1}$ and $\log~{\rm
L_X}$$>$43.0\,erg~s$^{-1}$.  While radio luminosities of this magnitude are
similar to those exhibited by bright B2 and 3CR radio galaxies
\cite[]{b2_2,3cr,b2_1}, the X-ray luminosity of these RGB sources is 1-3
orders of magnitude larger \cite[]{rg_xray1,rg_xray2}.  There is precedent for
this kind of object, including 3C\,264 \cite[]{3c264_1,3c264_2} and the
Einstein source J2310$-$43.  \cite{tanabaum} conclude the latter object is
likely an AGN similar to BL~Lacs which, either because of orientation or
certain intrinsic properties of the AGN itself, lack both strong emission
lines and significant nonthermal emission in the optical band.

Alternatively, it is also possible these objects are ``optically passive X-ray
galaxies'' \cite[]{passive,2sigma,tanabaum}, extraordinary early-type galaxies
which emit at least an order of magnitude more radio and X-ray luminosity than
other galaxies with similar optical properties.  \cite{2sigma} found 8 such
objects in the Einstein Two-Sigma catalog.  They concluded, on the basis of
followup radio observations, that the objects were unlikely to be hidden AGN
whereas optical imaging showed that they may be members of previously unknown
clusters or small groups of galaxies.  Alternatively, these sources may be
early type galaxies with an extraordinarly hot ISM.  Similar objects have also
been found in selected deep ROSAT PSPC fields and the study of them has led to
the suggestion that a mini-cooling flow in a small cluster or group of
galaxies may trigger AGN activity in these objects \cite[]{passive}.

Without additional information, especially deep optical imaging, we cannot
make any conclusions about the true nature of the galaxies presented here.  We
therefore assign the ``Galaxy'' classification to them, but note that these
objects are not typical early-type galaxies and that followup observations are
warranted.

\subsection{BL Lacs} \label{sect:bllacs}
The classical definition of a BL~Lac is an AGN with a highly variable,
linearly polarized, nonthermal continuum which contains no optical emission
lines \cite[]{stein1}.  It is now generally recognized that objects which
violate some of these criteria generally exhibit BL~Lac-like properties and
should be considered part of the class \nocite{vista} (see Kollgaard [1994]
for a summary).  For example, long-term, high S/N monitoring of BL~Lacs has
shown that many exhibit low luminosity emission lines with the most striking
example being BL~Lac itself \cite[]{millerhawley,corbett}.  A standard
definition is that BL~Lacs are AGN with emission lines whose equivalent width
does not exceed 5\,\AA\ \cite[]{stocke}.  However, many observational criteria
exist which have added additional constraints to the classical definition
\cite[]{remillard,schwartz,stickel,stocke,slew,NA}.

While these new criteria originally delineated clear boundaries between
BL~Lacs and other classes of AGN, it is becoming clear that they are more
arbitrary than originally suspected.  The resulting selection affects are only
now being addressed \cite[e.g.,][]{marcha1,PGcat,marcha2,scarpa}.  Because
X-ray$-$selected BL~Lacs (XBLs/HBLs) typically exhibit properties intermediate
between radio galaxies and radio-selected BL~Lacs (RBLs/LBLs), the issues are
particularly relevant in X-ray$-$selected samples which the RGB at least 
partially is.  New spectroscopic observations also show the division between
BL~Lacs and weak-lined AGN and elliptical galaxies is becoming increasingly
blurred \cite[]{marcha2,scarpa}.  It is therefore necessary to develop a
BL~Lac definition which encompasses objects which exhibit the same {\it
intrinsic\/} (not observed) characteristics.  This is particularly difficult
because much of the observed radiation in BL~Lacs is dominated by
orientation-dependent beamed radiation.

We consider the approach proposed by \cite{marcha2}, who present a set of
self-consistent orientation-independent rules for discriminating BL~Lacs from
ordinary quasars, Seyfert, radio and elliptical galaxies.  The \cite{marcha2}
definition relies on the results of \cite{ebreak} who show that less than 5\%
of elliptical galaxies have a \ion{Ca}{2}~H\&K ($\lambda\lambda$3933,3968)
break contrast, Br$_{4000}$, below 0.4.  Here, the term ``break contrast''
refers to the relative depression of the continuum blueward of the
\ion{Ca}{2}~H\&K lines (see below).  It is therefore likely that any object
with a weaker break contrast has an additional blue component present.  It is
also true that many objects with Br$_{4000}$$\ge$0.4 may have a similar
continuum present, but it is impossible to determine this on the basis of the
break contrast or any other property that correlates with it.  This effect
will obviously become more important for the intrinsically most luminous host
galaxies with the weakest blue components, assuming that the two are
uncorrelated.  However, even modest host galaxy luminosities can be sufficient
to obscure contributions from underlying blue continua \cite[]{newBL}.

In addition to these complications, it is important to note that there are two
possible sources of the additional component: an AGN or star formation.
Ellipticals must be very young in order to have enough blue light present to
significantly weaken the break contrast \cite[]{jjg}.  The RGB objects with
weak break contrasts are at redshifts that correspond to a lookback time of
only $\sim$1\,Gyr and it is therefore unlikely these objects constitute a
young population of starforming ellipticals.  Merger induced star formation is
a far more plausible scenario.  \cite{starform} find that $\sim$15\% of
galaxies with abnormally weak break contrasts show significant evidence of
star formation where up to 60\% of the nuclear light at 4000\,\AA\ originates
in hot young stars.  However, these objects typically also show Balmer lines
whose strength varies with the assumed contribution from hot young stars.  We
therefore assert that galaxies with weak break contrasts which lack spectral
evidence for a burst of recent star formation in the form of Balmer lines, are
likely old elliptical galaxies which harbor a weak BL~Lac nucleus.  The BL~Lac
classification is also strengthened by the presence of high luminosity radio
and X-ray emission consistent with the AGN hypothesis.

We adopt the same definition of break contrast used by \cite{ebreak}, namely:
\begin{equation}\label{eqn:break}
{\rm Br}_{4000} = \frac{f^+ - f^-}{f^+},
\end{equation}
where $f^+$ is the average flux in the range 4050$-$4250\,\AA\ and $f^-$ is
the average flux in the range 3750$-$3950\,\AA.  All wavelengths are in the
rest frame.  Two caveats deserve mention.  First, $f^+$ and $f^-$ are {\em
average\/} fluxes which are good measures only if the slopes in each band are
flat.  For most sources this appears to be true.  Second, the wavelength range
for $f^-$ includes the \ion{Ca}{2}~H, but not the \ion{Ca}{2}~K, line.  This
will decrease $f^-$ slightly compared to the value it would have if the region
around 3933\,\AA\ were excluded.  However, this produces a change which is
less than 1$-$2\% in our reported break strengths and is less than the
uncertainties introduced by observational effects.  The value of the break
strength in those objects for which we could detect it are given in the notes
to Table \ref{tab:obs_times} and we show a histogram of our measured break
strengths in Figure \ref{fig:ca}.  We note that the RGB objects observed here
exhibit a continuous range of break contrasts, ranging from 0\%-50\%, and that
no clear point exists where a division between BL~Lacs and normal ellipticals
can be made.  Nevertheless, our formal definition presented in
\S\ref{sect:bllacs} makes an arbitrary cut at Br$_{4000}$$=$40\%.  We note 
that we are following up with optical polarimetry, high resolution (nuclear)
optical imaging and X-ray and radio studies all RGB BL~Lacs and galaxies {\em
regardless\/} of break strength in order to determine the presence of any
nuclear BL~Lac component.  It is therefore possible that some new RGB galaxies
may be reclassified as BL~Lacs in the future, but we have no definitive 
evidence at present to warrant such reclassification.

Break contrast alone is not enough to unambiguously distinguish BL~Lacs from
other types of low luminosity extragalactic objects.  An emission line
criterion must be established to exclude both radio and Seyfert galaxies which
also show absorption features.  A maximum permitted equivalent width has
traditionally been set at W$_{\lambda}$$=$5\,\AA\ \cite[]{stickel,stocke},
although some definitions have used the rest and others the observed frame.
In either case, there is mounting evidence that this may be too stringent a
limit \cite[e.g.,][]{slew,marcha2,mg2} and that the only differences in the
line properties of BL~Lacs and highly polarized beamed quasars (HPQ blazars)
are those introduced by the arbitrary 5\,\AA\ equivalent width criterion
\cite[]{scarpa}.  In both these cases, the measured equivalent width depends
on the strength of beamed optical emission in the sense that an object very
close to the line-of-sight will exhibit smaller equivalent widths than one
further from the line-of-sight because continuum emission will be enhanced,
but the intrinsic line luminosity will remain the same.

\cite{marcha2} suggest criteria which take this effect into account.  Using
examples from their `200\,mJy' radio-selected sample and models which vary the
orientation of 3C\,371 (RGB\,J1806+698) to the line-of-sight, they
simultaneously characterize the effects beaming produce in the break contrast
$-$ equivalent width plane \cite[the Br$_{4000}$$-$W$_{\lambda}$ plane; see
Figure 6 in][]{marcha2}.  All objects which are more `BL~Lac-like'' than
3C\,371 (i.e., have weaker lines for a given break strength) and {\em also\/}
have Br$_{4000}$$<$40\% are considered BL~Lacs.  The \cite{marcha2} criteria
encompass nearly all objects which would be classified as BL~Lacs using the
traditional criteria \cite[W$_{\lambda}$$\le$5\,\AA\ and
Br$_{4000}$$\le$25\%;][]{stocke} but cover a larger region of the
Br$_{4000}$$-$W$_{\lambda}$ plane.  While these criteria are arbitrary, they
are a useful first step toward unambiguously defining the BL~Lac class.

The RGB BL~Lac classification criteria we adopt are summarized in Figure
\ref{fig:flow_chart} and are as follows:
\begin{itemize}
\item{If the spectrum is featureless or the only features observed are
emission lines with W$_{\lambda}$$\le$5\,\AA, the object is classified as a
BL~Lac. All measurements of equivalent width are taken in the rest-frame.}

\item{If absorption features are present and Br$_{4000}$$<$25\%, we classify
the object as a BL~Lac, provided any emission lines present have
W$_{\lambda}$$\le$5\,\AA\ and no strong Balmer absorption lines are observed.
Objects with strong Balmer absorption would have been classified as
starforming galaxies but no such objects were found.  However, four RGB
BL~Lacs exhibit very weak Balmer absorption lines.  They are RGB\,J0656+426,
RGB\,J1516+293, RGB\,J2250+384 and RGB\,J2322+346 (Figures 1.23, 1.89, 1.162
\& 1.173).  All have been classified as ``BL~Lacs?''.}

\item{If the \ion{Ca}{2} break contrast is between 25-40\%, we classify the
object as a possible BL~Lac if any emission line present also has an
equivalent width smaller than that required by the \cite{marcha2} criterion
(specified by the symbol W$_{\lambda}^{\rm M}$ in Table \ref{tab:obs_times})
for that particular break strength.  These tentative RGB BL~Lacs are denoted
by a question mark after their classification.  As described above, any
objects with strong Balmer absorption lines would have been classified as
starforming galaxies.  We note that because the spectroscopic setup was
designed to obtain Br$_{4000}$ for z$\simeq$0$-$0.5, many spectra do not
extend far enough to the red to determine the H$\alpha$ emission line
strength.  For this reason, the equivalent width criterion we use for
classification is not as uniform as that employed by \cite{marcha2}, whose
equivalent width measurements are generally based on H$\alpha$.  Followup
spectroscopy in the region where H$\alpha$ is expected would be useful in
standardizing the classification of these objects.  We therefore place the
note ``H$\alpha$ useful'' in Table \ref{tab:obs_times} for these sources.}

\item{If the \ion{Ca}{2} break contrast is $>$40\% we classify the object as a
galaxy because of the lack of spectroscopic evidence for an AGN.  We note,
however, that a low luminosity BL~Lac nucleus may be present in some of these
objects.  We again emphasize that the distribution of \ion{Ca}{2} break
contrasts shows no clear point at which the spectroscopic division between
galaxies and BL~Lacs should be made (Figure \ref{fig:ca}).  We are pursuing
higher S/N observations, studies which characterize the strength of H$\alpha$,
optical polarimetry and spatially resolved spectroscopy of the core of these
objects since they may help distinguish the two classes.}
\end{itemize}

The formal RGB BL~Lac definition is given by the above criteria.  However,
because of either poor quality data or variations in spectroscopic setup, we
define two further criteria:
\begin{itemize}
\item{When our spectra did not extend far enough to the blue to measure
Br$_{4000}$, we made a tentative classification based the criteria defined
above and, in the absence of Br$_{4000}$, on the O$-$E colors.  Objects with
weak (or no) emission lines and colors bluer than O$-$E$=$2.0\,mag are
tentatively classified as BL~Lacs.  In Table \ref{tab:obs_times} we report
both the colors and the note ``Br$_{4000}$ needed'' for these objects which
we then denote with the ``BL~Lac?'' classification.}

\item{We classify one object (RGB\,J1012+424) which appears to have two
emission lines in its spectrum (W$_{\lambda}(\lambda 4554)$$=$4\,\AA\ \&
W$_{\lambda}(\lambda 4380)$$=$10\,\AA) and no measured \ion{Ca}{2} break
contrast as a possible BL~Lac (``BL~Lac?'' in all tables).  Although the lines
violate the criteria given above, the measured wavelengths of the emission
features correspond to no obvious emission line pairs and therefore may not be
real.  Further observations are required.}
\end{itemize}

The variety of criteria used to classify BL~Lacs raises concerns regarding the
completeness of various surveys.  Even catalogs which are completely
identified may be missing BL~Lacs because of the misclassification of objects,
or even contain additional objects which are not BL~Lacs at all.  This latter
point was raised most recently by \cite{mg2} who suggest the 1\,Jy RBL sample
contains a number of lensed radio-loud quasars.  On the other hand, there is
evidence that differences in classification criteria have not led to serious
errors in the characterization of the bulk properties of the XBL subclass
\cite[]{lmheao}, although existing samples may be incomplete, particularly at
low luminositiies \cite[]{marcha1,newBL}.  Analyses which rely on the space
density and shape of the luminosity function or LogN$-$LogS distribution will
be particularly sensitive to these issues and care must therefore be taken to
address them \cite[]{rgbbllacs}.

\section{Comments on Individual Sources} \label{sect:indiv}
Before presenting the multiwavelength properties of the RGB sources, we
briefly discuss some individual sources.

{\em RGB\,J0044+193.} $-$ The presence of both broad Balmer lines and Fe
emission at $\sim$5300\,\AA\ in this object is similar to that seen in
Narrow-lined Seyfert~I galaxies.  Unlike this object, however, Narrow-lined
Seyfert~I galaxies are typically radio-quiet \cite[]{UAG}.  Therefore, because
this object is both radio-loud and optically bright (M$_{\rm
O}$$\le$$-$22.0\,mag), we tentatively classify it as a radio-loud quasar,
although it is unusual for radio-loud quasars to exhibit strong Fe emission
\cite[]{fe_rqq}.  However, the subclass of radio-loud AGN which are also
strong IRAS sources, as this object is, have been known exhibit extremly
strong Fe emission \cite[]{lipari}.

{\em RGB\,J1248+514.} $-$  This object is an M4 ($\pm$1) star with an implied
X-ray luminosity of 10$^{29}$$-$10$^{30}$\,erg~s$^{-1}$.  Stars of this type
typically exhibit strong Balmer emission \cite[]{lxmstar}, but Figure 1.58
shows only aborption lines.  Additionally, the implied radio luminosity of
this source is a factor of $\sim$10$^3$ larger than the most radio luminous
dMe stars known \cite[]{radiomstar}.  We therefore believe the optical
counterpart to this source is a spurious identification
(\S\ref{candidate_selection}).

{\em RGB\,J1413+436.} $-$ This object may be a radio$-$ and X-ray$-$luminous
galaxy (\S\ref{sect:galaxies}) or a low luminosity AGN.  Our spectrum does not
extend far enough to the blue to measure Br$_{4000}$, making it difficult to
determine the fraction of nonthermal emission.  The FWHM of the
H$\alpha$/[NII] blend is approximately 1900\,km\,s$^{-1}$, which is too broad
for a galaxy classification.  We therefore classify this object as a BLRG.

{\em RGB\,J1808+468.} $-$ The RASS and VLA positions of this source differ by
69$\arcsec$, greater than our 40$\arcsec$ radio/X-ray offset criterion
(\S\ref{candidate_selection}).  However, new HRI observations show the radio
and X-ray positions differ by only 2$\arcsec$ \cite[]{hri}.

{\em RGB\,J1811+584.} $-$ The spectrum of this object is that of a K star or
very low redshift elliptical galaxy (z$=$0.002).  New data obtained at Lick
observatory (not presented here) show that this object has zero redshift and
is therefore likely a spurious identification (see
\S\ref{candidate_selection}).

{\em RGB\,J1922+691.} $-$ This object is notable for the strength of its
forbidden O[II] and O[III] lines and concurrent lack of any detectable Balmer
lines.  Similar objects have been found in other surveys \cite[e.g., the
EMSS;][]{stocke}.  Observations of H$\alpha$ would be useful to determine
whether this object simply has a large Balmer decrement or if the Balmer lines
are intrinsically weak.

\section{Multiwavelength Properties}\label{sect:mult}

In Table \ref{tab:flux} we reproduce for convenience the coordinates and
multiband fluxes given in \cite{rgb} and \cite{siebert}.  For those objects
with measured break contrasts, we also estimate the separate contributions of
the galaxy and AGN to the total optical flux (see below).  Table
\ref{tab:flux} also lists the radio-loudness parameter ($\log{\rm S}_{\rm
r}/{\rm S}_{\rm opt}$), redshift and classification for all sources.  The
radio-loudness parameter has been K-corrected assuming $\alpha_{\rm
r,core}$$=$$0.0$ and $\alpha_{\rm opt}$$=$$1.0$, although individual fluxes in
Table \ref{tab:flux} have not been K-corrected.  Sources with unknown
redshifts are K-corrected using the median redshift for the class (0.16 for
BL~Lacs, 0.770 and 0.231 for those quasars and BLRGs, respectively, for which
a redshift could not be determined).  These objects are all at high Galactic
latitude (b$>$25$^{\circ}$) and have not been corrected for extinction.  Such
corrections would be small compared with the $\sim$0.5~mag uncertainty in the
POSS derived POSS O magnitudes \cite[]{thesis}.  X-ray fluxes have been
computed from the RASS count rates by assuming an average spectral index of
$\alpha_{\rm x}$$=$1.2 and Galactic {\sl N}$_{\rm H}$ absorption
\cite[]{nh1,nh2,Brinkmann}.   

\placetable{tab:flux}
\begin{table}
\dummytable \label{tab:flux}
\end{table}

Table \ref{tab:lum} contains the multiband luminosities of the newly
identified RGB objects grouped by spectroscopic class and lists the logarithm
of the 5\,GHz radio power, the absolute O magnitude and the logarithm of the
X-ray luminosity for all objects which have a measured redshift.  All
luminosities have been K-corrected.

\placetable{tab:lum}
\begin{table}
\dummytable \label{tab:lum}
\end{table}

Decomposition of the optical magnitude into separate AGN and galaxy components
was performed only for those sources for which a \ion{Ca}{2} break contrast
was measured.  The separate contributions are estimated by assuming the host
galaxy has an intrinsic break contrast of Br$_{4000}$$=$50\% and that
contrasts smaller than this are a result of an underlying blue powerlaw
continuum.  The AGN magnitudes are therefore likely upper limits.  Since the
contrast is measured at $\sim$4000\,\AA\ and the effective wavelength of the
POSS\,I O magnitude is $\sim$4400\,\AA, the true fraction of light originating
in the galaxy does not change substantially in this 400\,\AA\ bandpass.  We
therefore calculate the AGN contribution as follows: the AGN continuum just
redward and blueward of 4000\,\AA\ is assumed to be a constant, A (valid
because the break at 4000\,\AA\ is very sharp).  Then,\\

\begin{equation}
\frac{R - B}{R} \equiv 0.5~~~{\rm and}~~~\frac{R+A - (B + A)}{R+A} = s,
\end{equation}

\noindent where A is the flux arising from the AGN component at 4000\,\AA, B
is the flux of the galaxy component blueward of the break, R is the galaxy
flux redward of the break, and s represents the measured break strength.  The
relative contribution of AGN to host galaxy light is then given by:\\

\begin{equation}\label{eqn:contam}
\frac{A}{R} = \frac{0.5}{s} - 1.
\end{equation}

\noindent While not as accurate as a full spectral deconvolution, our method
yields an AGN light fraction that differs by only 6\% from that obtained by a
full spectral deconvolution of the BL~Lac object E\,0336$-$248 \cite[]{newBL}.
This yields confidence that our estimates of the fraction of light arising
from the AGN are accurate to $\pm$10\% or better.  Generally we estimate that 
30-60\% of the total flux at 4000\,\AA\ is contributed by the AGN.

In Figures \ref{fig:lr}$-$\ref{fig:lx}, we show histograms of the multiband
luminosities of the various classes of sources.  Because the samples presented
here are incomplete (consisting only of those RGB objects which did not have a
previous spectroscopic classification), a detailed analysis of these data is
unwarranted.  However the figures provide a useful evaluation of many of the
classification criteria discussed here and we therefore limit our discussion
to these issues.  A full discussion of the properties of the all 118 RGB
BL~Lacs will be presented in \cite{rgbbllacs}.

The definitions of RGB BLRGs and quasars differ only by limits set on their
optical luminosities (\S\ref{sect:agn}).  Nevertheless, the RGB quasars
constitute the most luminous class of objects in all three wavebands although
both the radio and X-ray luminosities of the BLRGs (and NLRGs) overlap that of
the quasars.  The RGB classification criteria have the effect of abruptly
truncating both distributions at M$_{\rm O}$$=$$-22.0$\,mag (Figure
\ref{fig:lo}).  However, a small discontinuity in the combined absolute
optical magnitudes of the BLRGs and quasars exists at M$_{\rm
O}$$\simeq$$-22.0$\,mag.  Although it is not a well-defined break, this
indicates that our criteria do correspond to some intrinsic difference in the
quasar and BLRG populations.  However, Figure \ref{fig:lo} shows that many
transition objects undoubtedly exist and that detailed study of the RGB sample
would provide useful insight into the transition between these two AGN
classes.

As noted in \S\ref{sect:galaxies}, the RGB galaxies are overluminous in both
the radio and X-ray.  As discussed above, the X-ray emission may (at least
partially) originate in diffuse cluster emission, but the strength of the
radio emission is difficult to explain.  It is possible that many of these
objects belong to the new class of optically passive X-ray galaxies, although
other explanations exist (see \S\ref{sect:galaxies}).  

The newly discovered BL~Lacs show a wide range of radio, optical and X-ray
luminosities.  The broadband spectral energy distribution of these objects is
in fact much larger than expected, a point which will be discussed extensively
in \cite{rgbbllacs}.  Figure \ref{fig:lo} also shows the distributions of both
the nuclear and host galaxy absolute blue magnitudes for those 16 RGB BL~Lacs
with measured break contrasts (\S\ref{sect:mult}).  Both the host galaxy and
AGN absolute magnitudes peak near M$_{\rm O}$$=$$-19.5$\,mag and both
distributions are smooth and well-behaved.  Given an uncertainty of
$\sim$0.5\,mag in the apparent O magnitudes, the average estimated RGB BL~Lac
host galaxy luminosity of $\langle {\rm M}_{\rm V}\rangle$$=$$-23.5$\,mag is
consistent with more precise measurements obtained for samples of low redshift
BL~Lacs from the 1\,Jy sample \cite[$\langle {\rm M}_{\rm
V}\rangle$$=$$-22.9$;][]{SFK} and Fanaroff-Riley Type\,I radio galaxies
\cite[$\langle {\rm M}_{\rm V}\rangle$$=$$-23.1$;][]{FRhosts}, after
converting to the same cosmology and assuming a typical host galaxy has a
(B$-$V) color of $\sim$1.0.  This supports our assertion that the method
presented in \S\ref{sect:mult} is reasonably sound.

The full RGB sample is large and flux-limited (in both the radio and X-ray)
and as such can provide useful insight into the transitions between various
classes of AGN.  Of particular interest is the transition between RBLs and
XBLs.  Figure \ref{fig:ratio} shows the $\log ({\rm S}_{\rm x}/{\rm S}_{\rm
r})$ vs.\ $\log ({\rm S}_{\rm o}/{\rm S}_{\rm r})$ diagram for the newly
identified RGB objects \cite[see Figure 11 in][]{siebert}.  As noted in
\cite{siebert}, RGB objects generally span the intermediate region in this
diagram between those regions populated by radio-loud quasars and RBLs at one
end (low S$_{\rm x}$/S$_{\rm r}$ ratio) and XBLs at the other (high S$_{\rm
x}$/S$_{\rm r}$ ratios).  Figure \ref{fig:ratio} shows this ``zone of
avoidance'' in which very few previously known BL~Lacs reside.  While it is
clear the fraction of RGB BL~Lacs increases with increasing S$_{\rm
x}$/S$_{\rm r}$, more than 20\% of the newly discoverd RGB BL~Lacs reside in
the region previously unpopulated by BL~Lacs.  This has important implications
for BL~Lac search strategies since it shows that RASS BL~Lac samples will be
severely incomplete if candidates are selected to have extreme S$_{\rm
x}$/S$_{\rm r}$ flux ratios.

\section{Summary}\label{sect:summary}

We present spectra of 169 sources and their spectroscopic classifications.
These sources belong to the RGB (RASS-Green Bank) sample which is based on
spatial coincidences between the ROSAT All-Sky Survey, Green Bank 5\,GHz, and
APM POSS\,I catalogs.  The criteria used to establish various classifications
are discussed in detail.  Of the 169 objects presented here, we find 53
BL~Lacs, 66 radio-loud quasars, 33 BLRGs, 5 NLRGs, 1 Seyfert~I galaxy and 11
galaxies or galaxies in clusters.  (The optical counterparts of two objects
are also identified as spurious; see \S\ref{sect:galaxies}).  For the vast 
majority of the objects (83\%), the data we present here yield for the first
time a spectroscopic classification or a redshift for an object previously
classified by other means.

Including the results of this work, approximately 70\% of the RGB objects with
POSS\,I counterparts and over 80\% of the objects with counterparts brighter
than O$=$18\,mag now have reliable spectroscopic classifications (see
\cite{Brinkmann} and \cite{siebert} for the other identifications).  The
numerous quasars and radio galaxies in addition to the less numerous, but
intriguing radio and X-ray luminous early-type galaxies, form a useful
database for further study, particularly since the RGB sample is
simultaneously both an X-ray and radio flux-limited sample.

Our ultimate goal is the creation of a large unbiased sample of BL~Lacs and in
this study we report the discovery of 38 previously unknown BL~Lacs.  We also
present and analyze our observations of 15 previously known BL~Lacs, which
were reobserved because they lacked redshift information.  Particular
attention is paid to the spectral information obtained for all BL~Lac
candidates.  Many spectra show evidence for the underlying host galaxy.
Following \cite{marcha2}, we characterize the mixture of host galaxy and
nuclear BL~Lac light in terms of the \ion{Ca}{2}~H\&K break contrast
(Br$_{4000}$).  Break strengths range from 0\% to 50\%, with those object with
Br$_{4000}$$>$40\% being classified as galaxies although no clear division
between BL~Lacs and galaxies exists.  We present a simple method for
decomposing blue optical magnitudes into separate contributions from the host
galaxy and BL~Lac.  In the absence of high S/N spatially resolved spectra, we
show this method gives reasonable upper limits to the optical AGN luminosity.

The RGB sample constitutes a new well-defined sample which is significantly
larger and more sensitive than previous samples obtained from radio and
X-ray$-$selected surveys.  It thus has the potential to reveal new aspects of
AGN populations and the astrophysics which govern radio-loud active galaxies.
In particular, we show that the RGB sample includes many BL~Lacs with
properties between those of traditional radio$-$ and X-ray$-$selected BL~Lacs.
In an associated paper \cite[]{rgbbllacs}, we will present the full RGB sample
of 118 BL~Lacs and discuss their properties in detail in addition to
implications this sample has for the unified scheme.

\acknowledgments
We extend our thanks to Dr.\ Richard McMahon who provided information on the
RGB optical counterparts prior to publication, to Dr.\ Ed Moran for useful
discussions on the spectroscopic classification of the emission line AGN
presented here and to Dr.\ Joe Pesce for useful commentary on the manuscript.
This work was partially supported by NASA Grant NAGW-2120 to EDF and partially
by the Department of Energy at the Lawrence Livermore National Lab under
contract W-7405-ENG-48.  RIK acknowledges support from Fermi National
Accelerator Laboratory.  We have made use of the NASA/IPAC Extragalactic
Database, operated by the Jet Propulsion Laboratory, California Institute of
Technology, under contract with NASA.  SALM also acknowledges partial support
from the NASA Space Grant Consortium through their Space Grant Fellow program.

\newpage



\clearpage

\begin{figure}
\caption{Figure 1 at end of paper.\label{fig:all_spectra}}
\end{figure}
\begin{figure}
\vspace{-0.3in}
\hspace{-0.3in}
\begin{minipage}[t]{6.3in}
\psfig{file=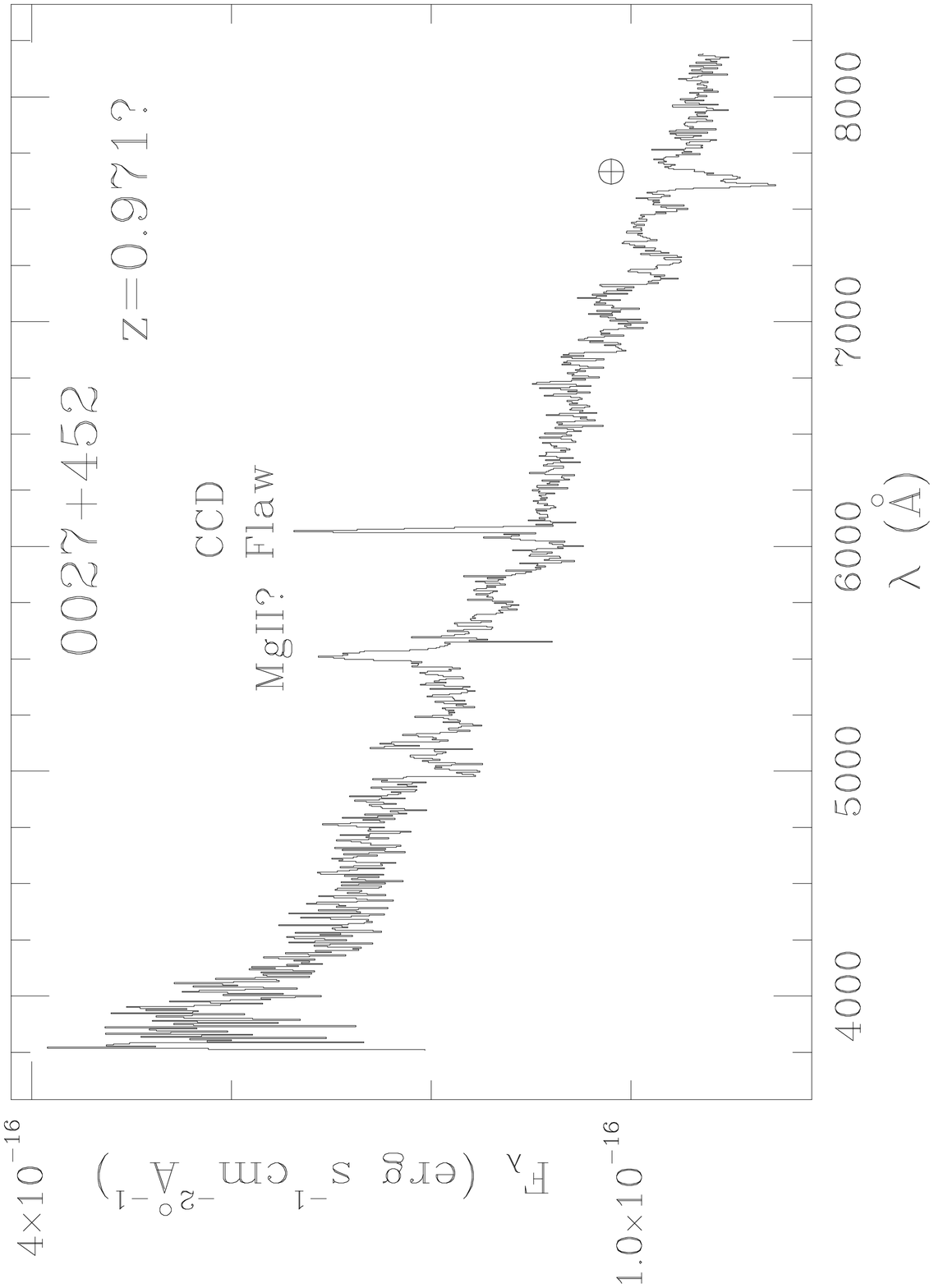,height=7.1cm,width=6.3cm}
\vspace{0.25in}
\psfig{file=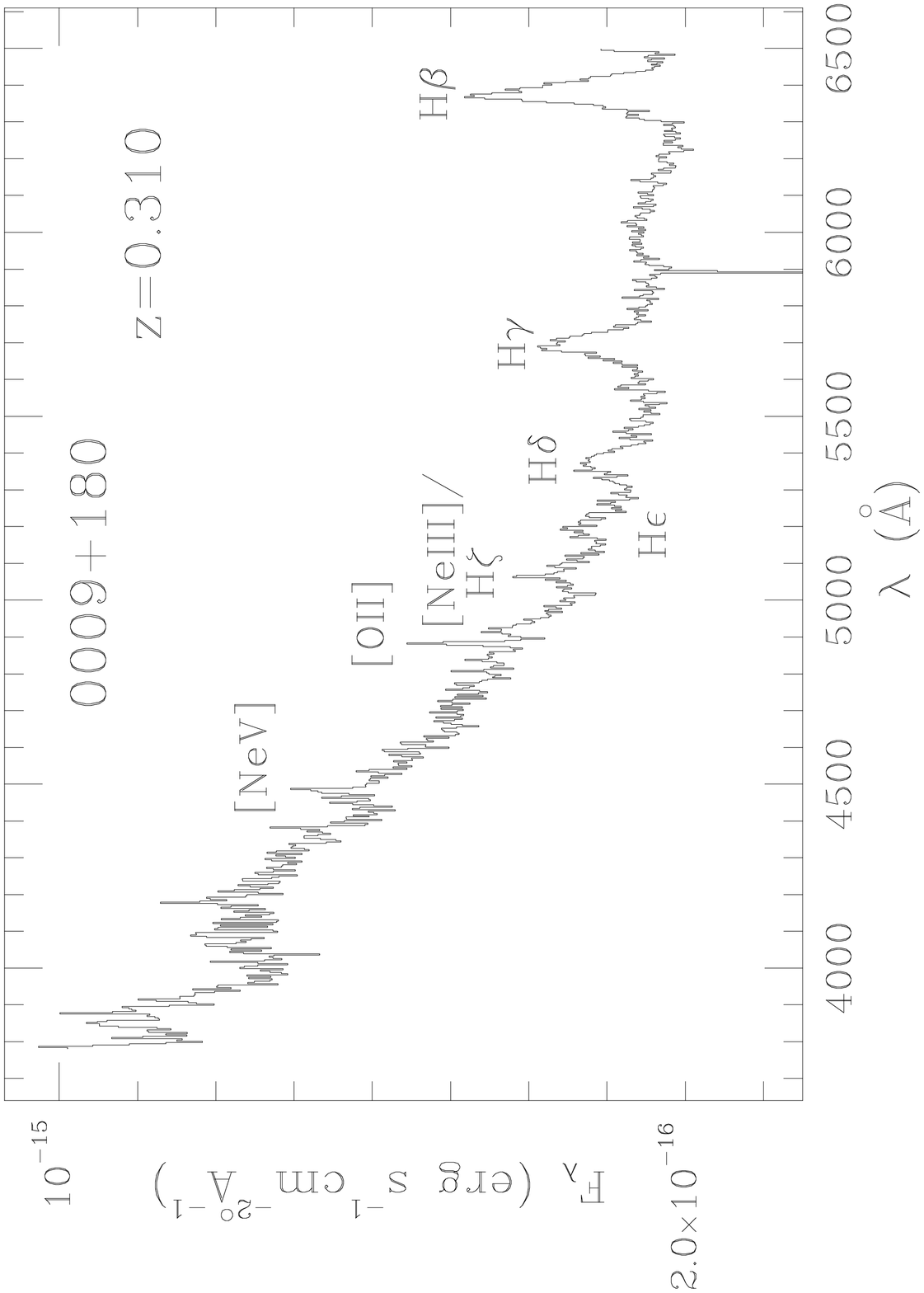,height=7.1cm,width=6.3cm}
\vspace{0.25in}
\psfig{file=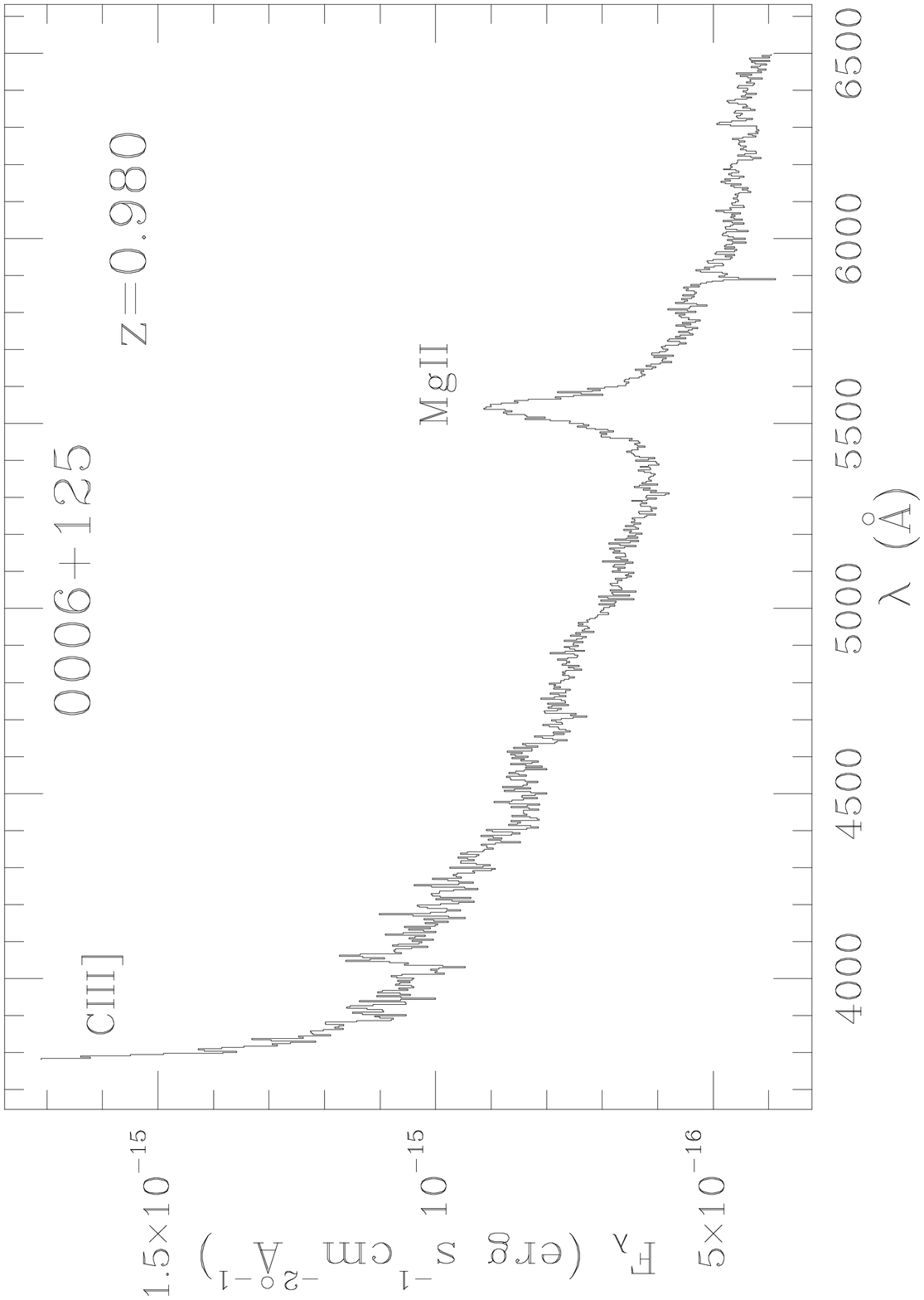,height=7.1cm,width=6.3cm}
\end{minipage}
\hspace{0.3in}
\begin{minipage}[t]{6.3in}
\psfig{file=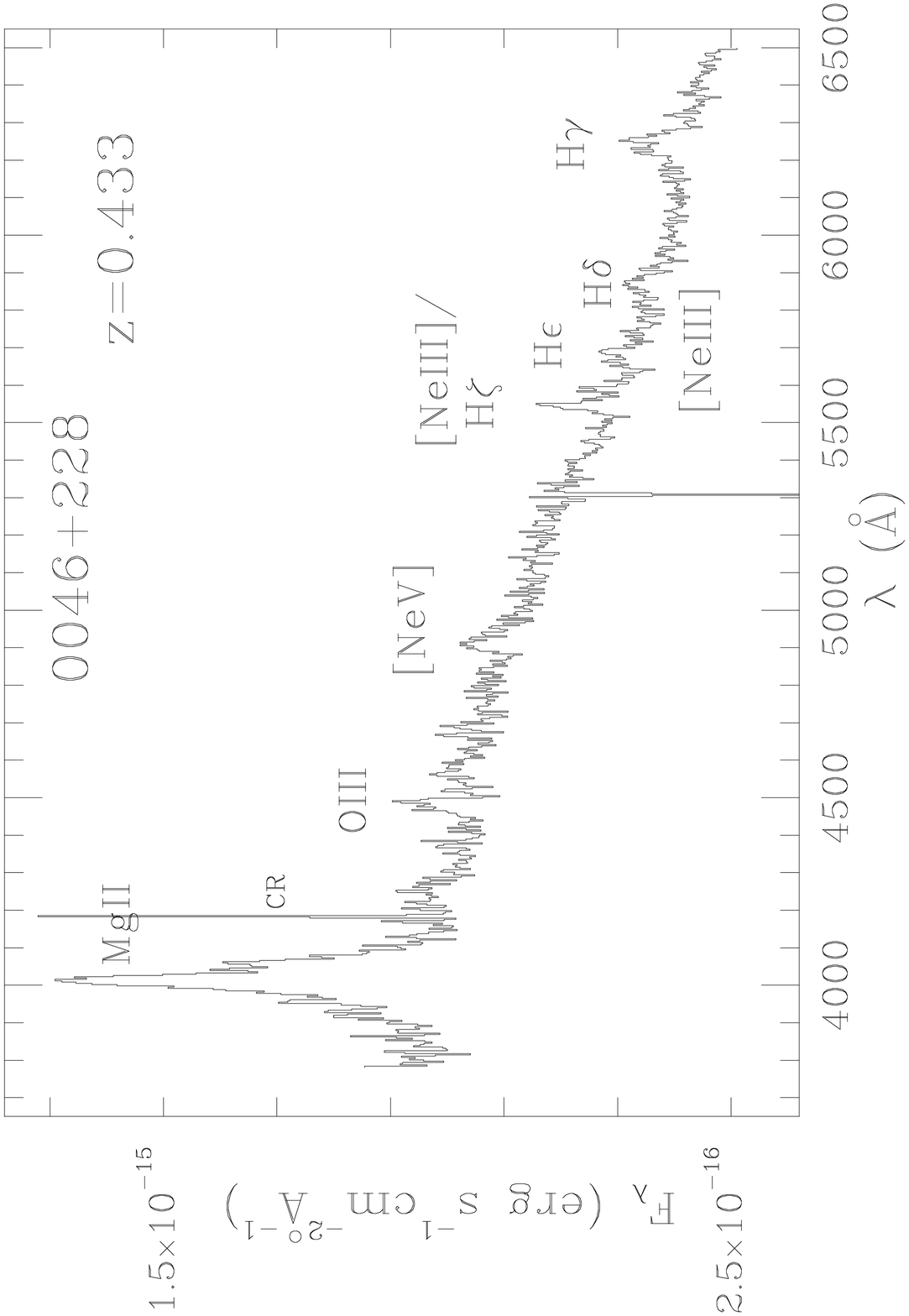,height=7.1cm,width=6.3cm}
\vspace{0.25in}
\psfig{file=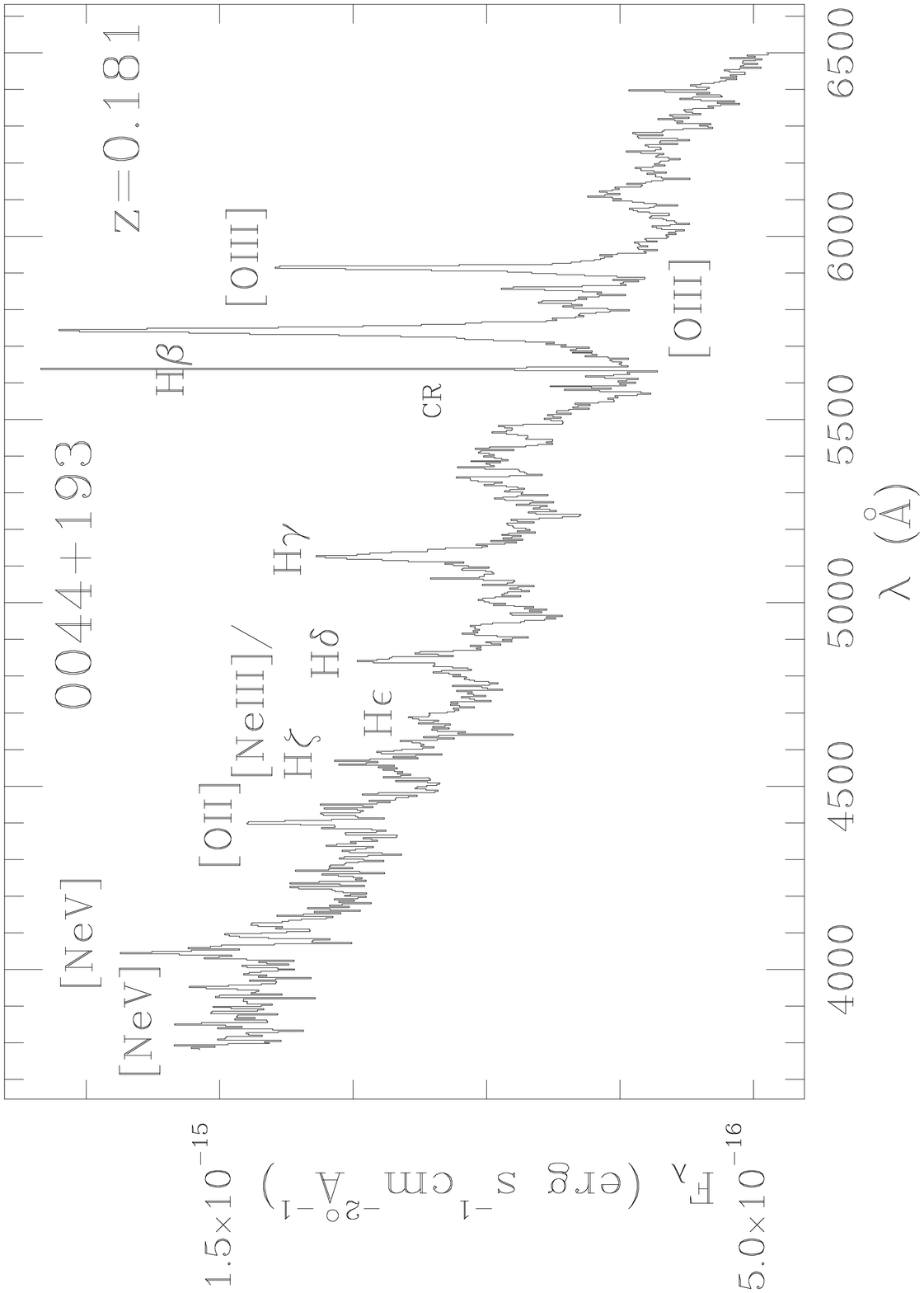,height=7.1cm,width=6.3cm}
\vspace{0.25in}
\psfig{file=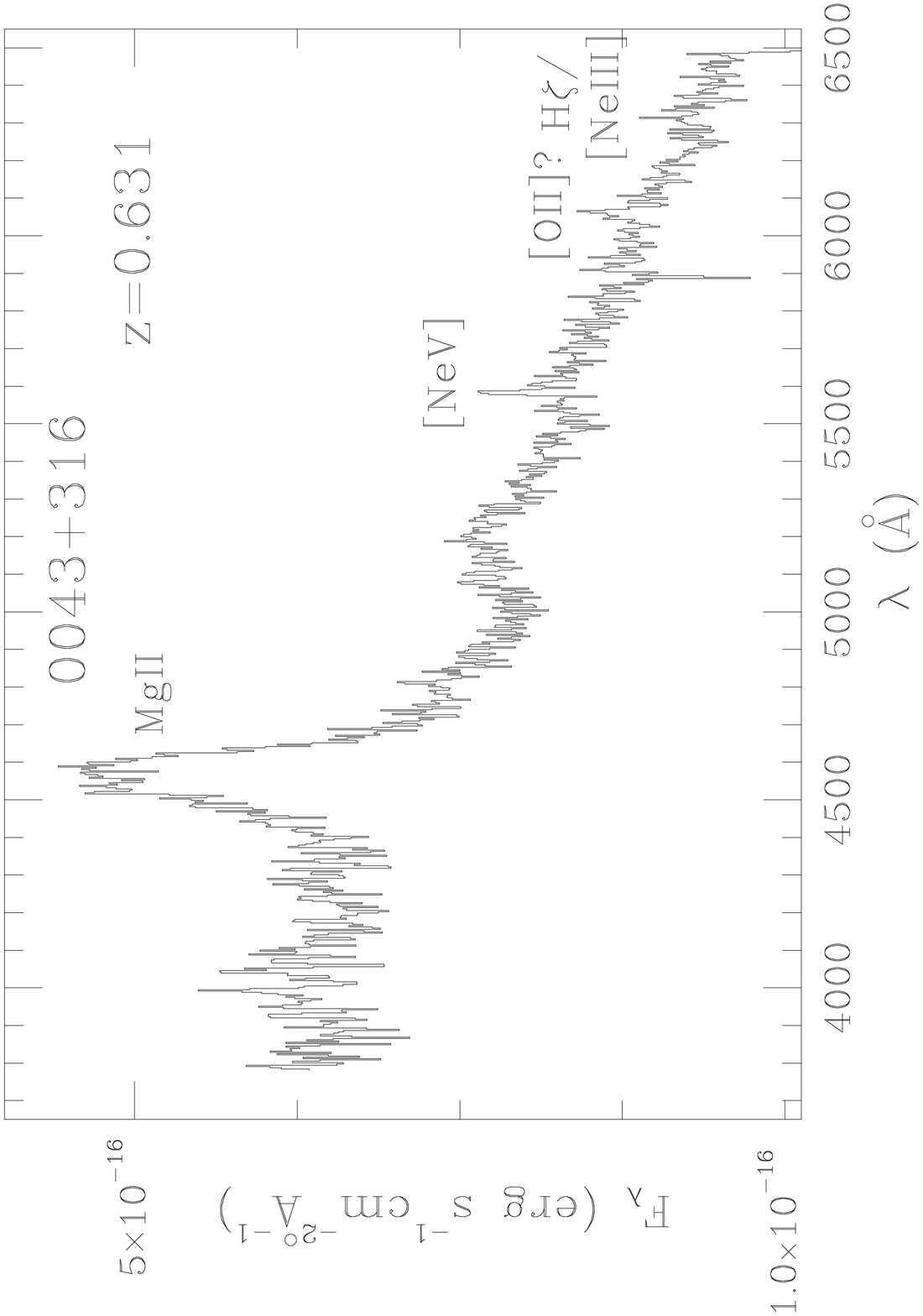,height=7.1cm,width=6.3cm}
\end{minipage}
\hfill
\begin{minipage}[t]{0.3in}
\vfill
\begin{sideways}
Figure 1.1 $-$ 1.6: Spectra of RGB objects
\end{sideways}
\vfill
\end{minipage}
\end{figure}
\clearpage

\begin{figure}
\vspace{-0.3in}
\hspace{-0.3in}
\begin{minipage}[t]{6.3in}
\psfig{file=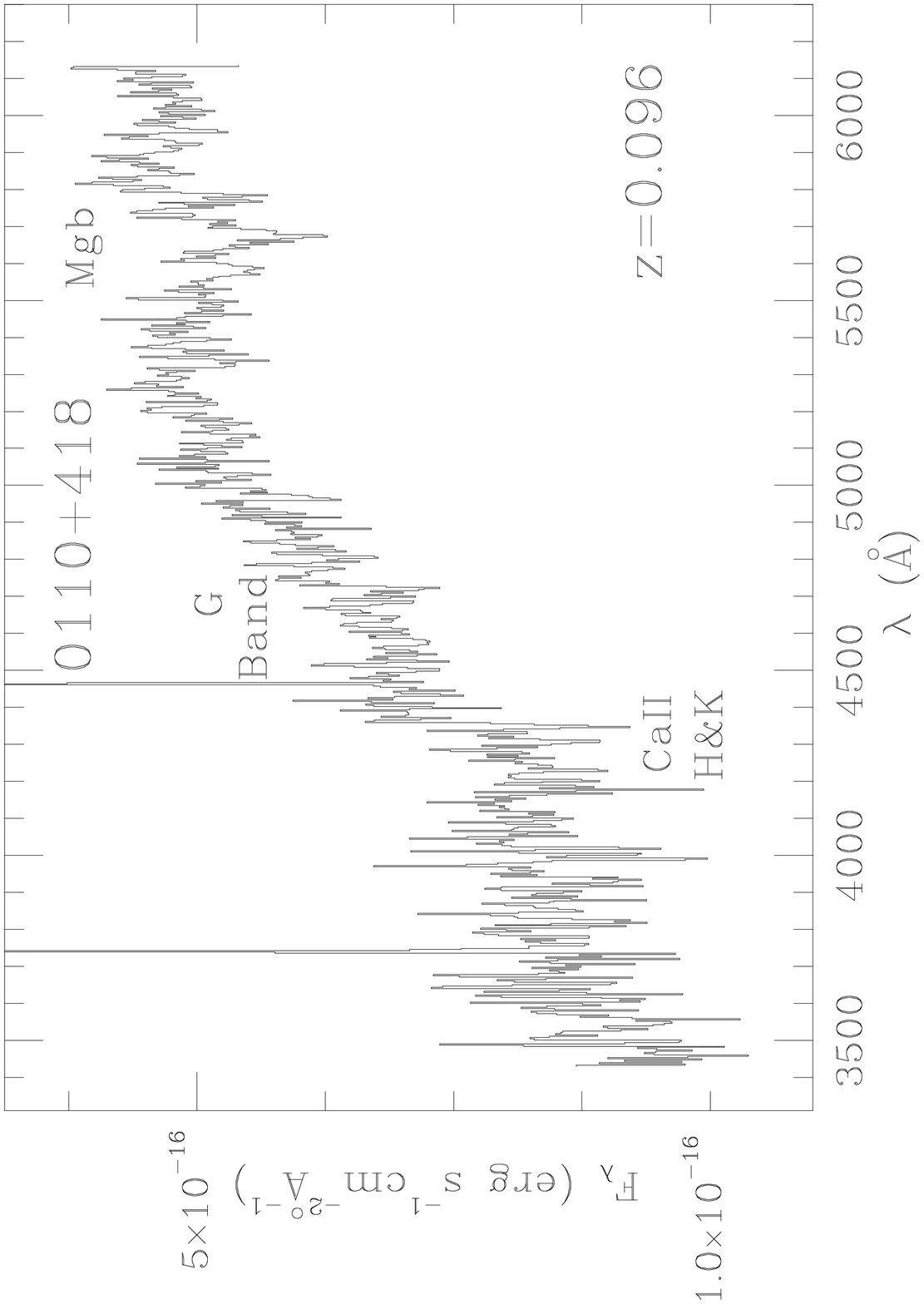,height=7.1cm,width=6.3cm}
\vspace{0.25in}
\psfig{file=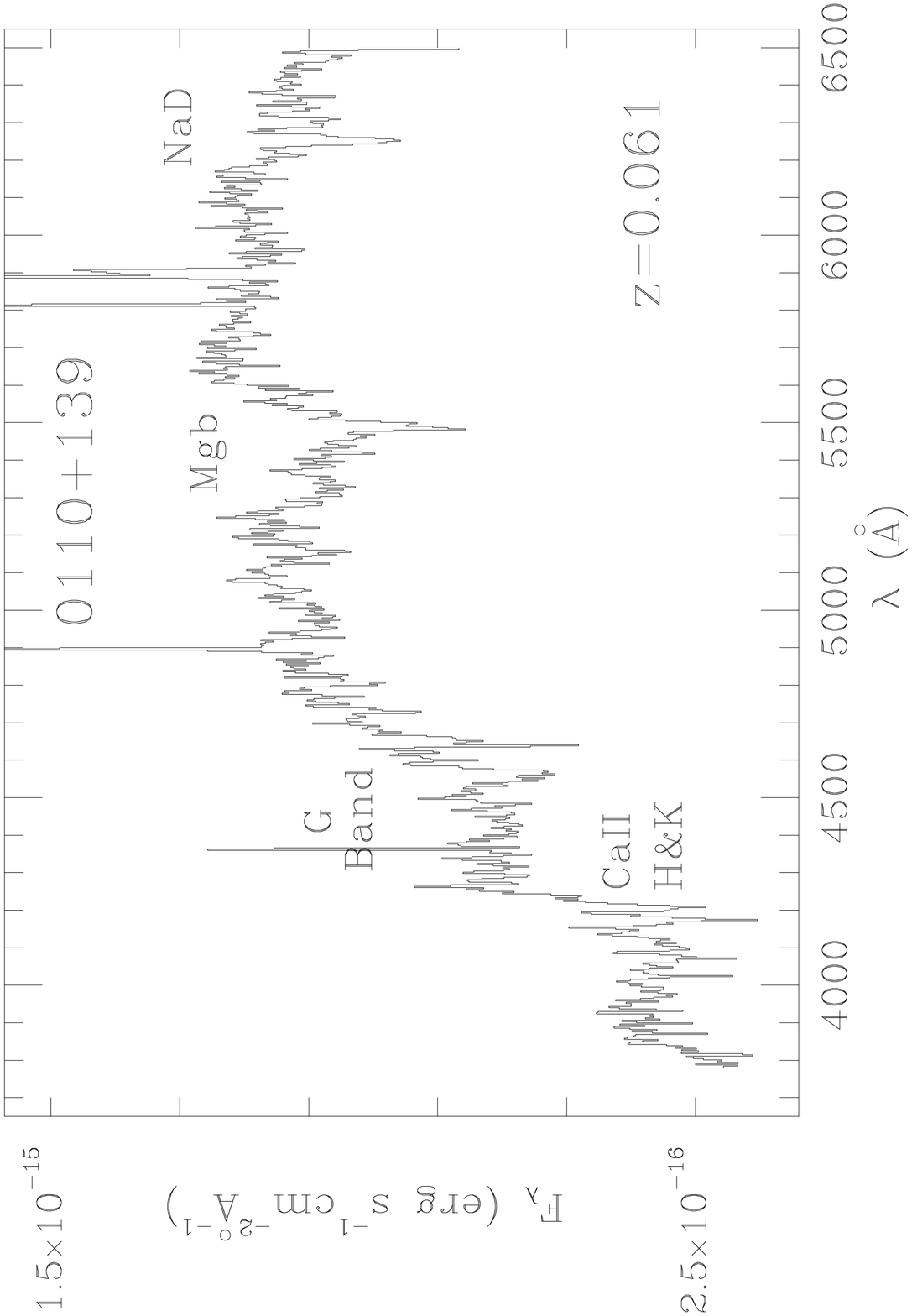,height=7.1cm,width=6.3cm}
\vspace{0.25in}
\psfig{file=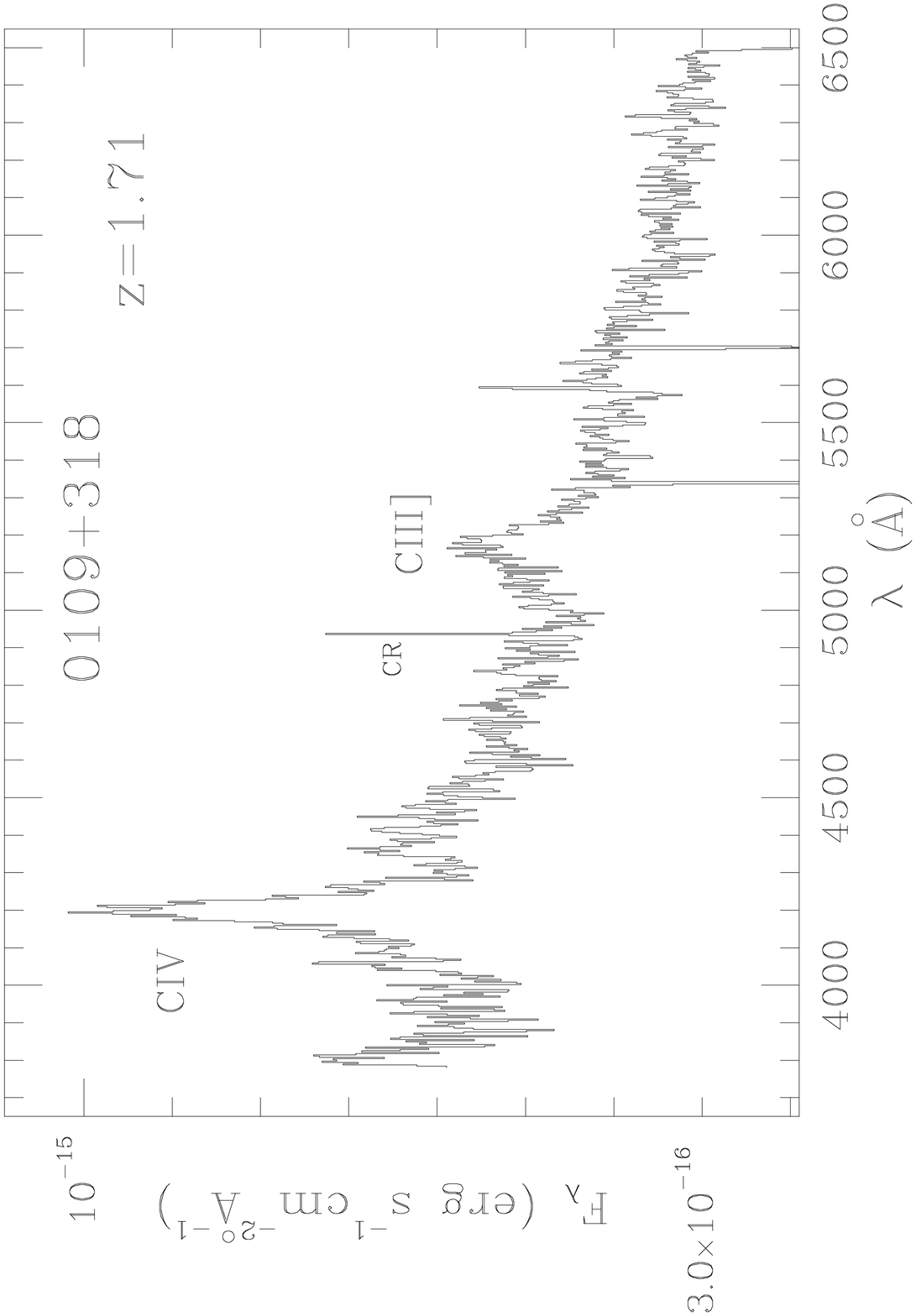,height=7.1cm,width=6.3cm}
\end{minipage}
\hspace{0.3in}
\begin{minipage}[t]{6.3in}
\psfig{file=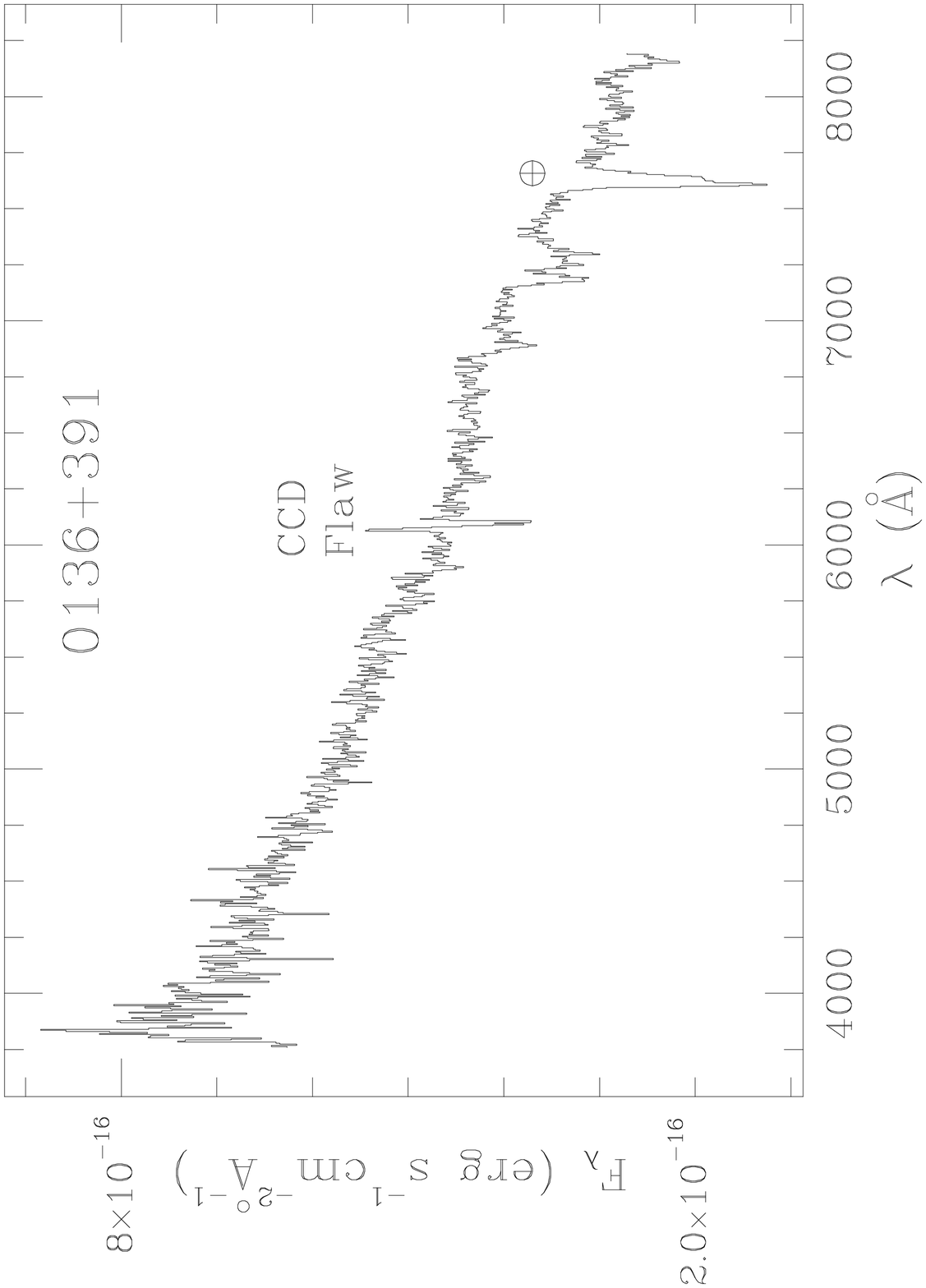,height=7.1cm,width=6.3cm}
\vspace{0.25in}
\psfig{file=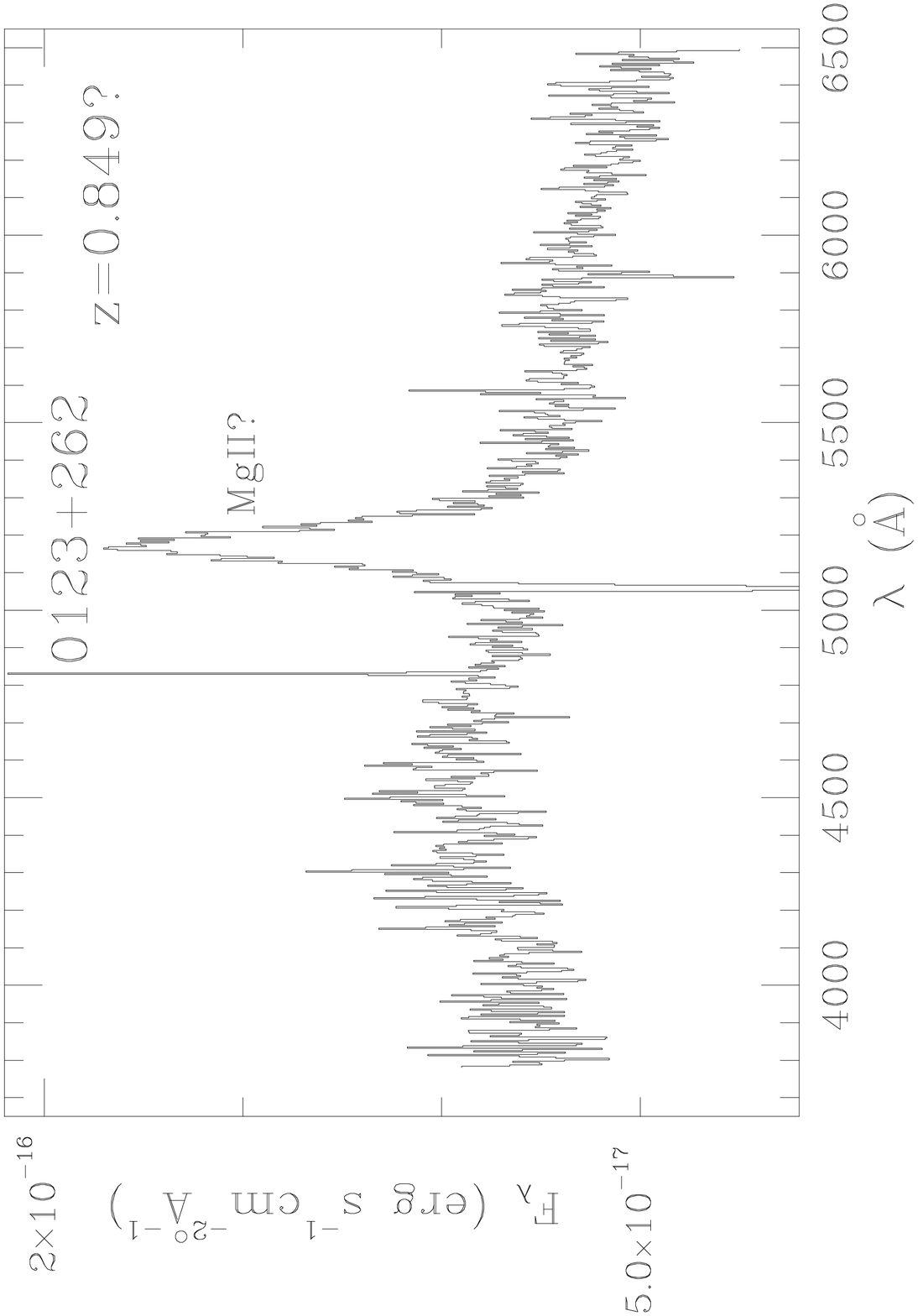,height=7.1cm,width=6.3cm}
\vspace{0.25in}
\psfig{file=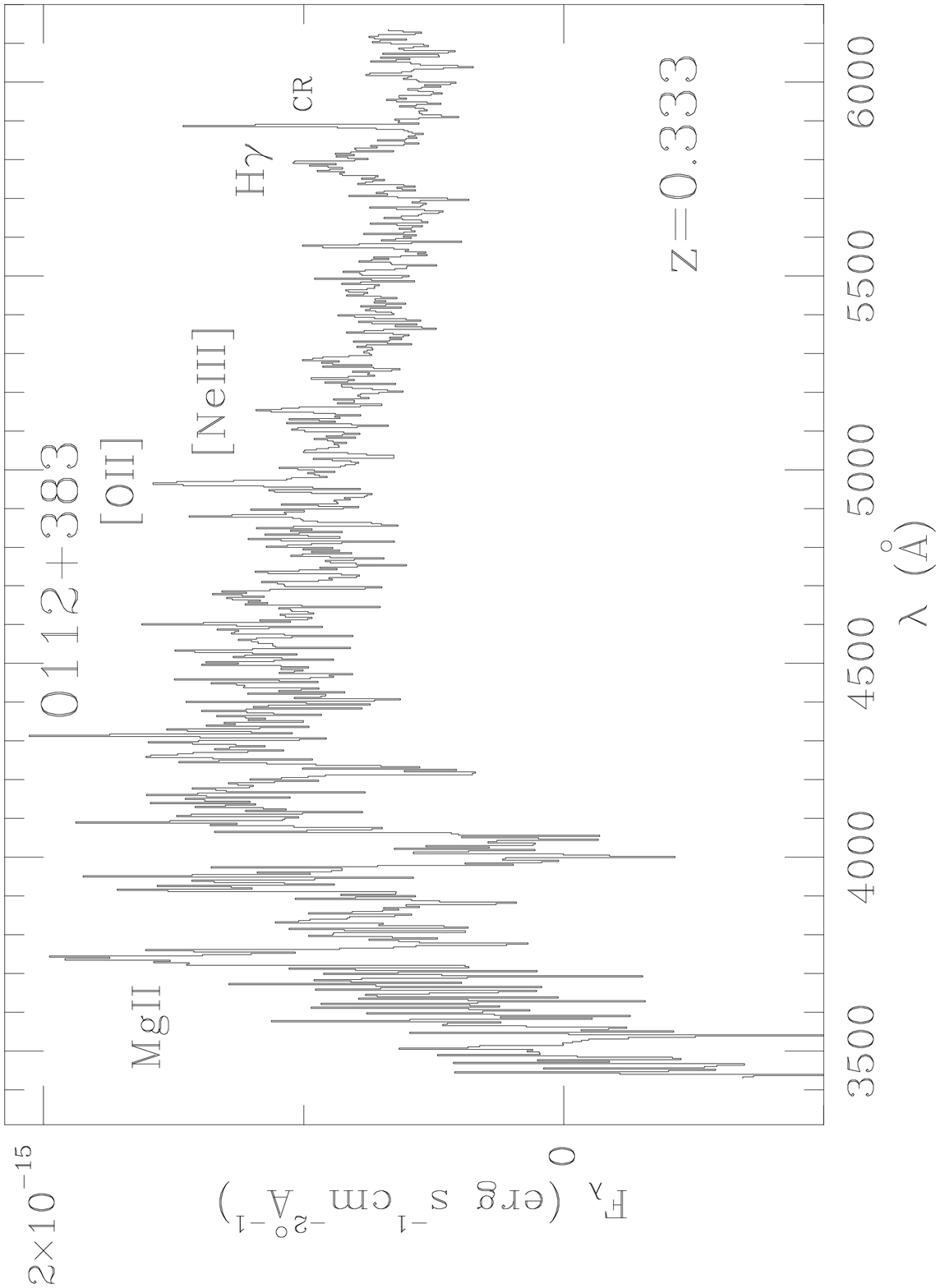,height=7.1cm,width=6.3cm}
\end{minipage}
\hfill
\begin{minipage}[t]{0.3in}
\vfill
\begin{sideways}
Figure 1.7 $-$ 1.12: Spectra of RGB Sources ({\it continued})
\end{sideways}
\vfill
\end{minipage}
\end{figure}
\clearpage

\begin{figure}
\vspace{-0.3in}
\hspace{-0.3in}
\begin{minipage}[t]{6.3in}
\psfig{file=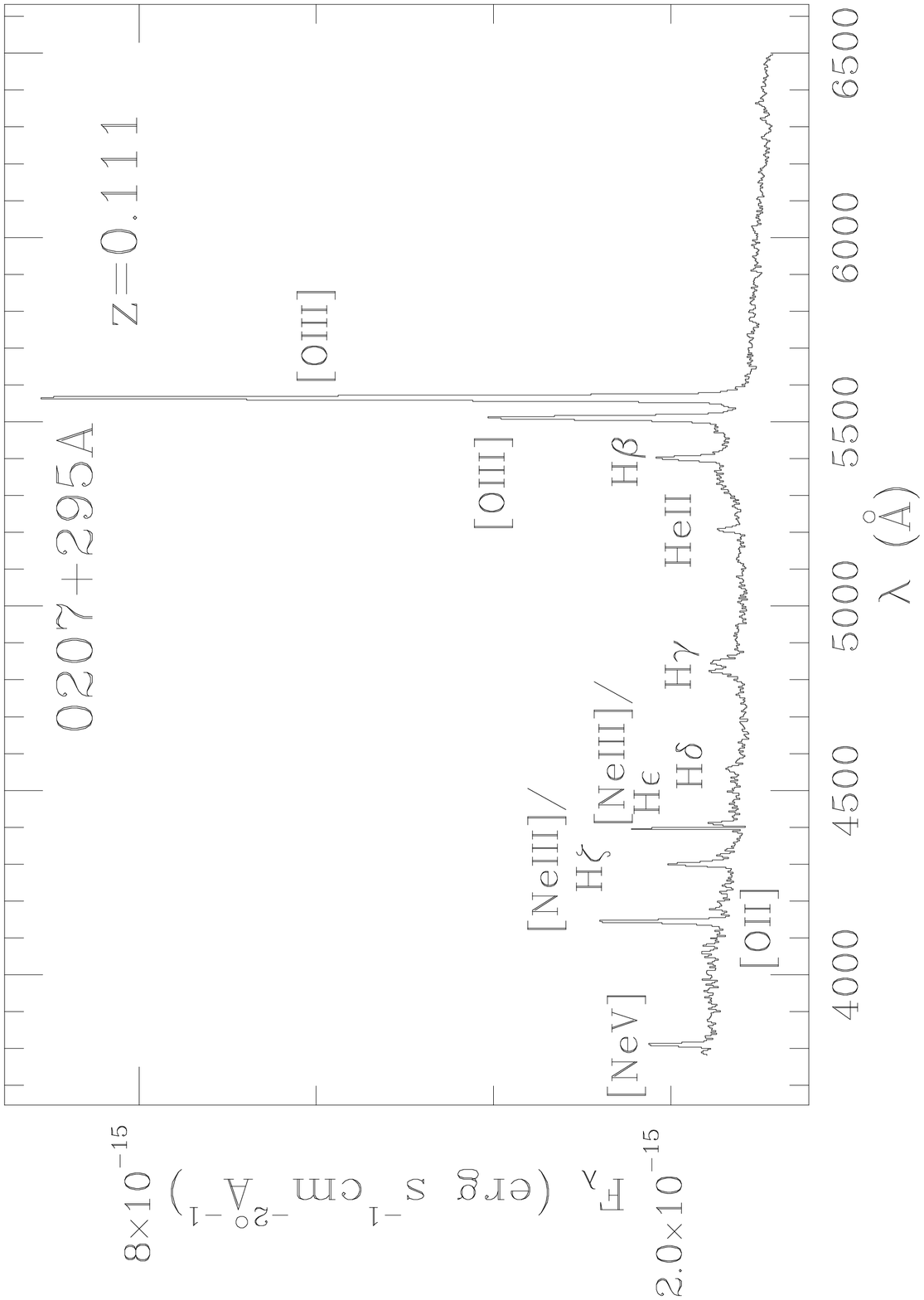,height=7.1cm,width=6.3cm}
\vspace{0.25in}
\psfig{file=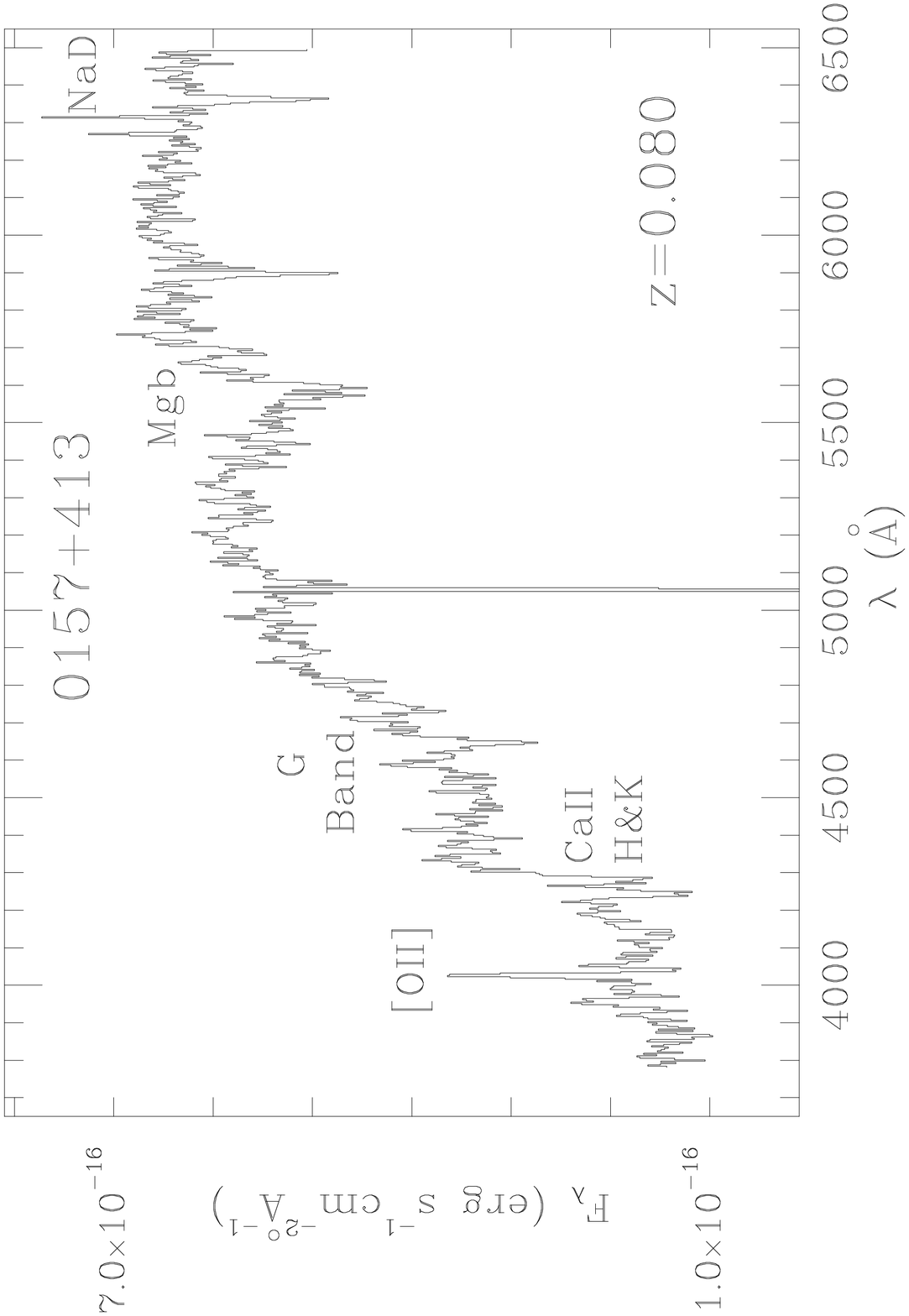,height=7.1cm,width=6.3cm}
\vspace{0.25in}
\psfig{file=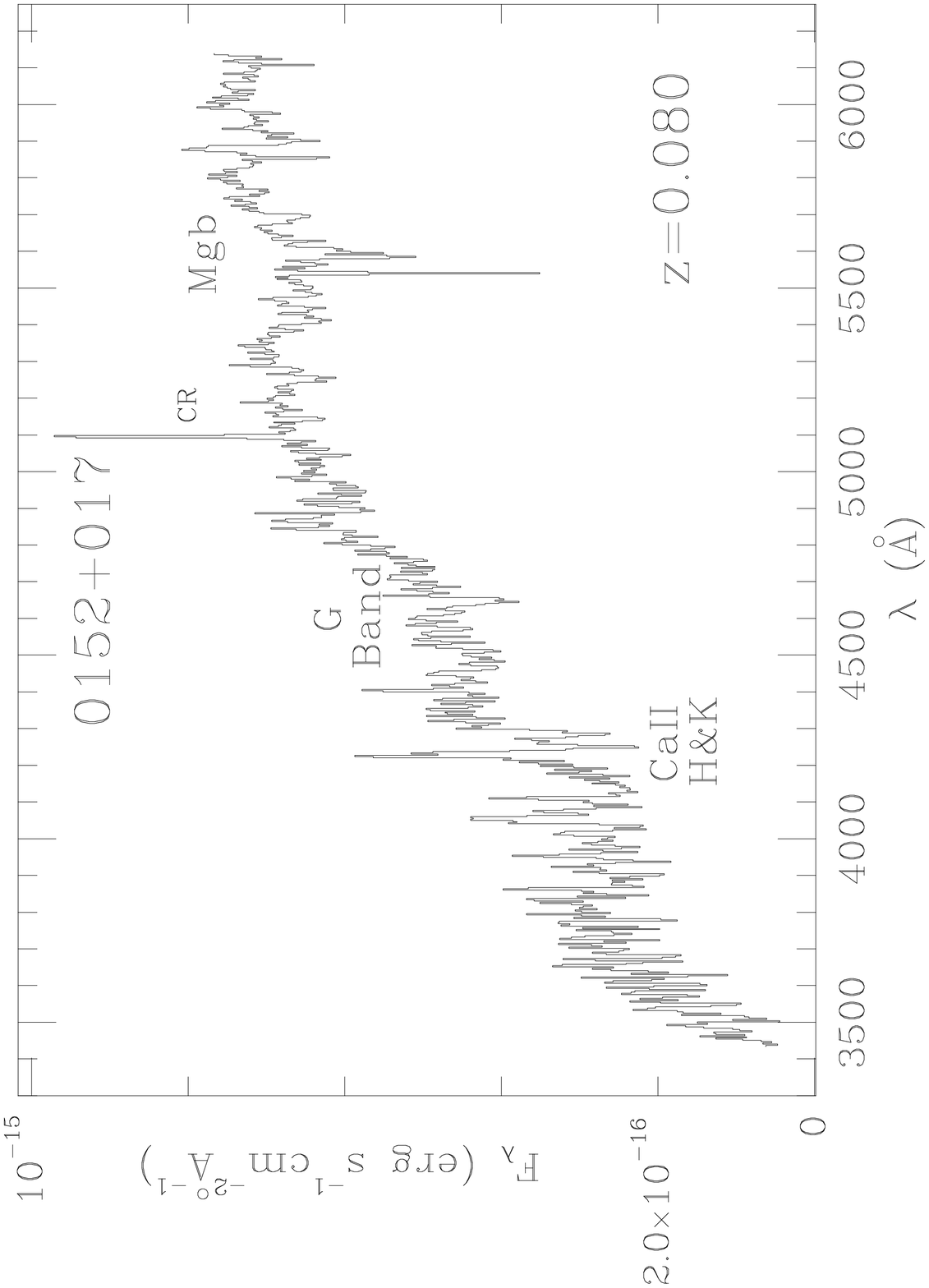,height=7.1cm,width=6.3cm}
\end{minipage}
\hspace{0.3in}
\begin{minipage}[t]{6.3in}
\psfig{file=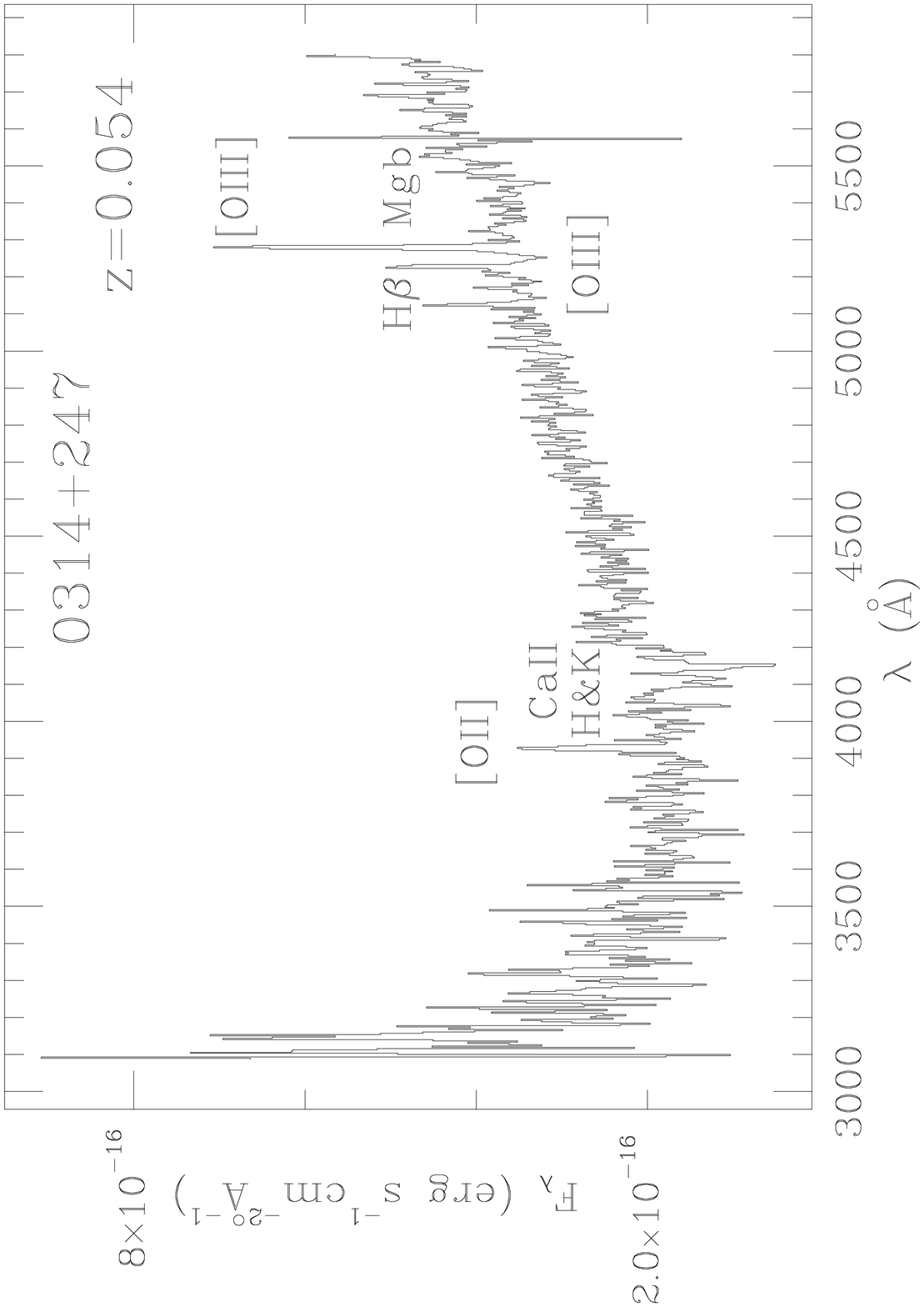,height=7.1cm,width=6.3cm}
\vspace{0.25in}
\psfig{file=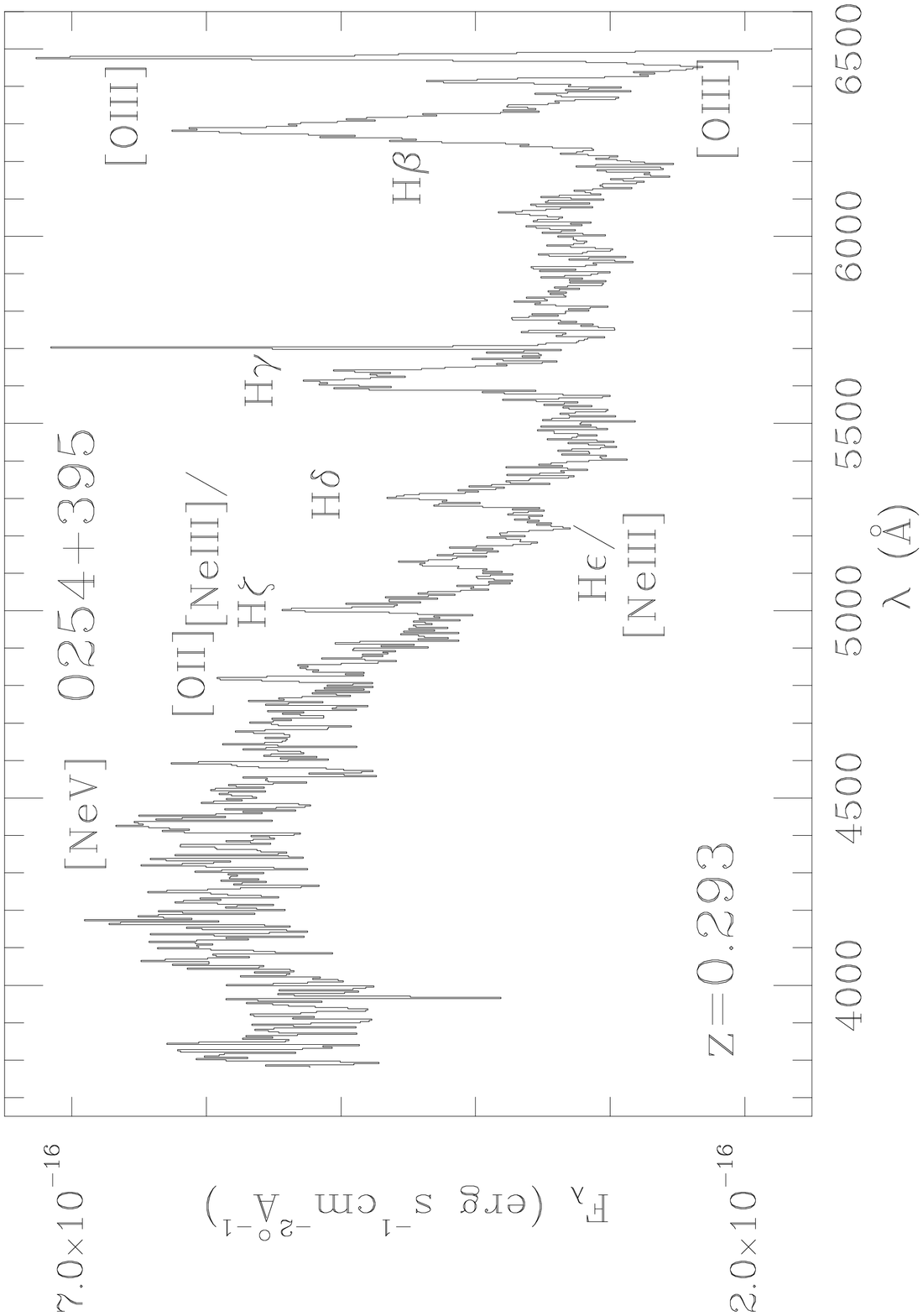,height=7.1cm,width=6.3cm}
\vspace{0.25in}
\psfig{file=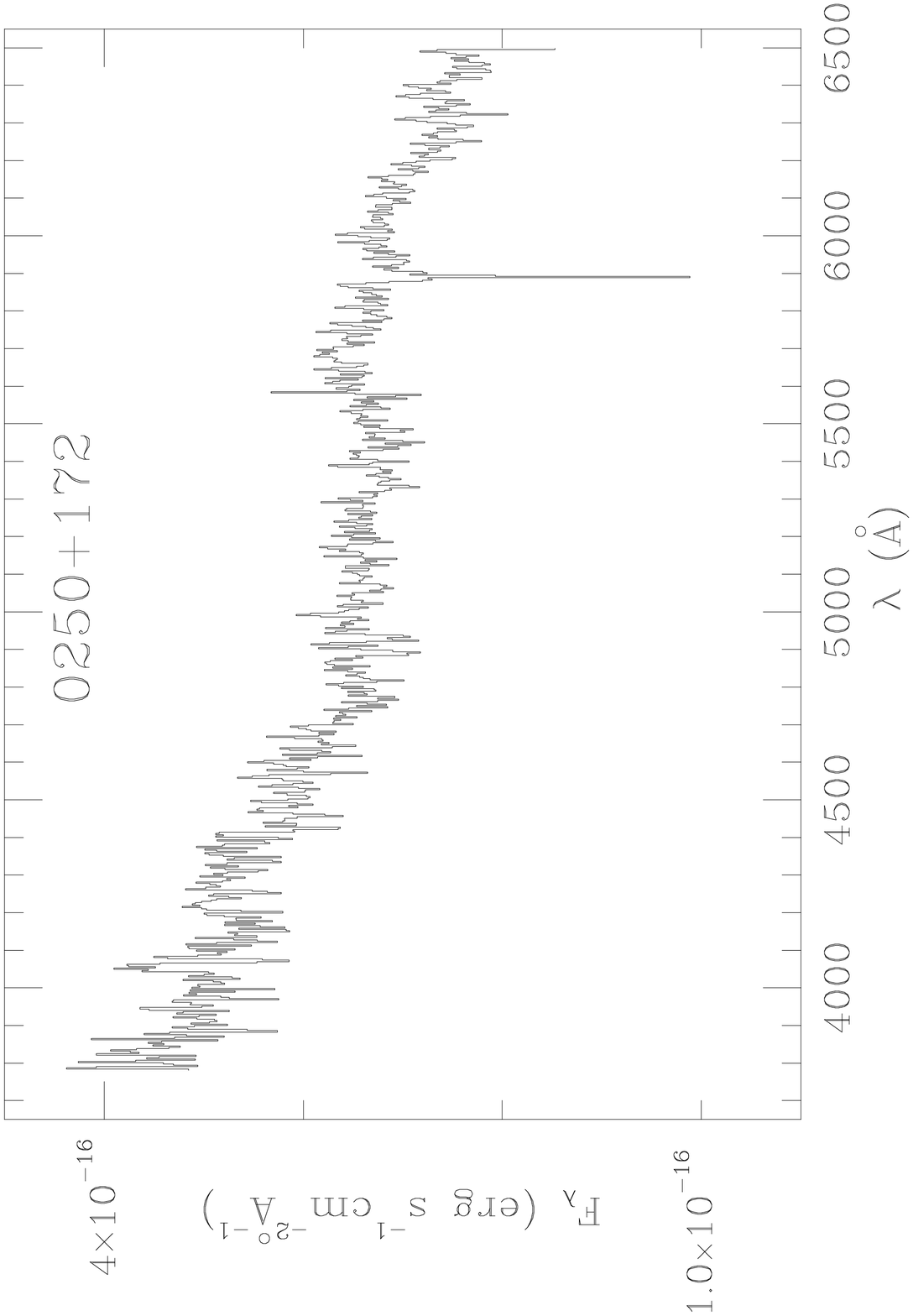,height=7.1cm,width=6.3cm}
\end{minipage}
\hfill
\begin{minipage}[t]{0.3in}
\vfill
\begin{sideways}
Figure 1.13 $-$ 1.18: Spectra of RGB Sources ({\it continued})
\end{sideways}
\vfill
\end{minipage}
\end{figure}
\clearpage

\begin{figure}
\vspace{-0.3in}
\hspace{-0.3in}
\begin{minipage}[t]{6.3in}
\psfig{file=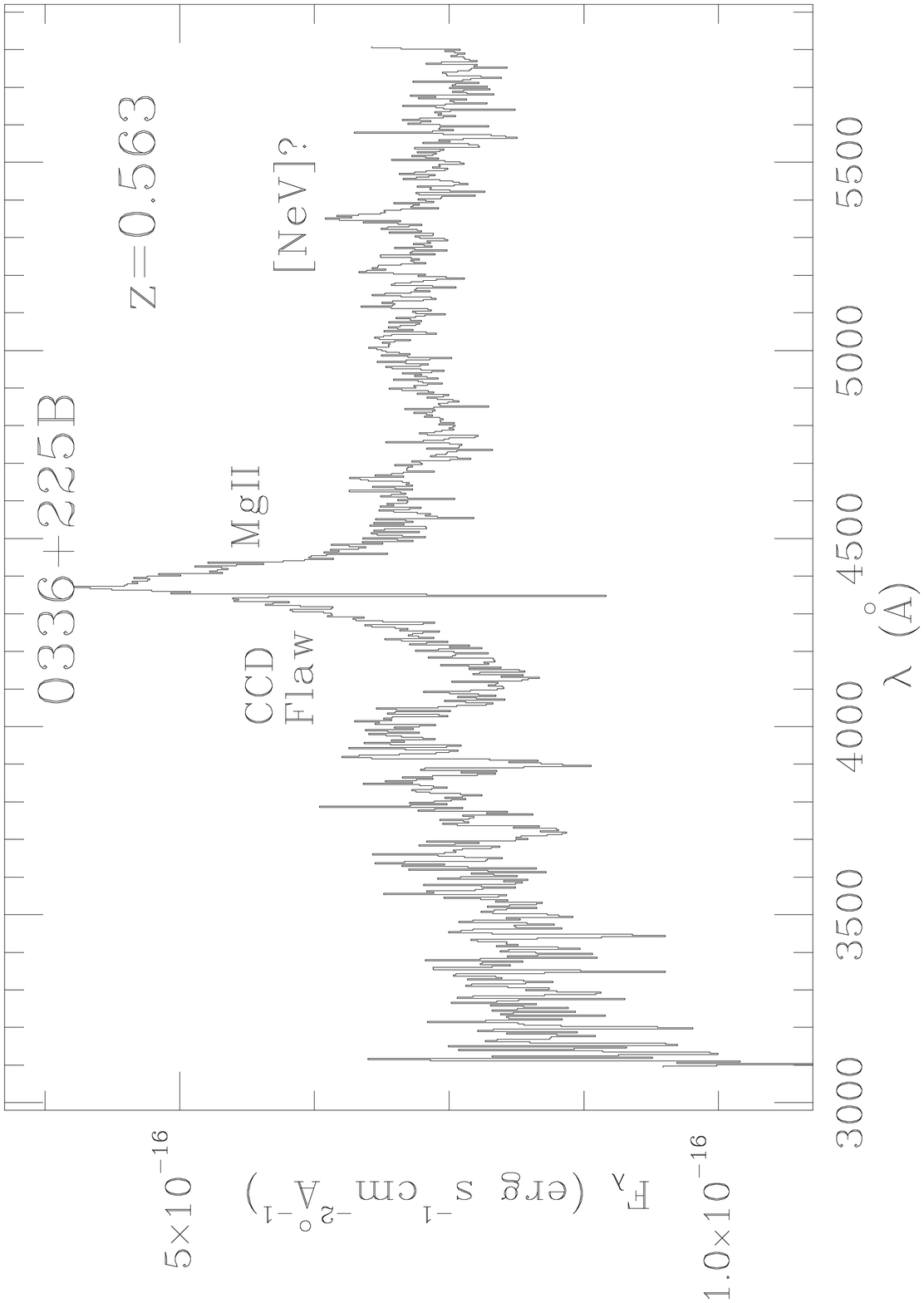,height=7.1cm,width=6.3cm}
\vspace{0.25in}
\psfig{file=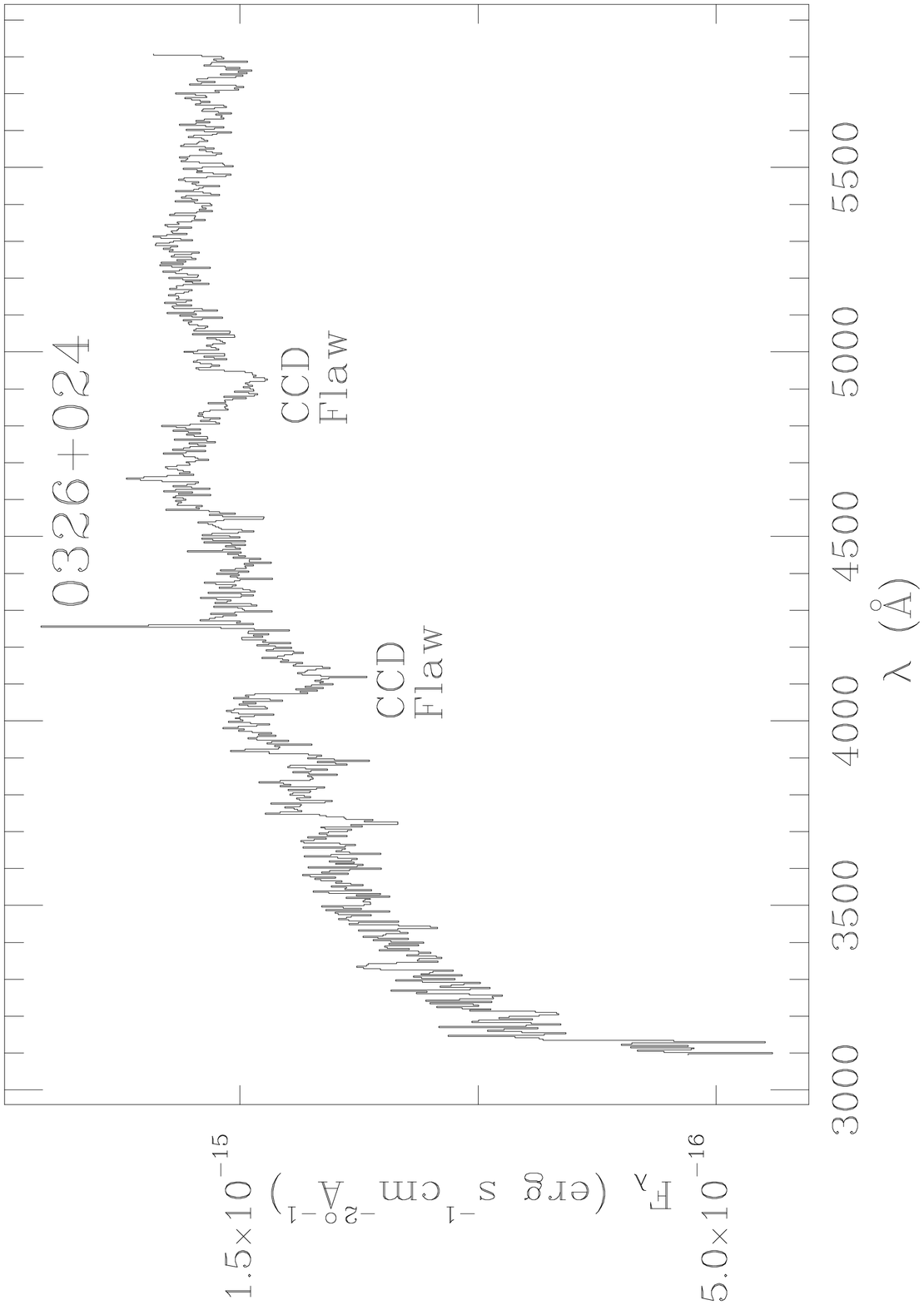,height=7.1cm,width=6.3cm}
\vspace{0.25in}
\psfig{file=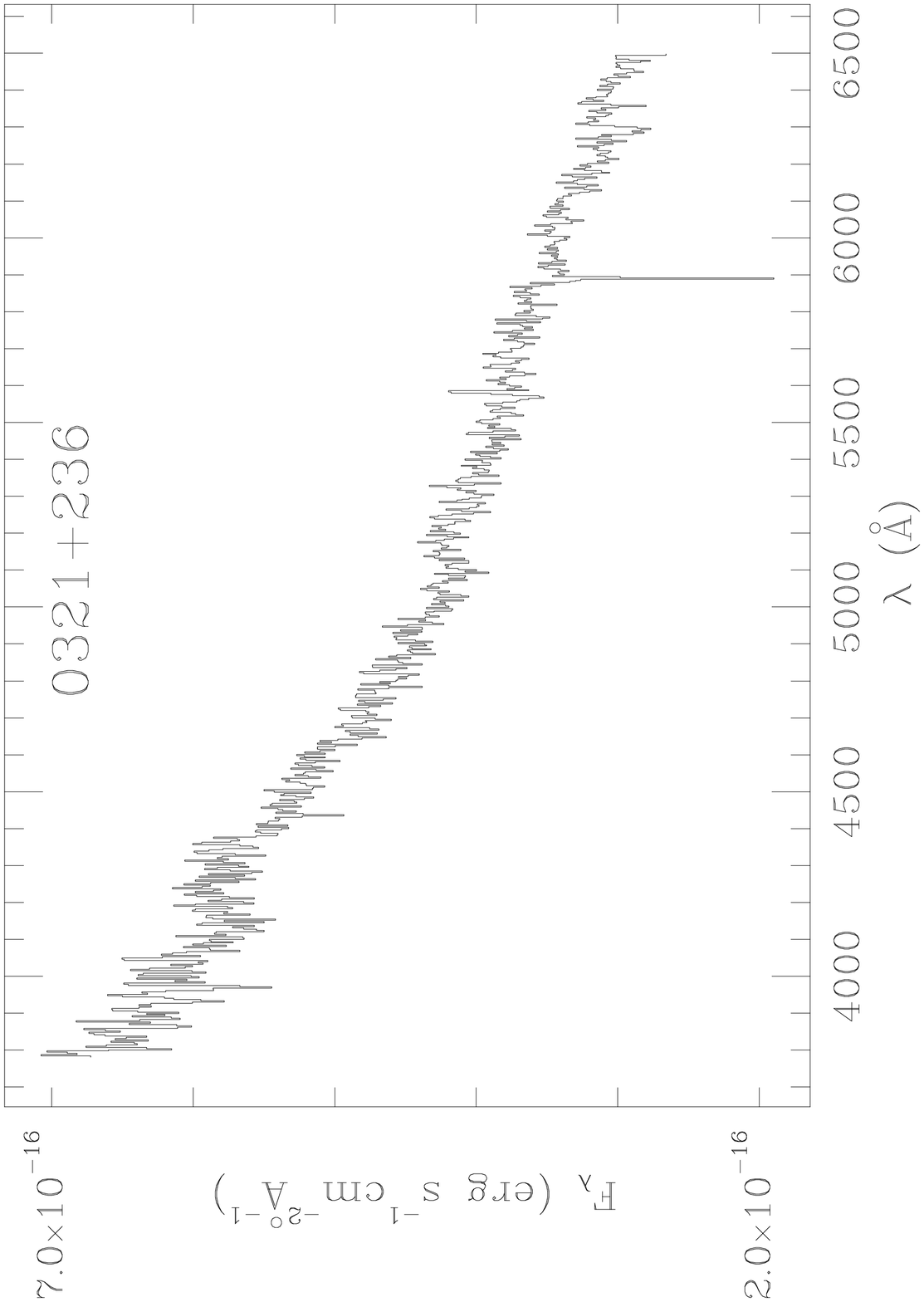,height=7.1cm,width=6.3cm}
\end{minipage}
\hspace{0.3in}
\begin{minipage}[t]{6.3in}
\psfig{file=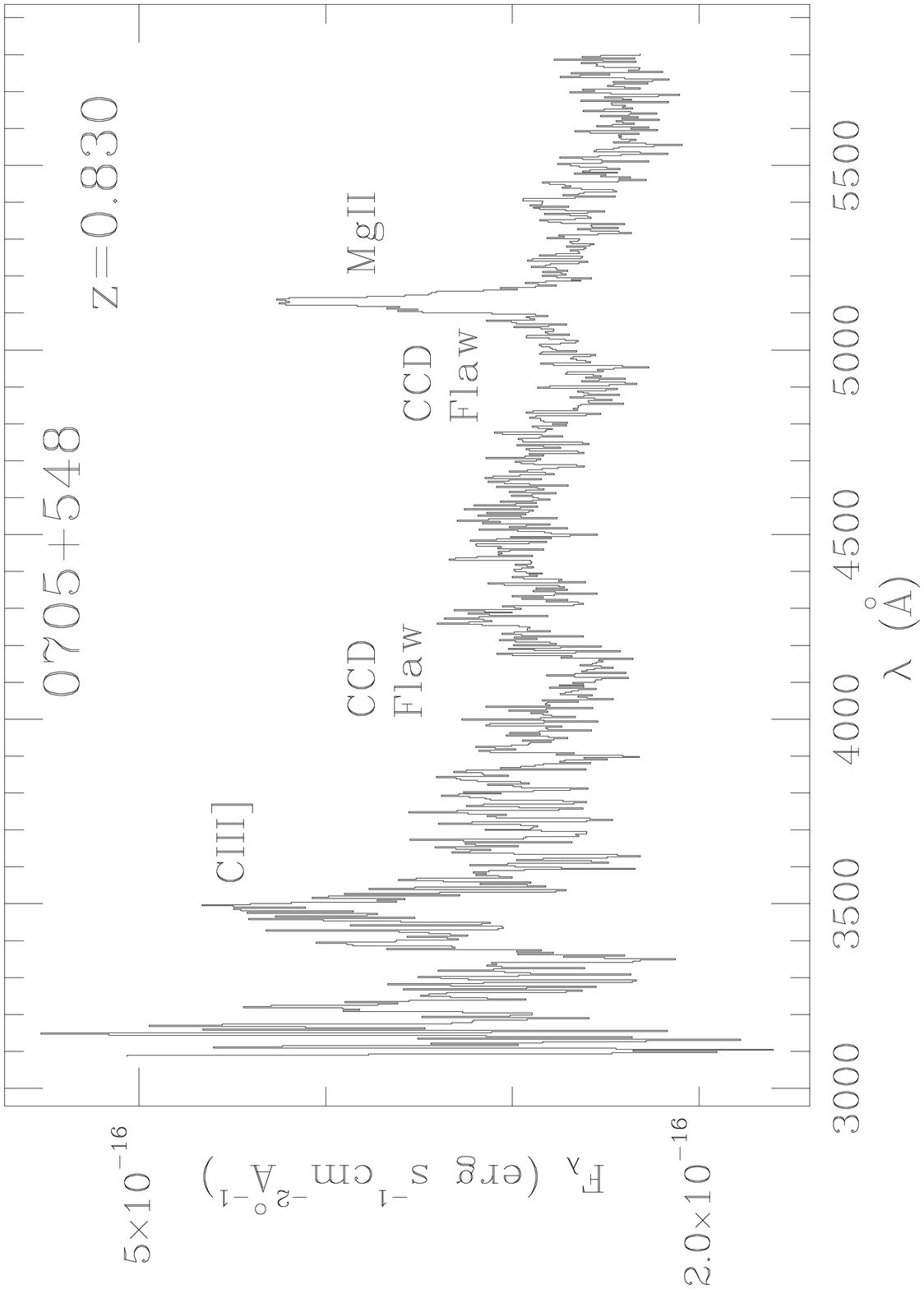,height=7.1cm,width=6.3cm}
\vspace{0.25in}
\psfig{file=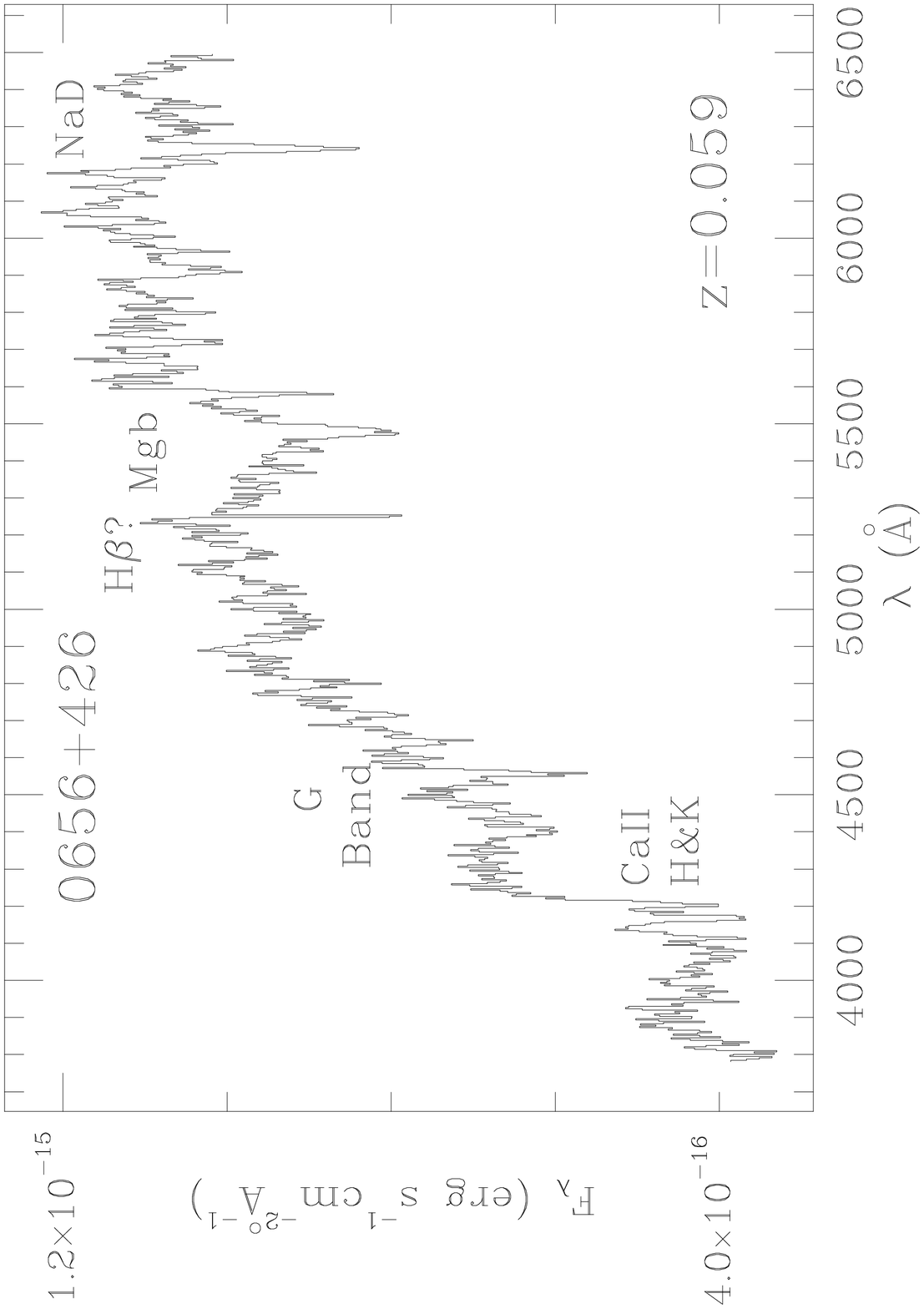,height=7.1cm,width=6.3cm}
\vspace{0.25in}
\psfig{file=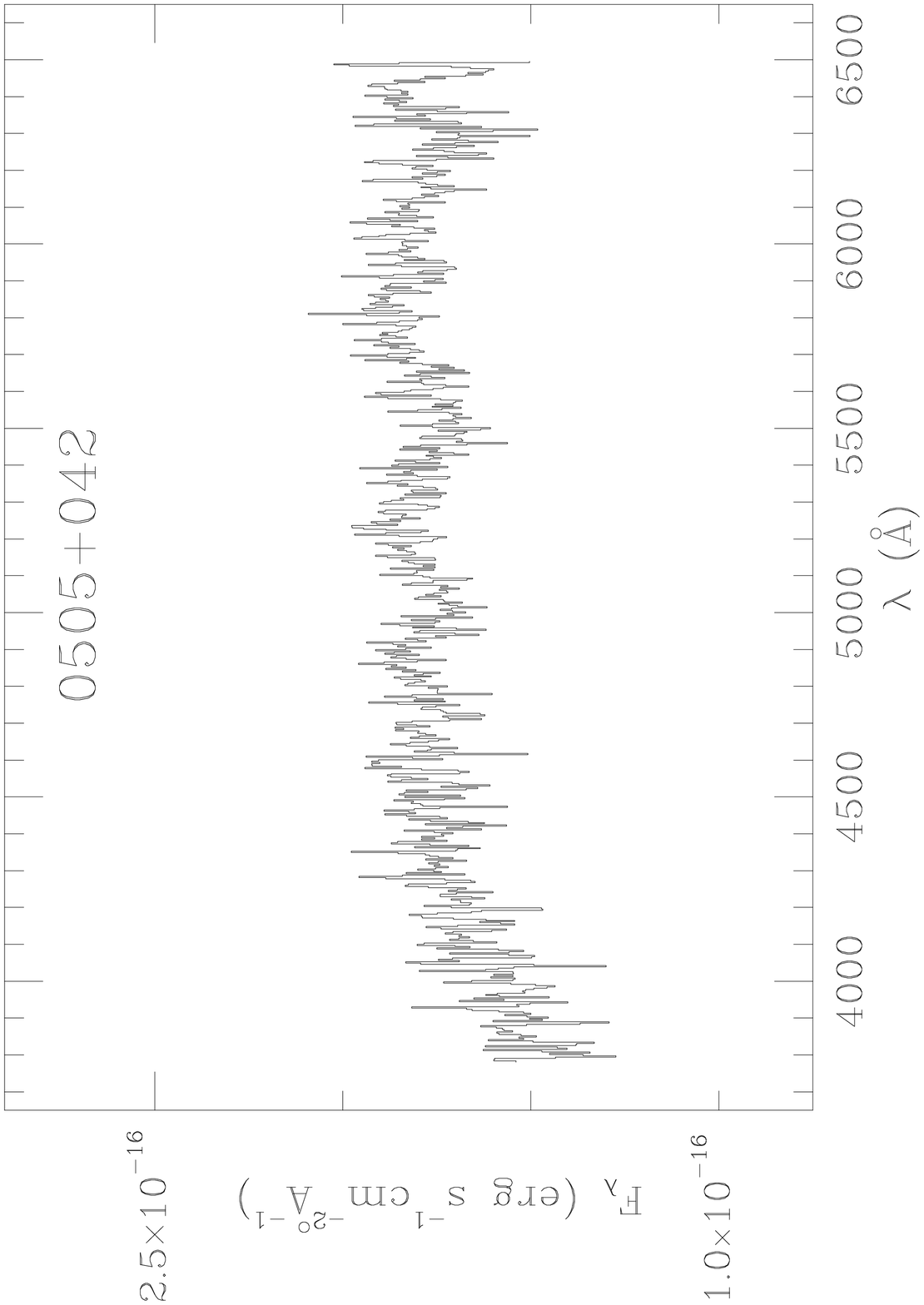,height=7.1cm,width=6.3cm}
\end{minipage}
\hfill
\begin{minipage}[t]{0.3in}
\vfill
\begin{sideways}
Figure 1.19 $-$ 1.24: Spectra of RGB Sources ({\it continued})
\end{sideways}
\vfill
\end{minipage}
\end{figure}
\clearpage

\begin{figure}
\vspace{-0.3in}
\hspace{-0.3in}
\begin{minipage}[t]{6.3in}
\psfig{file=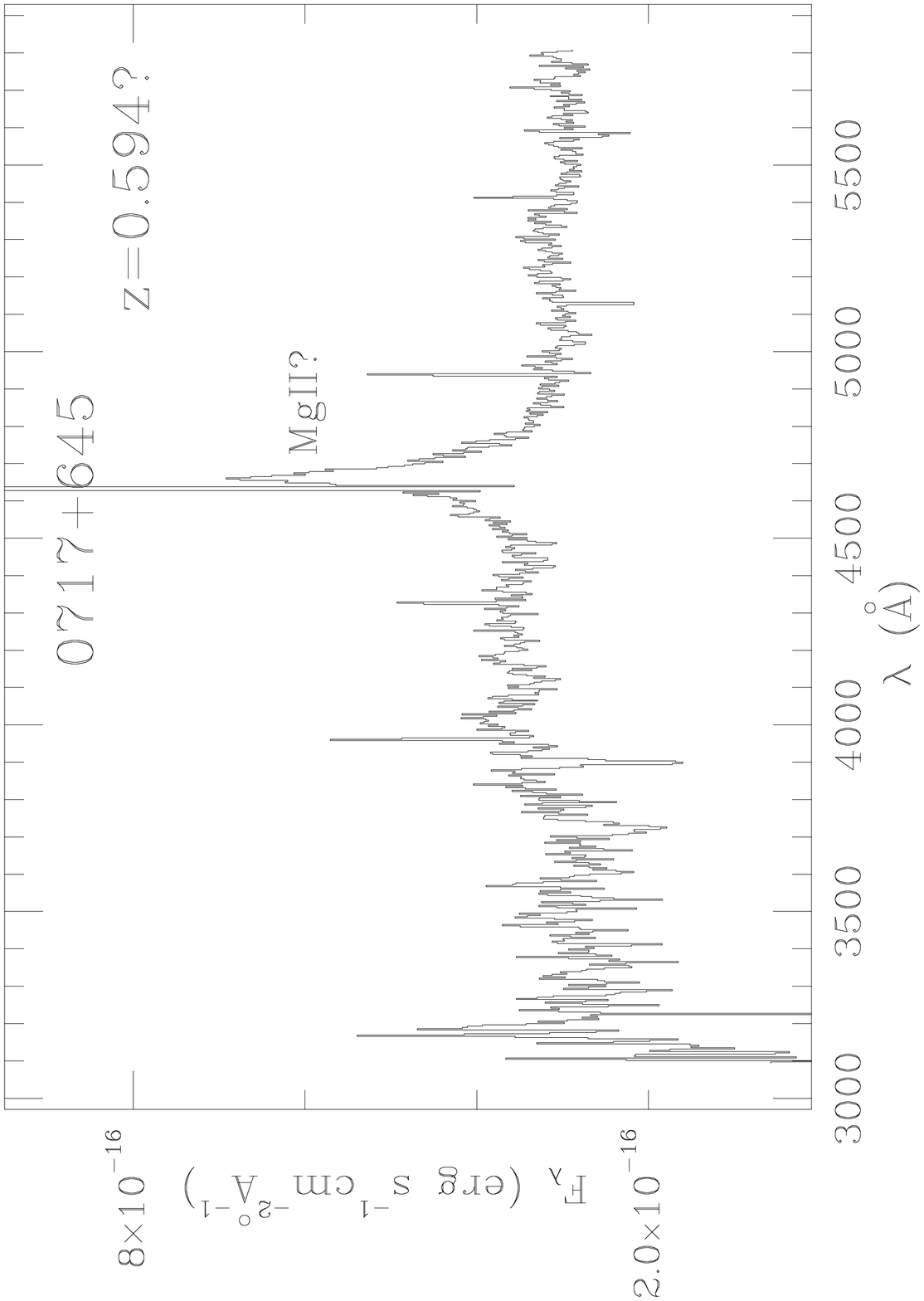,height=7.1cm,width=6.3cm}
\vspace{0.25in}
\psfig{file=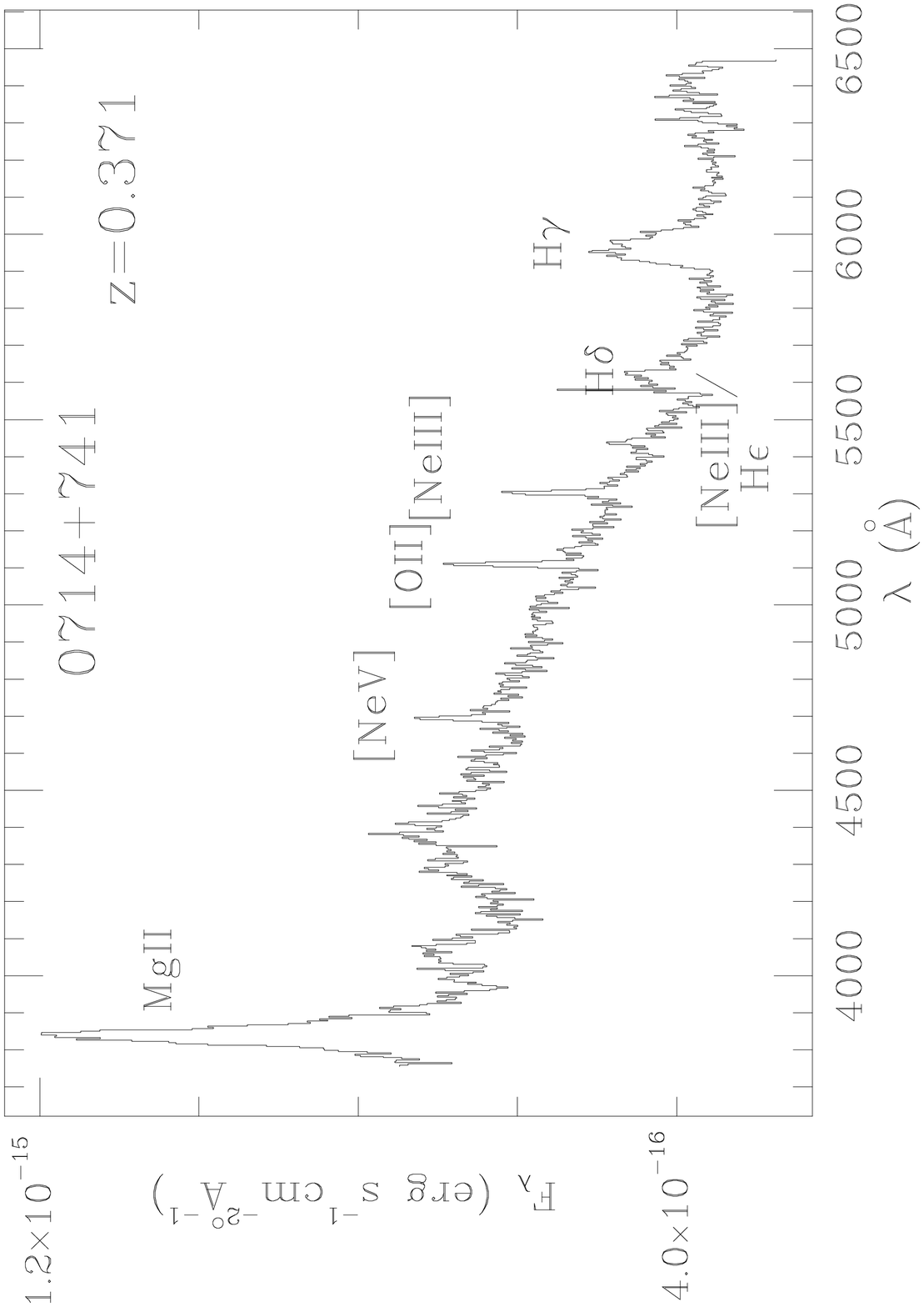,height=7.1cm,width=6.3cm}
\vspace{0.25in}
\psfig{file=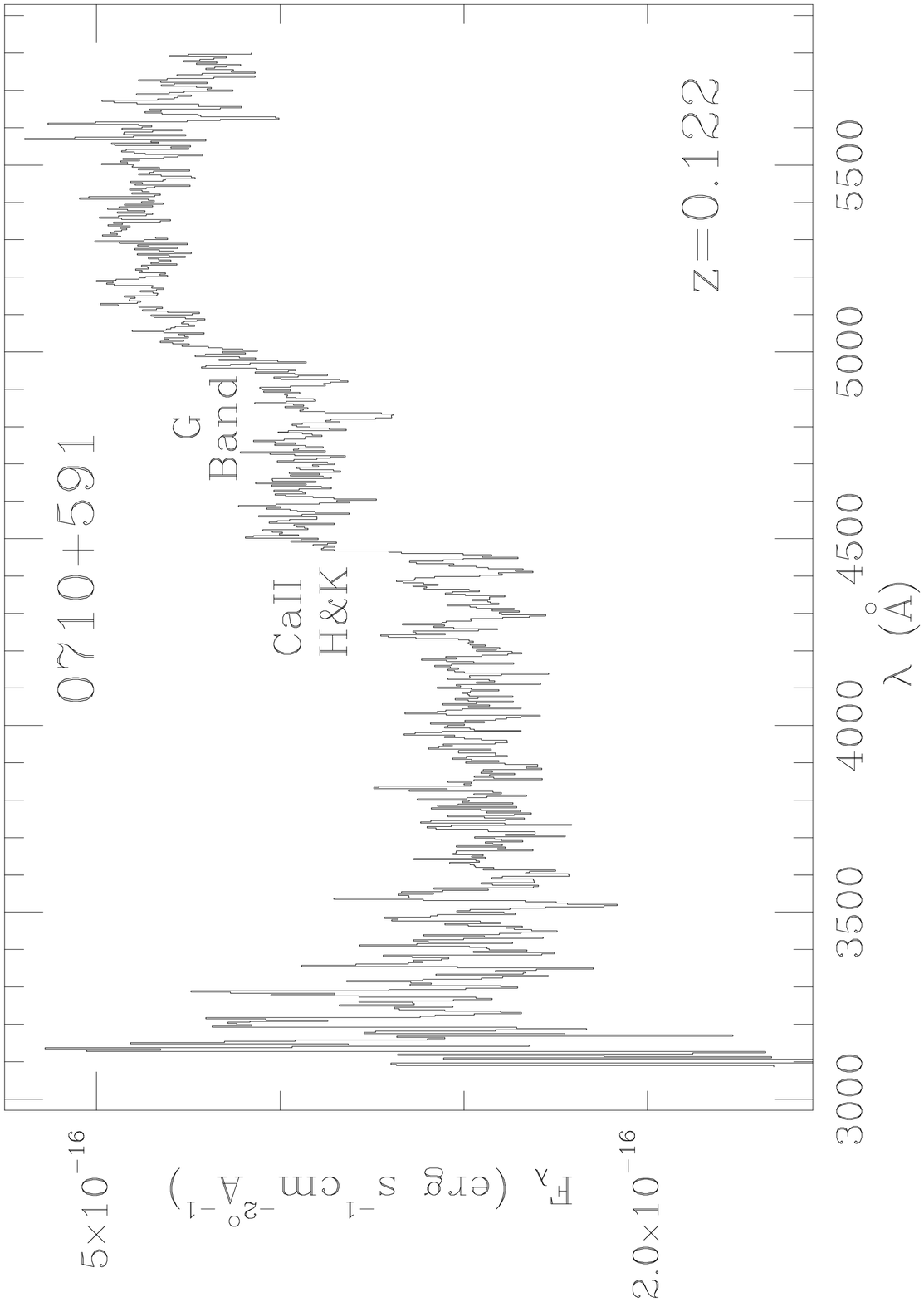,height=7.1cm,width=6.3cm}
\end{minipage}
\hspace{0.3in}
\begin{minipage}[t]{6.3in}
\psfig{file=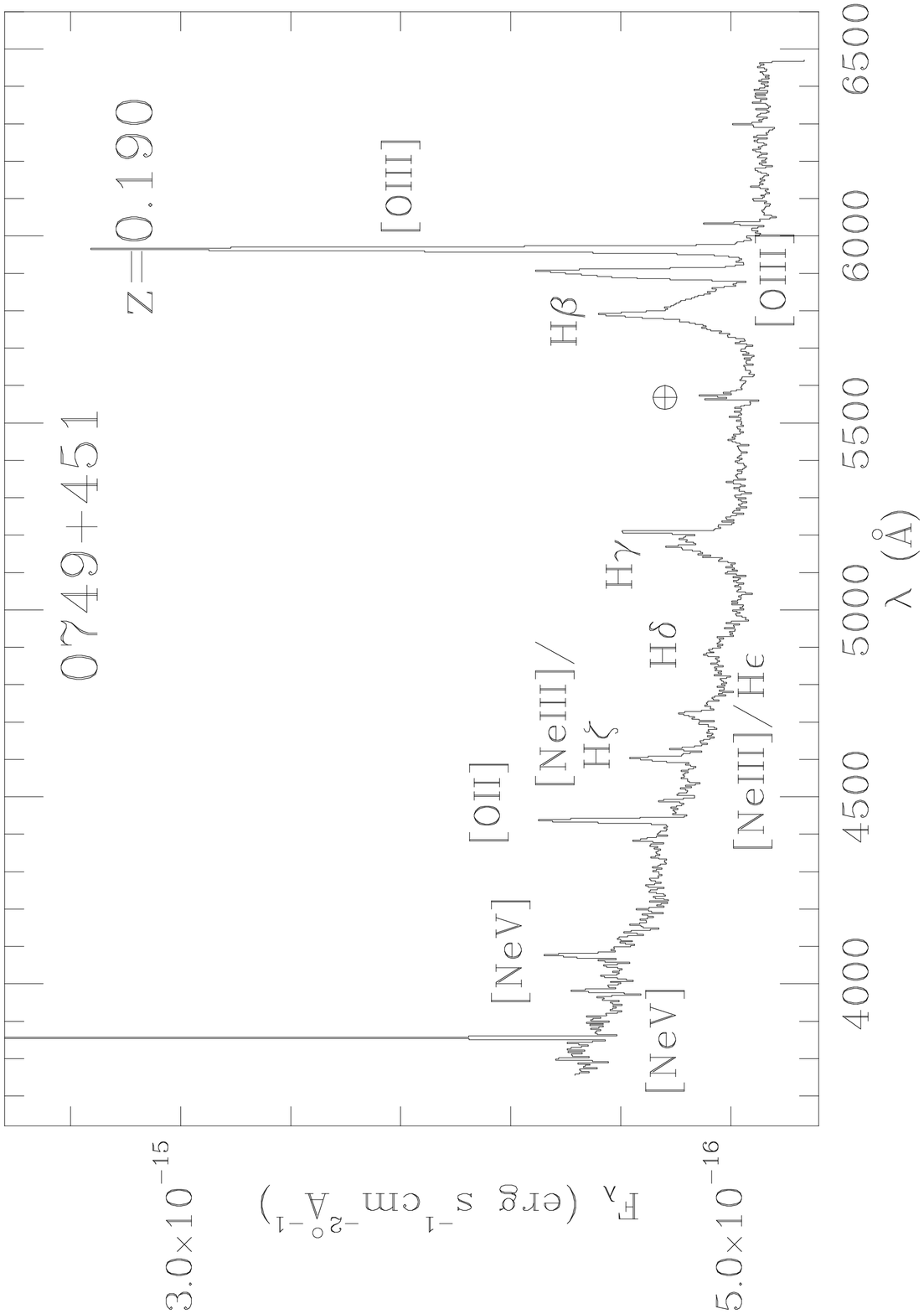,height=7.1cm,width=6.3cm}
\vspace{0.25in}
\psfig{file=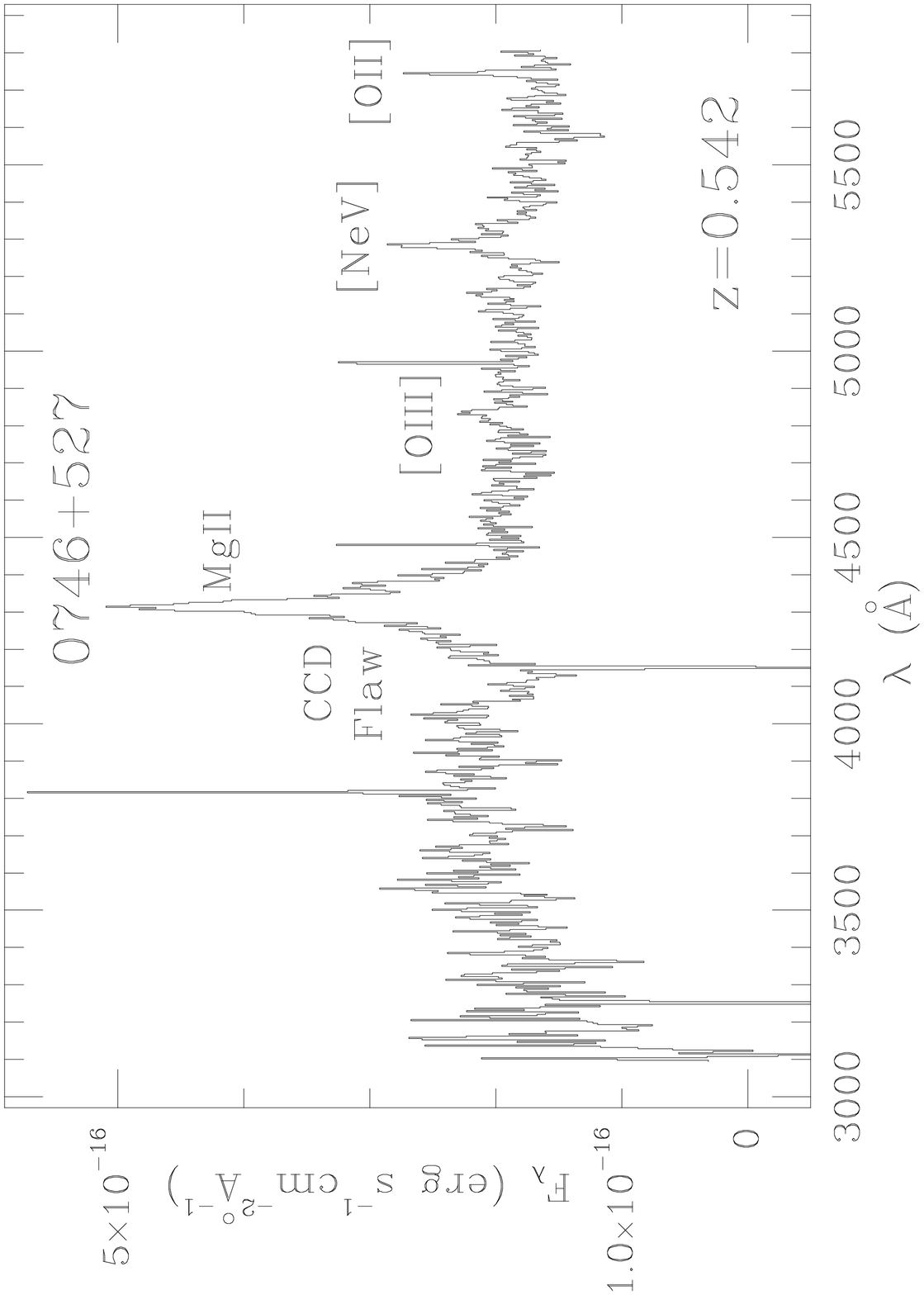,height=7.1cm,width=6.3cm}
\vspace{0.25in}
\psfig{file=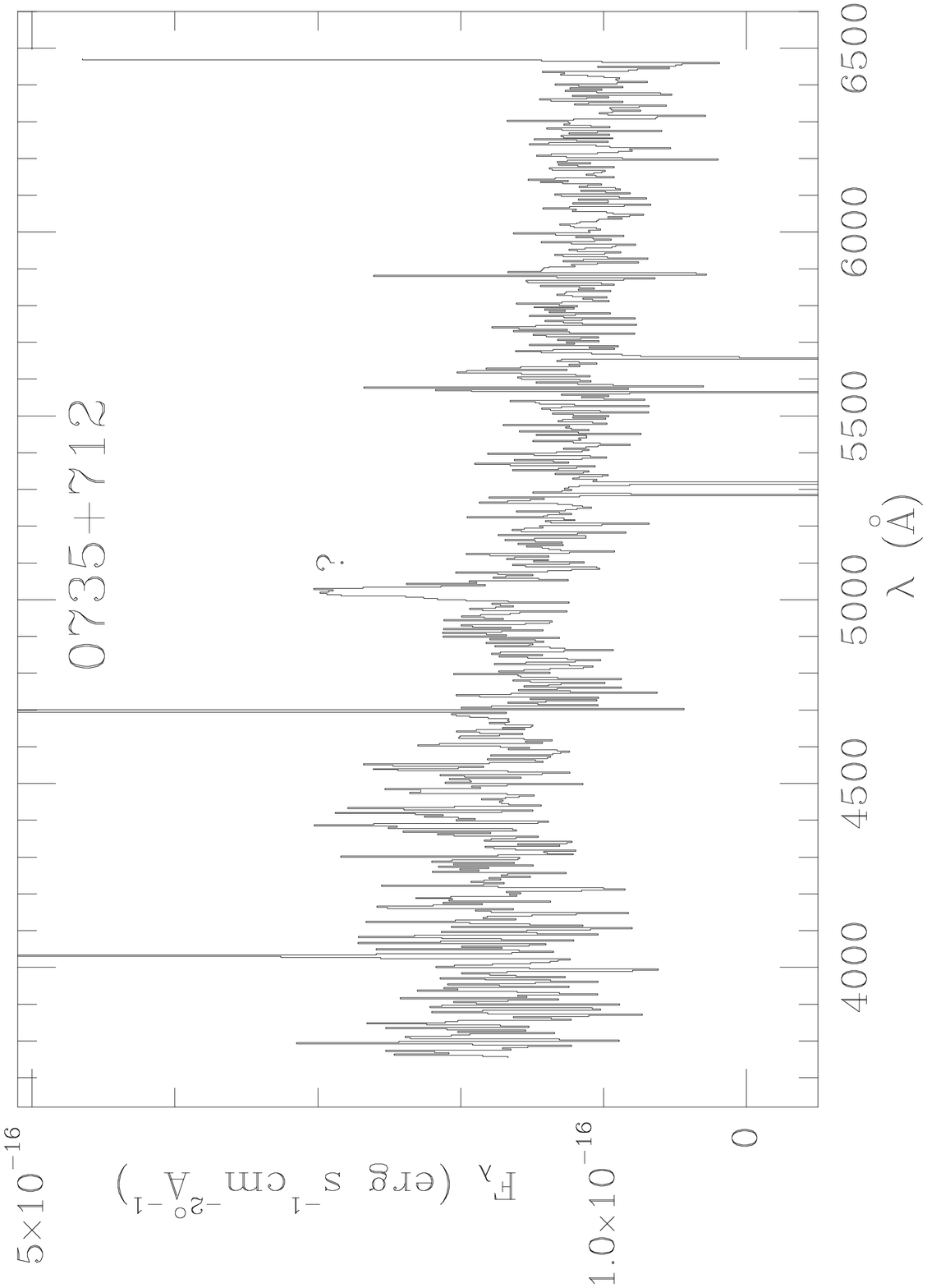,height=7.1cm,width=6.3cm}
\end{minipage}
\hfill
\begin{minipage}[t]{0.3in}
\vfill
\begin{sideways}
Figure 1.25 $-$ 1.30: Spectra of RGB Sources ({\it continued})
\end{sideways}
\vfill
\end{minipage}
\end{figure}
\clearpage

\begin{figure}
\vspace{-0.3in}
\hspace{-0.3in}
\begin{minipage}[t]{6.3in}
\psfig{file=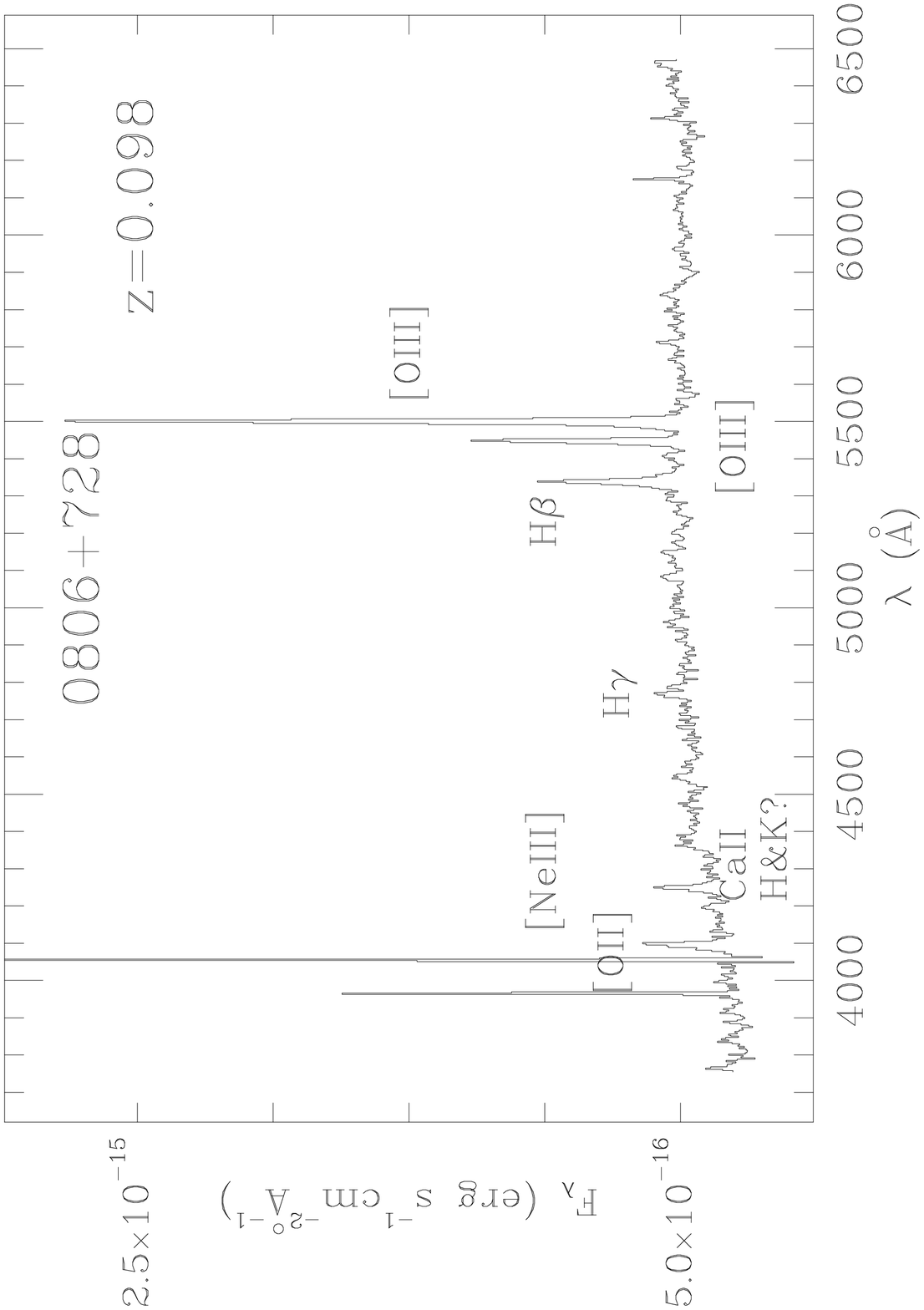,height=7.1cm,width=6.3cm}
\vspace{0.25in}
\psfig{file=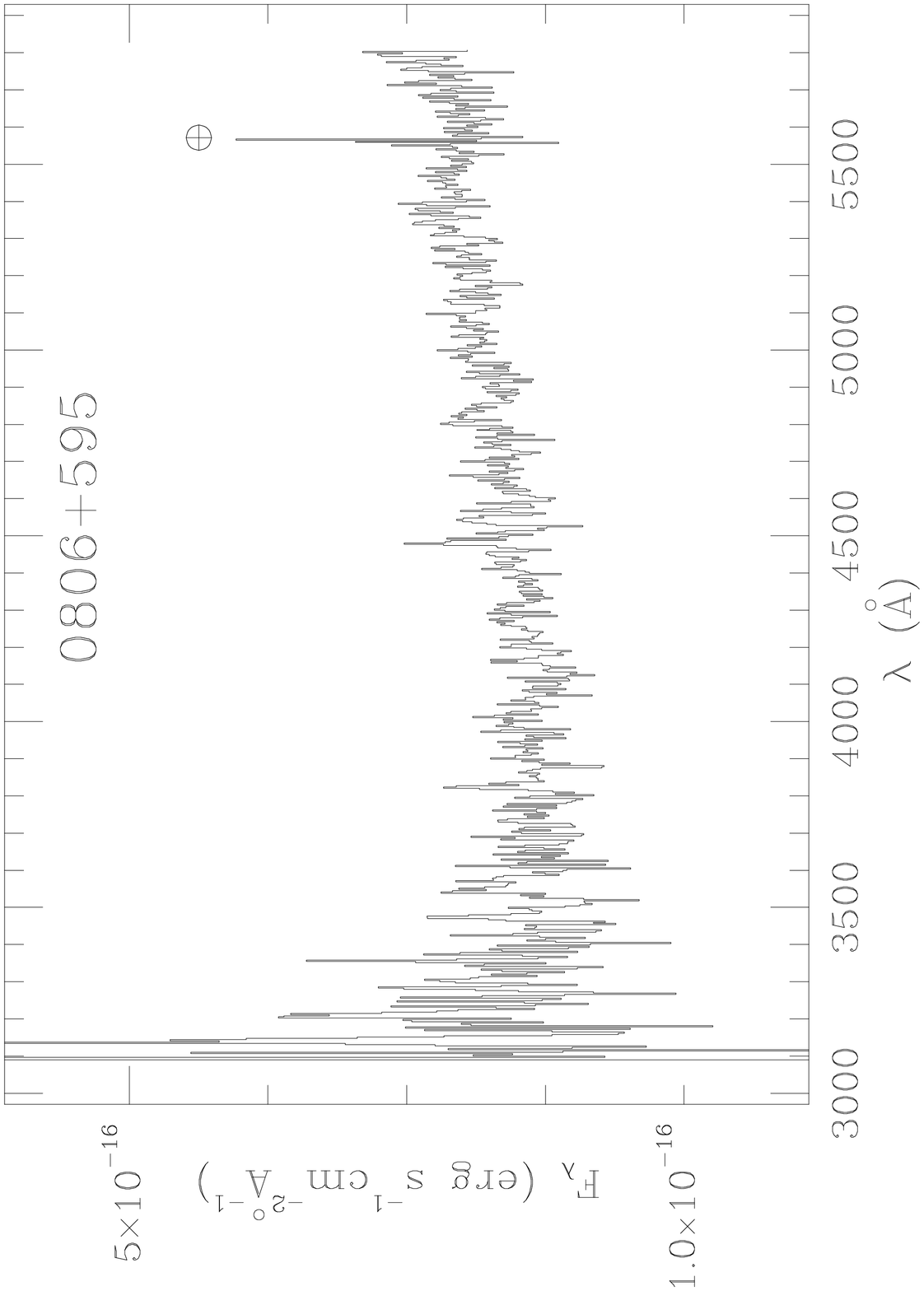,height=7.1cm,width=6.3cm}
\vspace{0.25in}
\psfig{file=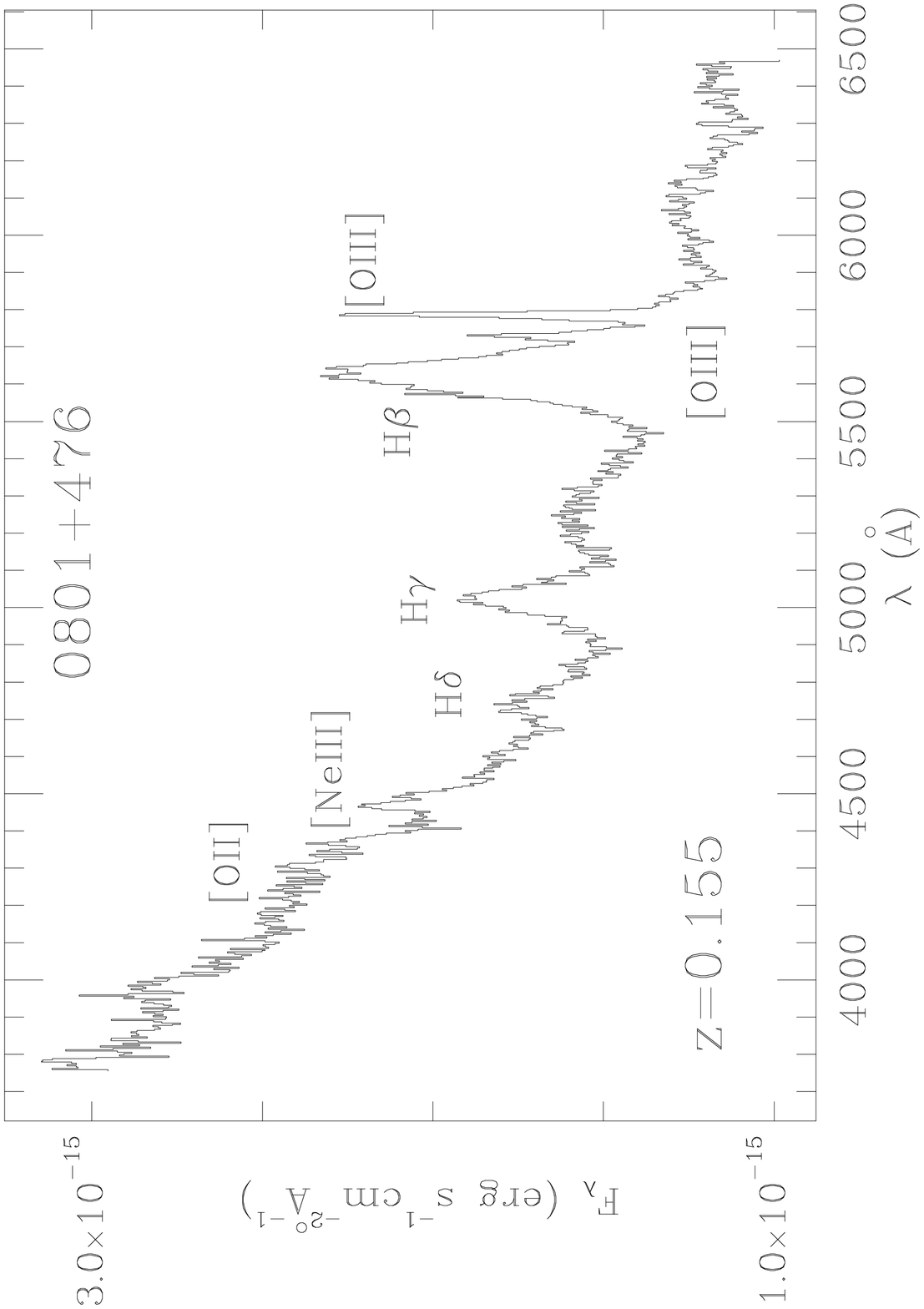,height=7.1cm,width=6.3cm}
\end{minipage}
\hspace{0.3in}
\begin{minipage}[t]{6.3in}
\psfig{file=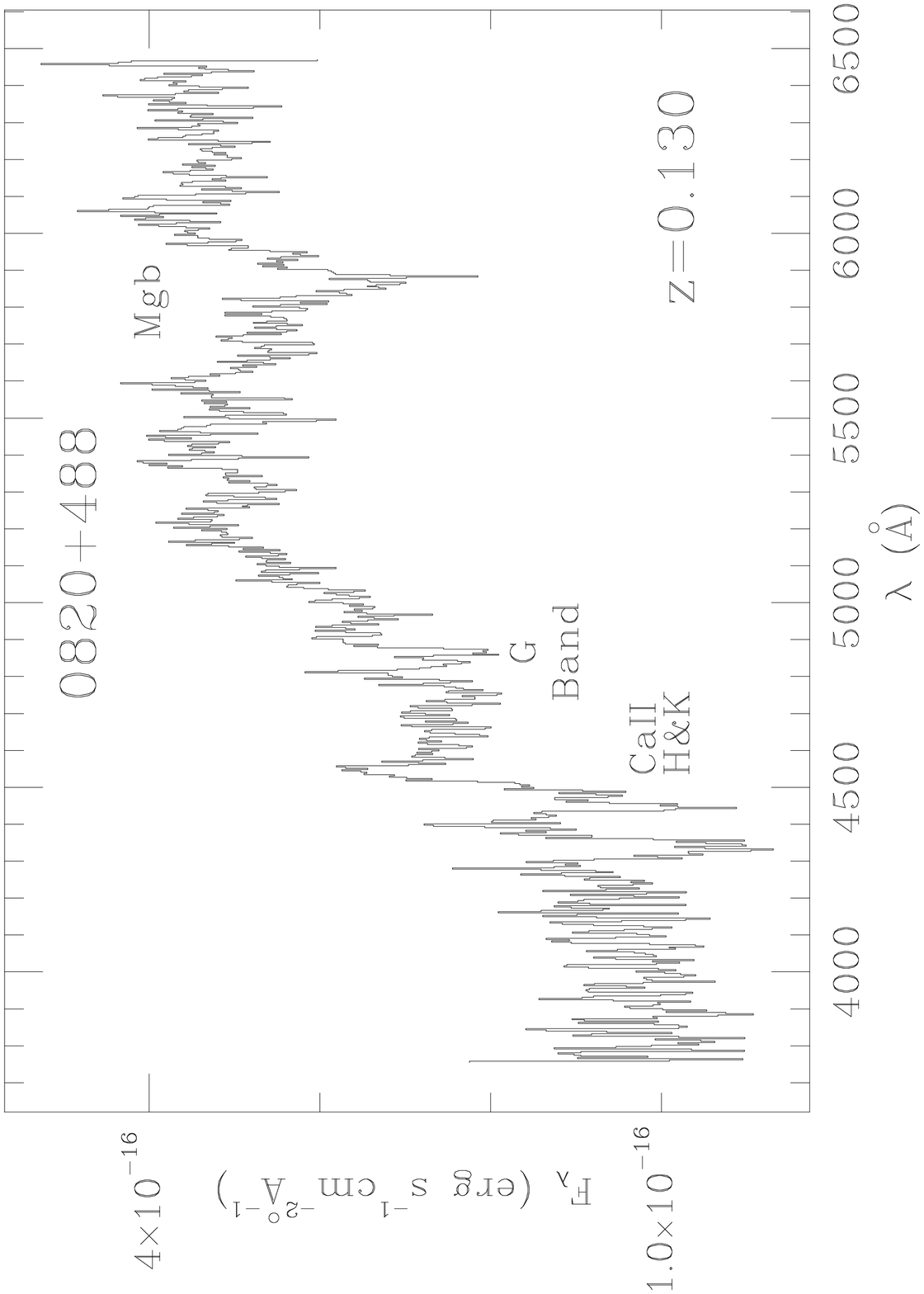,height=7.1cm,width=6.3cm}
\vspace{0.25in}
\psfig{file=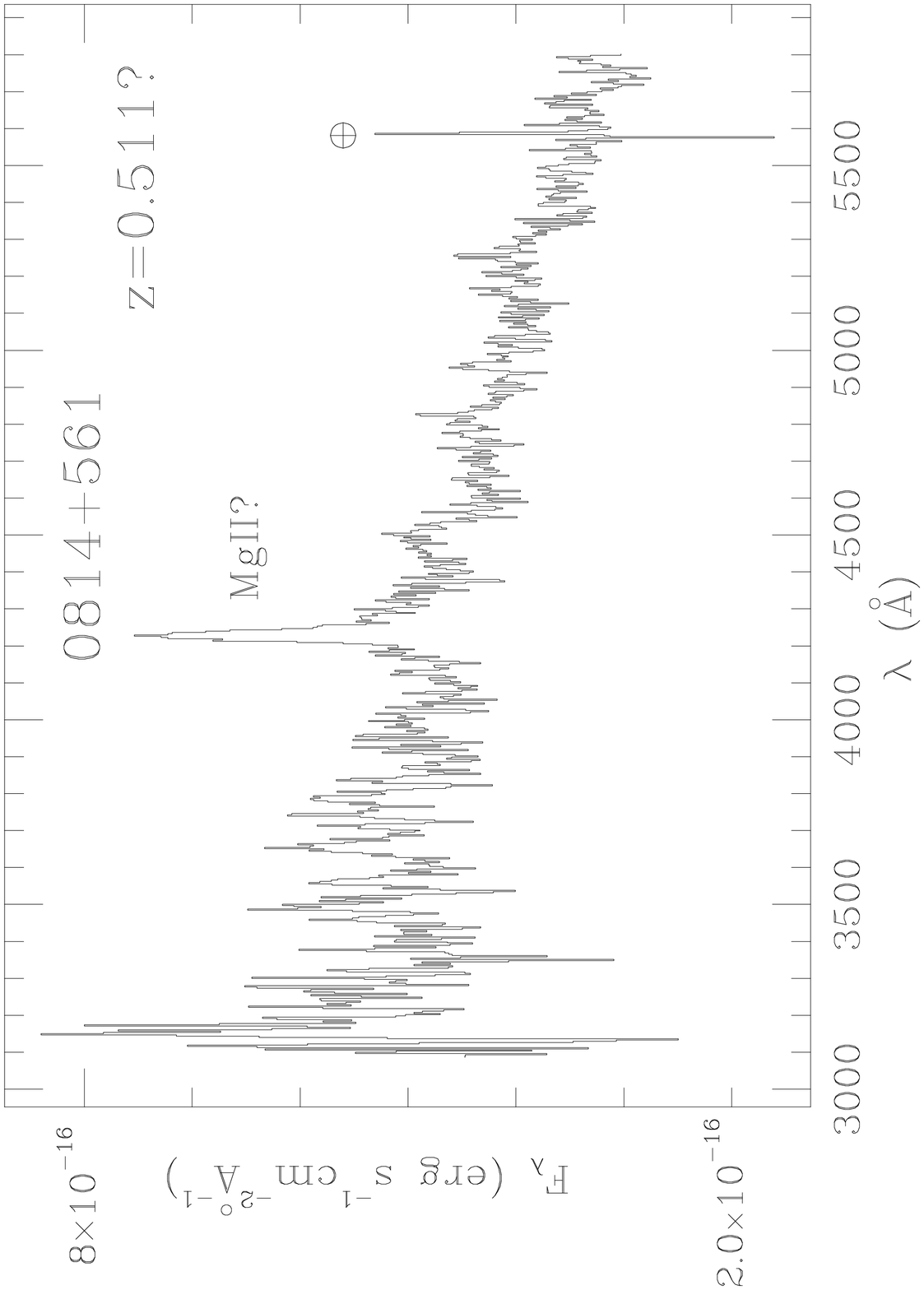,height=7.1cm,width=6.3cm}
\vspace{0.25in}
\psfig{file=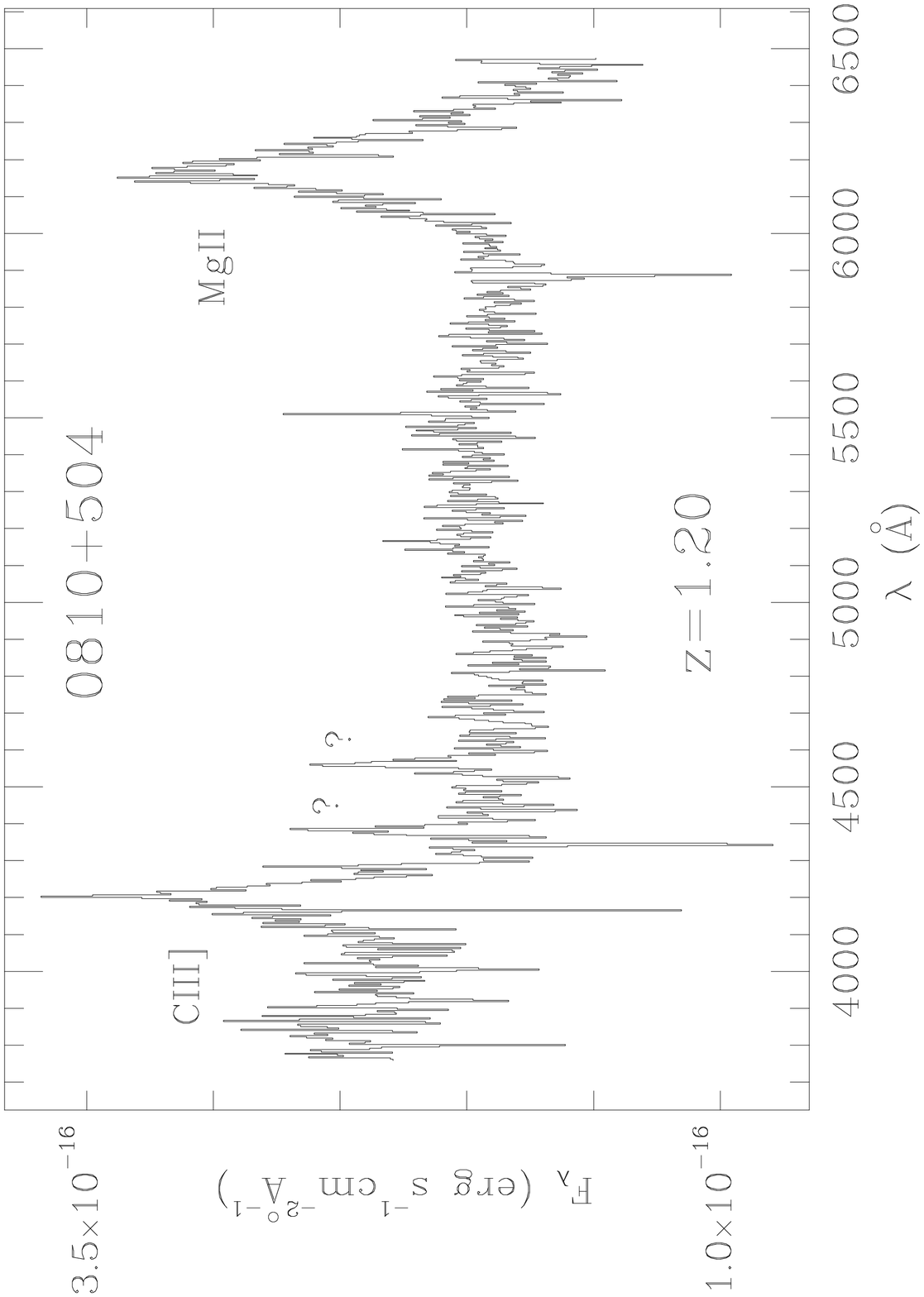,height=7.1cm,width=6.3cm}
\end{minipage}
\hfill
\begin{minipage}[t]{0.3in}
\vfill
\begin{sideways}
Figure 1.31 $-$ 1.36: Spectra of RGB Sources ({\it continued})
\end{sideways}
\vfill
\end{minipage}
\end{figure}
\clearpage
\begin{figure}
\vspace{-0.3in}
\hspace{-0.3in}
\begin{minipage}[t]{6.3in}
\psfig{file=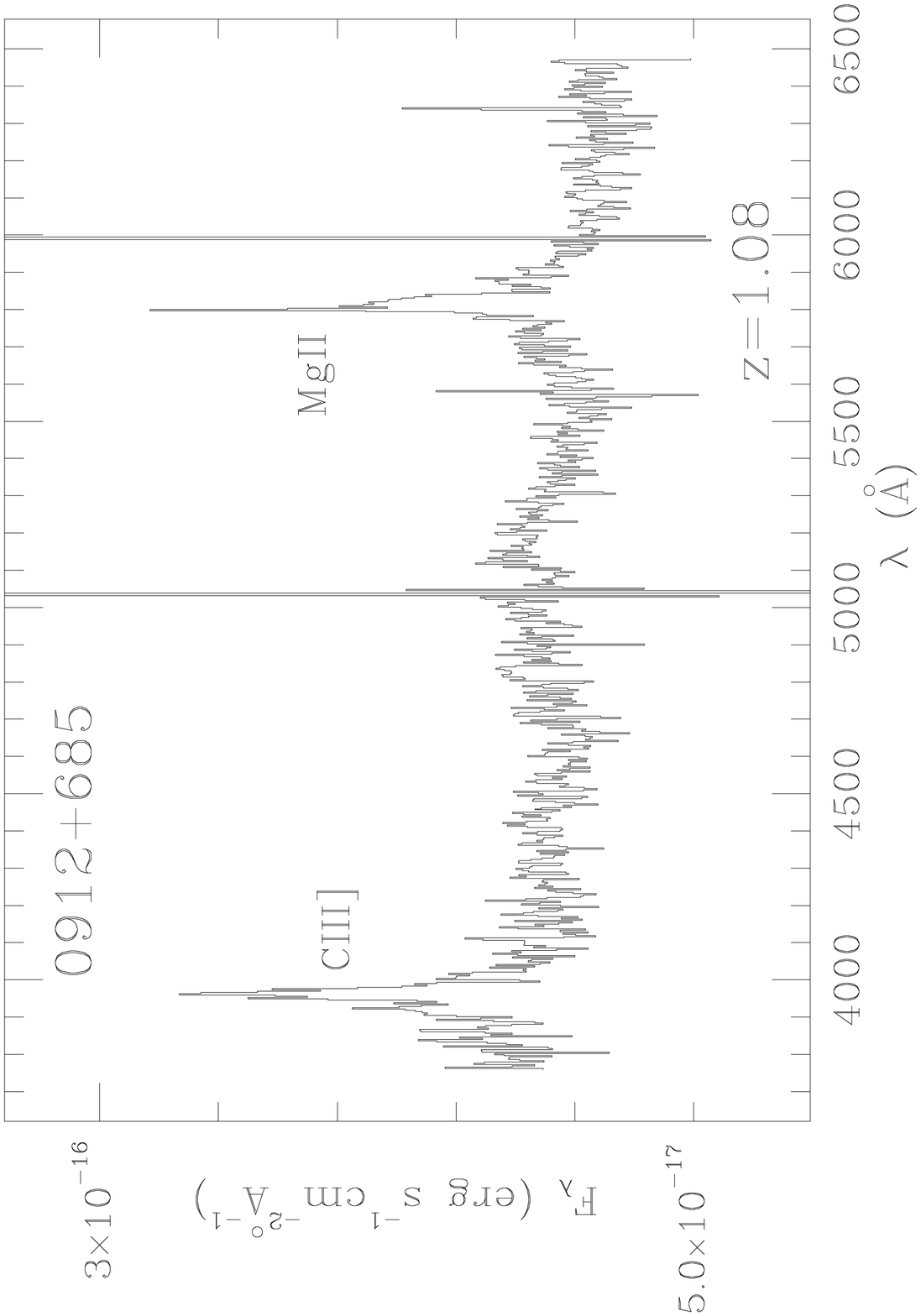,height=7.1cm,width=6.3cm}
\vspace{0.25in}
\psfig{file=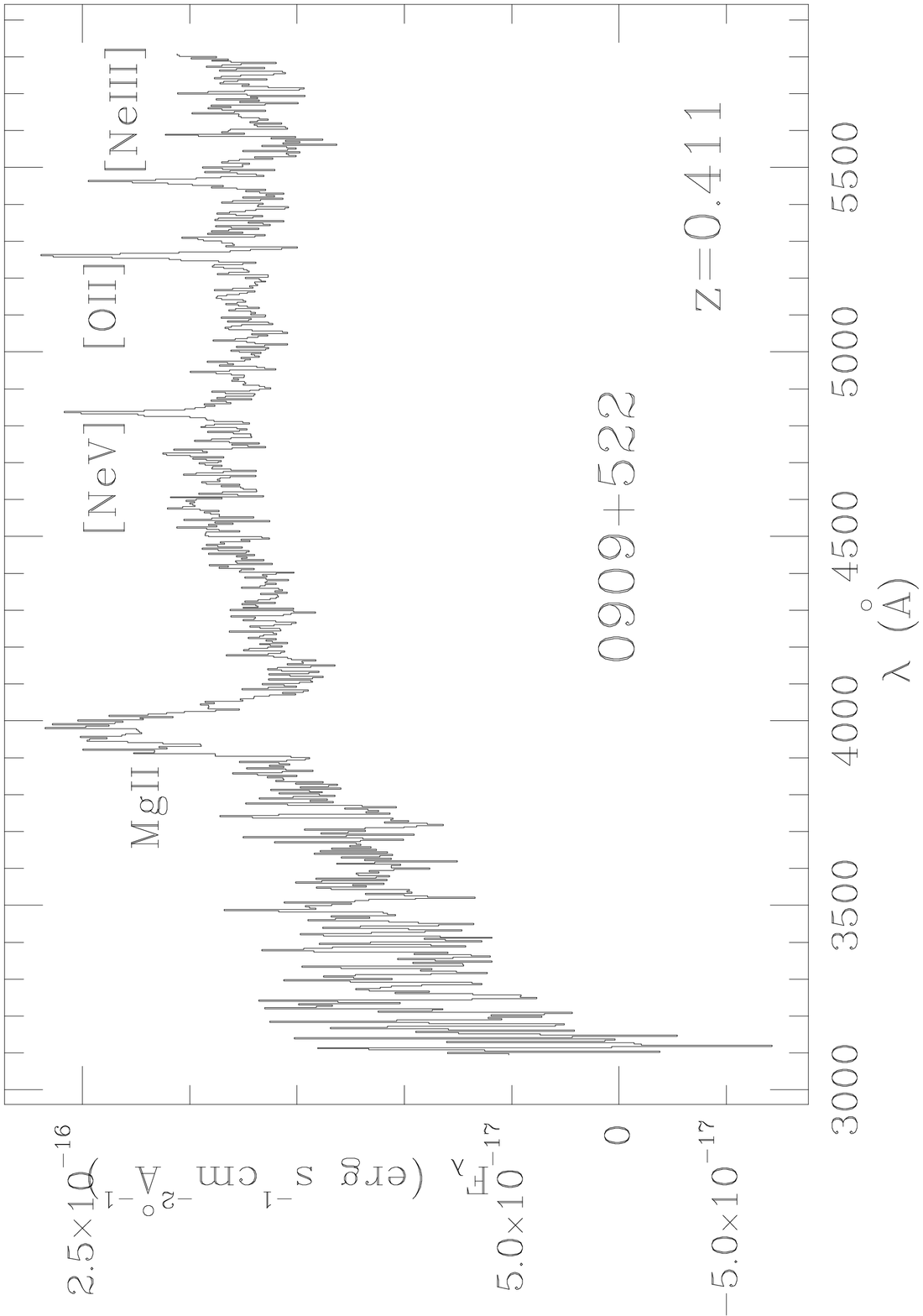,height=7.1cm,width=6.3cm}
\vspace{0.25in}
\psfig{file=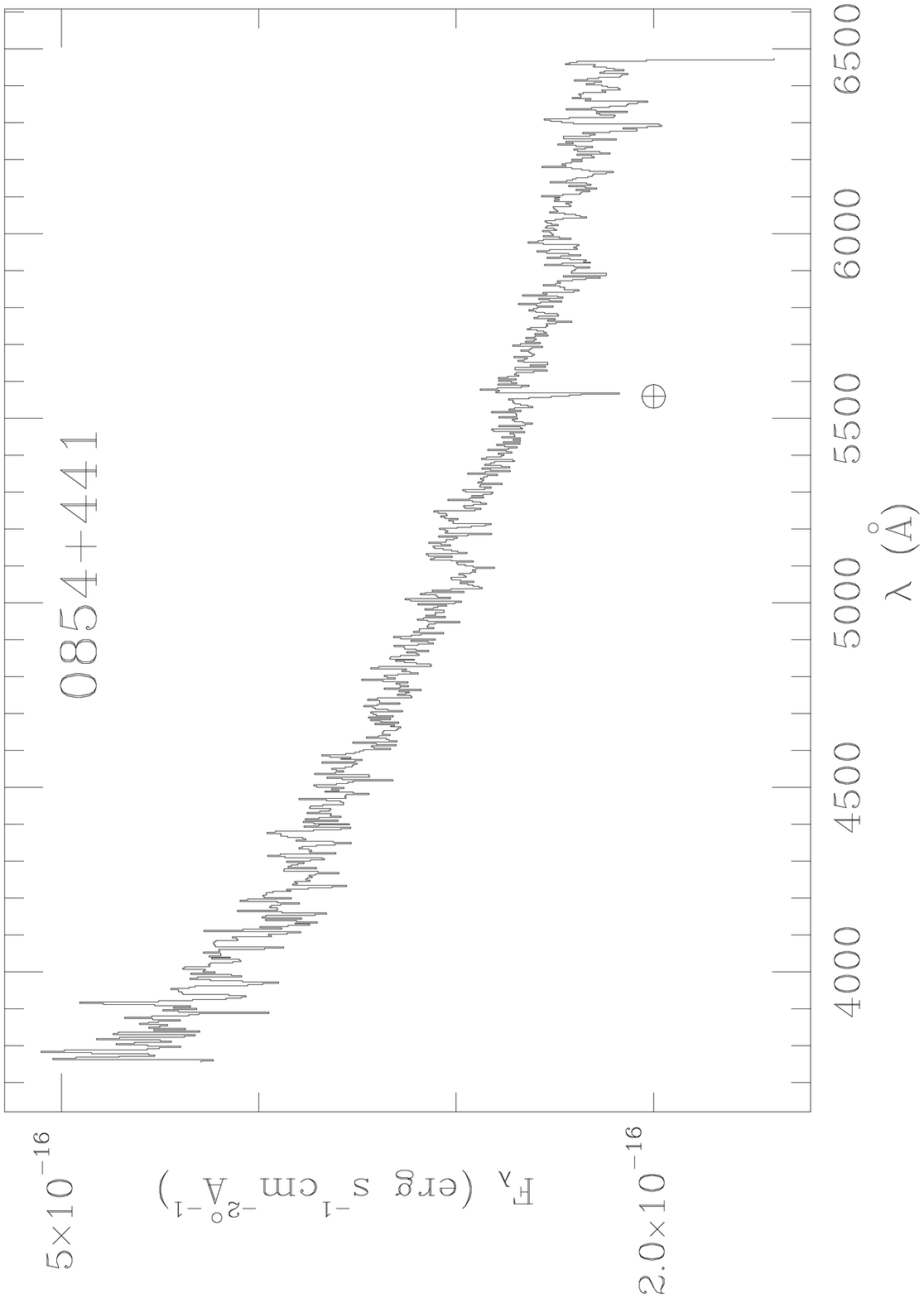,height=7.1cm,width=6.3cm}
\end{minipage}
\hspace{0.3in}
\begin{minipage}[t]{6.3in}
\psfig{file=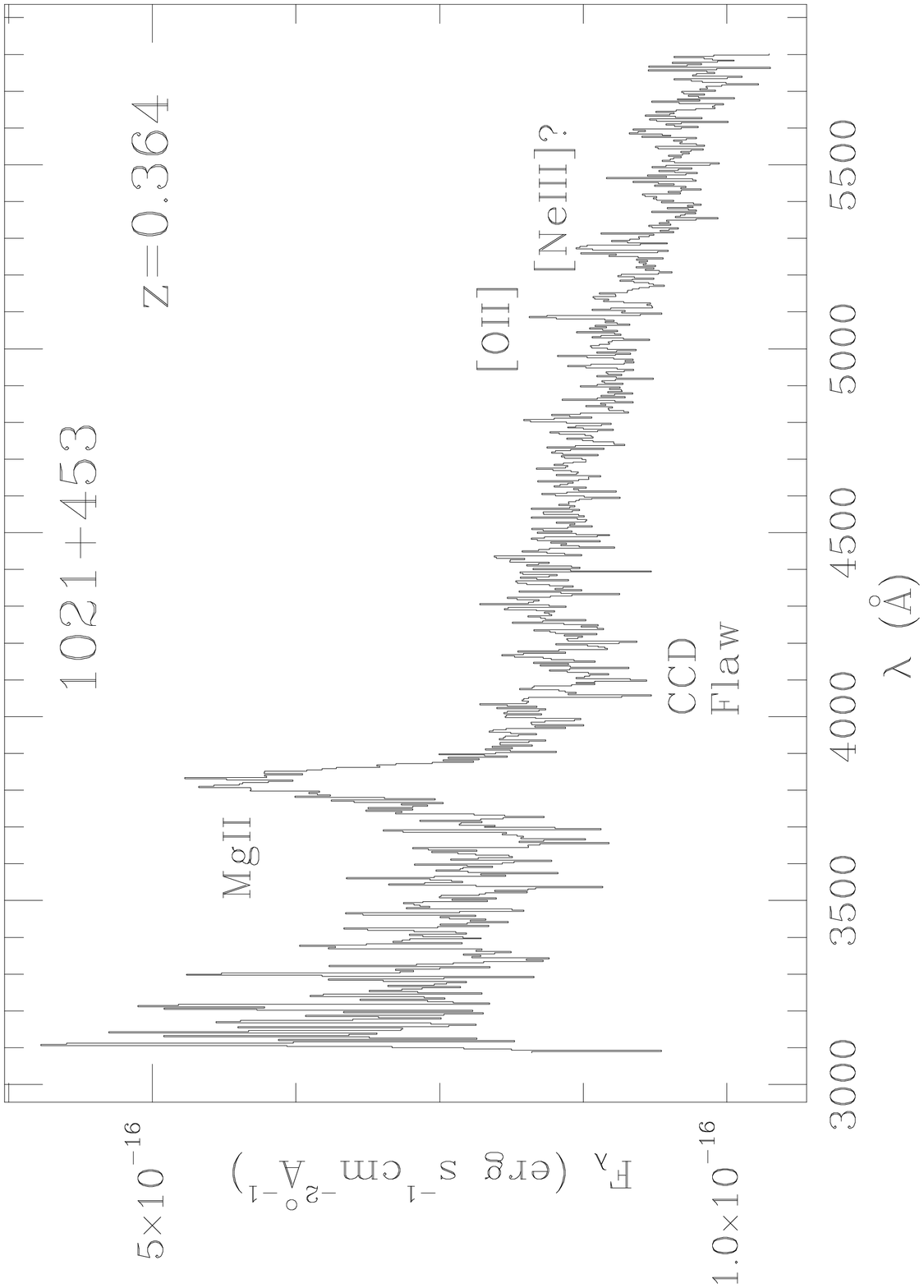,height=7.1cm,width=6.3cm}
\vspace{0.25in}
\psfig{file=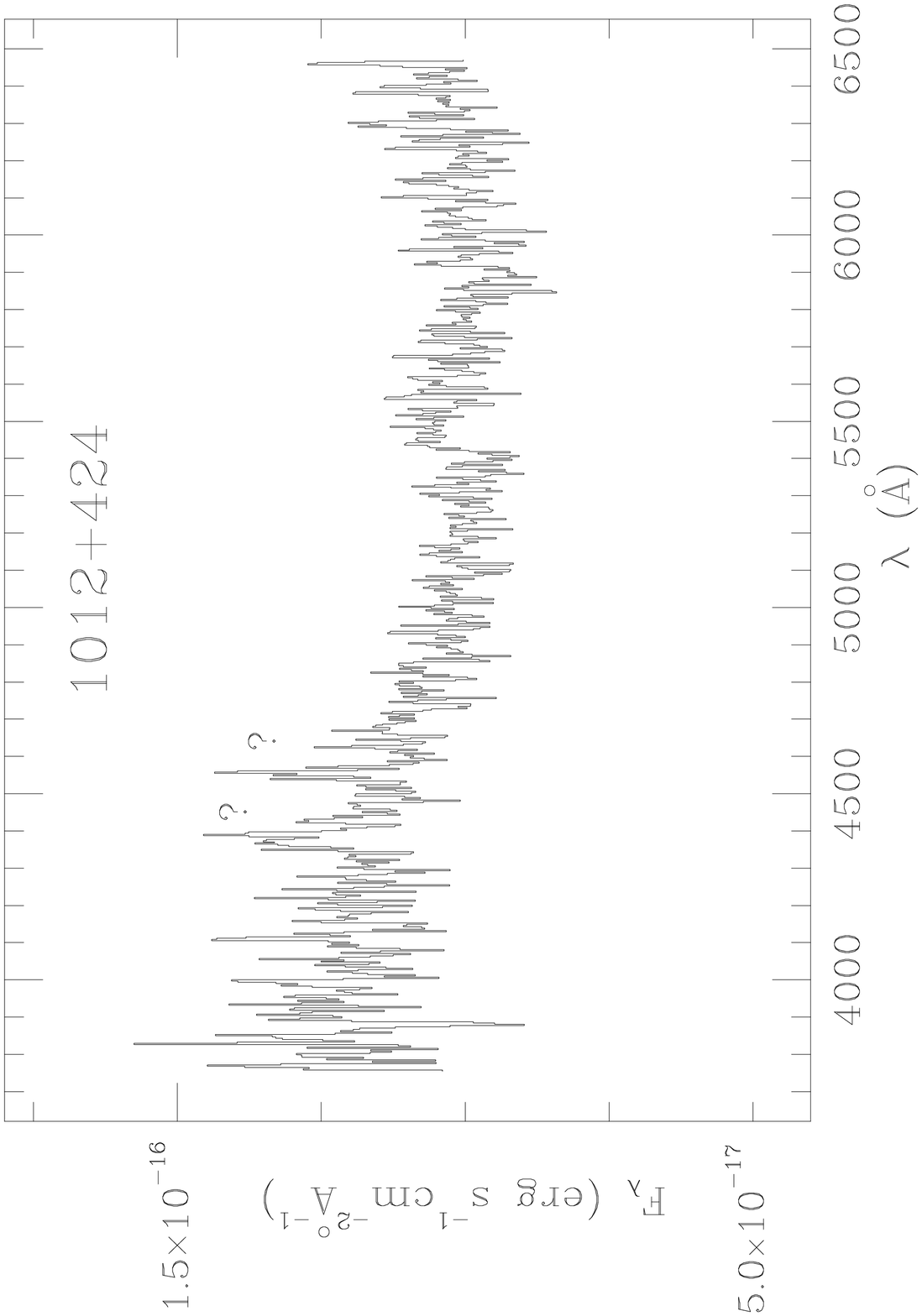,height=7.1cm,width=6.3cm}
\vspace{0.25in}
\psfig{file=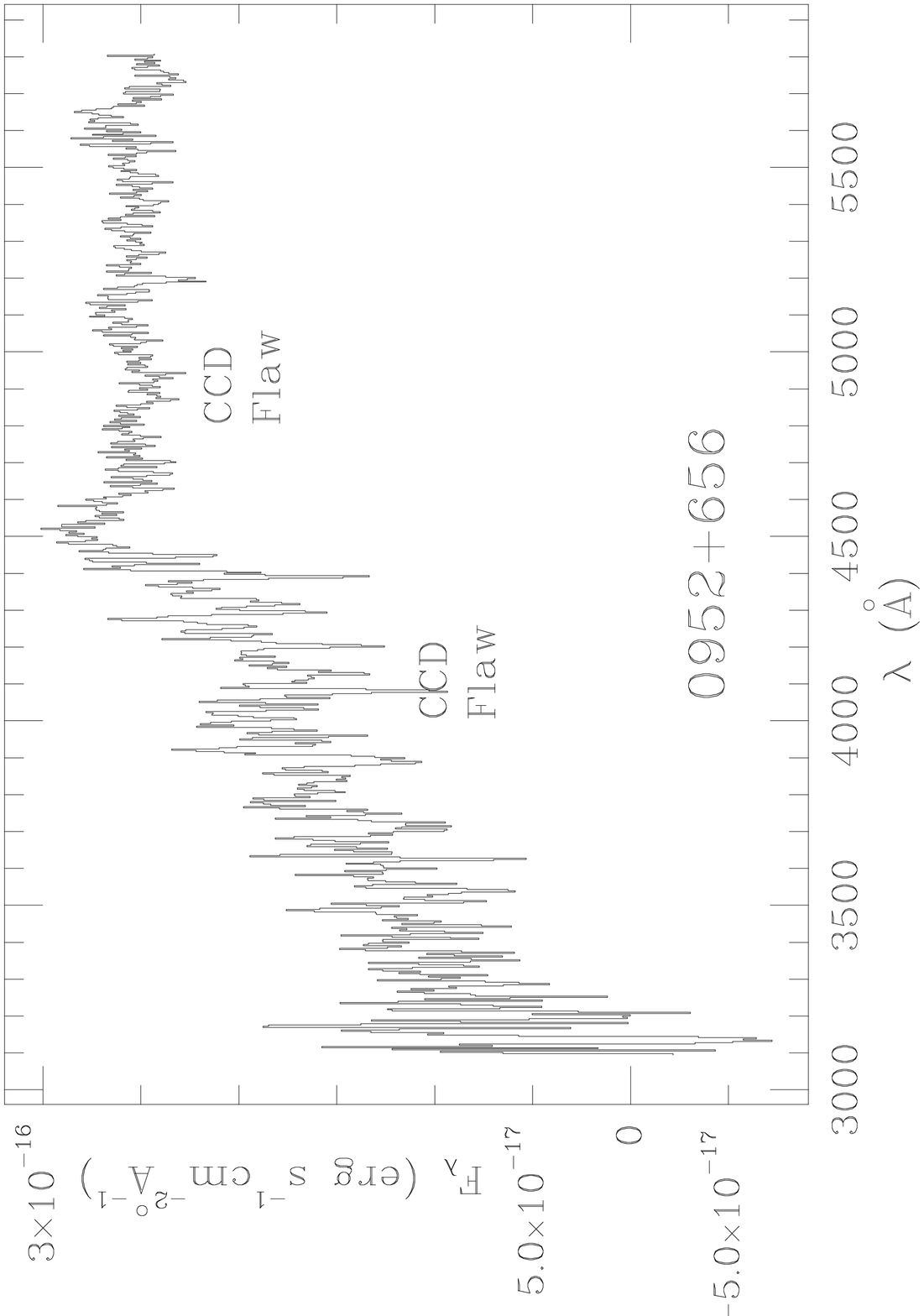,height=7.1cm,width=6.3cm}
\end{minipage}
\hfill
\begin{minipage}[t]{0.3in}
\vfill
\begin{sideways}
Figure 1.37 $-$ 1.42: Spectra of RGB Sources ({\it continued})
\end{sideways}
\vfill
\end{minipage}
\end{figure}
\clearpage
\begin{figure}
\vspace{-0.3in}
\hspace{-0.3in}
\begin{minipage}[t]{6.3in}
\psfig{file=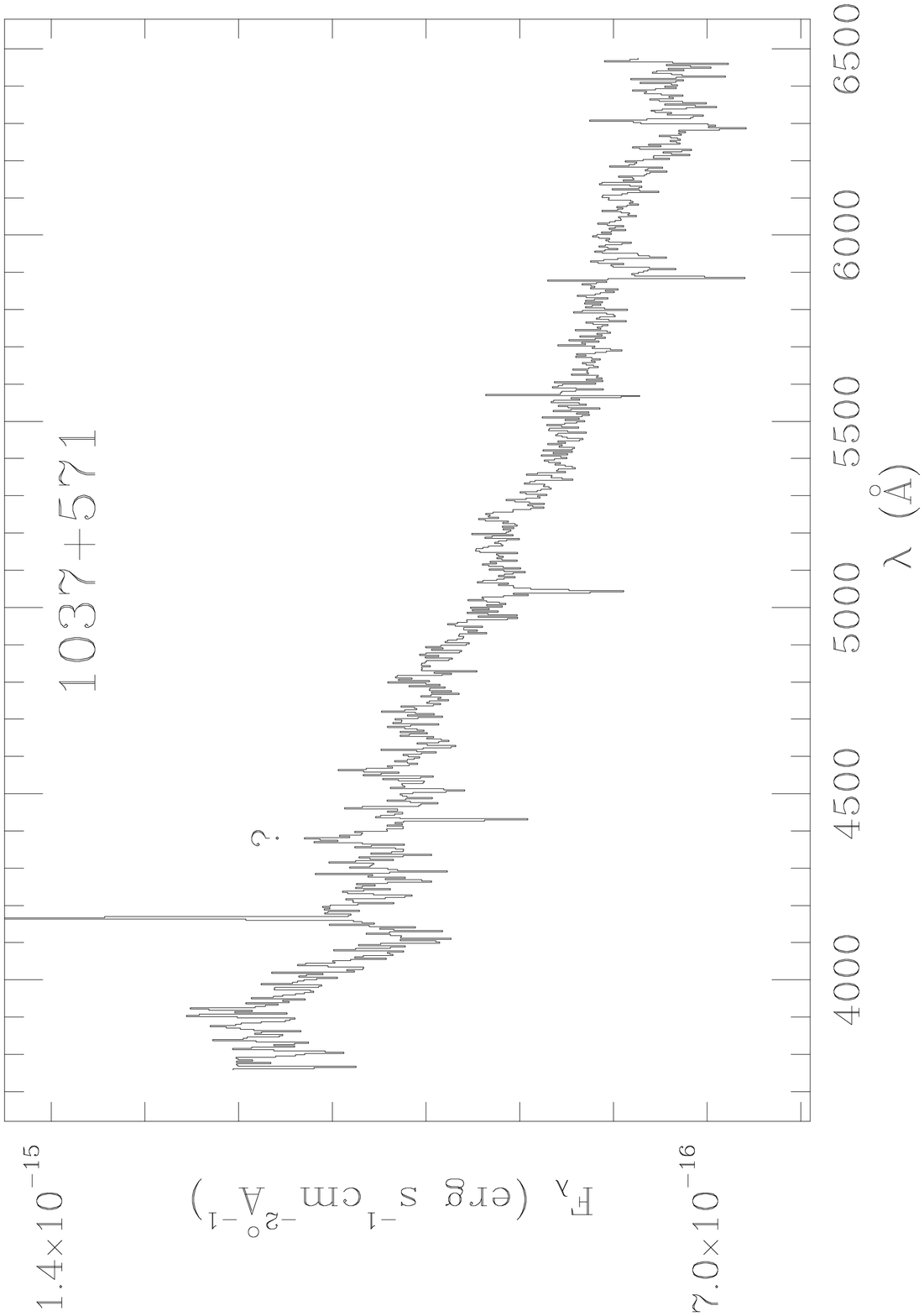,height=7.1cm,width=6.3cm}
\vspace{0.25in}
\psfig{file=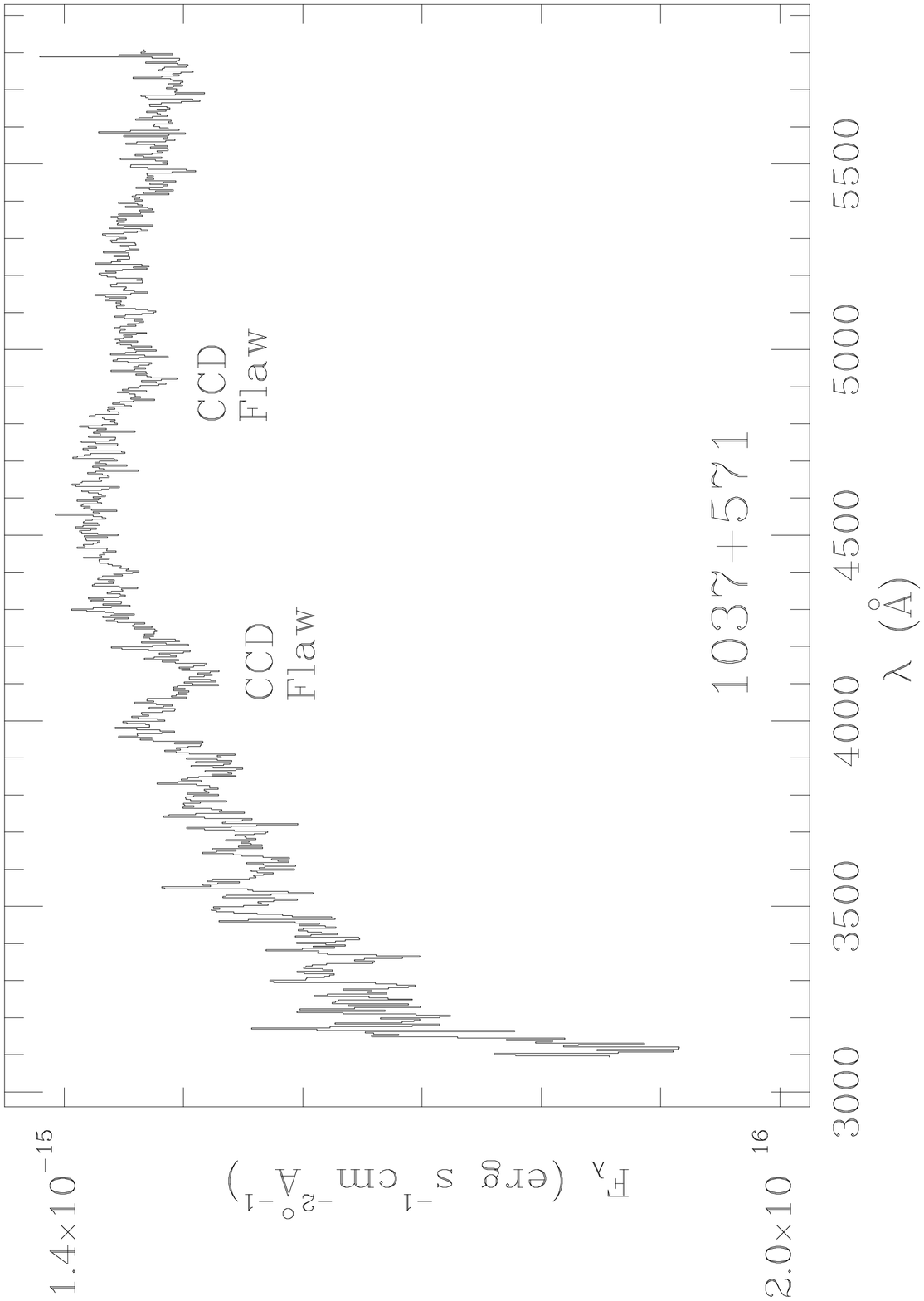,height=7.1cm,width=6.3cm}
\vspace{0.25in}
\psfig{file=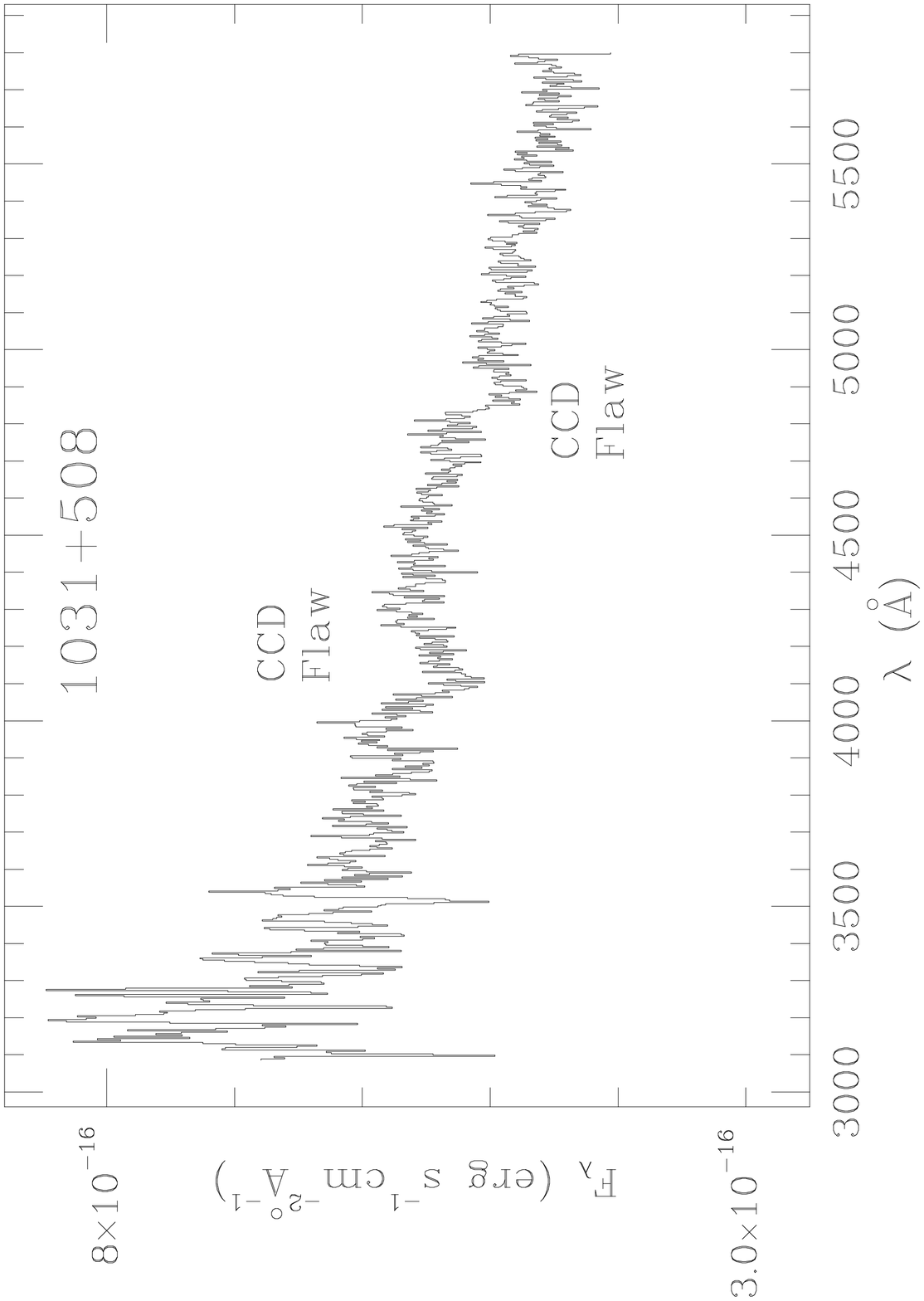,height=7.1cm,width=6.3cm}
\end{minipage}
\hspace{0.3in}
\begin{minipage}[t]{6.3in}
\psfig{file=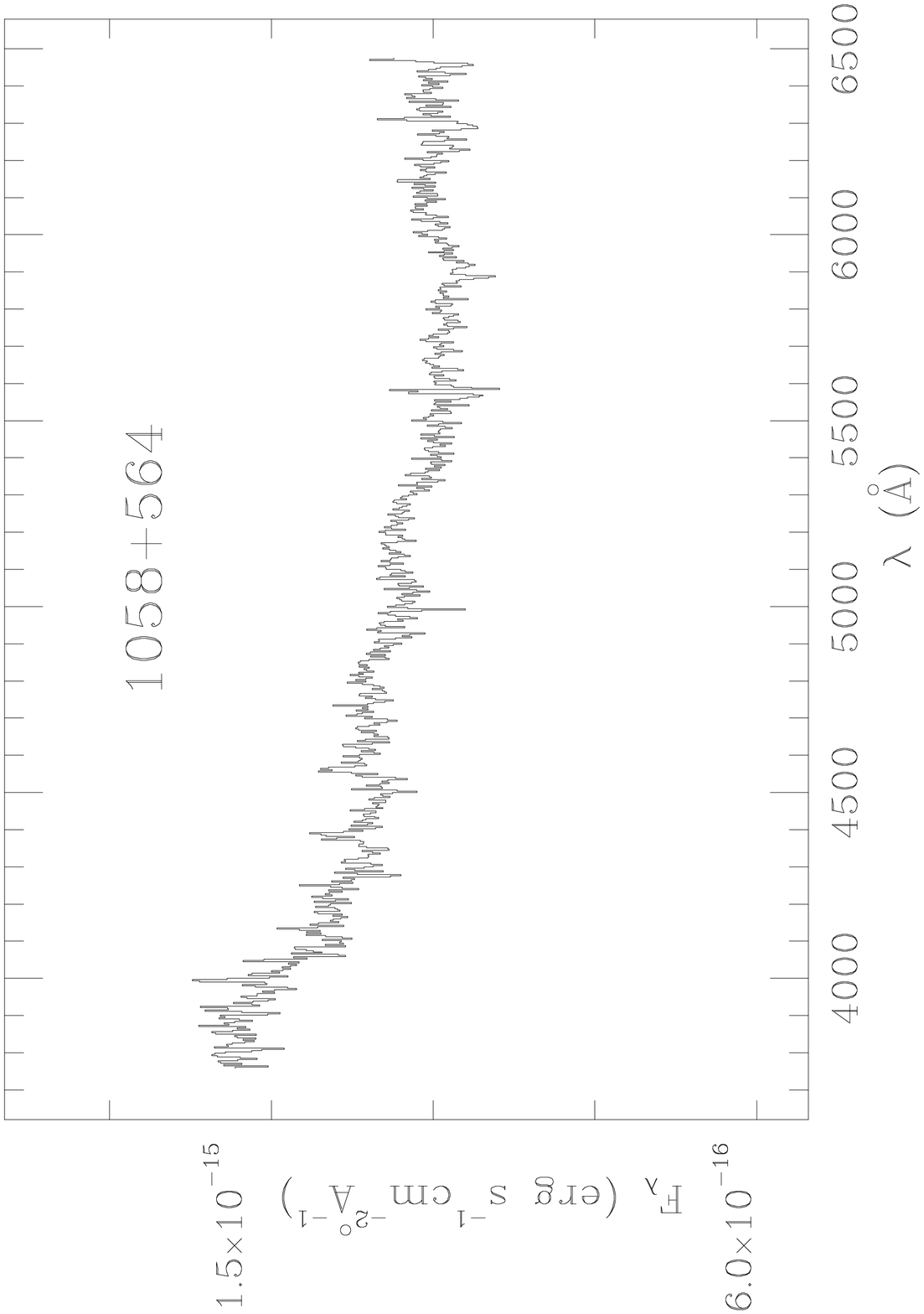,height=7.1cm,width=6.3cm}
\vspace{0.25in}
\psfig{file=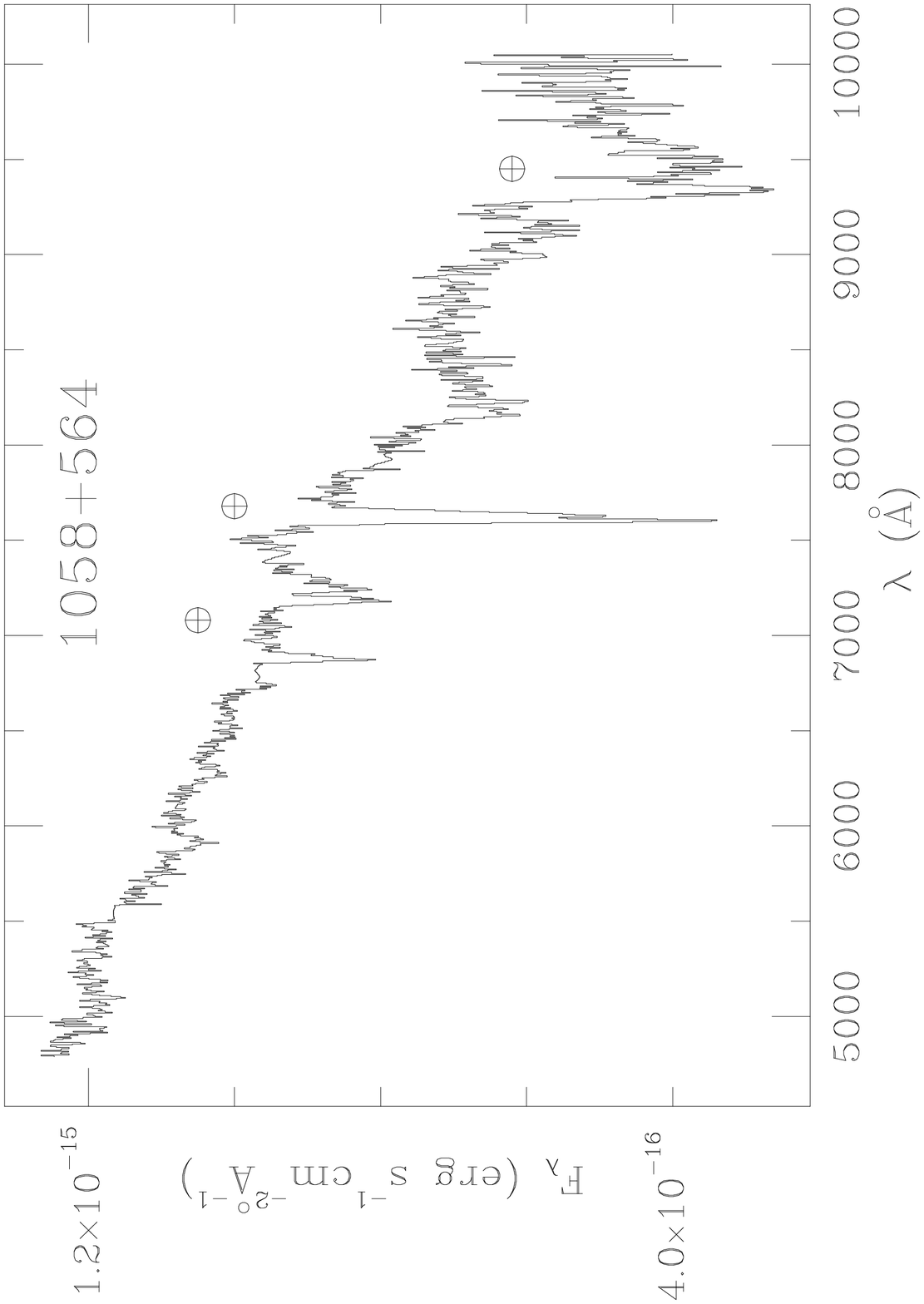,height=7.1cm,width=6.3cm}
\vspace{0.25in}
\psfig{file=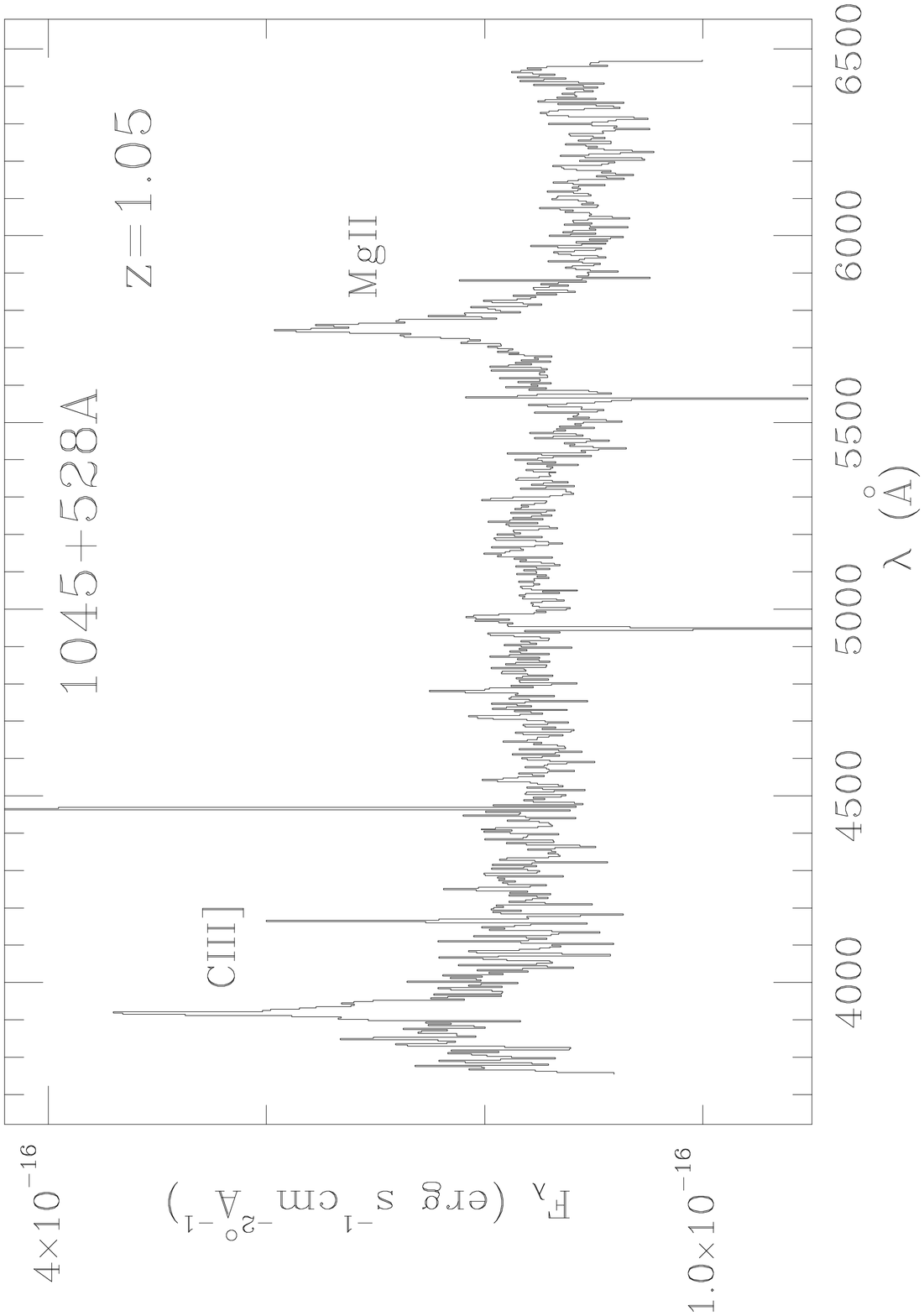,height=7.1cm,width=6.3cm}
\end{minipage}
\hfill
\begin{minipage}[t]{0.3in}
\vfill
\begin{sideways}
Figure 1.43 $-$ 1.48: Spectra of RGB Sources ({\it continued})
\end{sideways}
\vfill
\end{minipage}
\end{figure}
\clearpage

\begin{figure}
\vspace{-0.3in}
\hspace{-0.3in}
\begin{minipage}[t]{6.3in}
\psfig{file=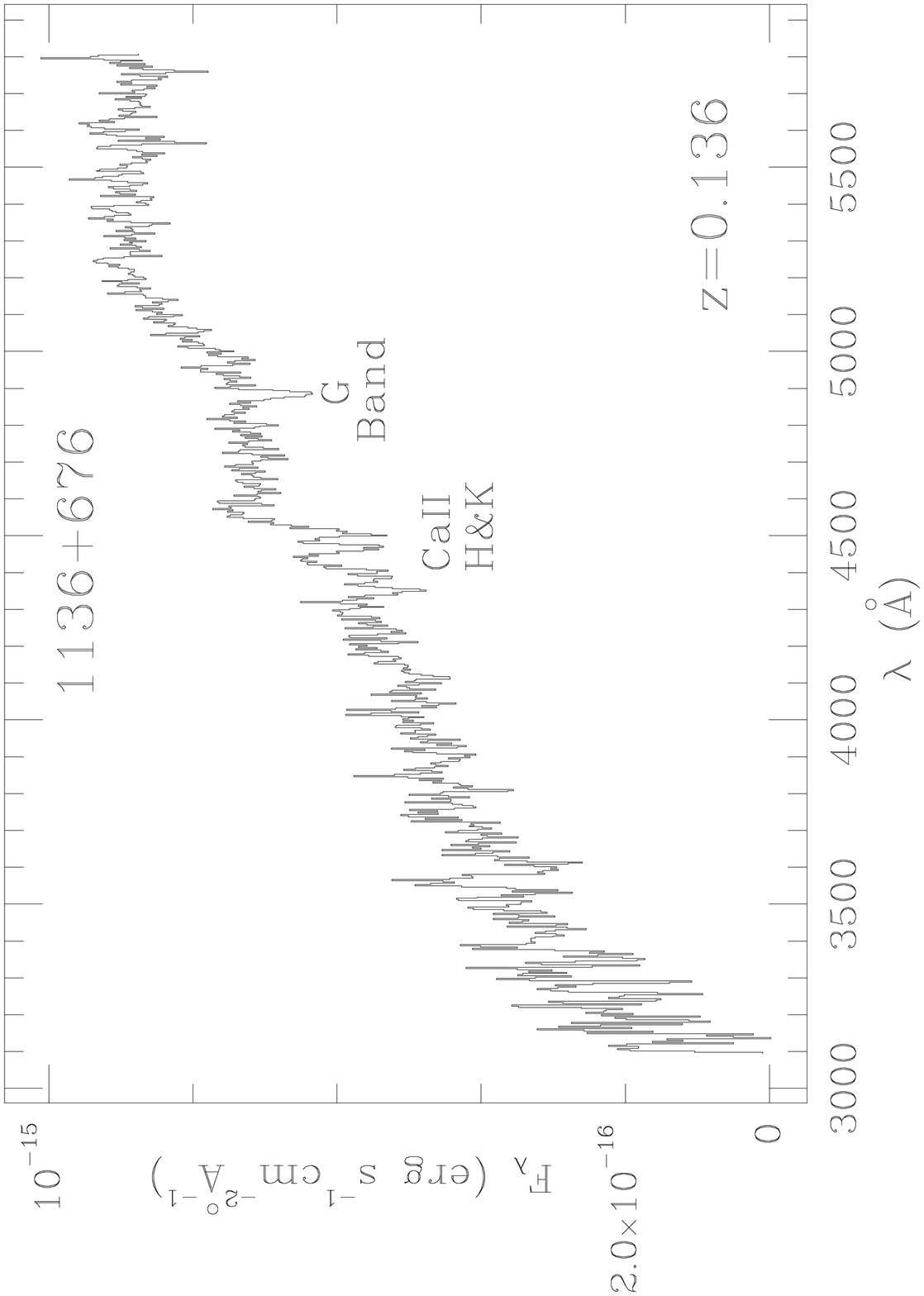,height=7.1cm,width=6.3cm}
\vspace{0.25in}
\psfig{file=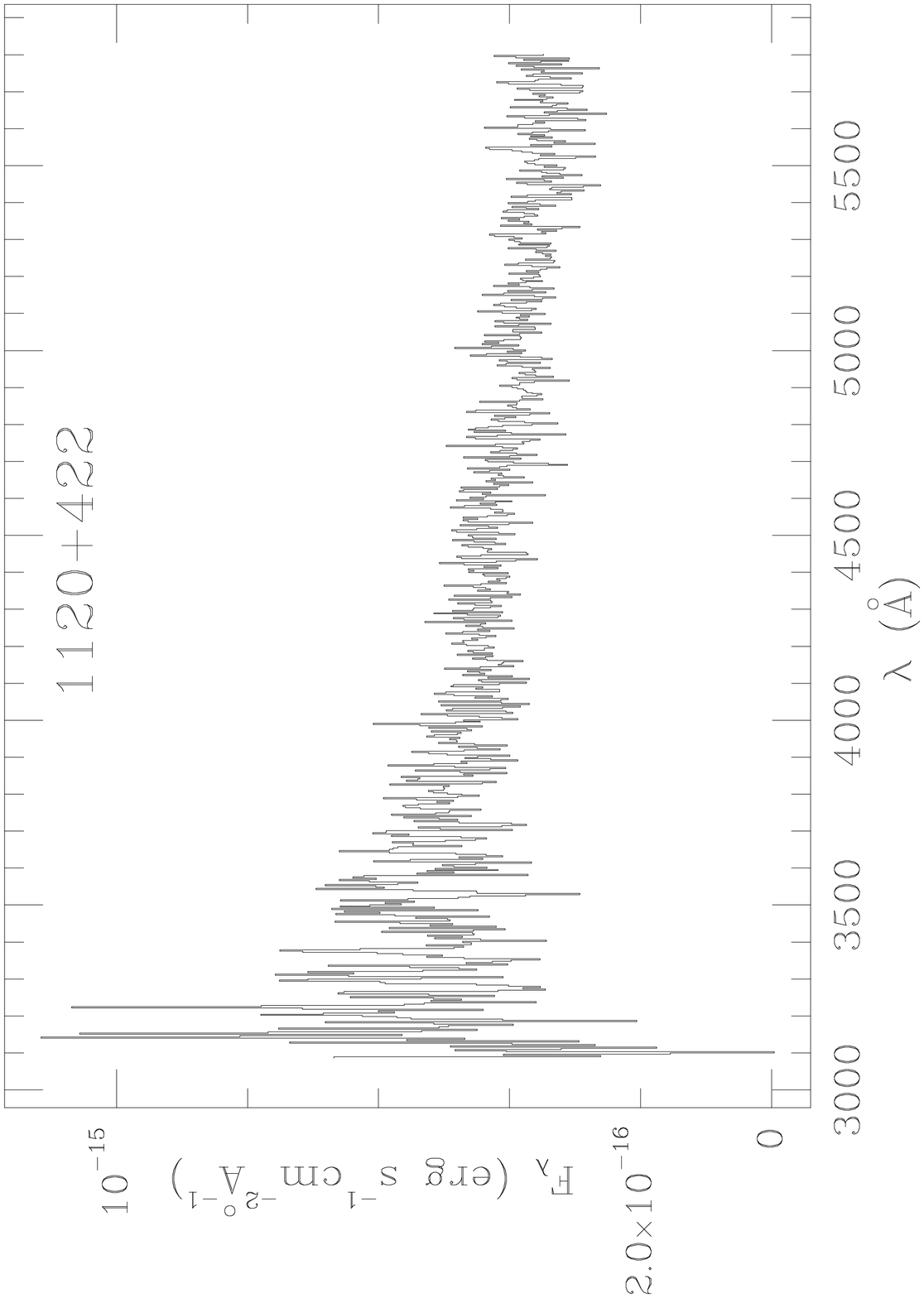,height=7.1cm,width=6.3cm}
\vspace{0.25in}
\psfig{file=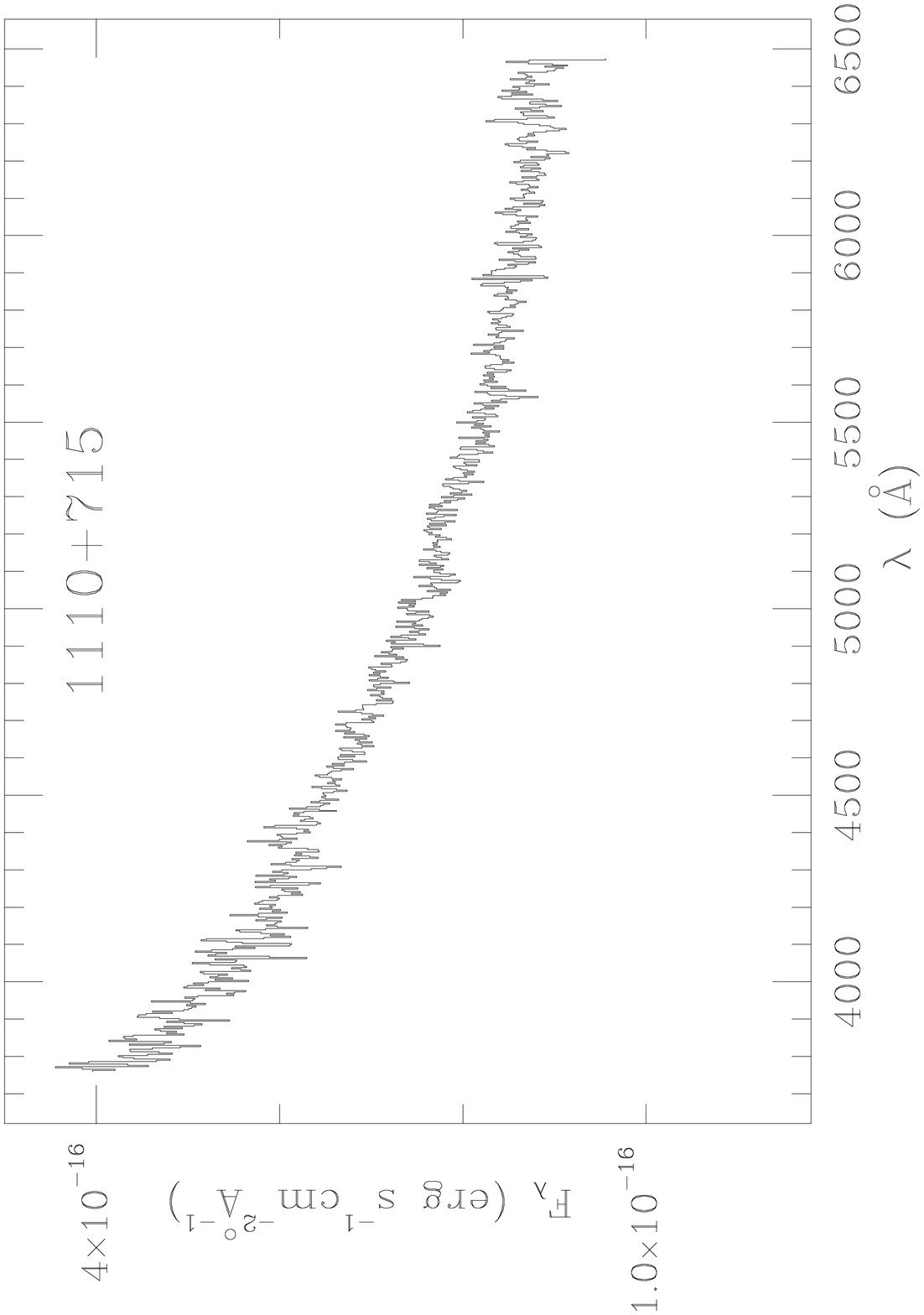,height=7.1cm,width=6.3cm}
\end{minipage}
\hspace{0.3in}
\begin{minipage}[t]{6.3in}
\psfig{file=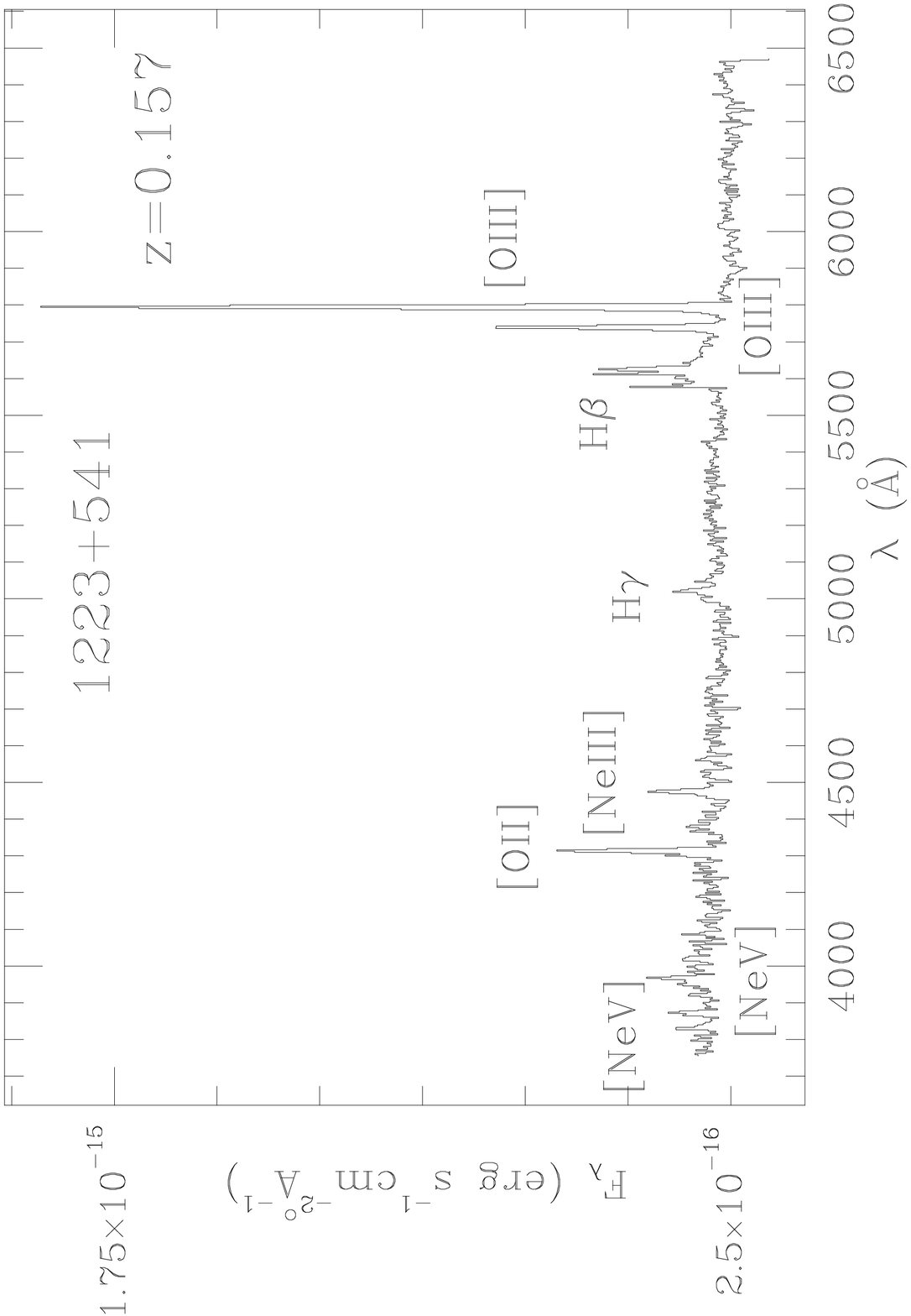,height=7.1cm,width=6.3cm}
\vspace{0.25in}
\psfig{file=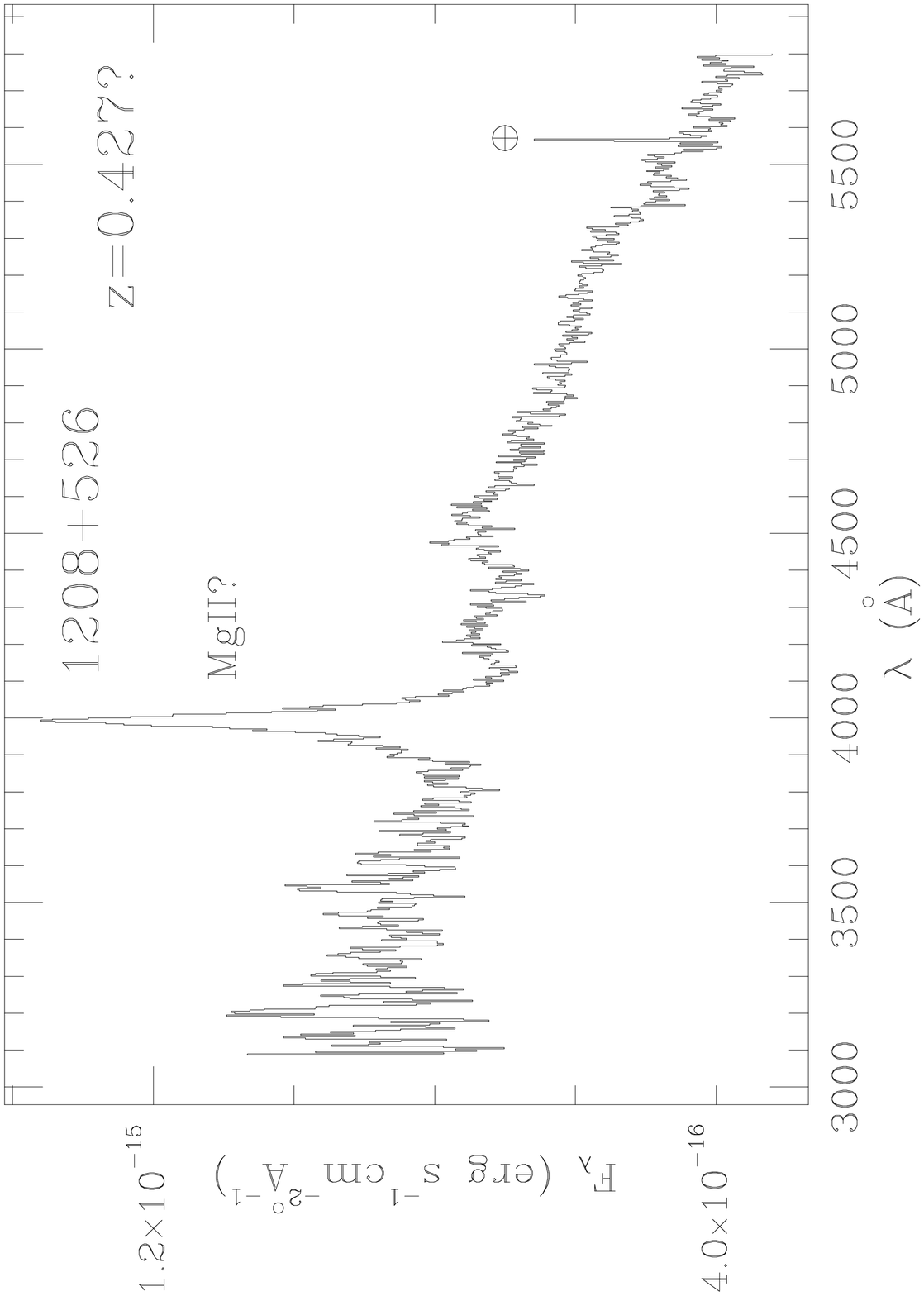,height=7.1cm,width=6.3cm}
\vspace{0.25in}
\psfig{file=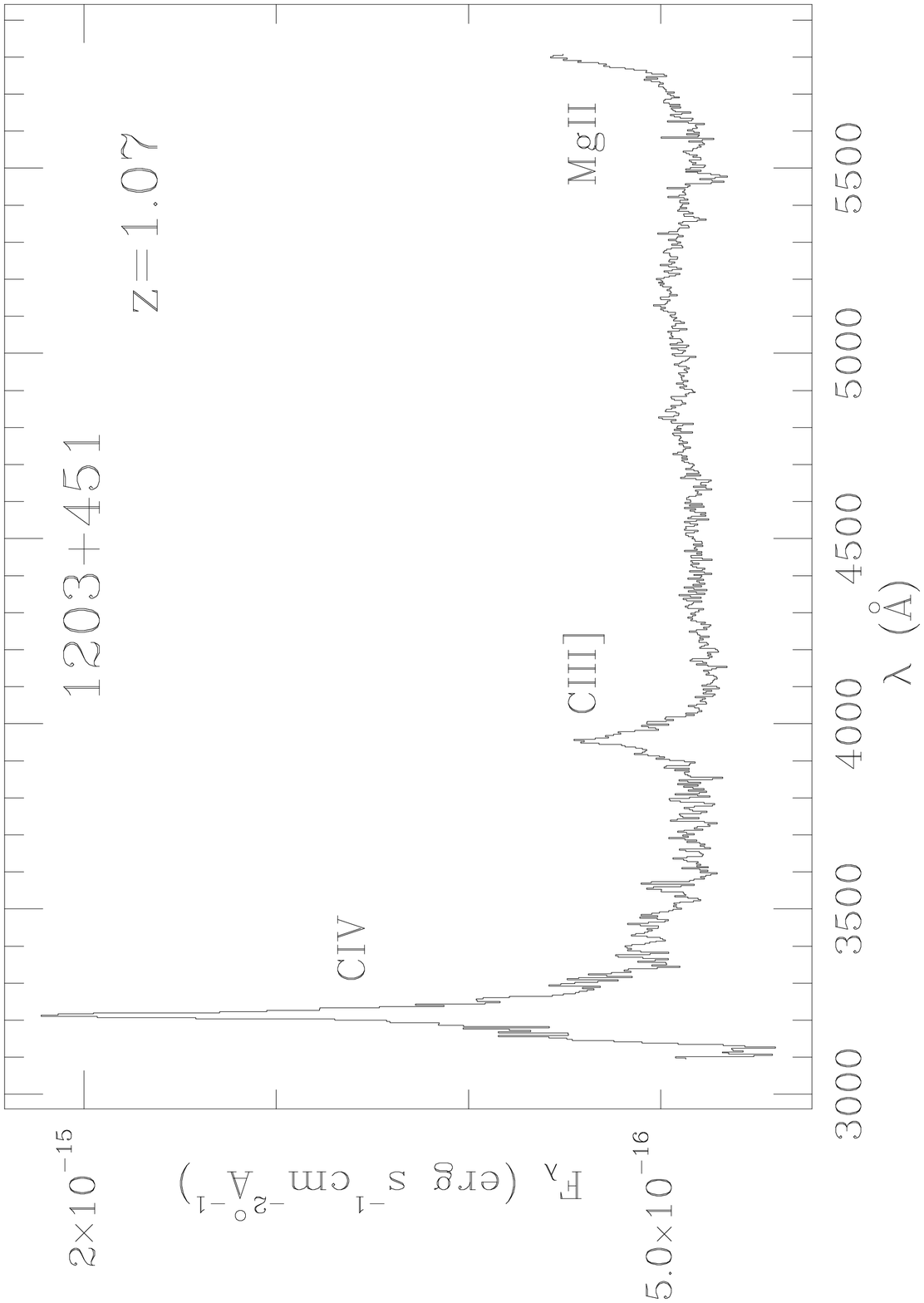,height=7.1cm,width=6.3cm}
\end{minipage}
\hfill
\begin{minipage}[t]{0.3in}
\vfill
\begin{sideways}
Figure 1.49 $-$ 1.54: Spectra of RGB Sources ({\it continued})
\end{sideways}
\vfill
\end{minipage}
\end{figure}

\clearpage
\begin{figure}
\vspace{-0.3in}
\hspace{-0.3in}
\begin{minipage}[t]{6.3in}
\psfig{file=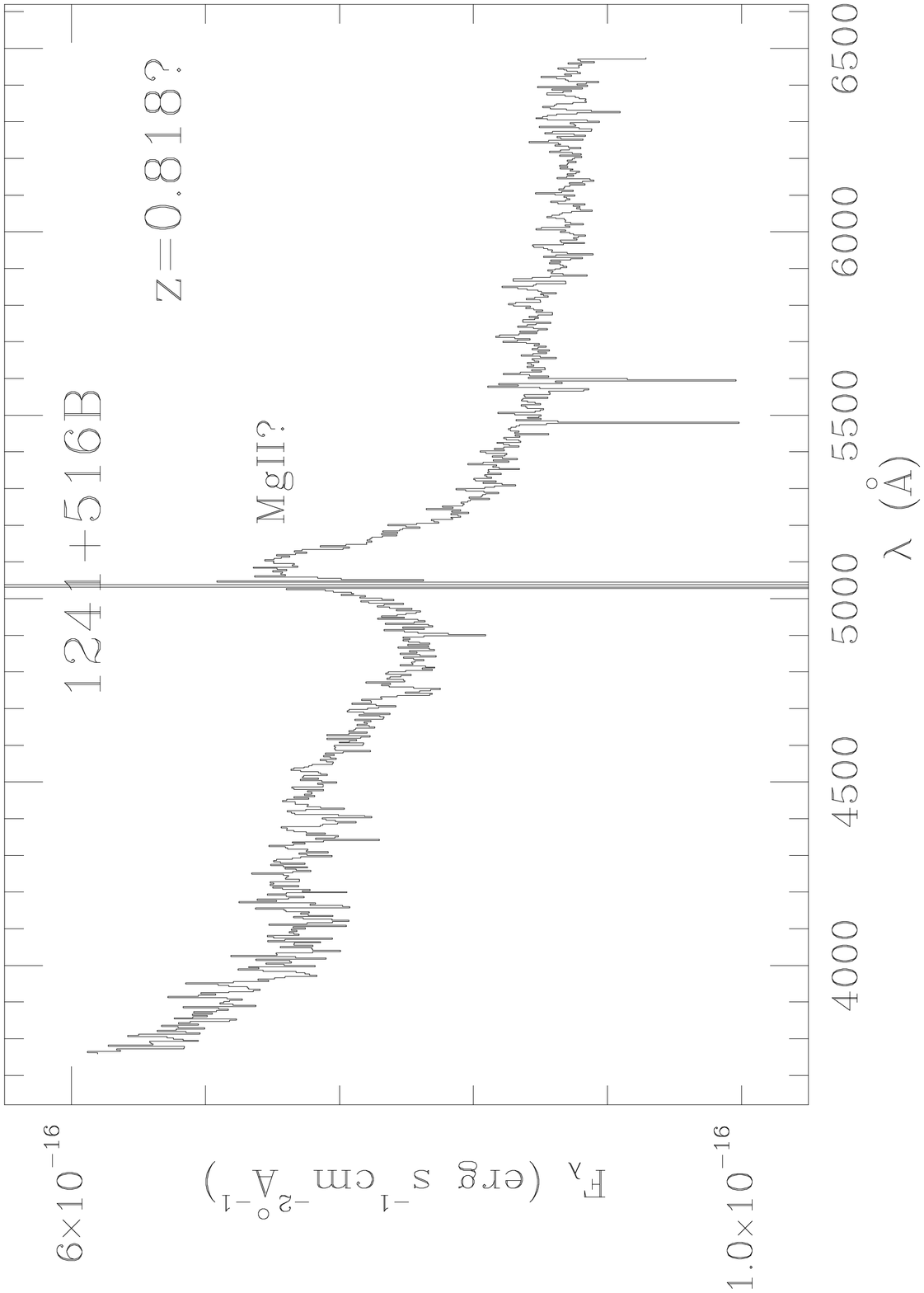,height=7.1cm,width=6.3cm}
\vspace{0.25in}
\psfig{file=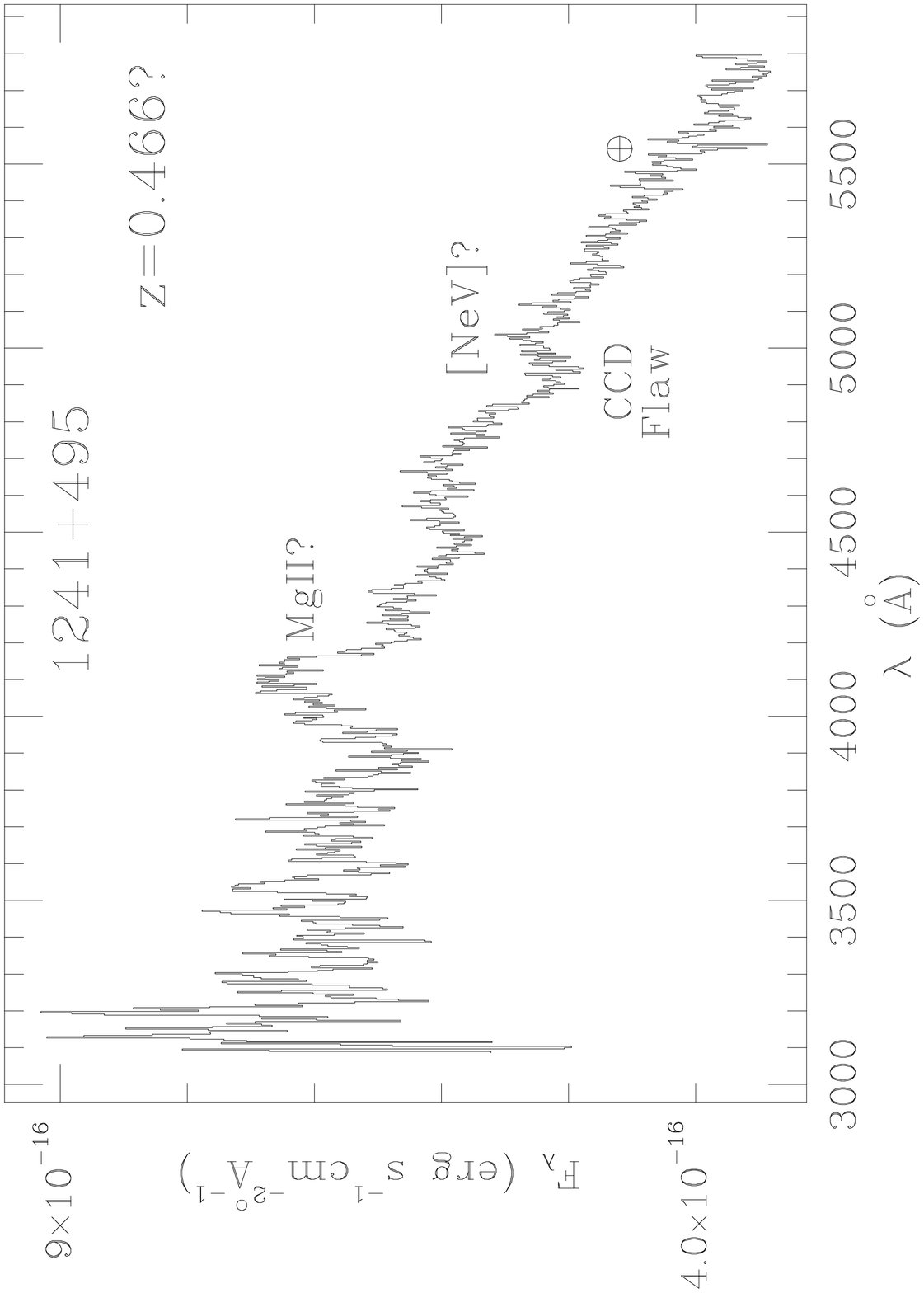,height=7.1cm,width=6.3cm}
\vspace{0.25in}
\psfig{file=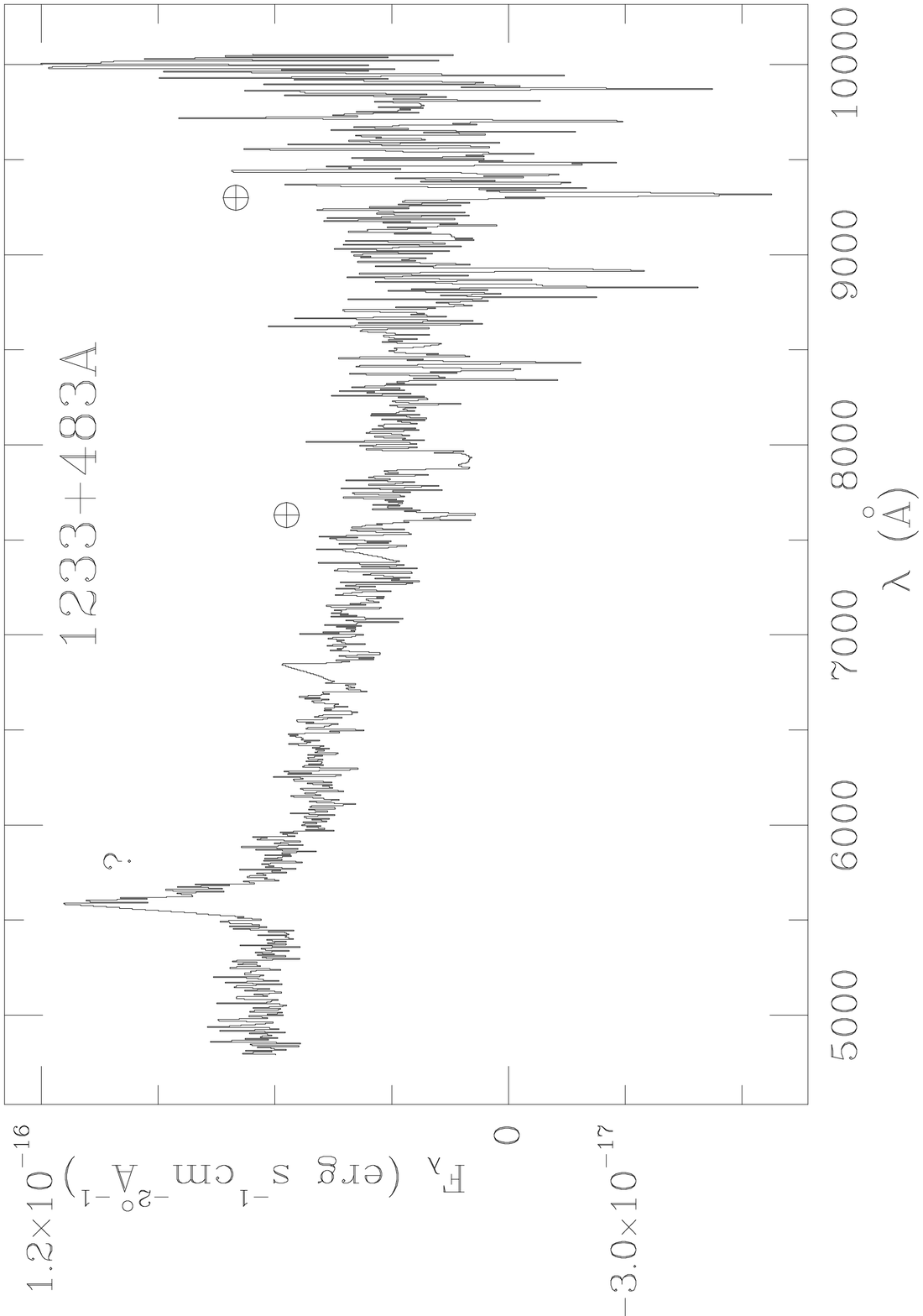,height=7.1cm,width=6.3cm}
\end{minipage}
\hspace{0.3in}
\begin{minipage}[t]{6.3in}
\psfig{file=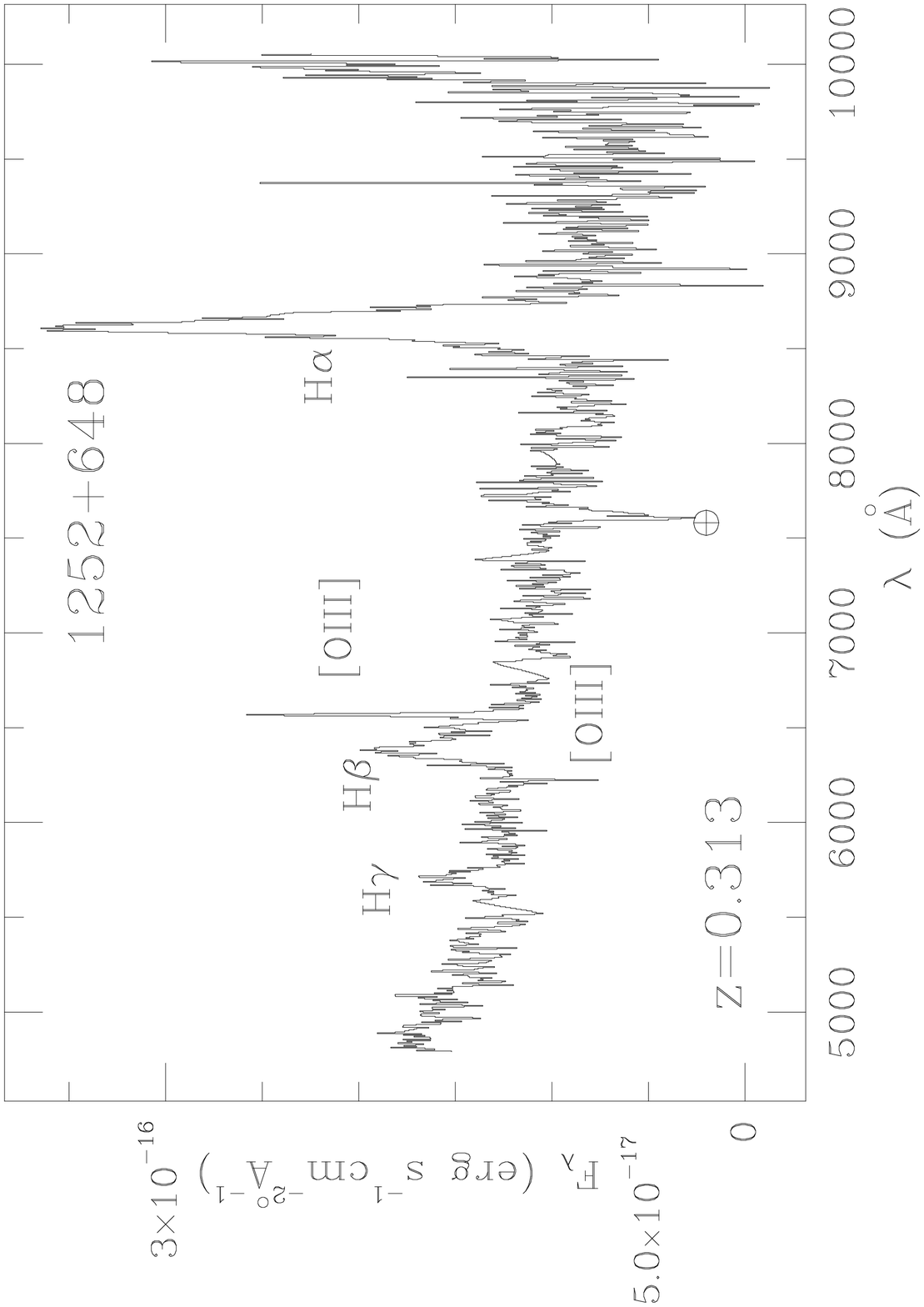,height=7.1cm,width=6.3cm}
\vspace{0.25in}
\psfig{file=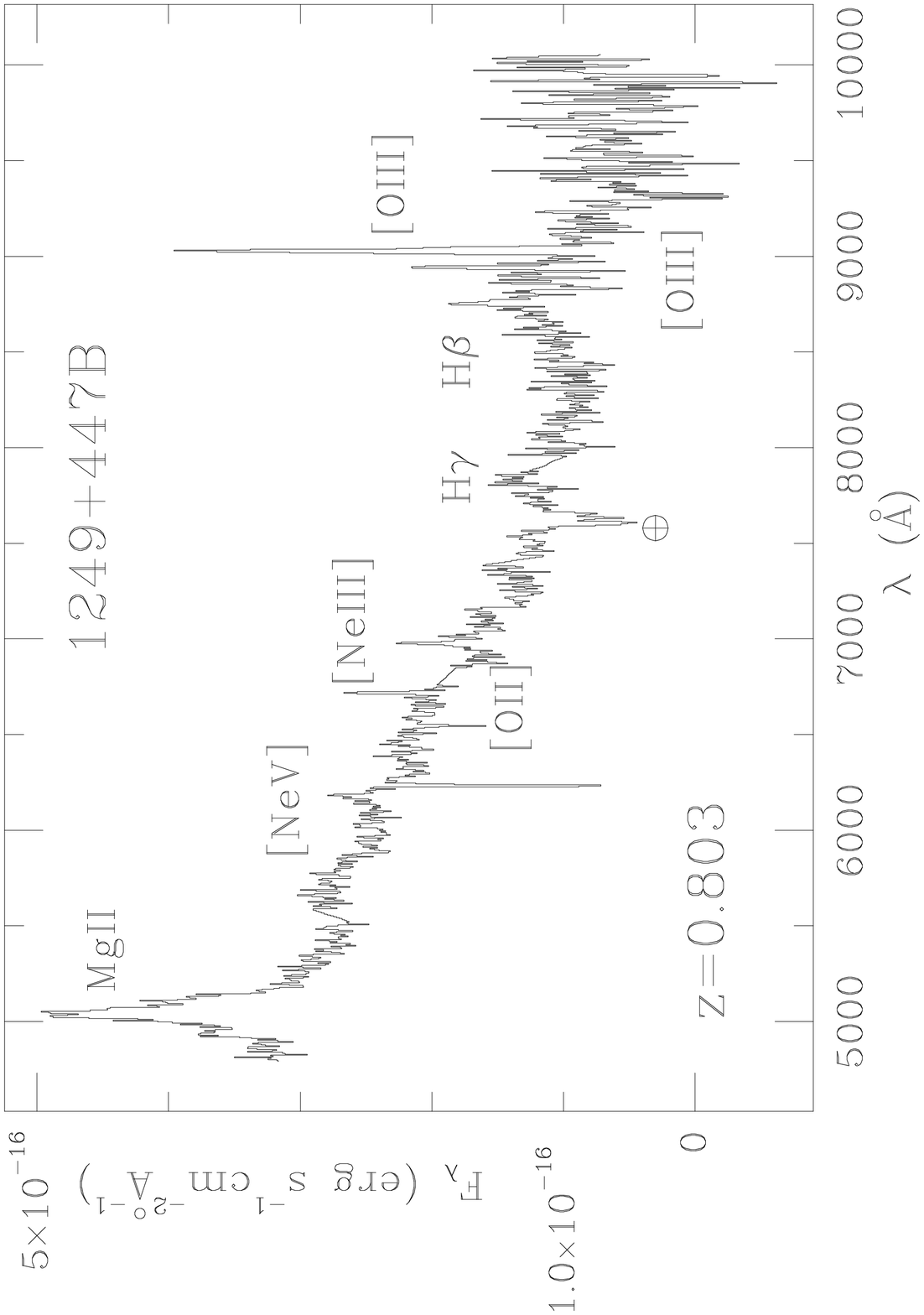,height=7.1cm,width=6.3cm}
\vspace{0.25in}
\psfig{file=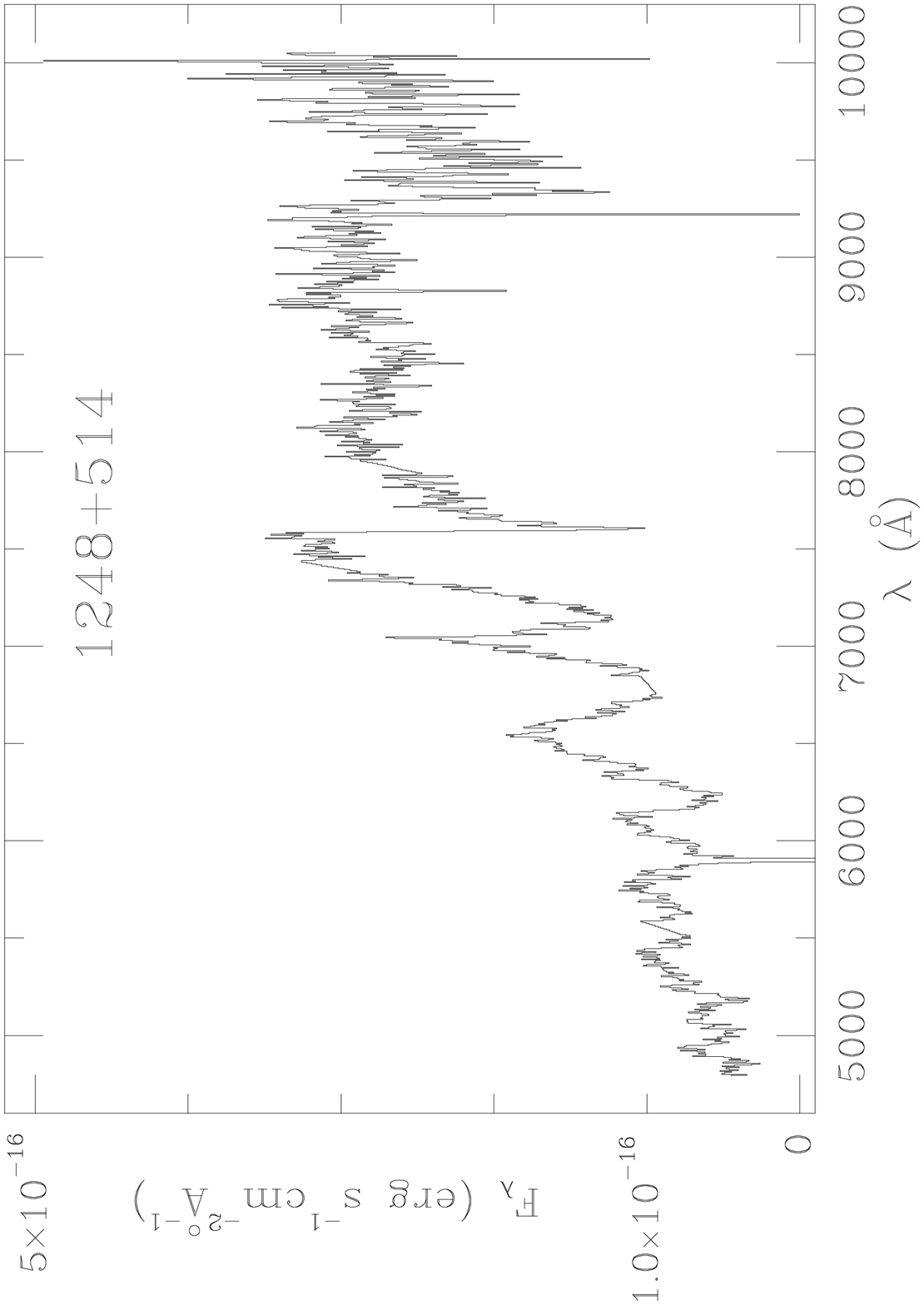,height=7.1cm,width=6.3cm}
\end{minipage}
\hfill
\begin{minipage}[t]{0.3in}
\vfill
\begin{sideways}
Figure 1.55 $-$ 1.60: Spectra of RGB Sources ({\it continued})
\end{sideways}
\vfill
\end{minipage}
\end{figure}

\clearpage
\begin{figure}
\vspace{-0.3in}
\hspace{-0.3in}
\begin{minipage}[t]{6.3in}
\psfig{file=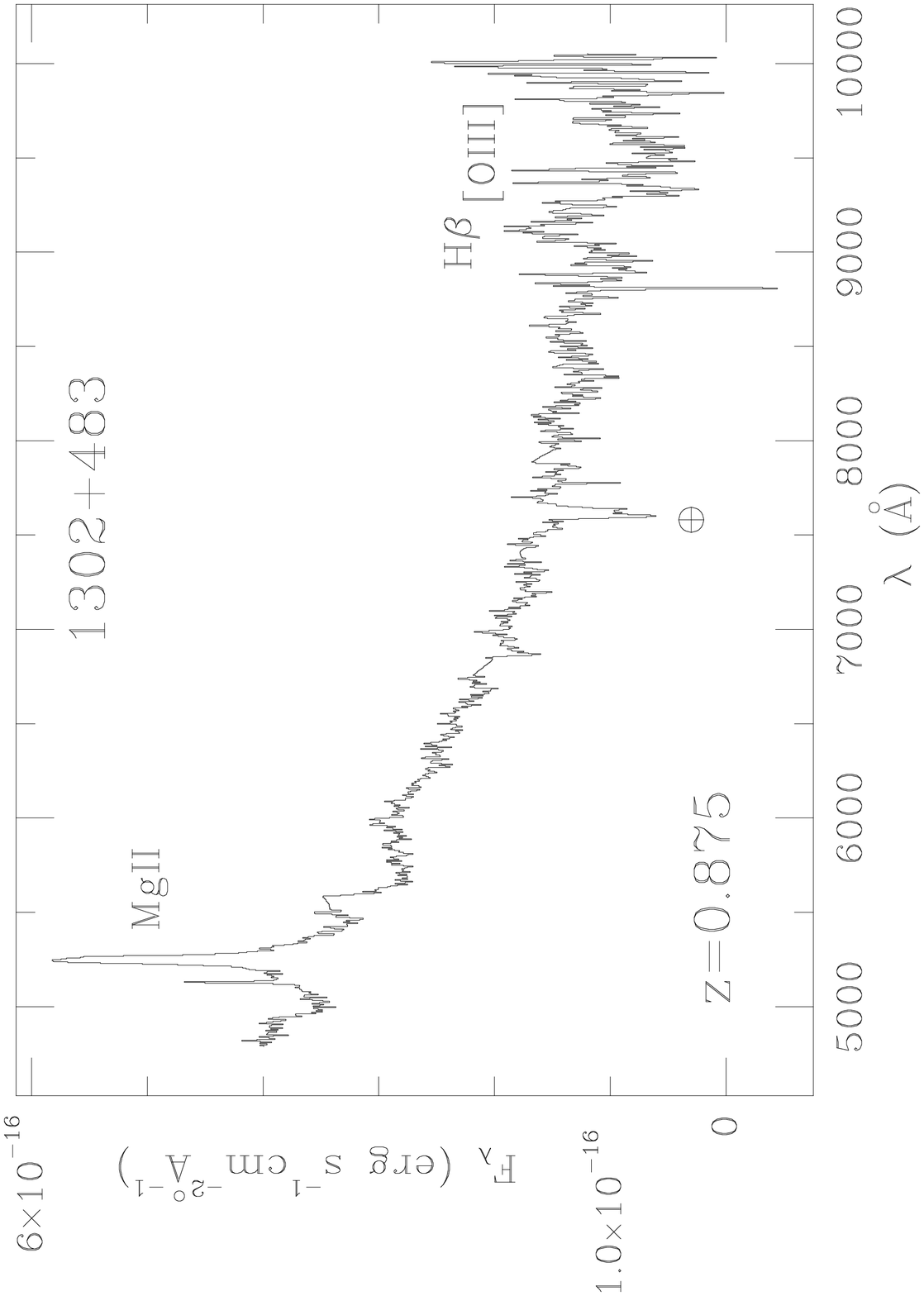,height=7.1cm,width=6.3cm}
\vspace{0.25in}
\psfig{file=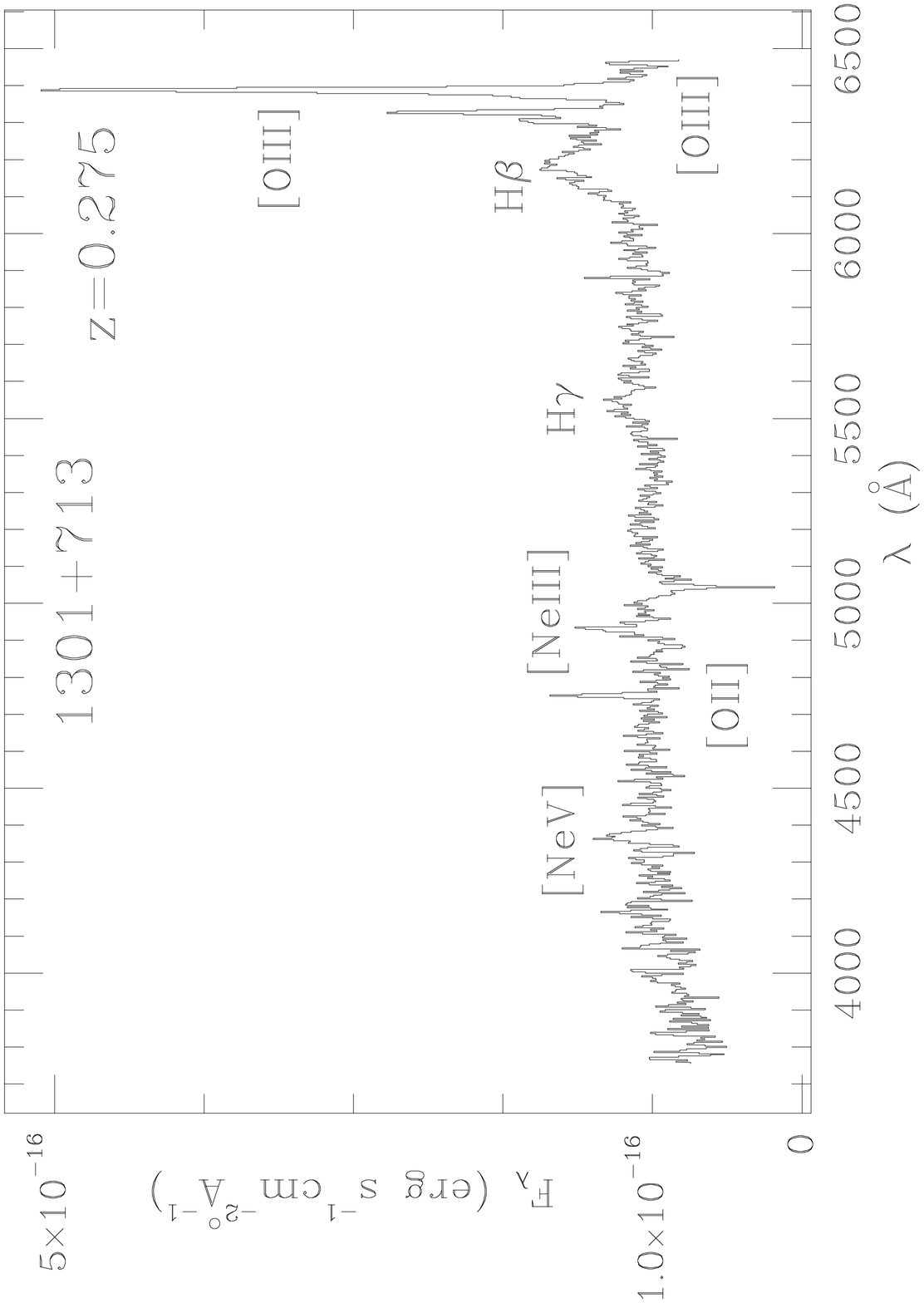,height=7.1cm,width=6.3cm}
\vspace{0.25in}
\psfig{file=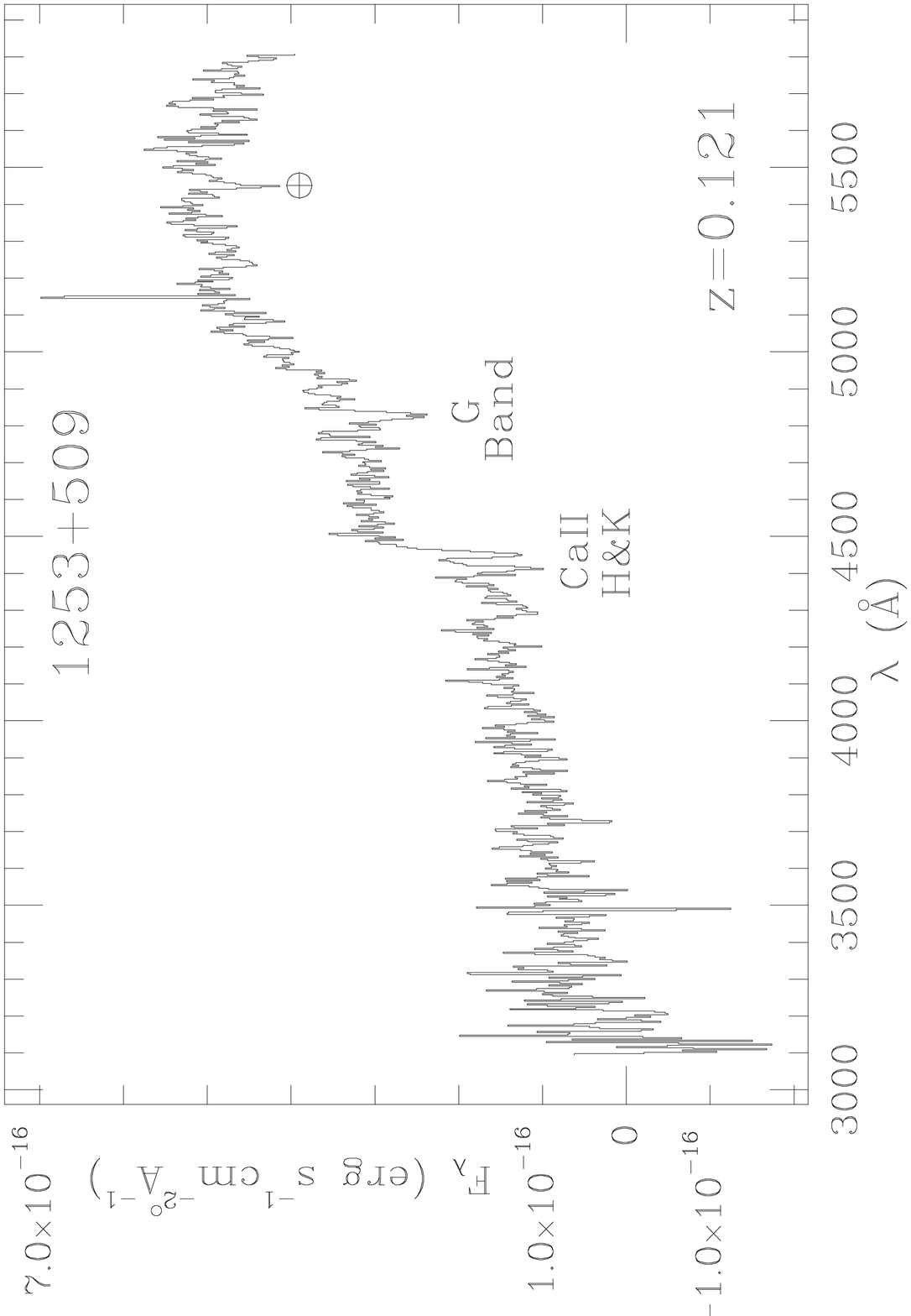,height=7.1cm,width=6.3cm}
\end{minipage}
\hspace{0.3in}
\begin{minipage}[t]{6.3in}
\psfig{file=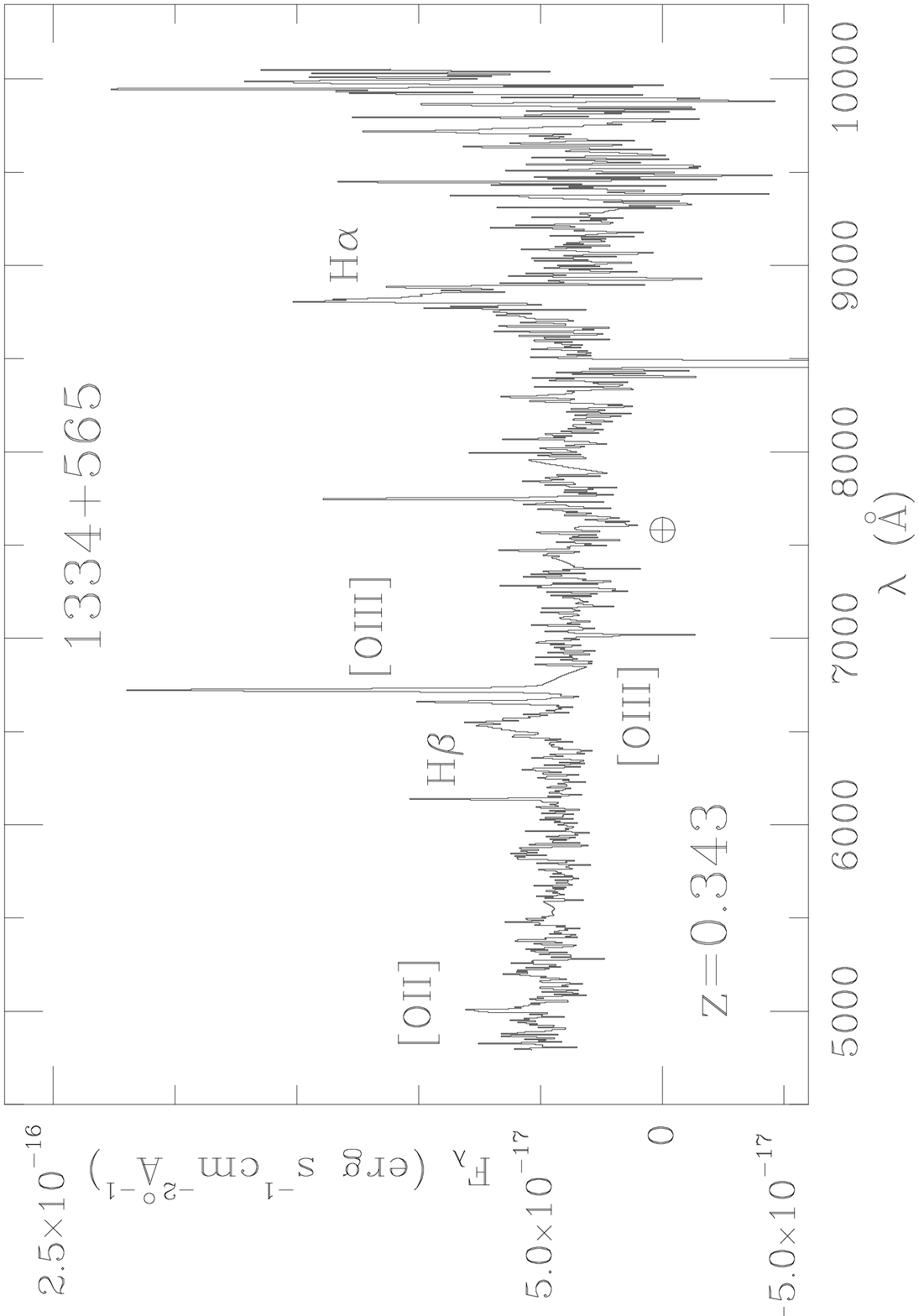,height=7.1cm,width=6.3cm}
\vspace{0.25in}
\psfig{file=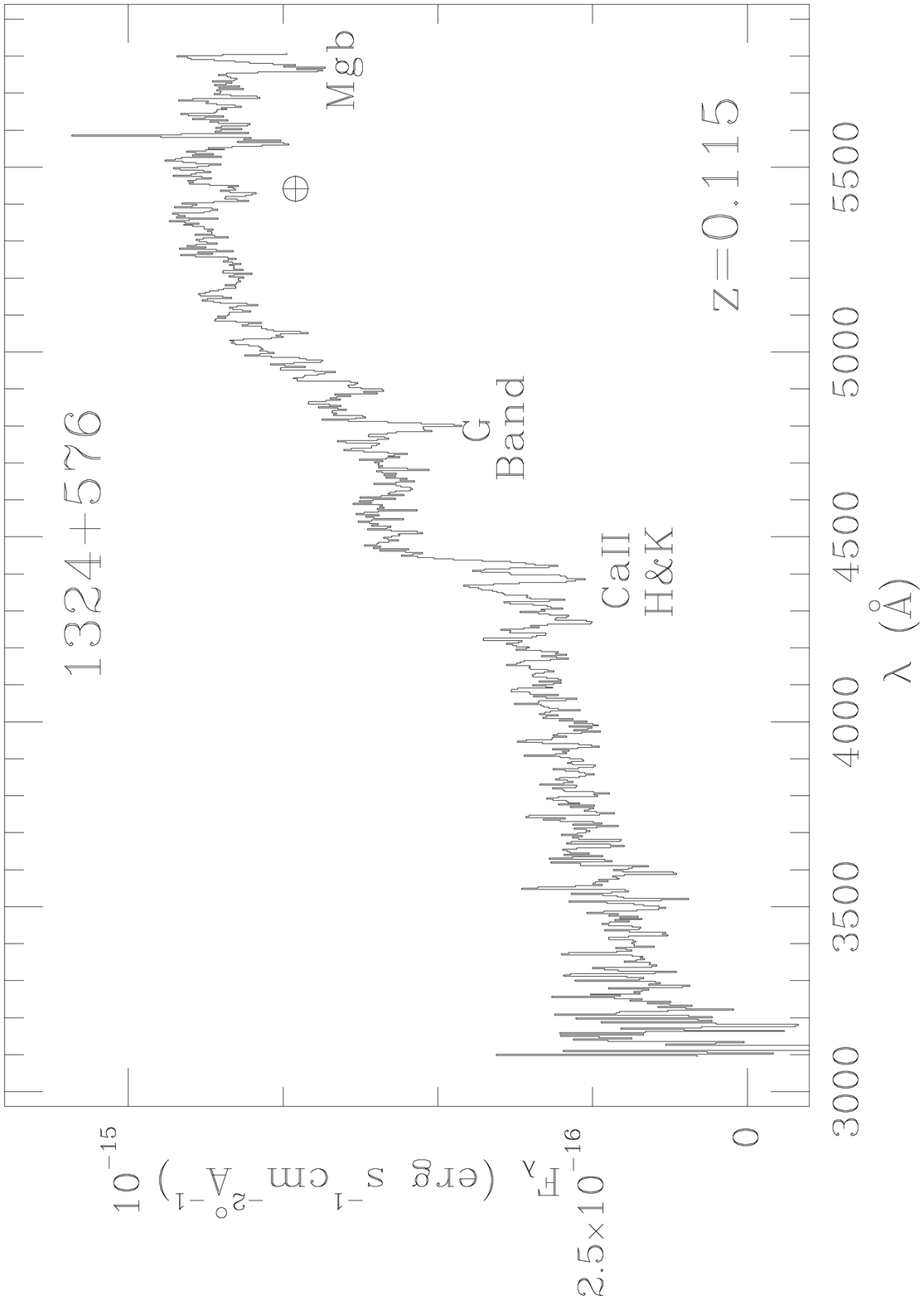,height=7.1cm,width=6.3cm}
\vspace{0.25in}
\psfig{file=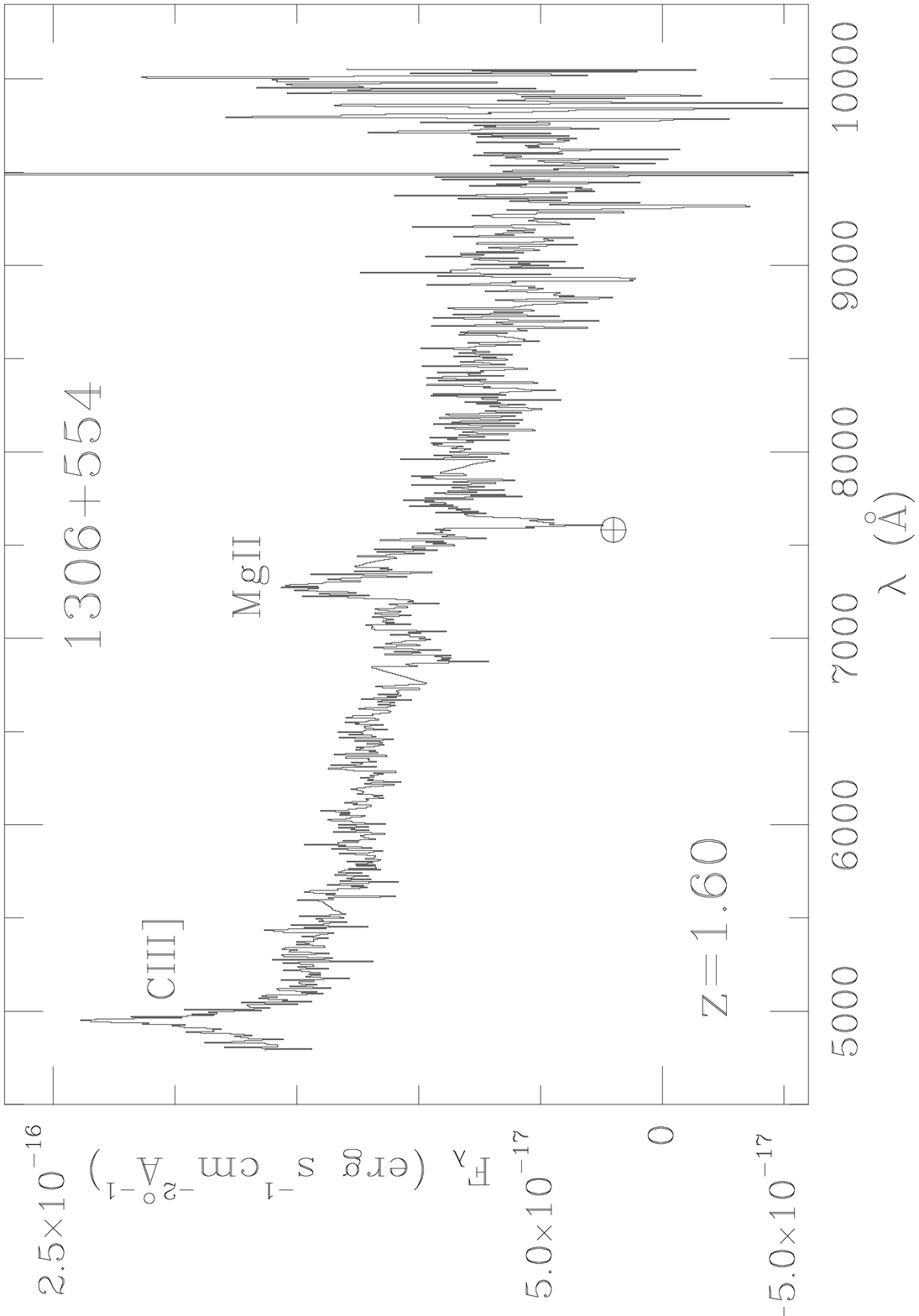,height=7.1cm,width=6.3cm}
\end{minipage}
\hfill
\begin{minipage}[t]{0.3in}
\vfill
\begin{sideways}
Figure 1.61 $-$ 1.66: Spectra of RGB Sources ({\it continued})
\end{sideways}
\vfill
\end{minipage}
\end{figure}

\clearpage
\begin{figure}
\vspace{-0.3in}
\hspace{-0.3in}
\begin{minipage}[t]{6.3in}
\psfig{file=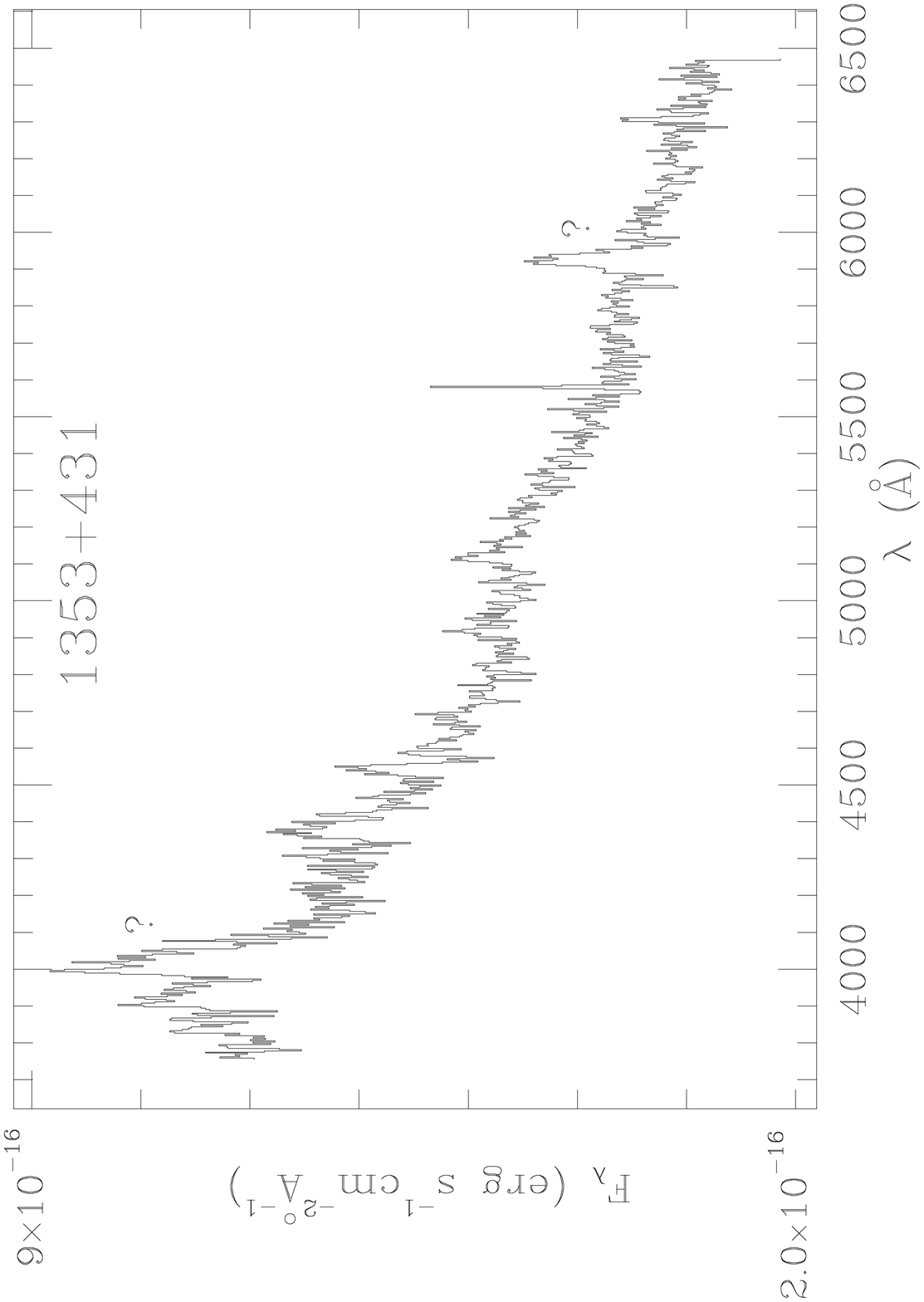,height=7.1cm,width=6.3cm}
\vspace{0.25in}
\psfig{file=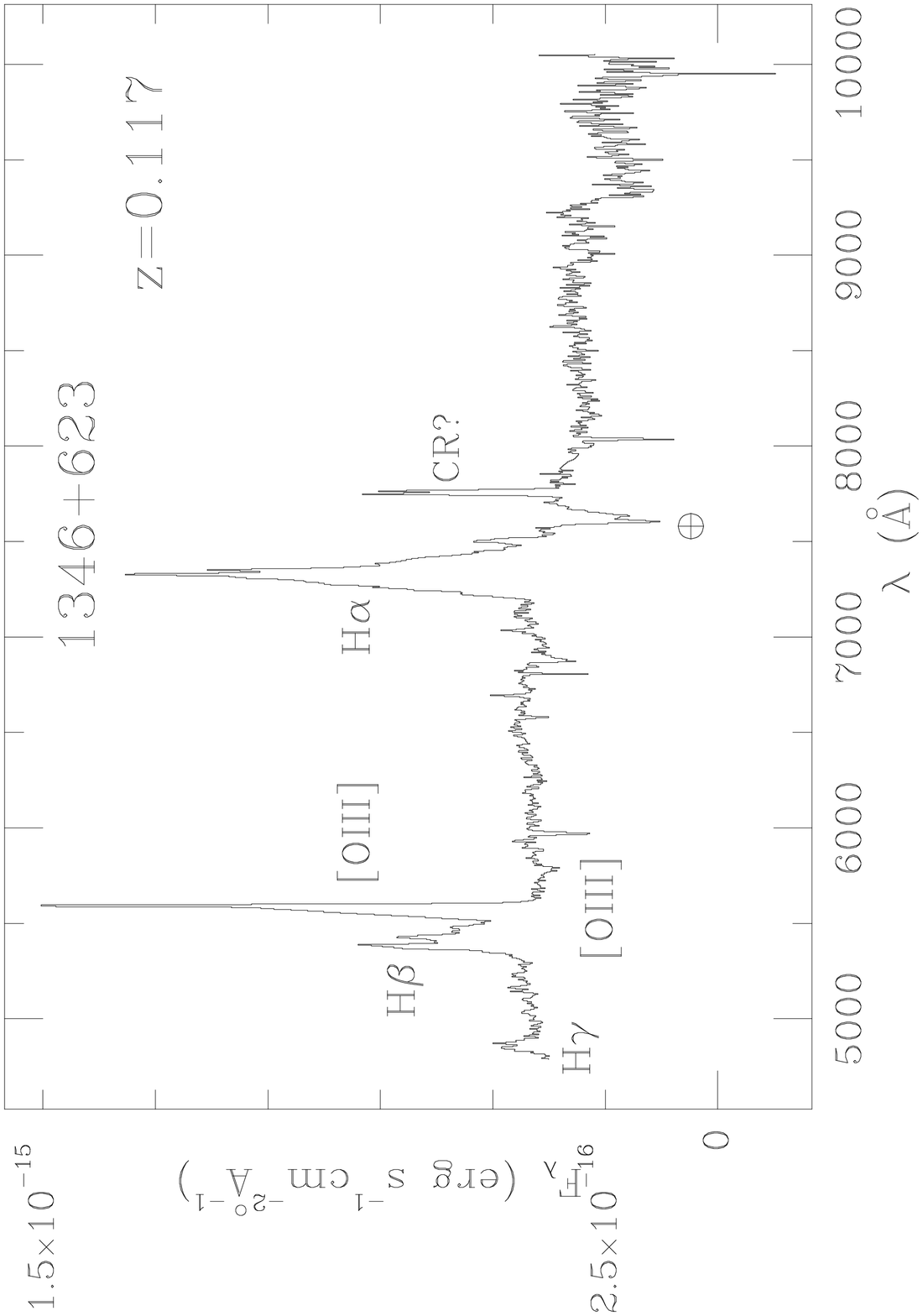,height=7.1cm,width=6.3cm}
\vspace{0.25in}
\psfig{file=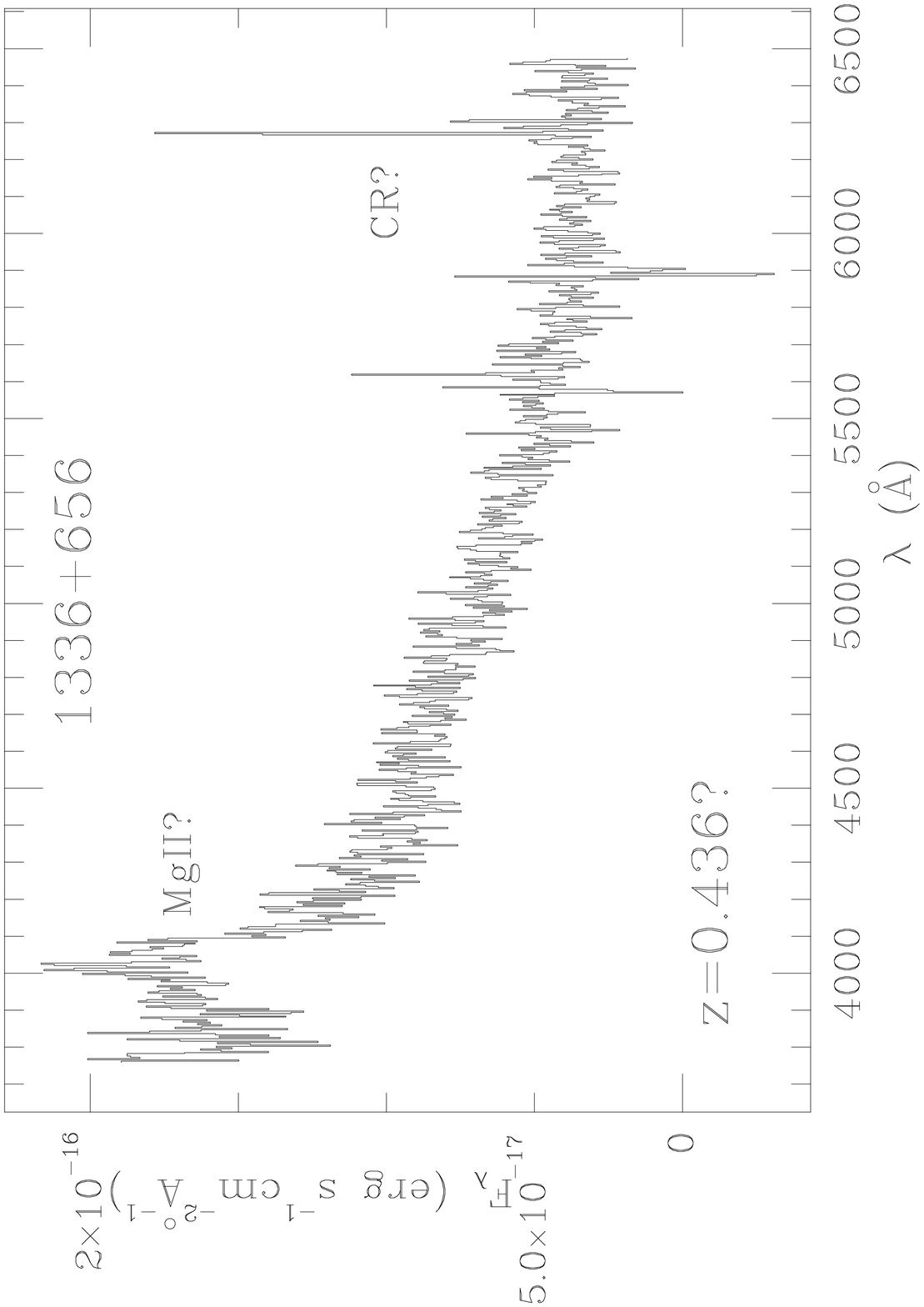,height=7.1cm,width=6.3cm}
\end{minipage}
\hspace{0.3in}
\begin{minipage}[t]{6.3in}
\psfig{file=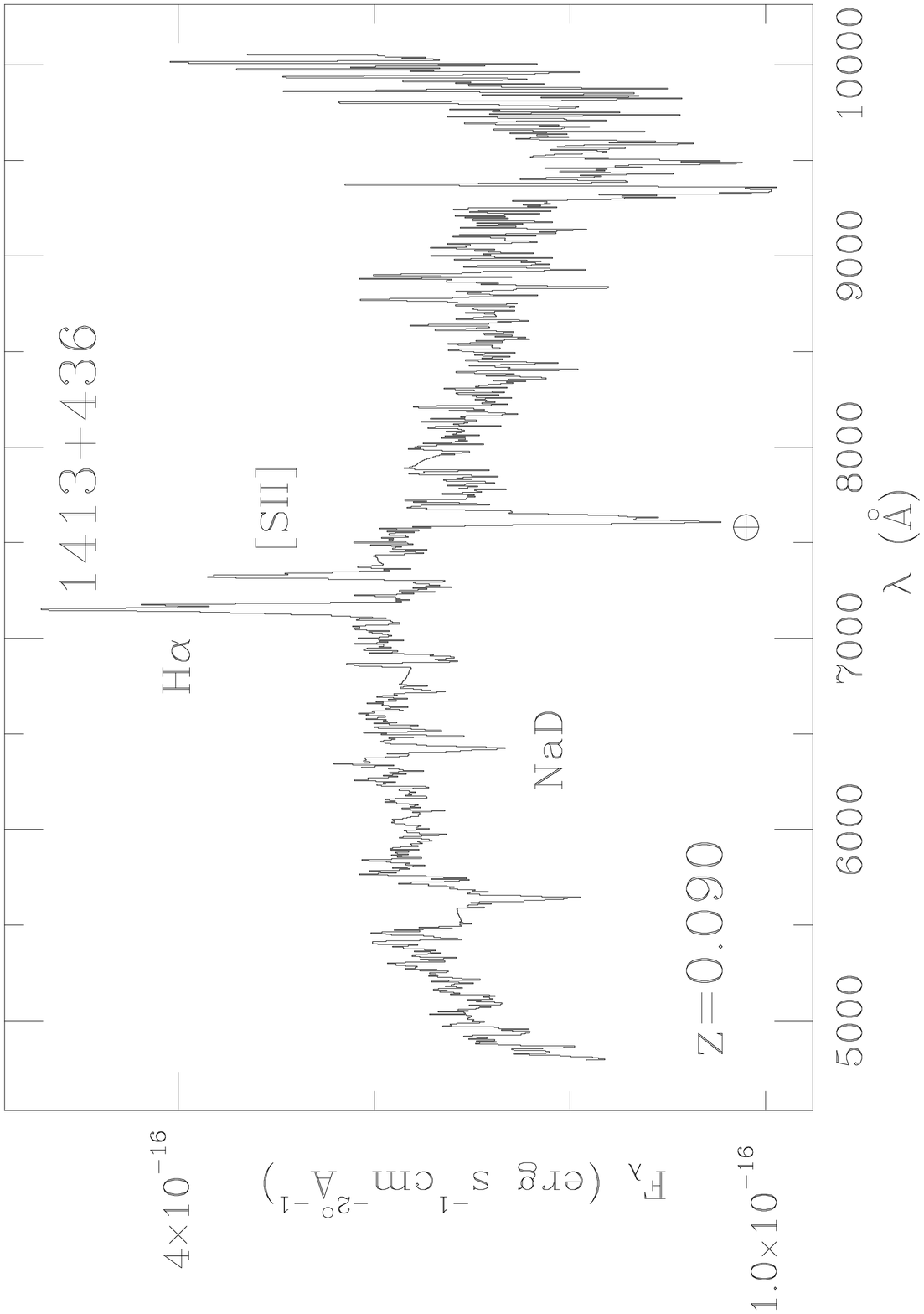,height=7.1cm,width=6.3cm}
\vspace{0.25in}
\psfig{file=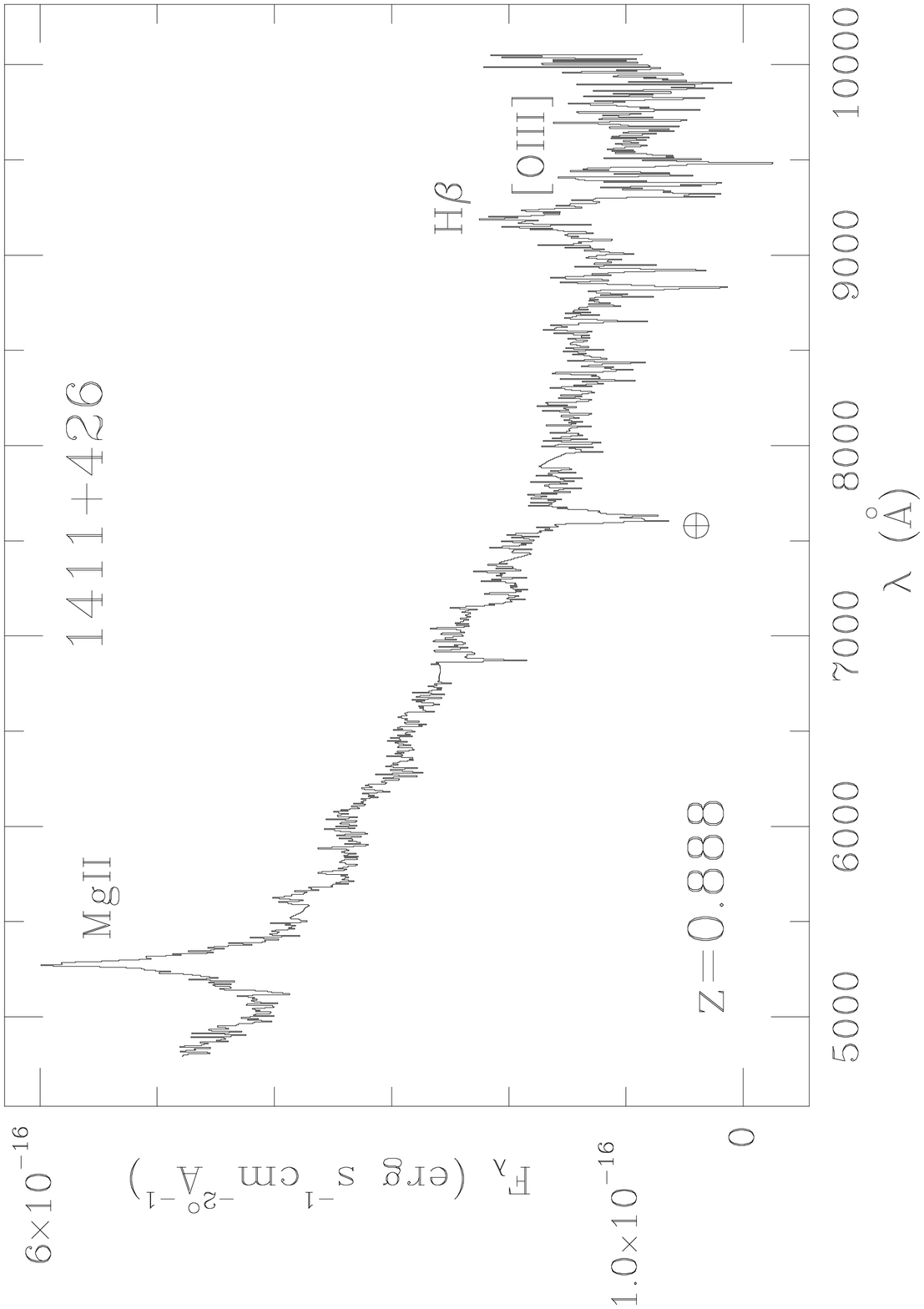,height=7.1cm,width=6.3cm}
\vspace{0.25in}
\psfig{file=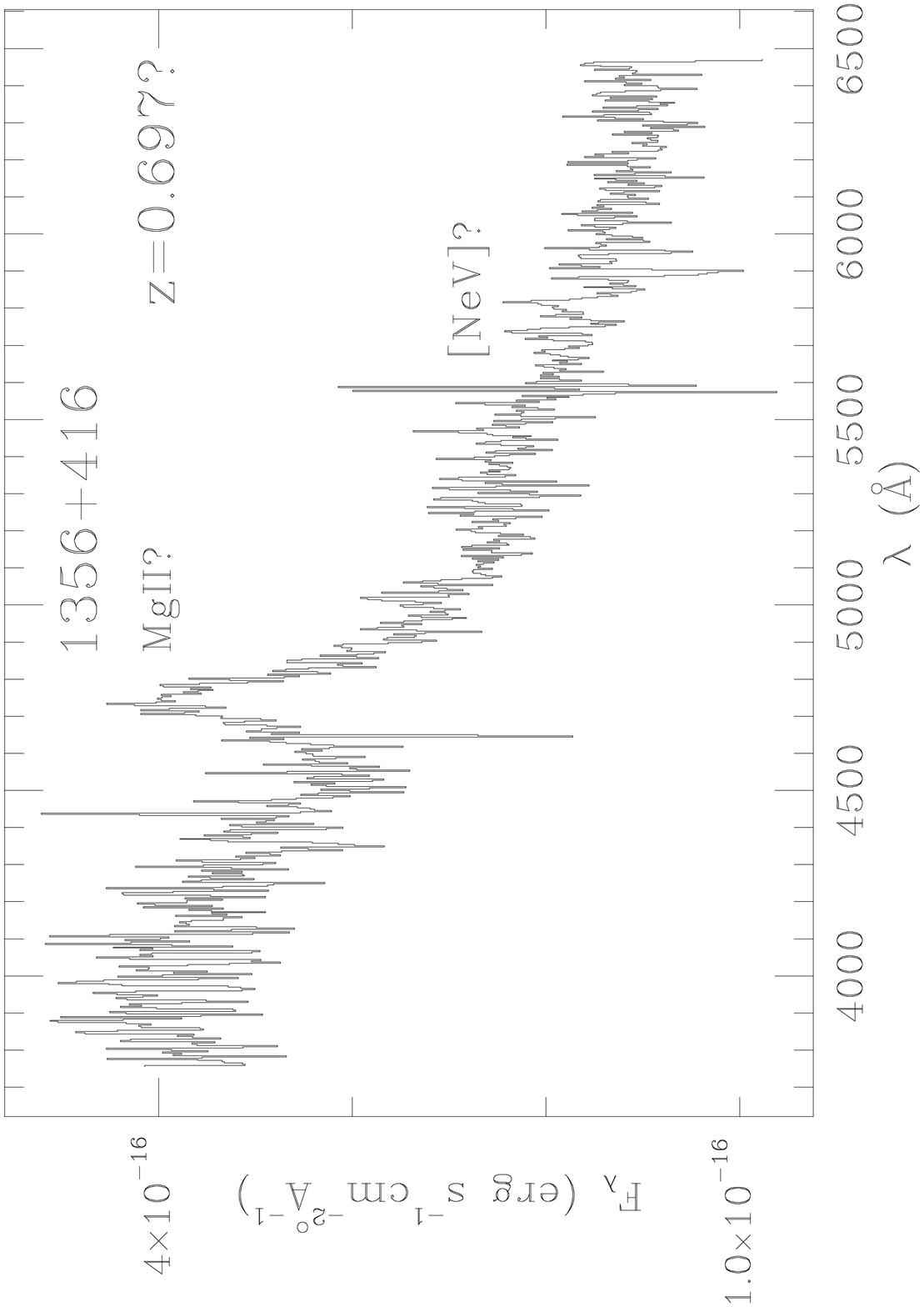,height=7.1cm,width=6.3cm}
\end{minipage}
\hfill
\begin{minipage}[t]{0.3in}
\vfill
\begin{sideways}
Figure 1.67 $-$ 1.72: Spectra of RGB Sources ({\it continued})
\end{sideways}
\vfill
\end{minipage}
\end{figure}

\clearpage
\begin{figure}
\vspace{-0.3in}
\hspace{-0.3in}
\begin{minipage}[t]{6.3in}
\psfig{file=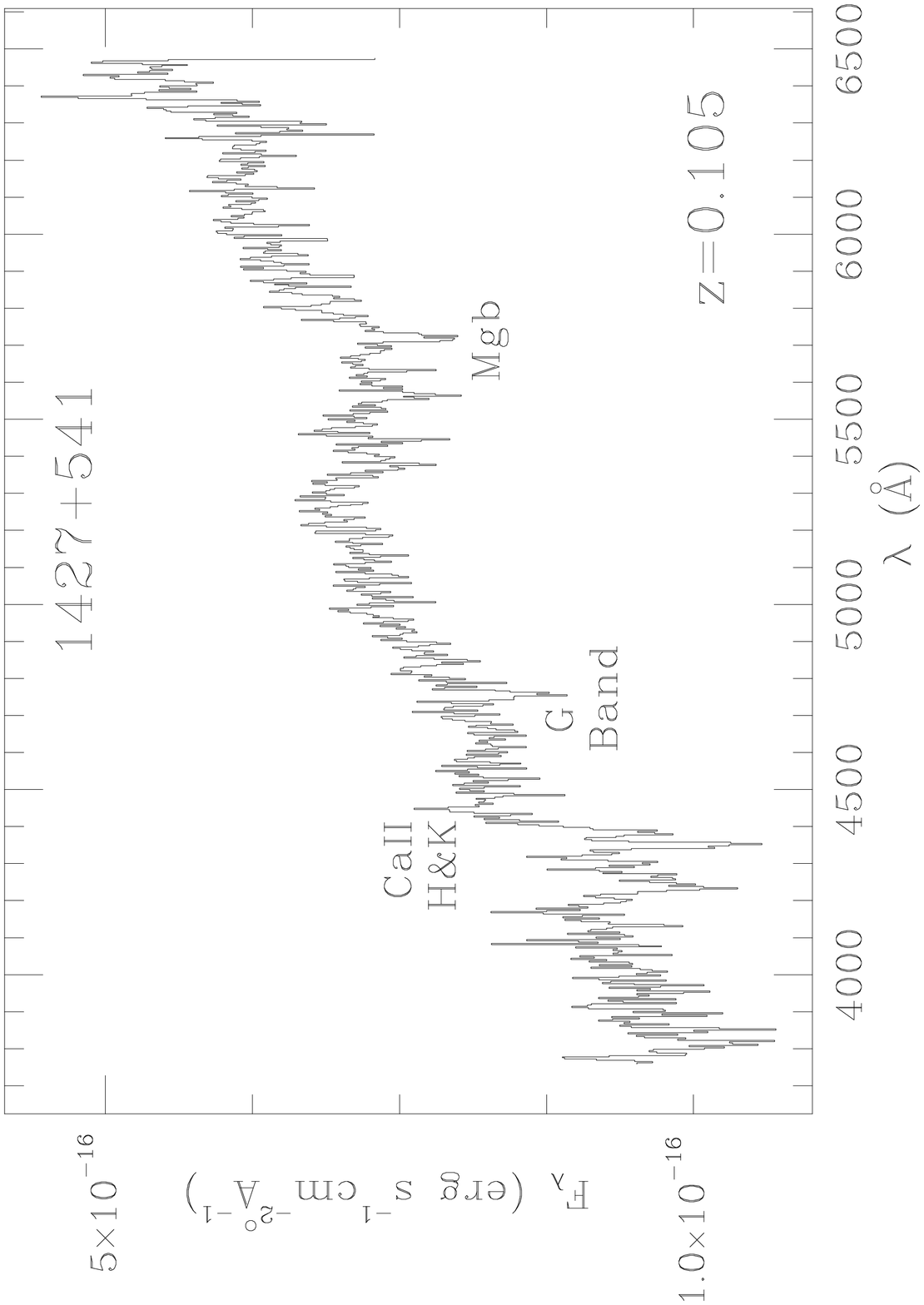,height=7.1cm,width=6.3cm}
\vspace{0.25in}
\psfig{file=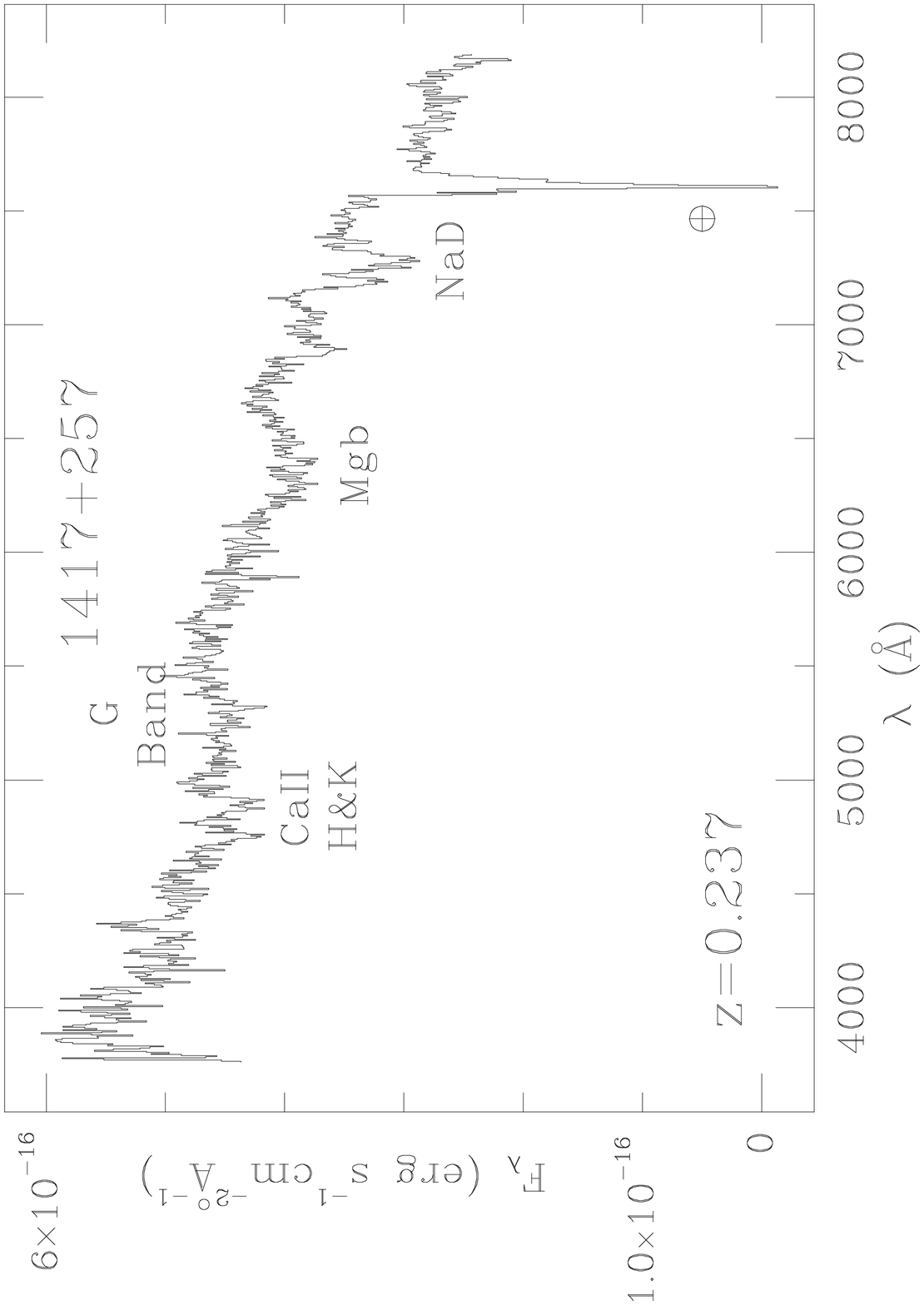,height=7.1cm,width=6.3cm}
\vspace{0.25in}
\psfig{file=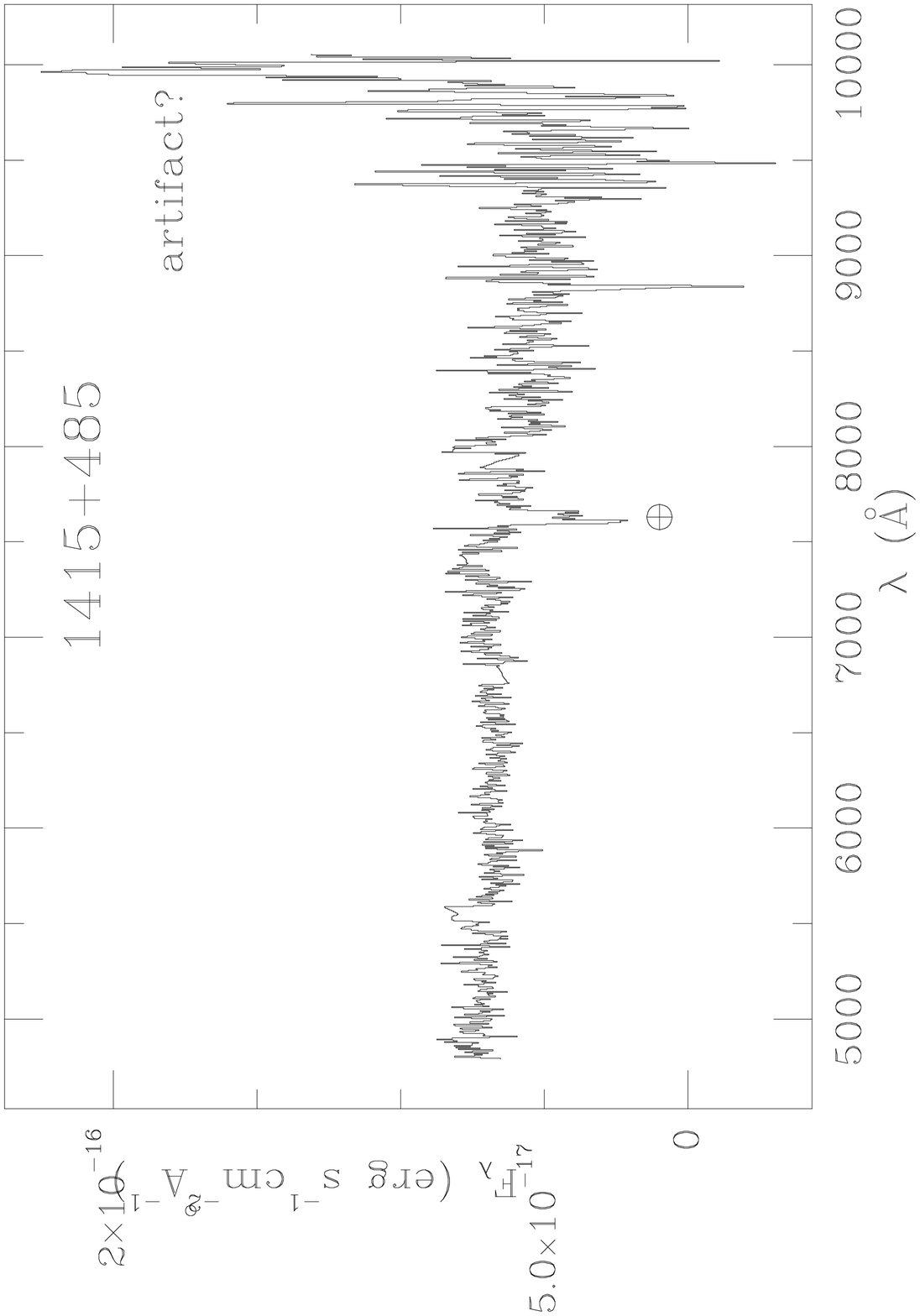,height=7.1cm,width=6.3cm}
\vspace{0.25in}
\end{minipage}
\hspace{0.3in}
\begin{minipage}[t]{6.3in}
\psfig{file=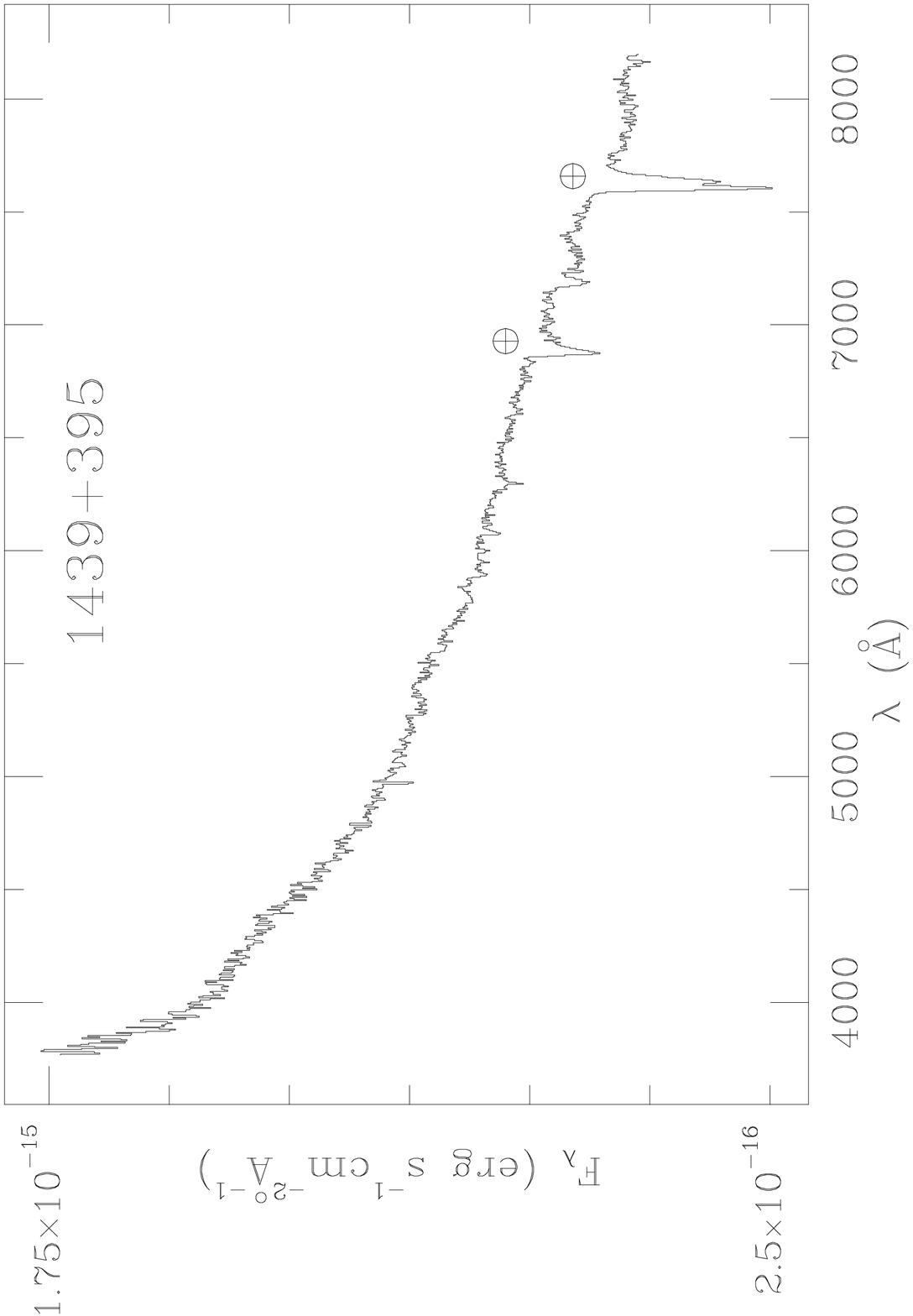,height=7.1cm,width=6.3cm}
\vspace{0.25in}
\psfig{file=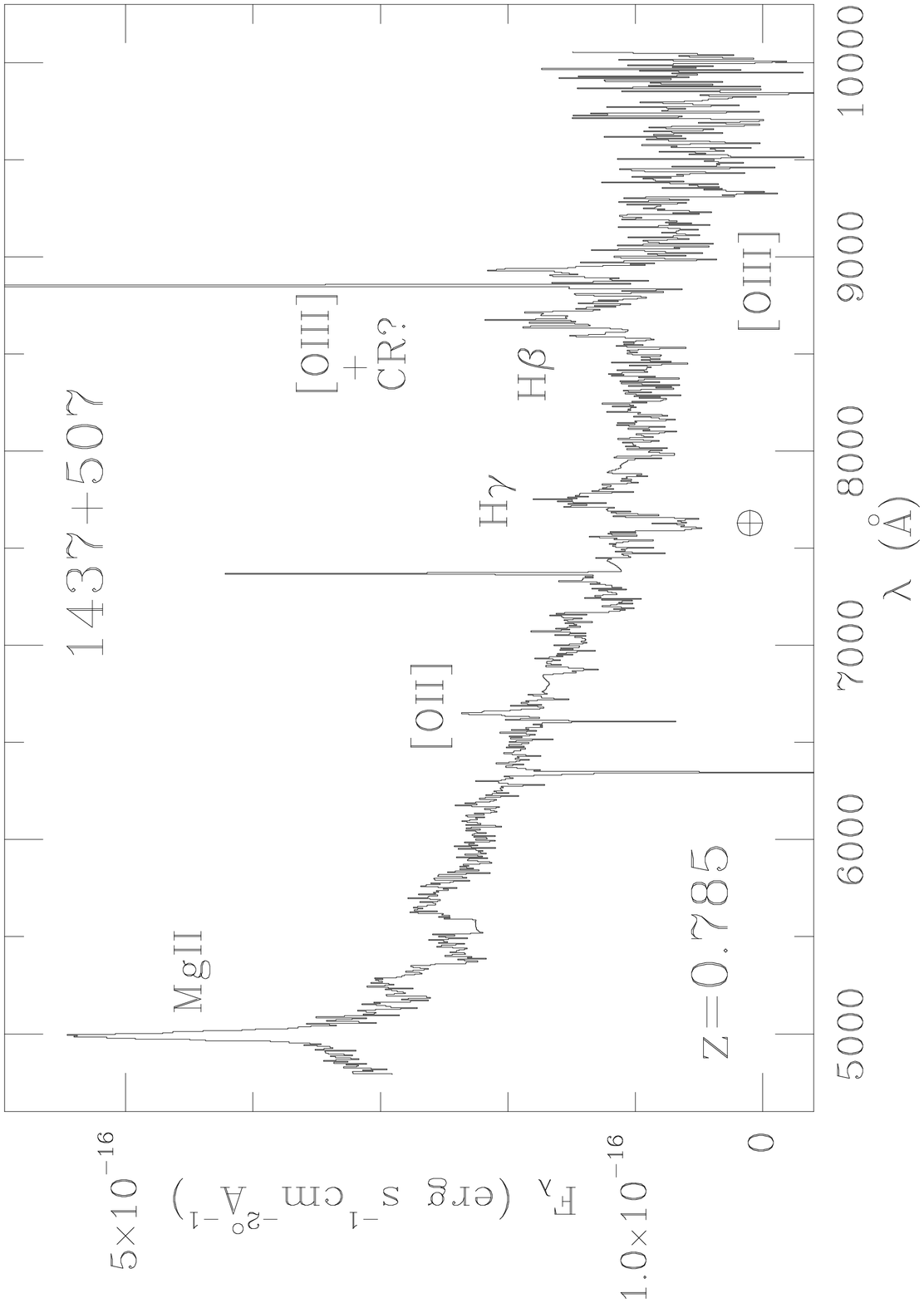,height=7.1cm,width=6.3cm}
\vspace{0.25in}
\psfig{file=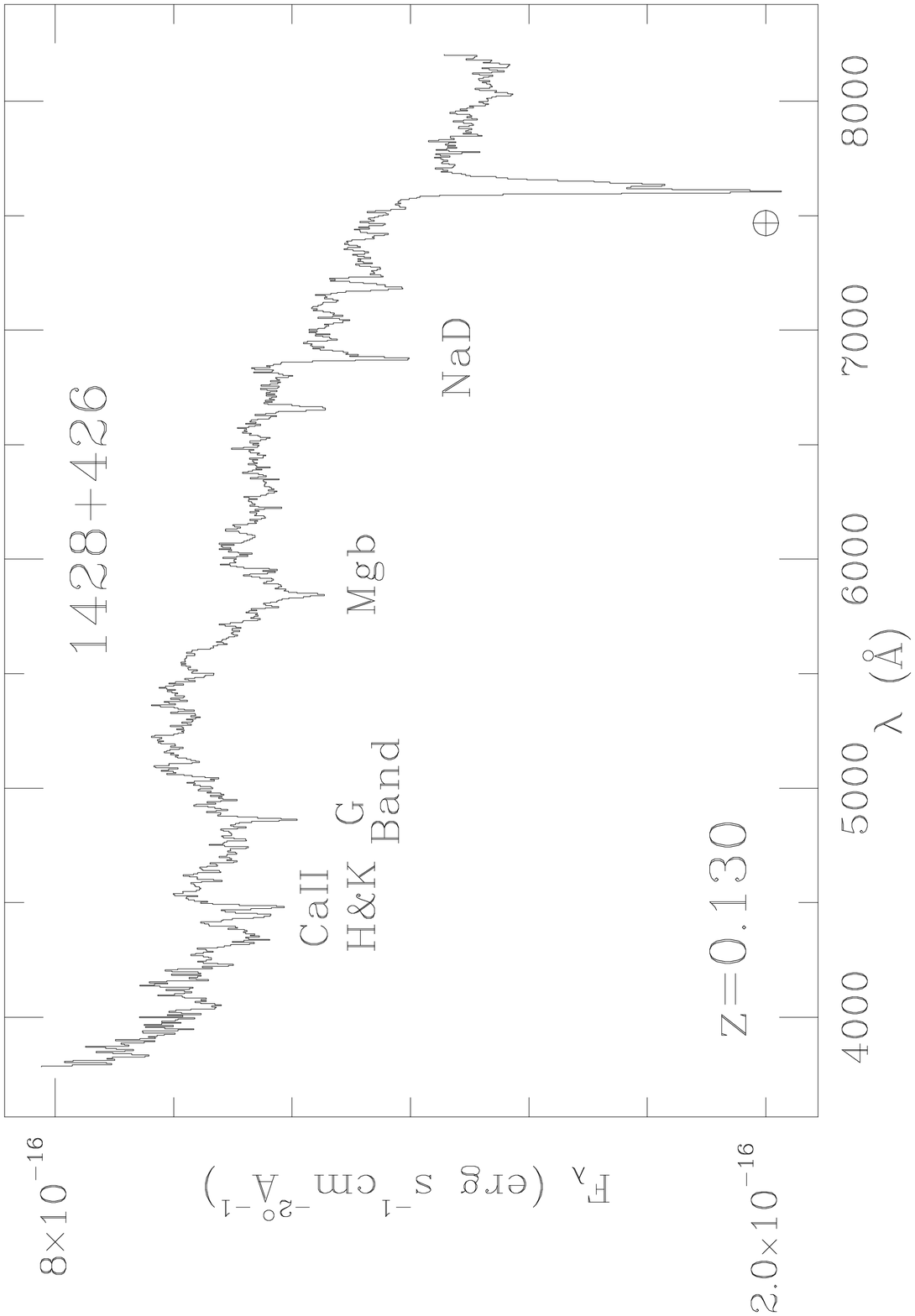,height=7.1cm,width=6.3cm}
\end{minipage}
\hfill
\begin{minipage}[t]{0.3in}
\vfill
\begin{sideways}
Figure 1.73 $-$ 1.78: Spectra of RGB Sources ({\it continued})
\end{sideways}
\vfill
\end{minipage}
\end{figure}

\clearpage
\begin{figure}
\vspace{-0.3in}
\hspace{-0.3in}
\begin{minipage}[t]{6.3in}
\psfig{file=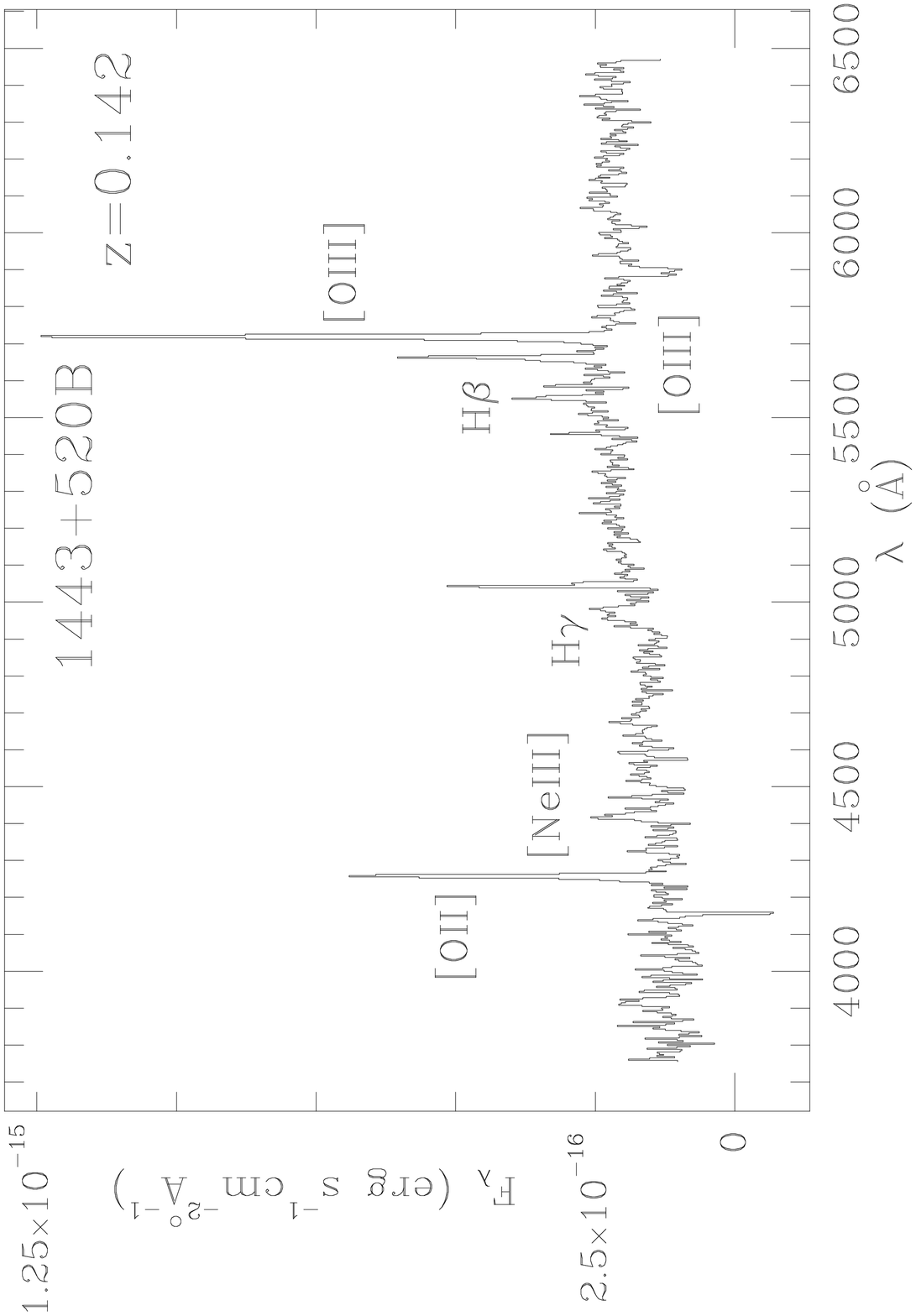,height=7.1cm,width=6.3cm}
\vspace{0.25in}
\psfig{file=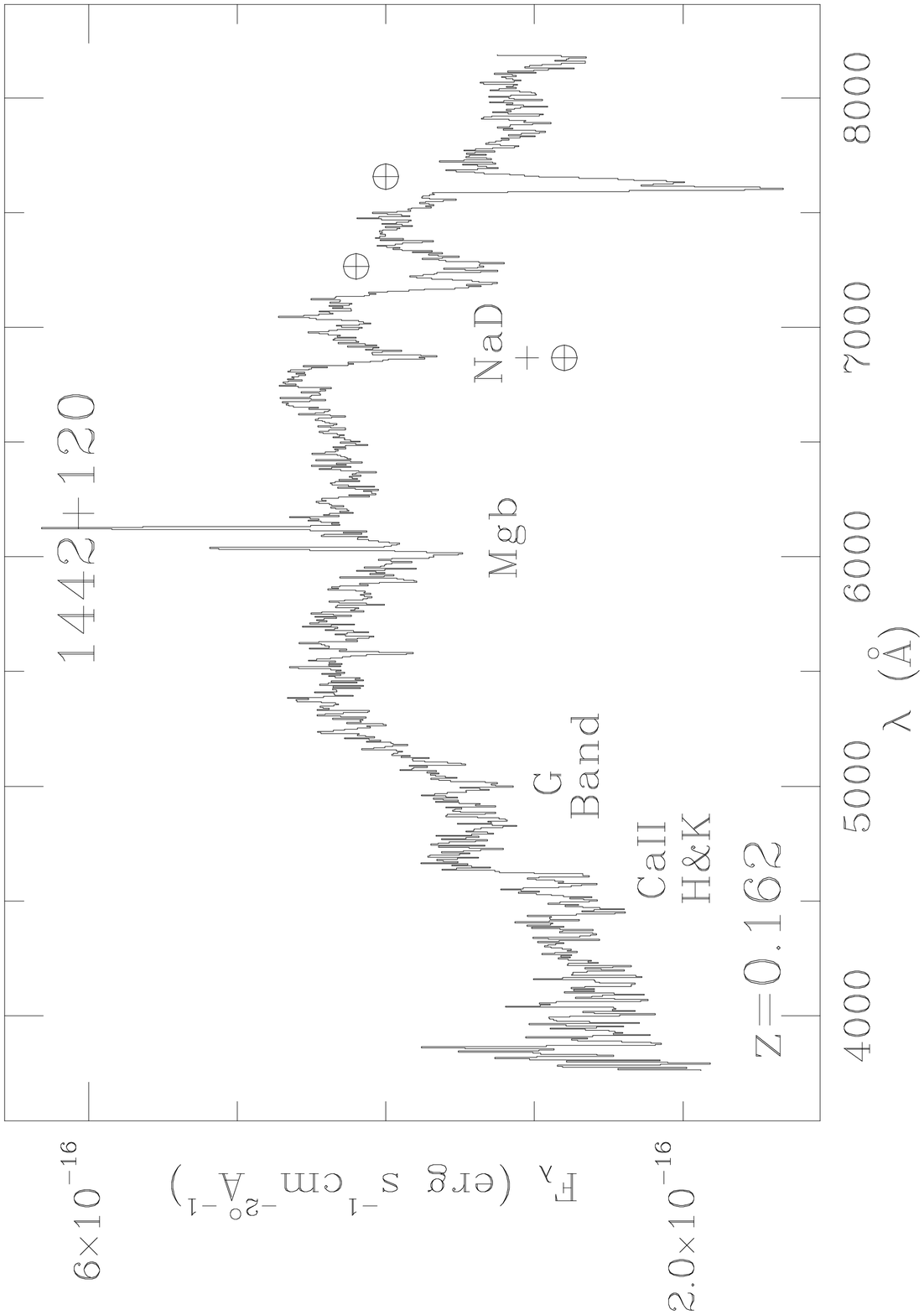,height=7.1cm,width=6.3cm}
\vspace{0.25in}
\psfig{file=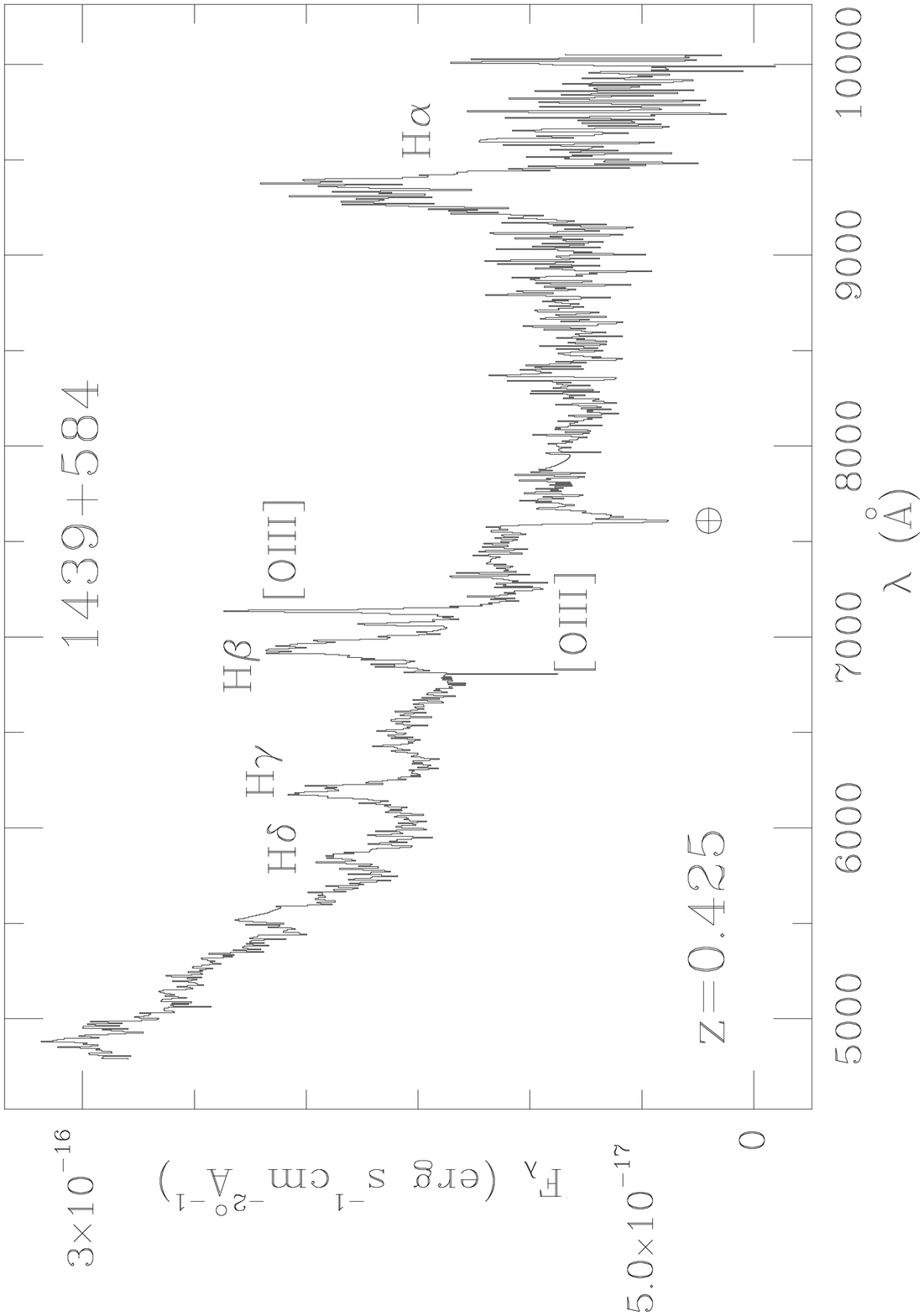,height=7.1cm,width=6.3cm}
\end{minipage}
\hspace{0.3in}
\begin{minipage}[t]{6.3in}
\psfig{file=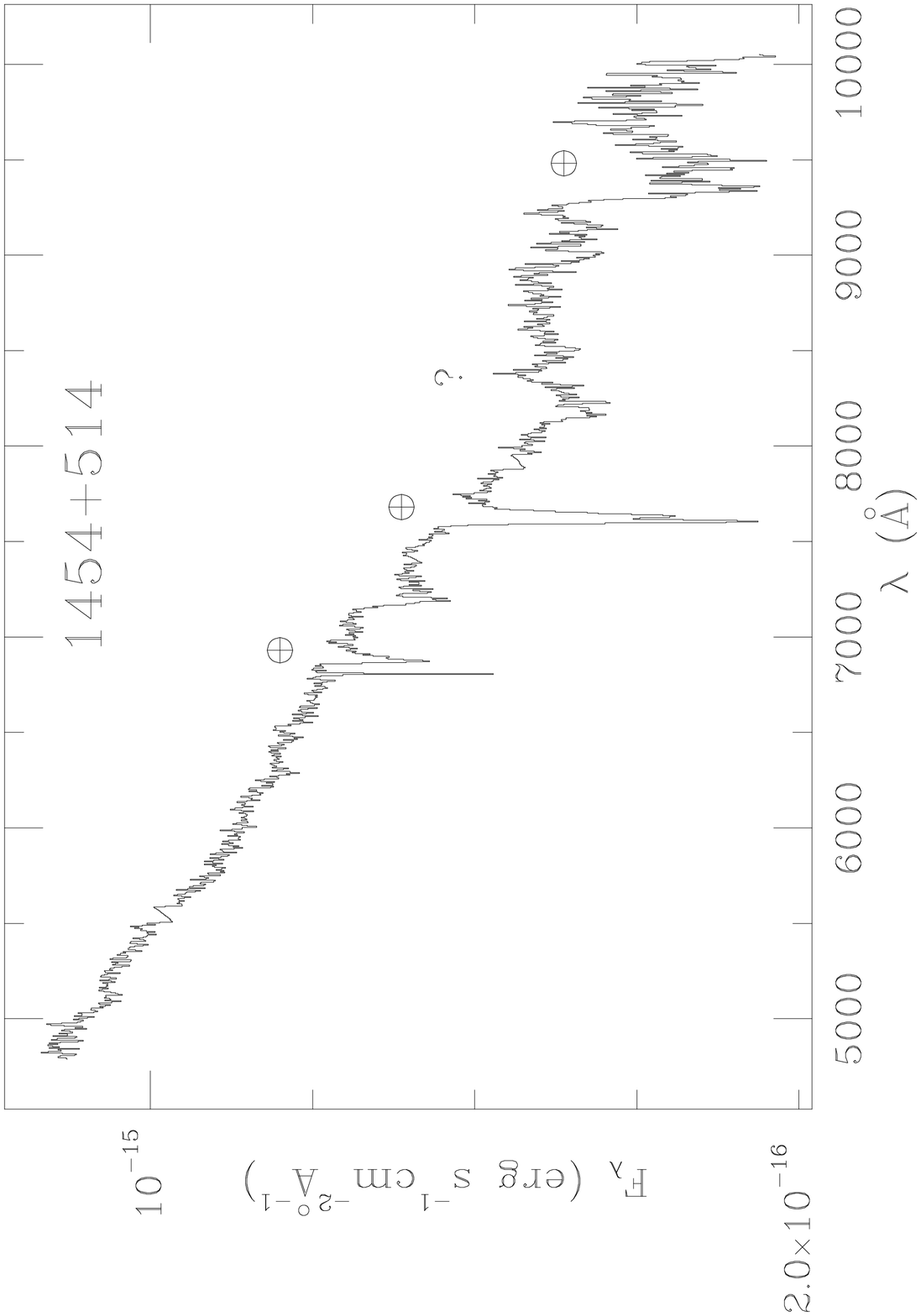,height=7.1cm,width=6.3cm}
\vspace{0.25in}
\psfig{file=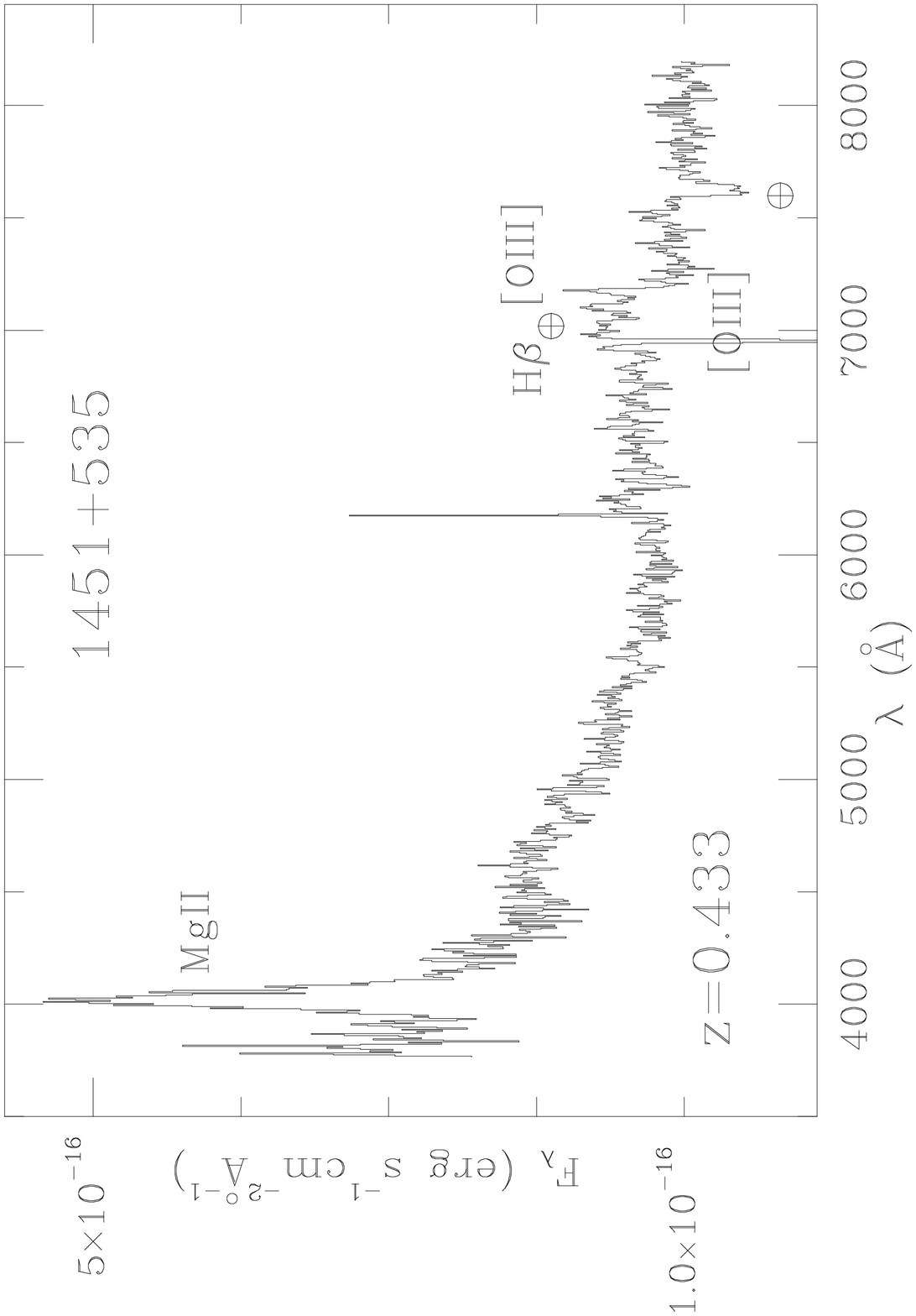,height=7.1cm,width=6.3cm}
\vspace{0.25in}
\psfig{file=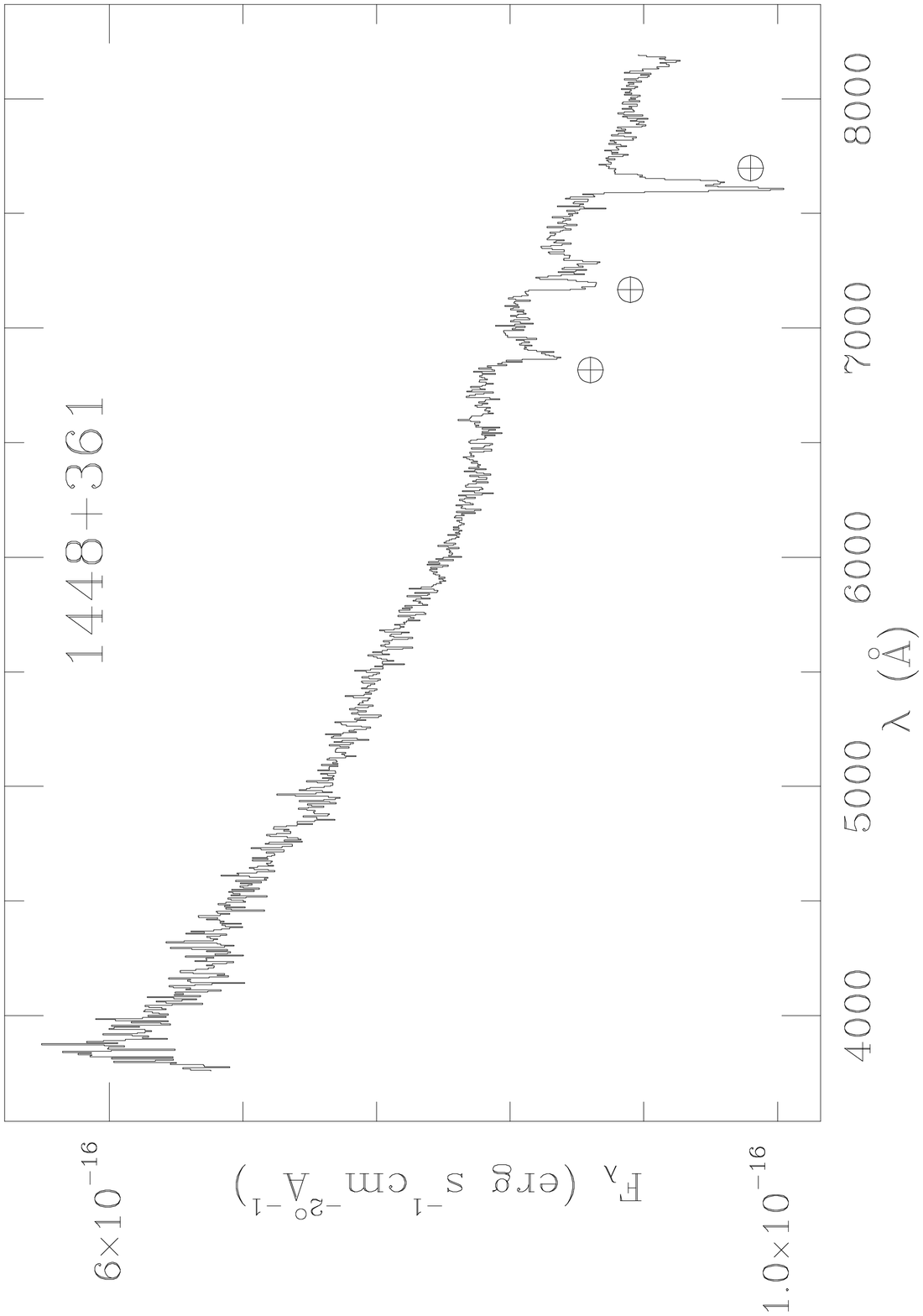,height=7.1cm,width=6.3cm}
\end{minipage}
\hfill
\begin{minipage}[t]{0.3in}
\vfill
\begin{sideways}
Figure 1.79 $-$ 1.84: Spectra of RGB Sources ({\it continued})
\end{sideways}
\vfill
\end{minipage}
\end{figure}

\clearpage
\begin{figure}
\vspace{-0.3in}
\hspace{-0.3in}
\begin{minipage}[t]{6.3in}
\psfig{file=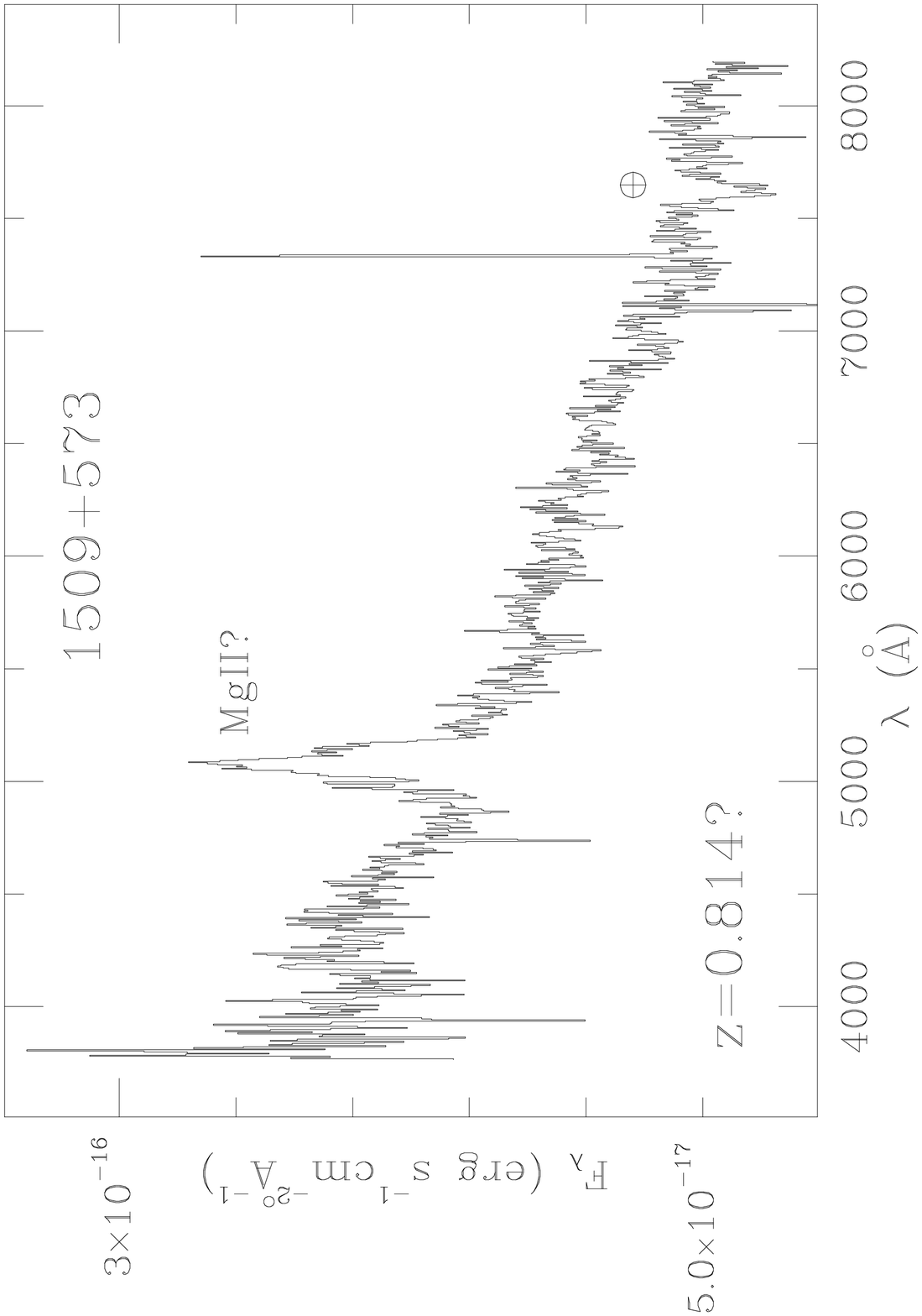,height=7.1cm,width=6.3cm}
\vspace{0.25in}
\psfig{file=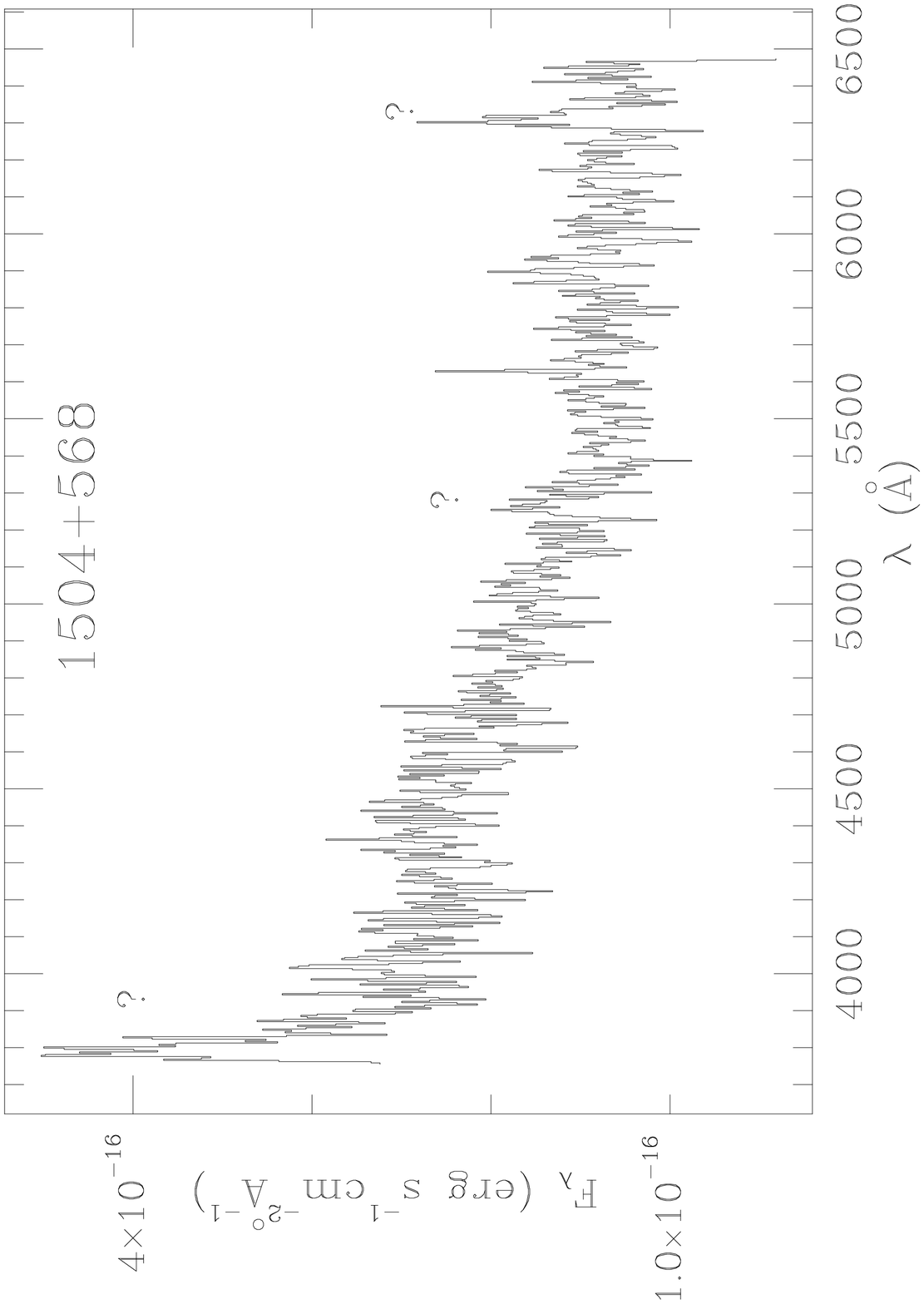,height=7.1cm,width=6.3cm}
\vspace{0.25in}
\psfig{file=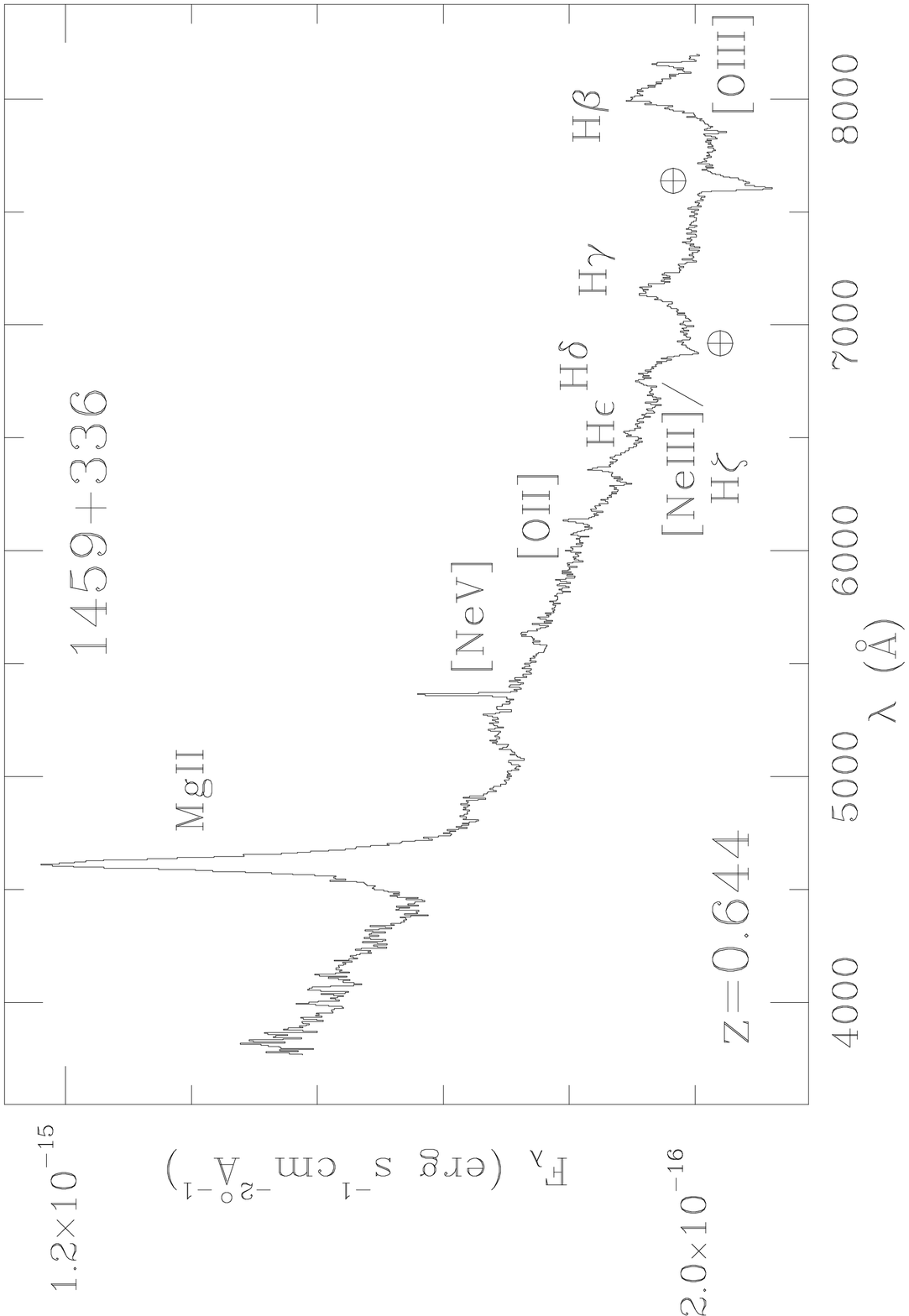,height=7.1cm,width=6.3cm}
\end{minipage}
\hspace{0.3in}
\begin{minipage}[t]{6.3in}
\psfig{file=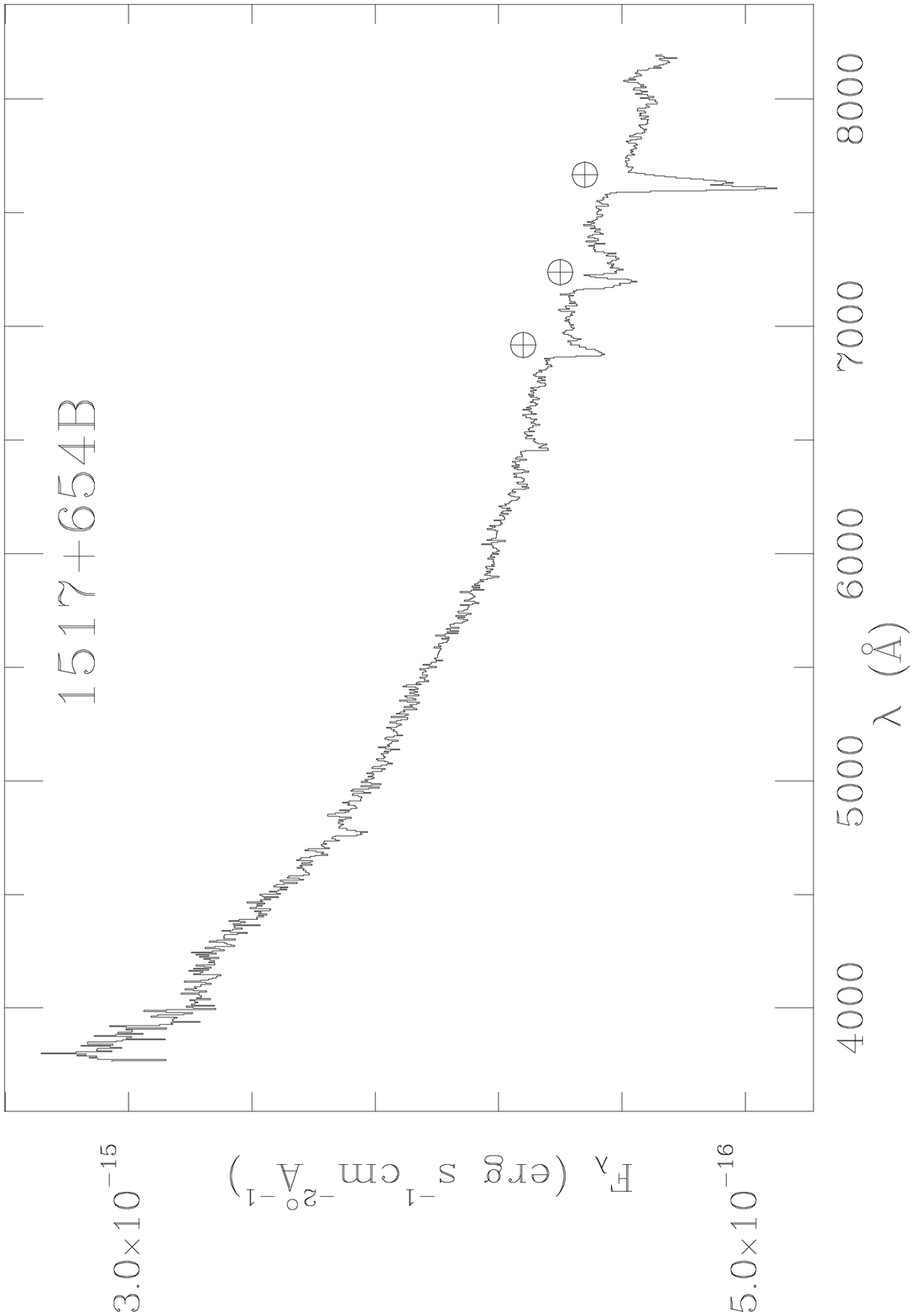,height=7.1cm,width=6.3cm}
\vspace{0.25in}
\psfig{file=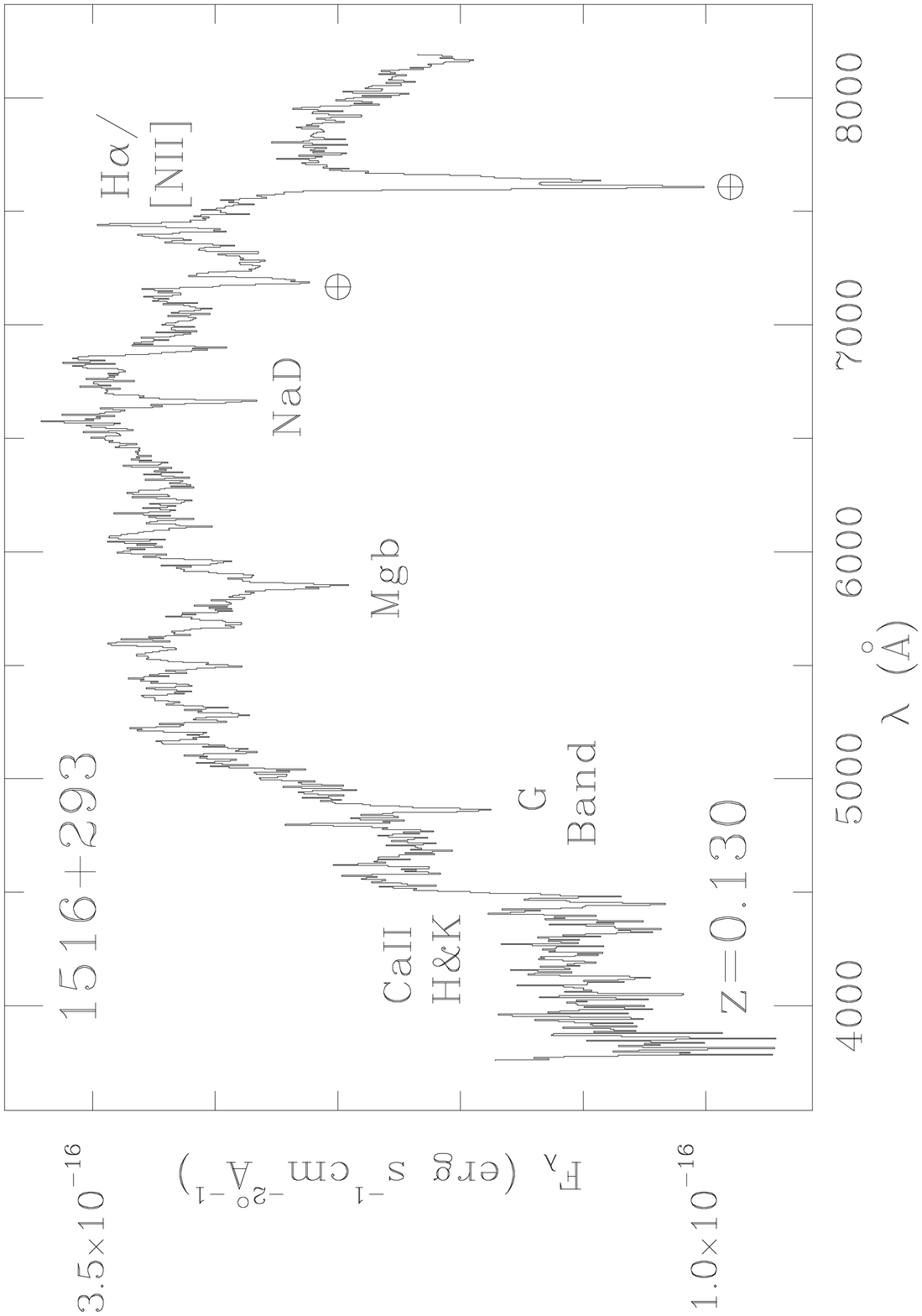,height=7.1cm,width=6.3cm}
\vspace{0.25in}
\psfig{file=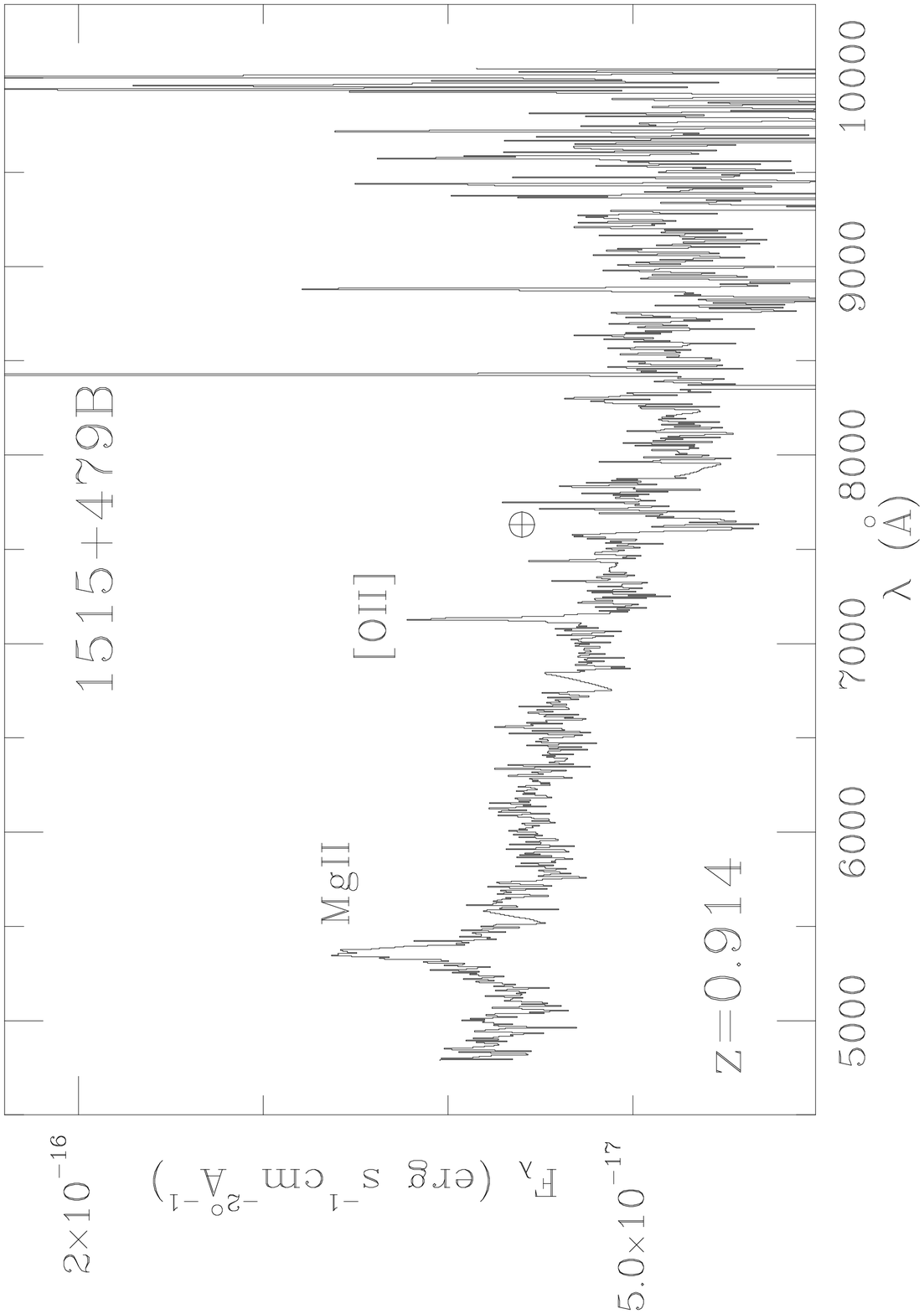,height=7.1cm,width=6.3cm}
\vspace{0.25in}
\end{minipage}
\hfill
\begin{minipage}[t]{0.3in}
\vfill
\begin{sideways}
Figure 1.85 $-$ 1.90: Spectra of RGB Sources ({\it continued})
\end{sideways}
\vfill
\end{minipage}
\end{figure}

\clearpage
\begin{figure}
\vspace{-0.3in}
\hspace{-0.3in}
\begin{minipage}[t]{6.3in}
\psfig{file=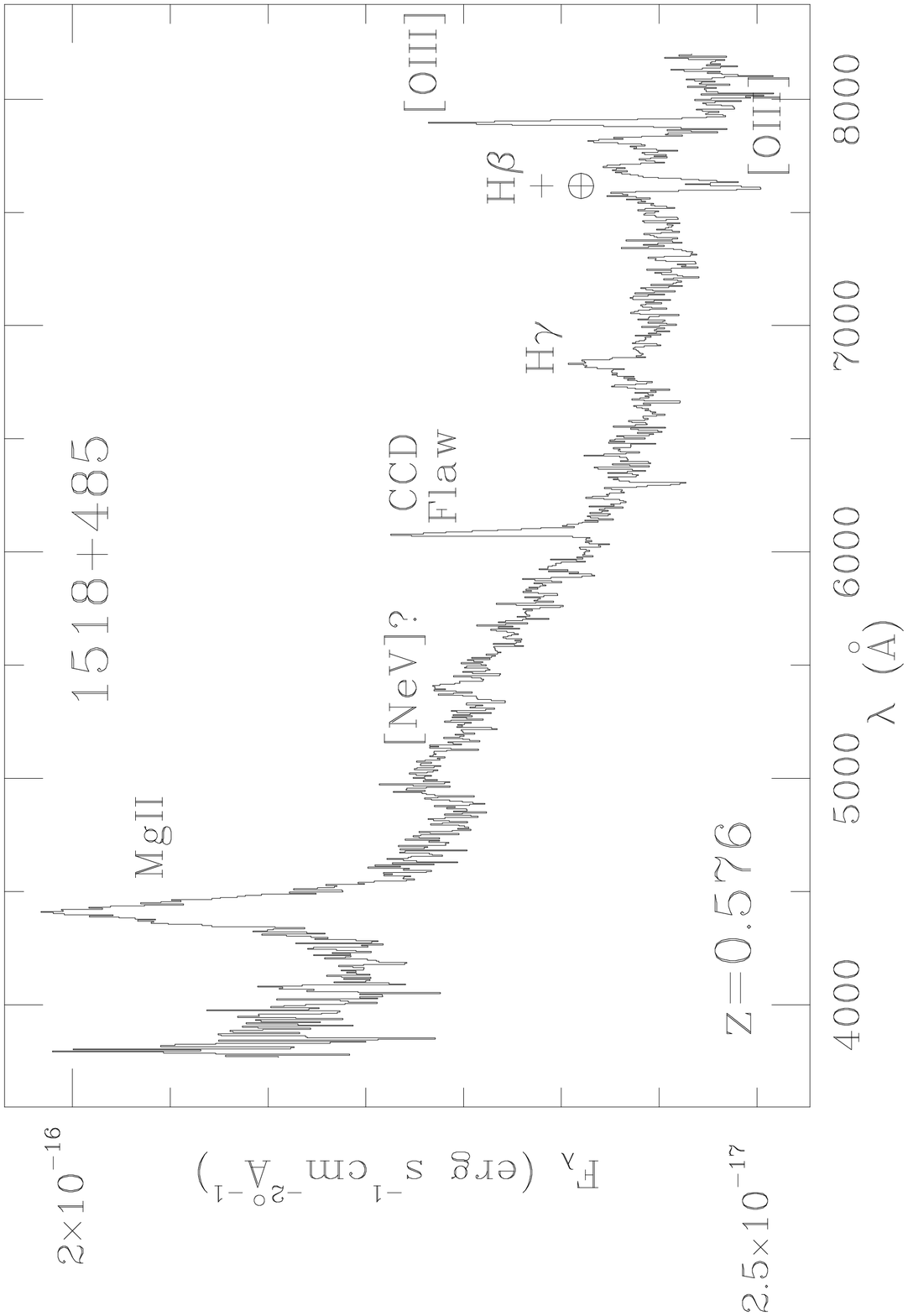,height=7.1cm,width=6.3cm}
\vspace{0.25in}
\psfig{file=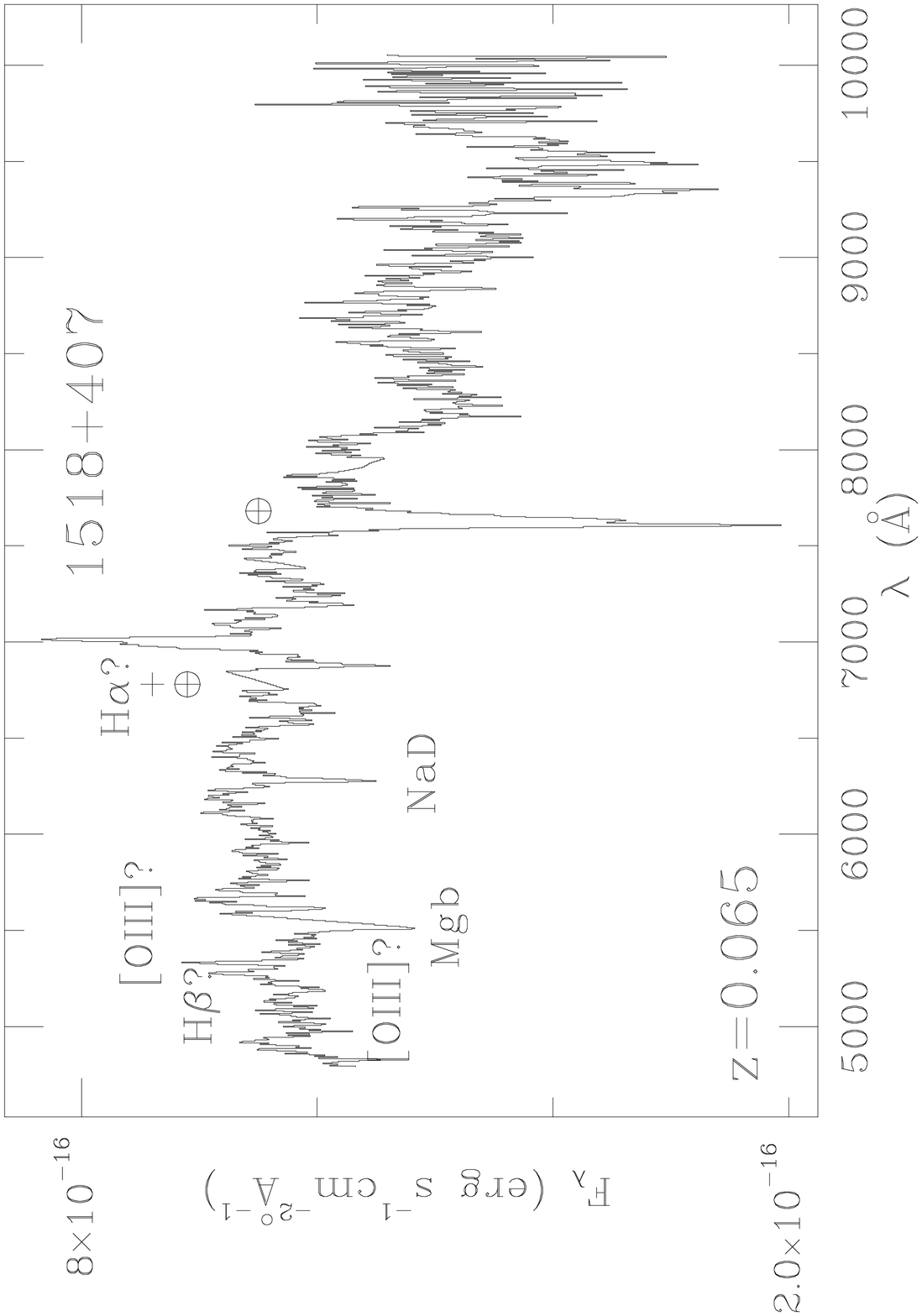,height=7.1cm,width=6.3cm}
\vspace{0.25in}
\psfig{file=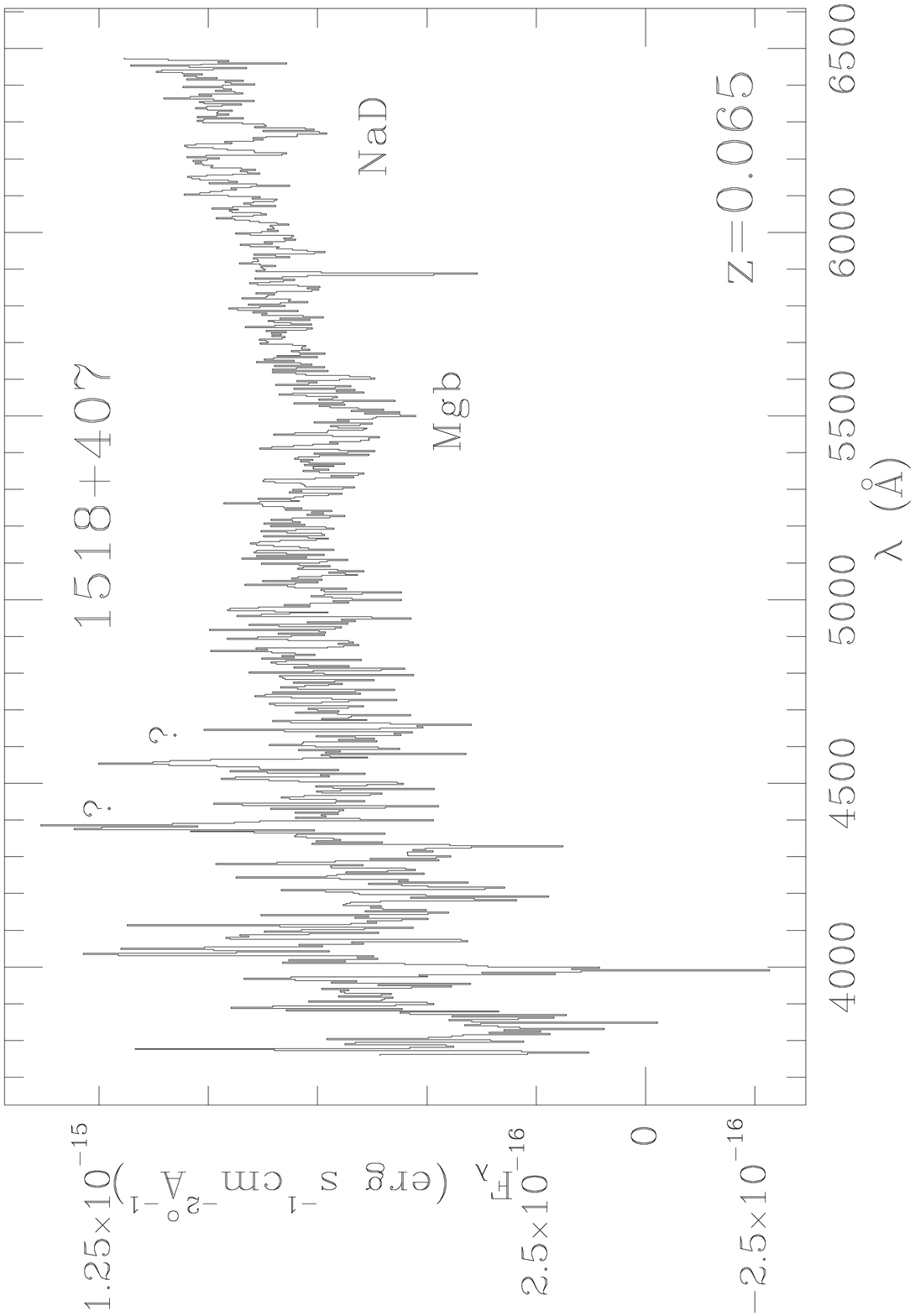,height=7.1cm,width=6.3cm}
\end{minipage}
\hspace{0.3in}
\begin{minipage}[t]{6.3in}
\psfig{file=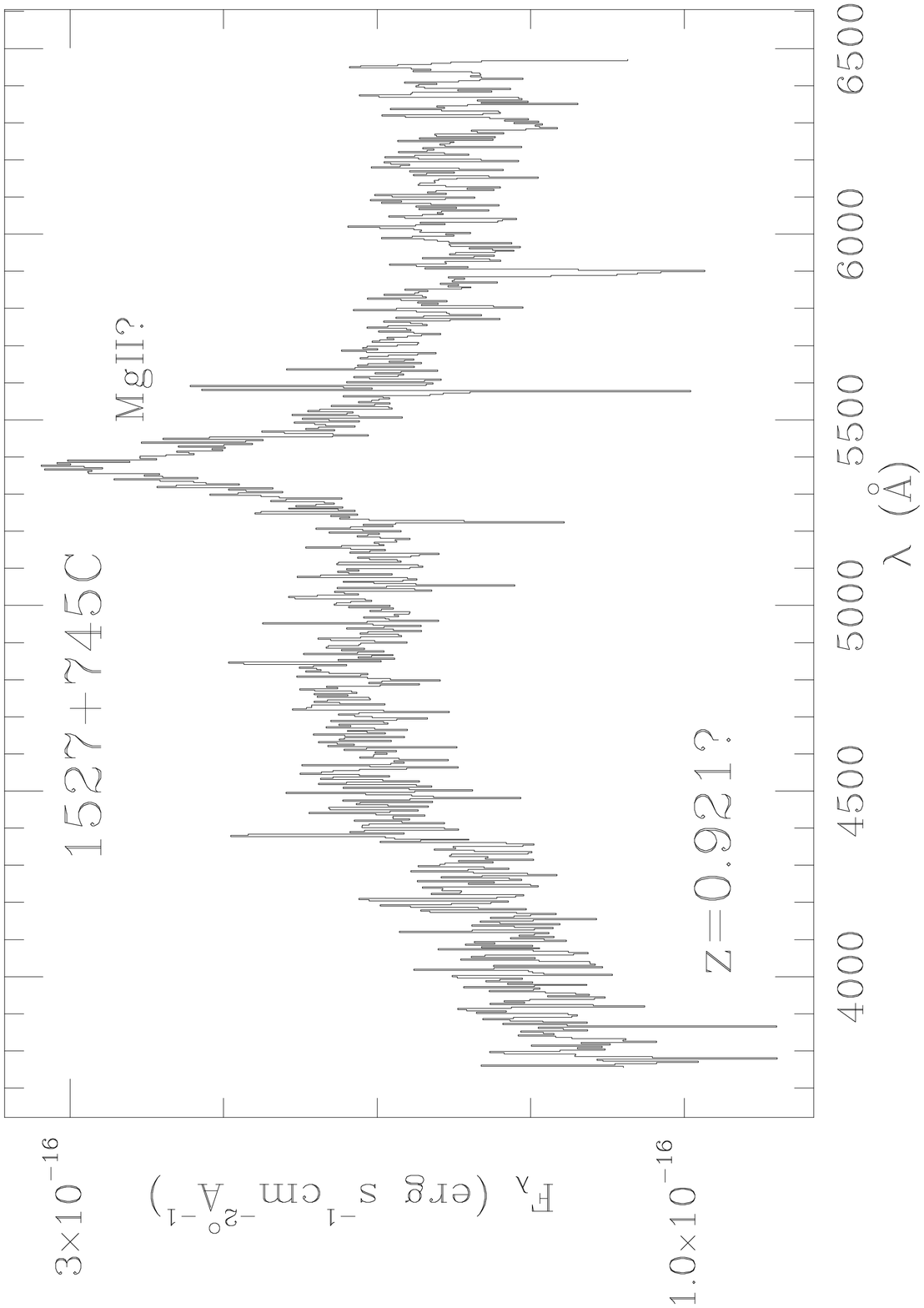,height=7.1cm,width=6.3cm}
\vspace{0.25in}
\psfig{file=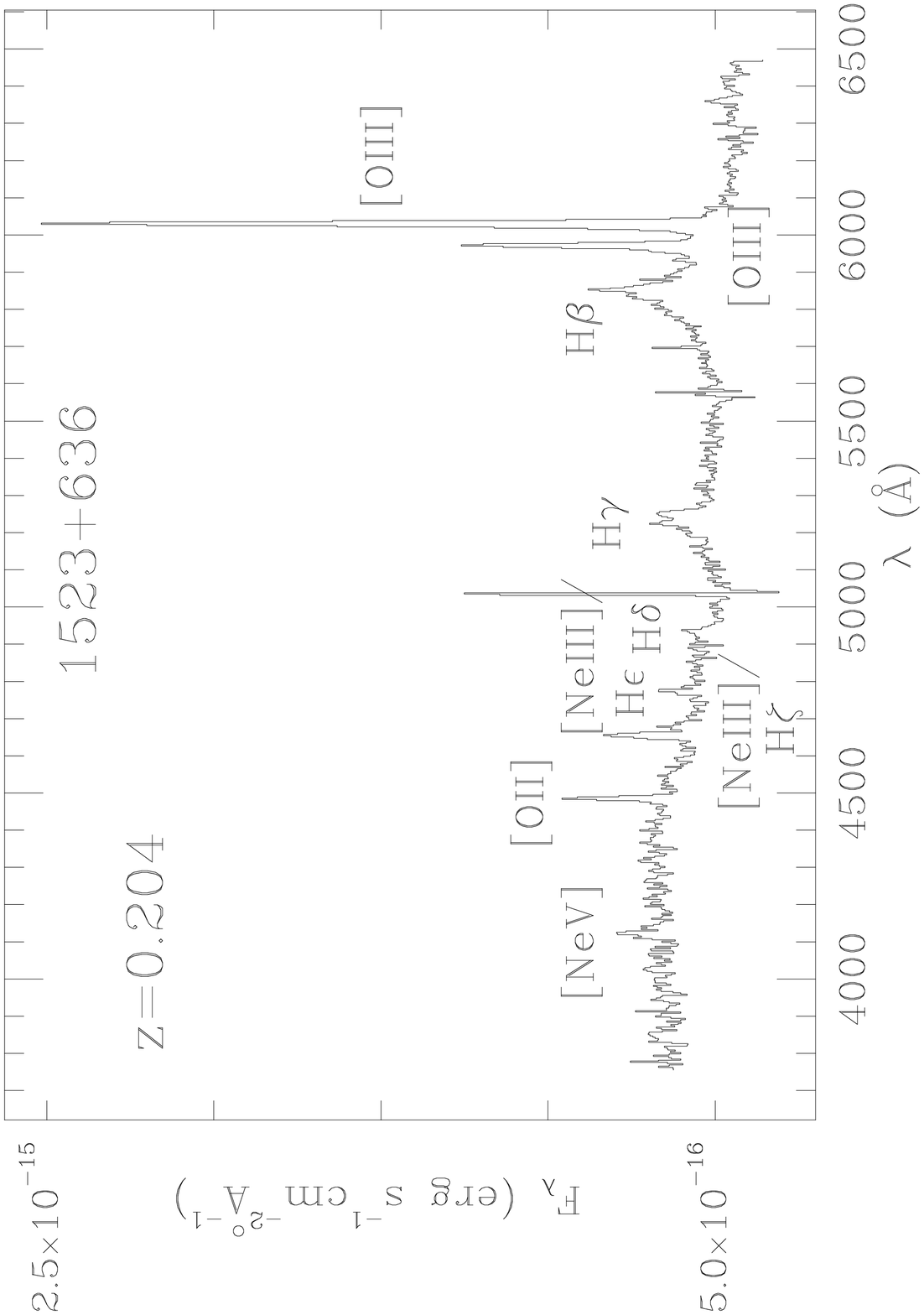,height=7.1cm,width=6.3cm}
\vspace{0.25in}
\psfig{file=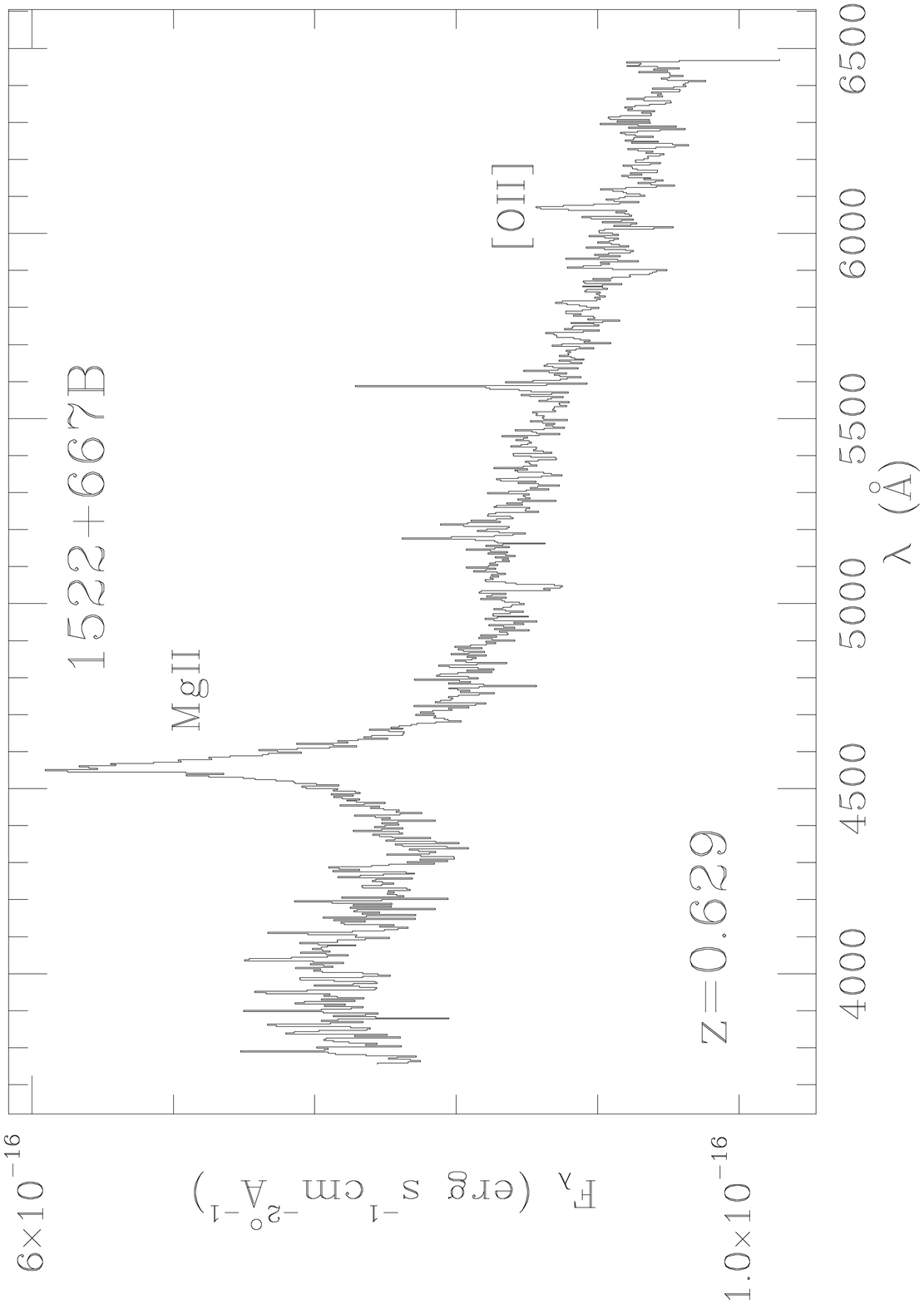,height=7.1cm,width=6.3cm}
\end{minipage}
\hfill
\begin{minipage}[t]{0.3in}
\vfill
\begin{sideways}
Figure 1.91 $-$ 1.96: Spectra of RGB Sources ({\it continued})
\end{sideways}
\vfill
\end{minipage}
\end{figure}

\clearpage
\begin{figure}
\vspace{-0.3in}
\hspace{-0.3in}
\begin{minipage}[t]{6.3in}
\psfig{file=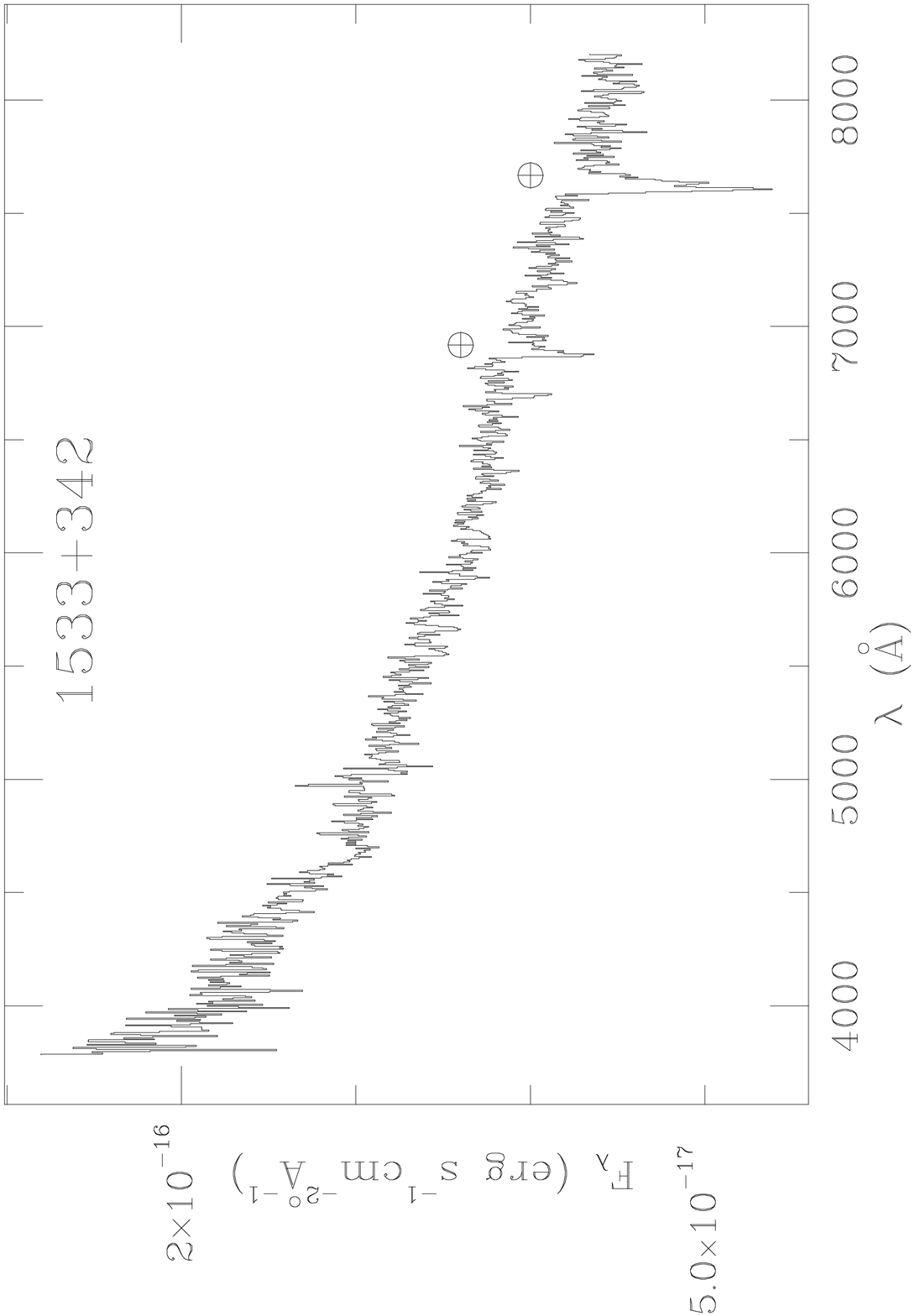,height=7.1cm,width=6.3cm}
\vspace{0.25in}
\psfig{file=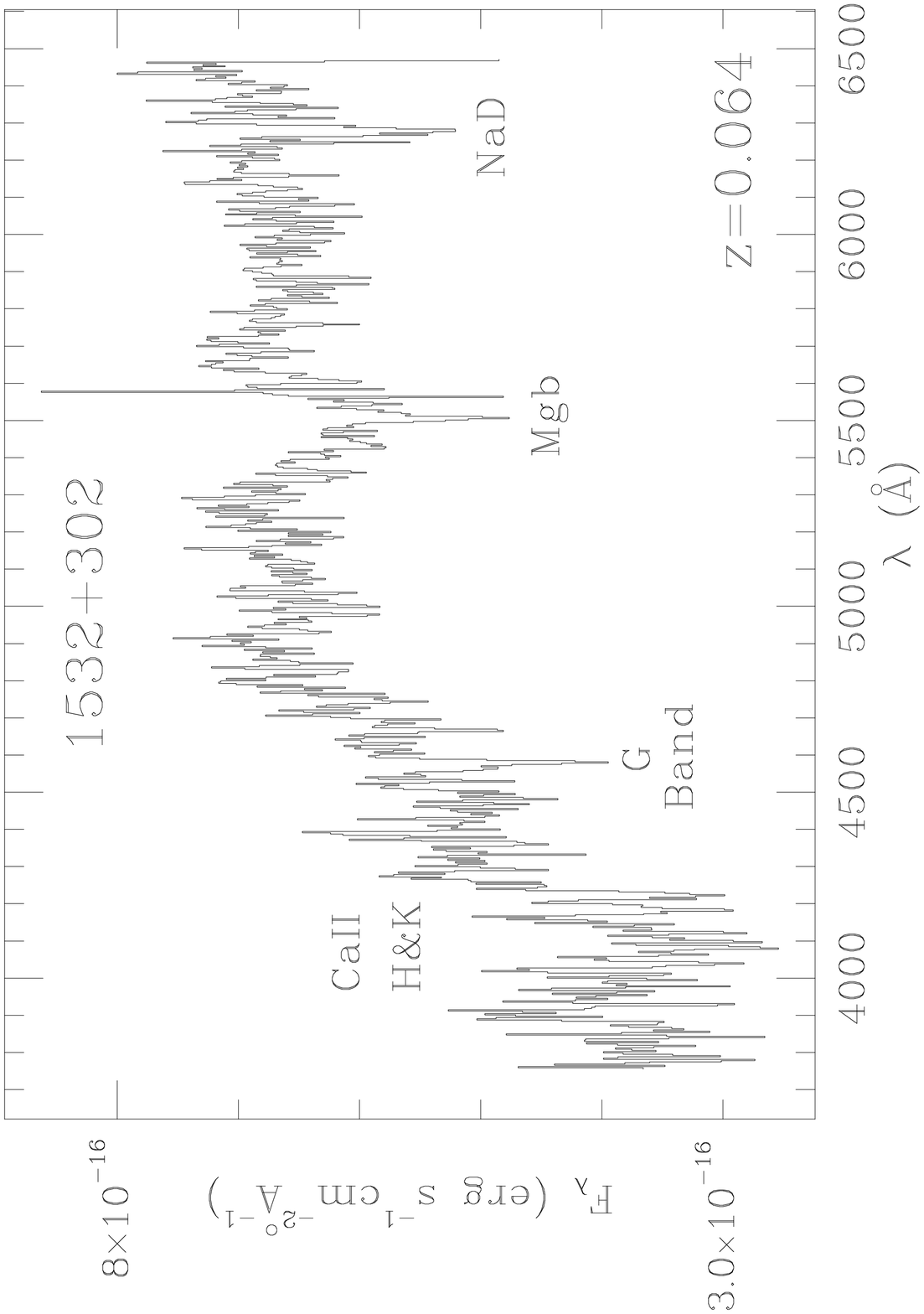,height=7.1cm,width=6.3cm}
\vspace{0.25in}
\psfig{file=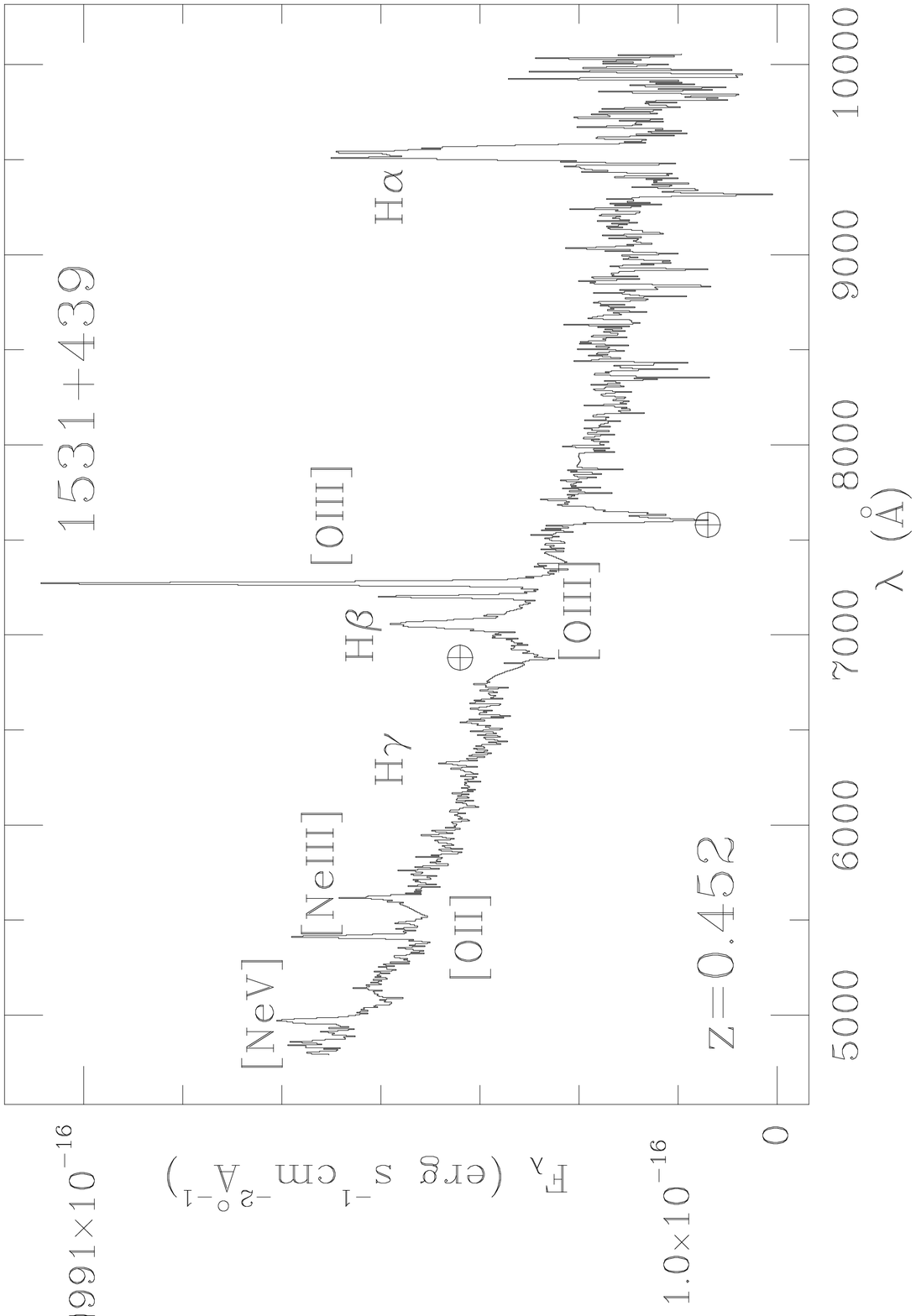,height=7.1cm,width=6.3cm}
\end{minipage}
\hspace{0.3in}
\begin{minipage}[t]{6.3in}
\psfig{file=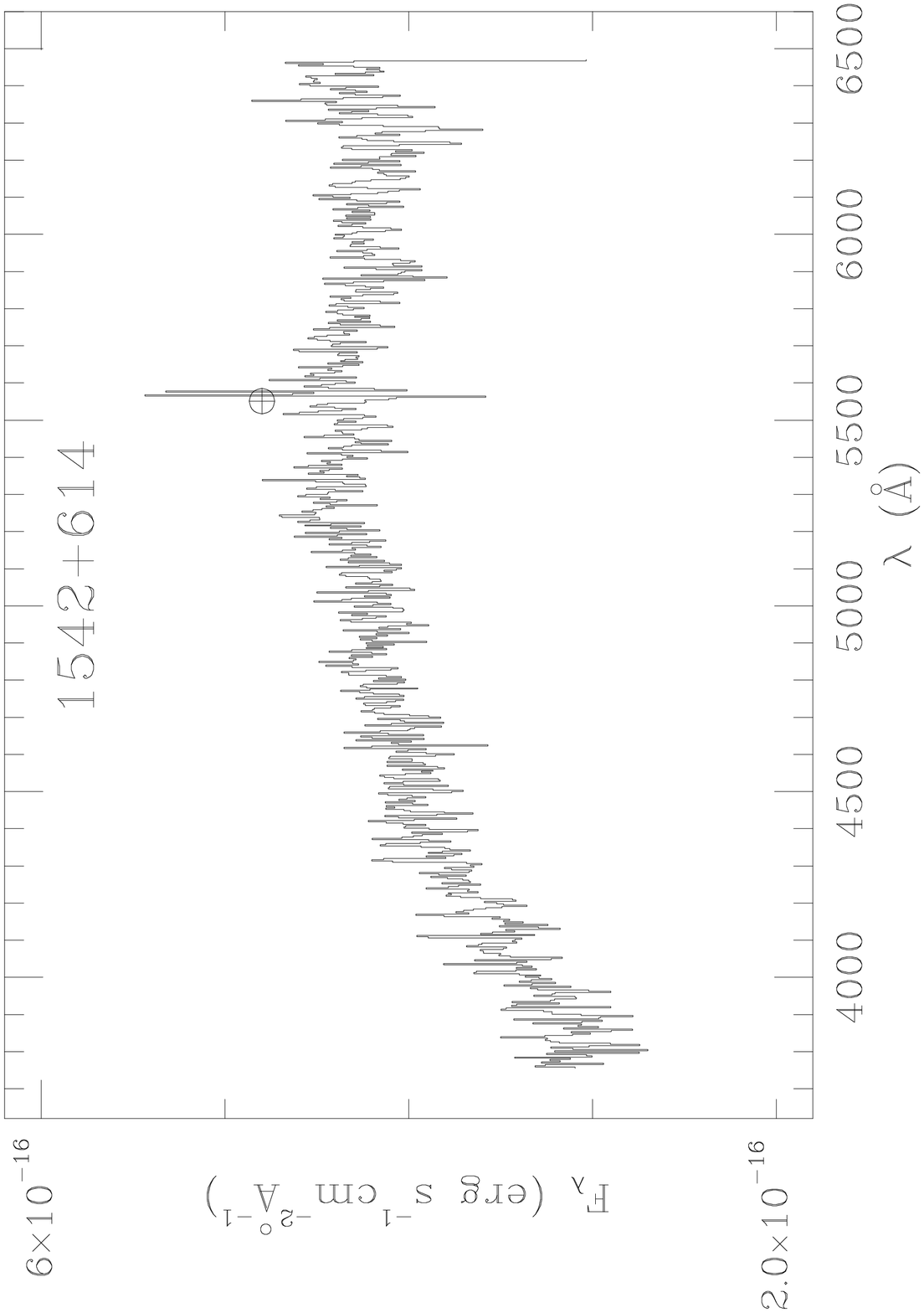,height=7.1cm,width=6.3cm}
\vspace{0.25in}
\psfig{file=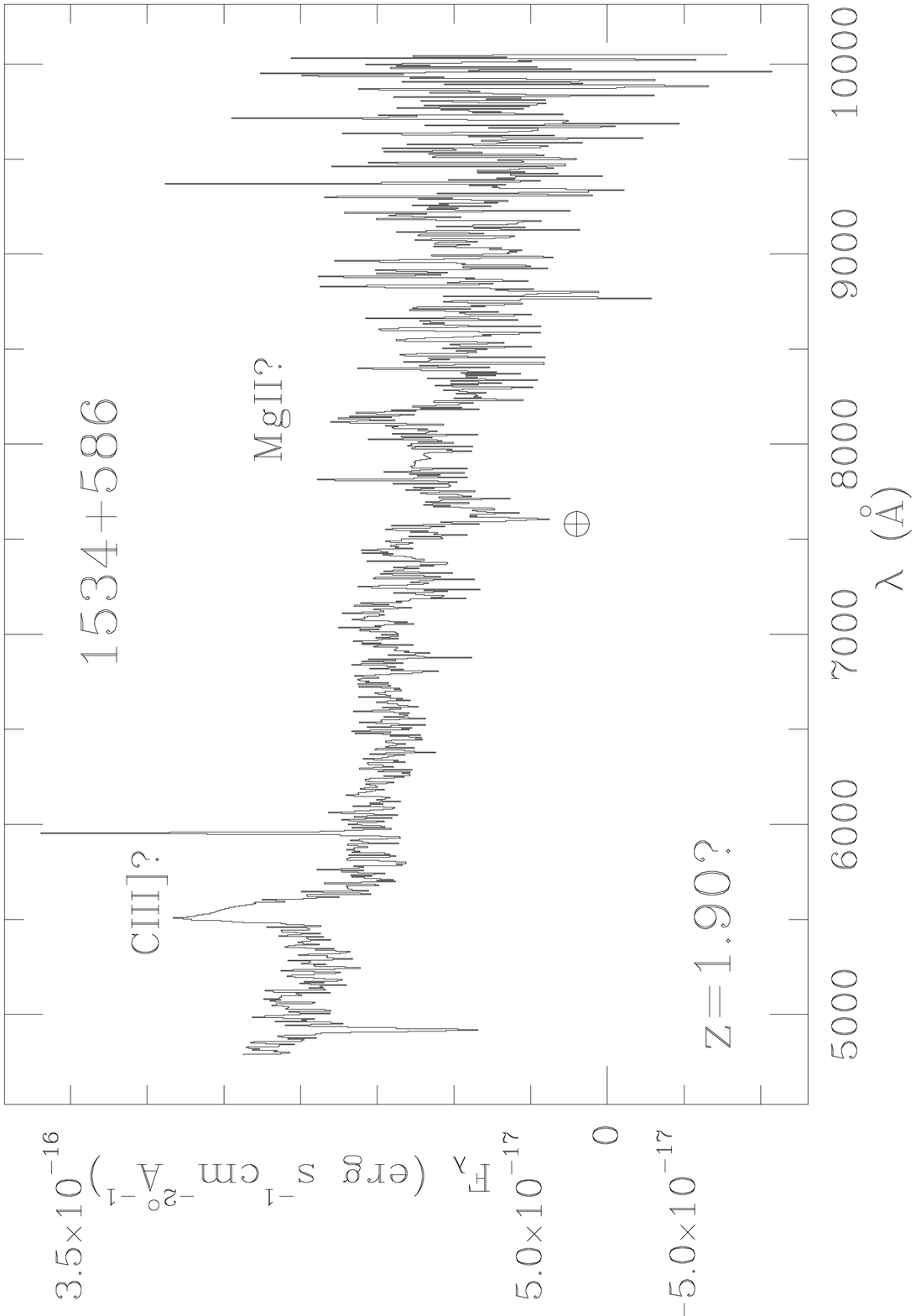,height=7.1cm,width=6.3cm}
\vspace{0.25in}
\psfig{file=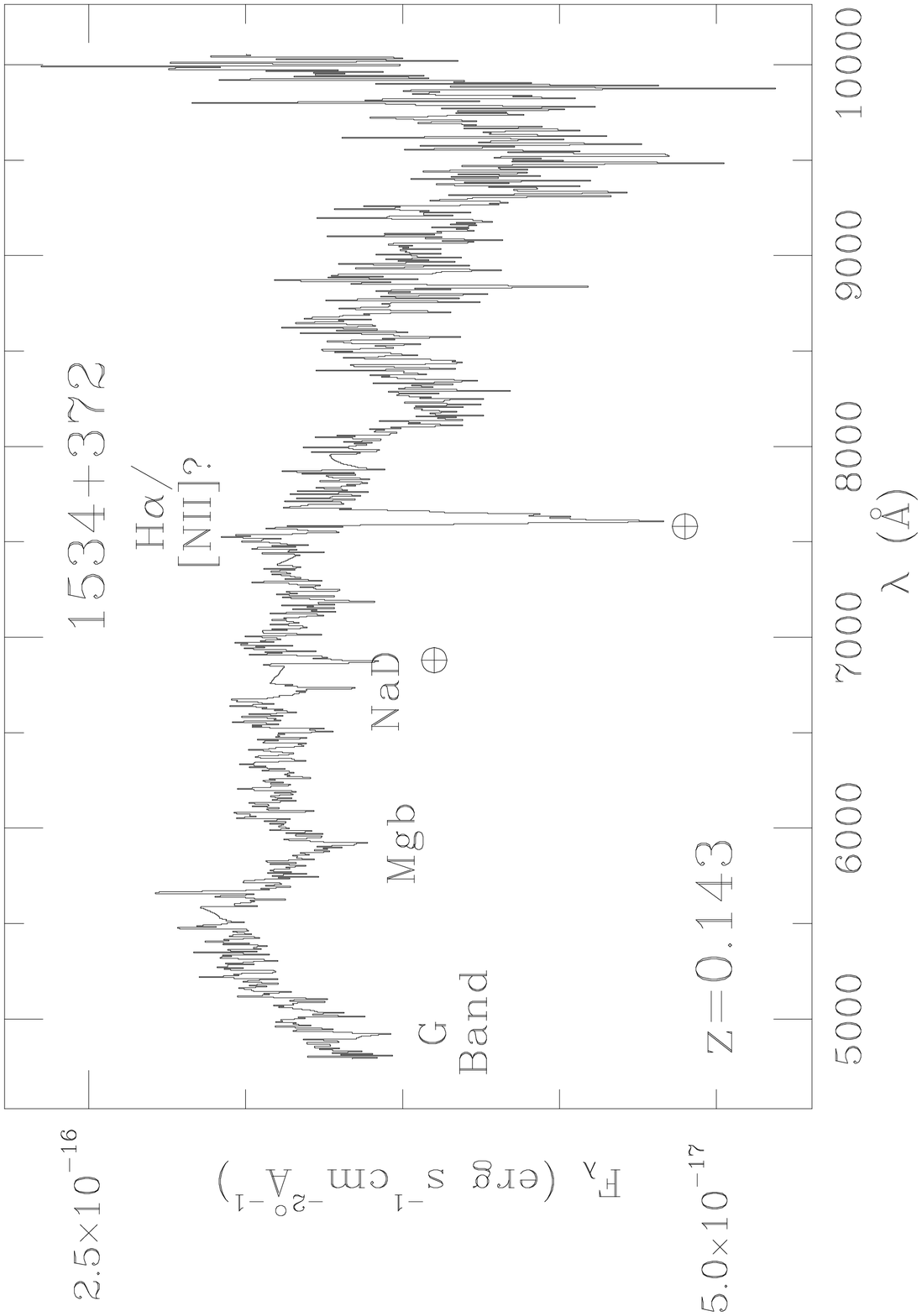,height=7.1cm,width=6.3cm}
\end{minipage}
\hfill
\begin{minipage}[t]{0.3in}
\vfill
\begin{sideways}
Figure 1.97 $-$ 1.102: Spectra of RGB Sources ({\it continued})
\end{sideways}
\vfill
\end{minipage}
\end{figure}

\clearpage
\begin{figure}
\vspace{-0.3in}
\hspace{-0.3in}
\begin{minipage}[t]{6.3in}
\psfig{file=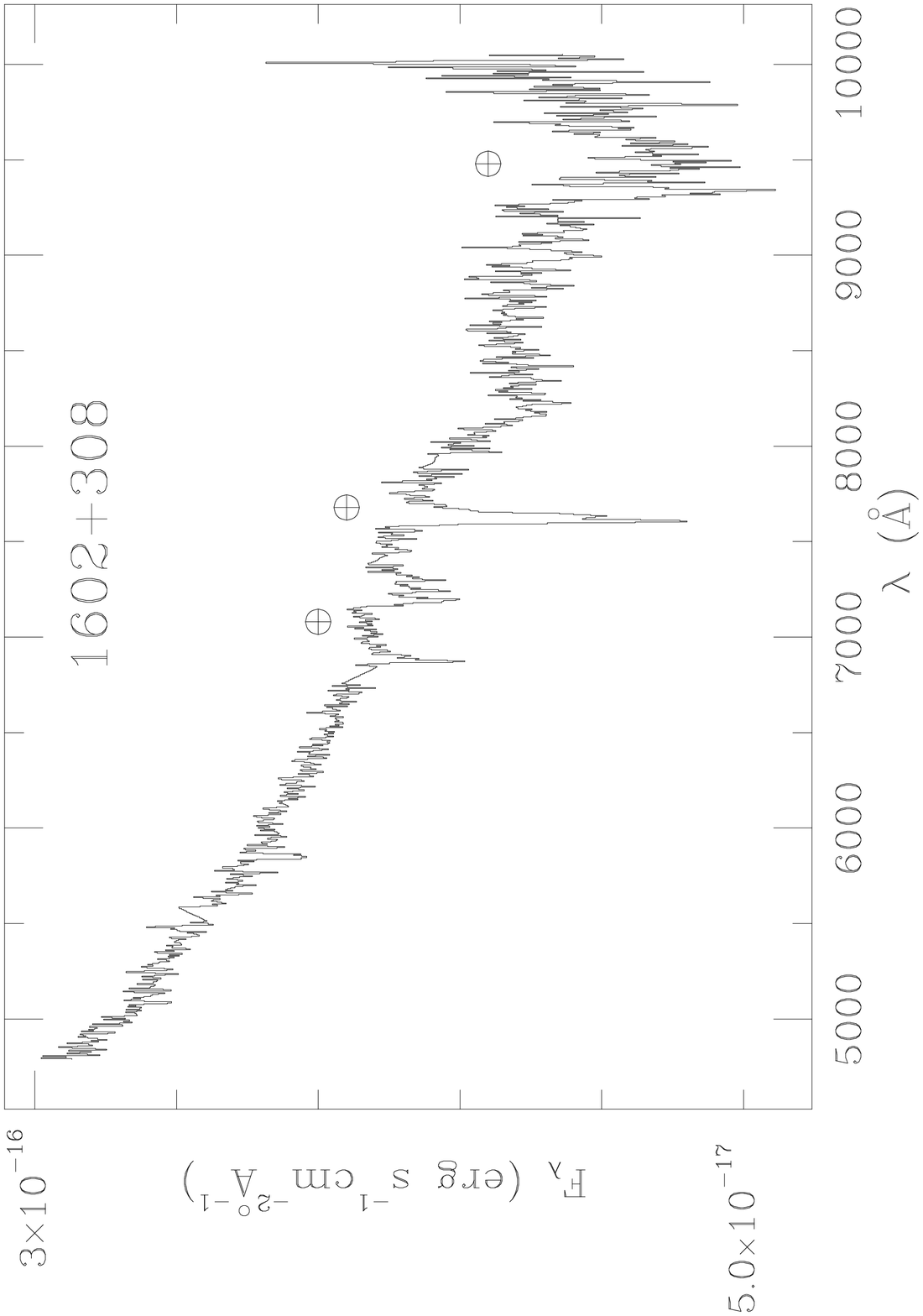,height=7.1cm,width=6.3cm}
\vspace{0.25in}
\psfig{file=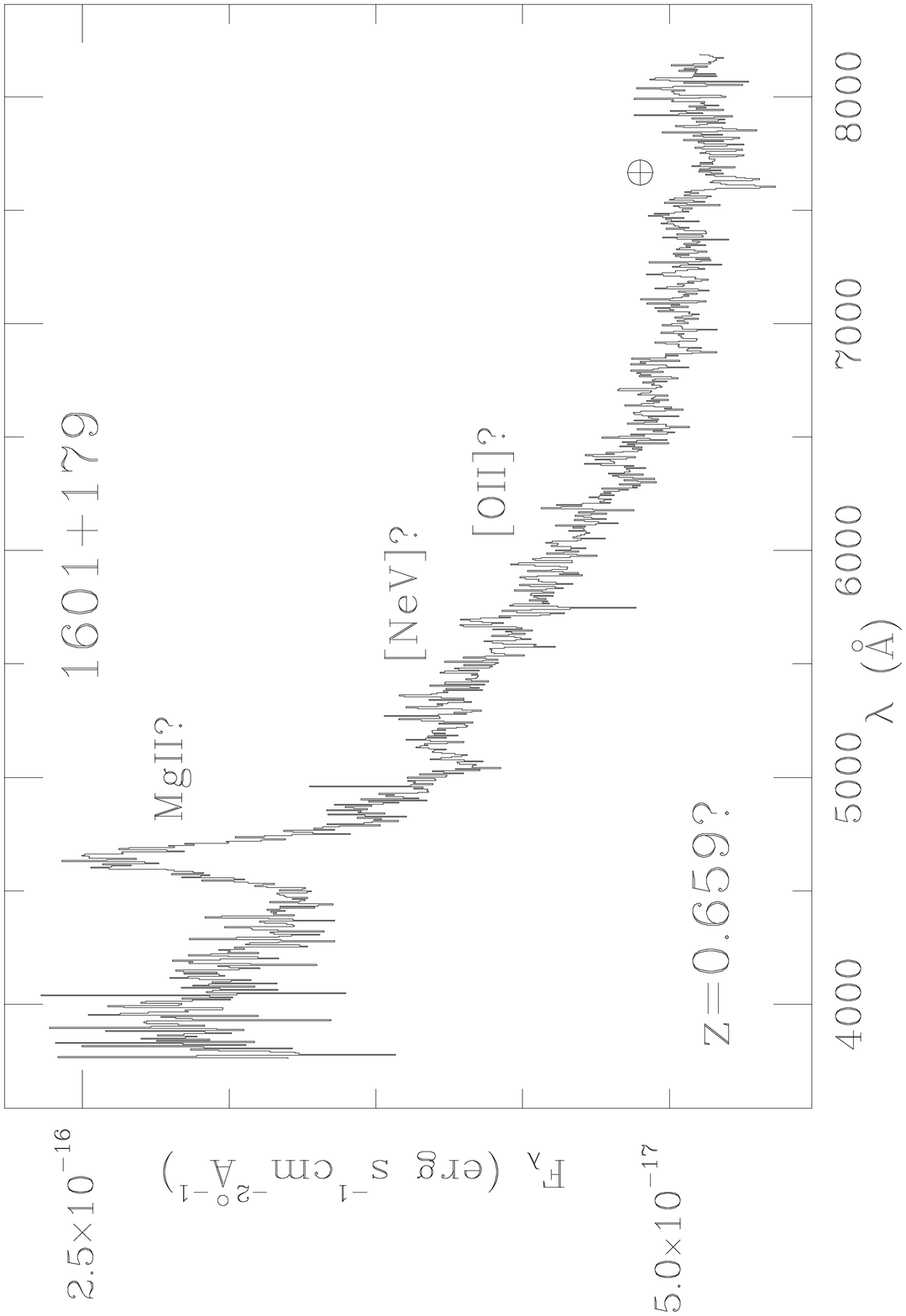,height=7.1cm,width=6.3cm}
\vspace{0.25in}
\psfig{file=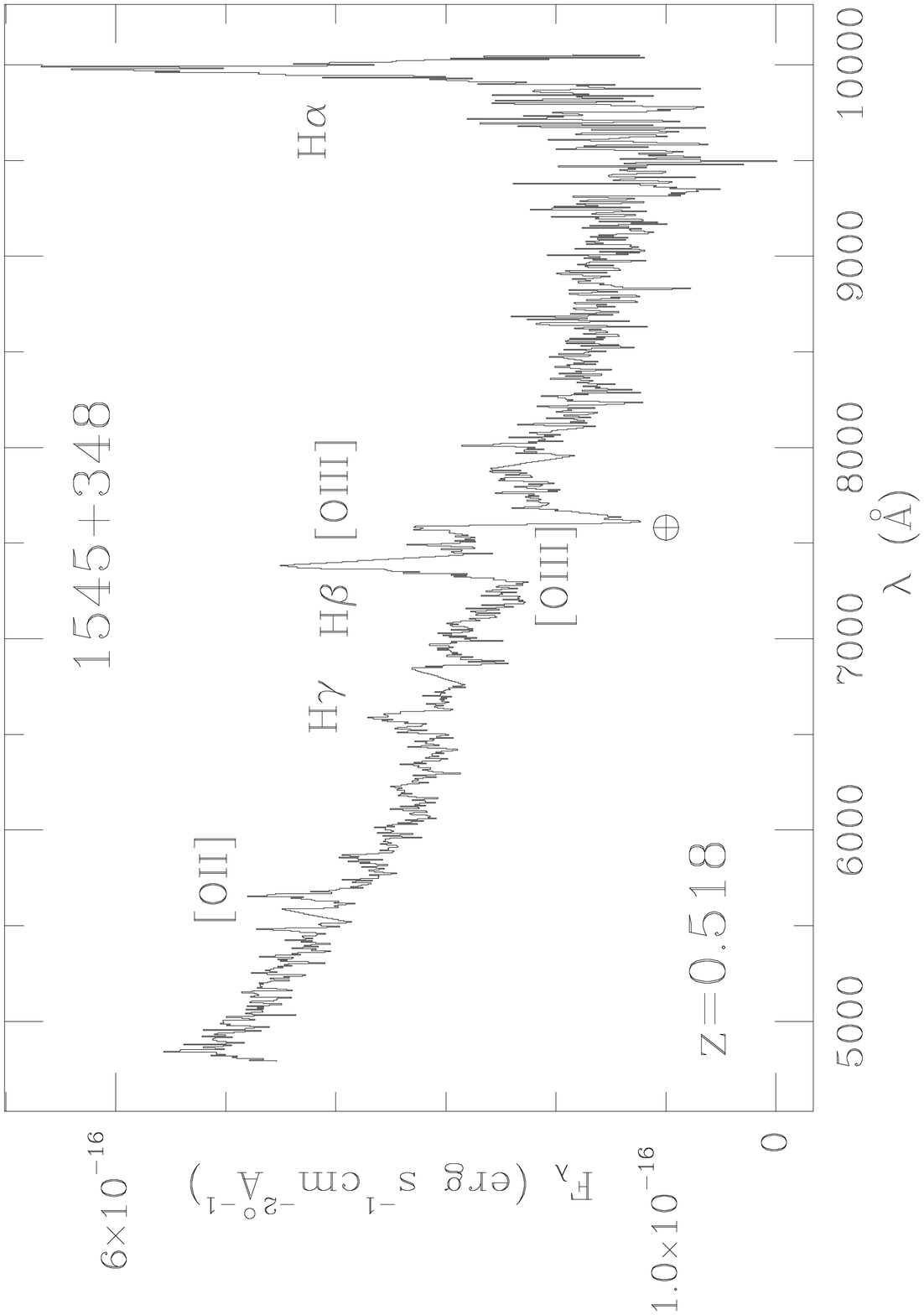,height=7.1cm,width=6.3cm}
\end{minipage}
\hspace{0.3in}
\begin{minipage}[t]{6.3in}
\psfig{file=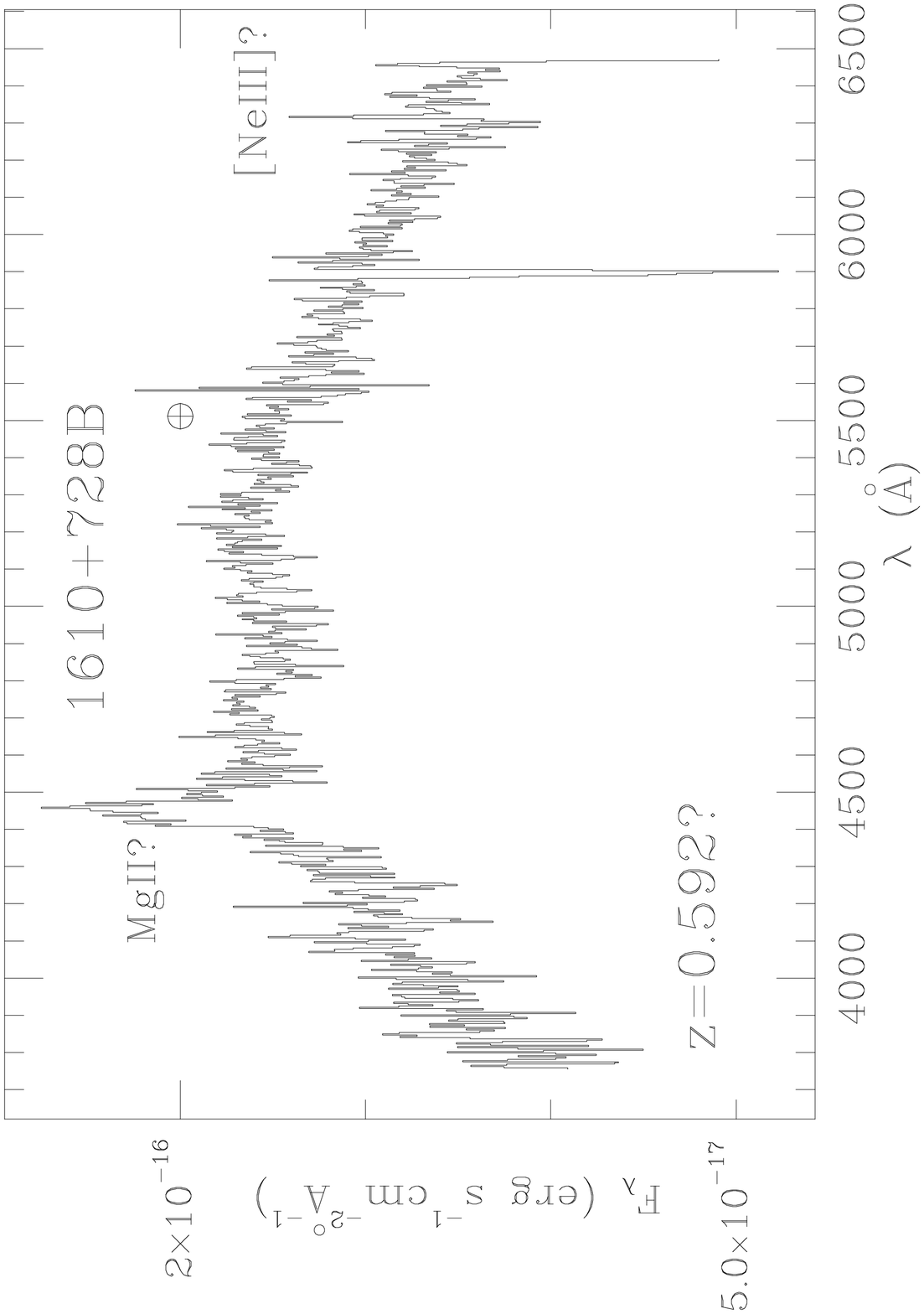,height=7.1cm,width=6.3cm}
\vspace{0.25in}
\psfig{file=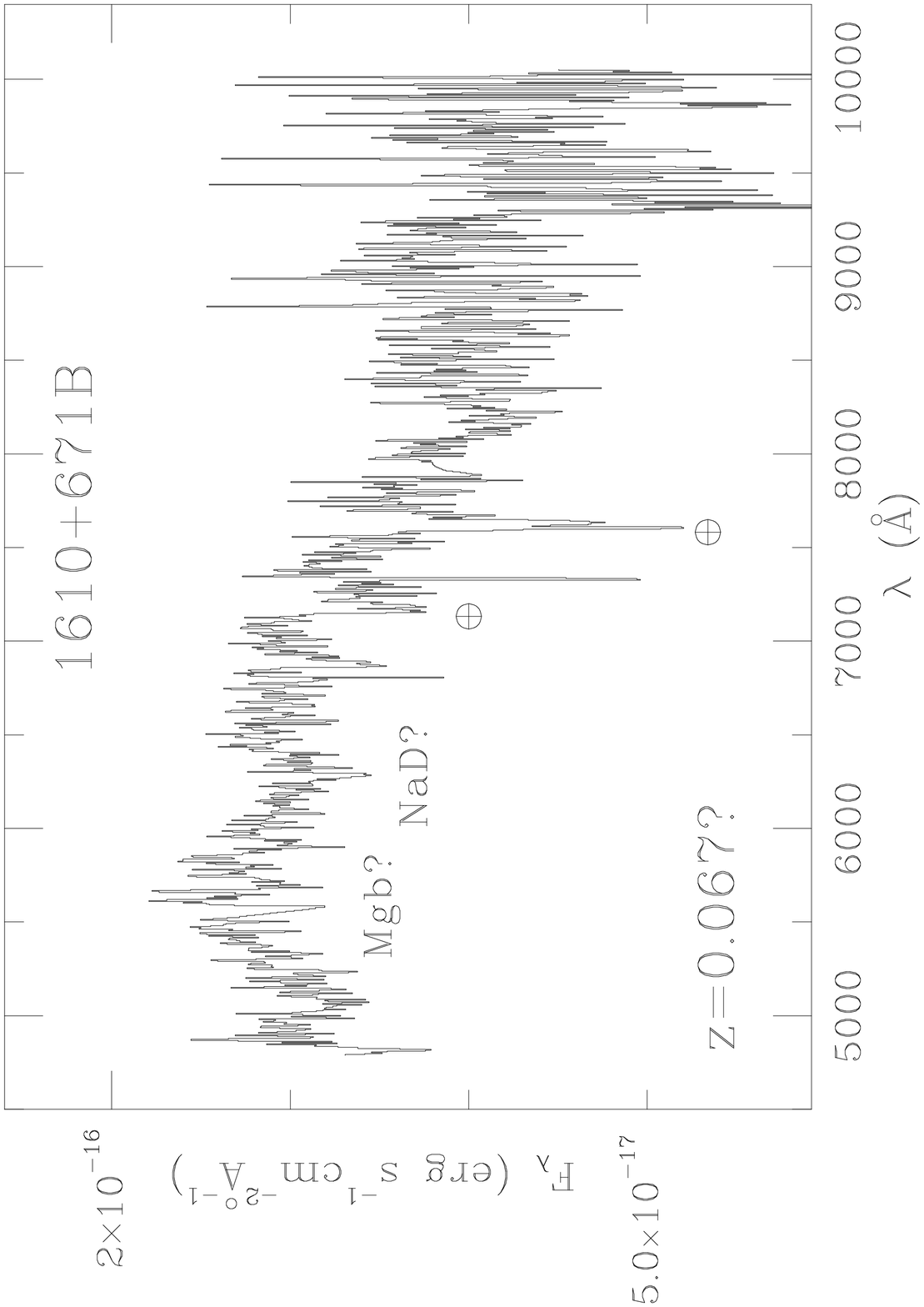,height=7.1cm,width=6.3cm}
\vspace{0.25in}
\psfig{file=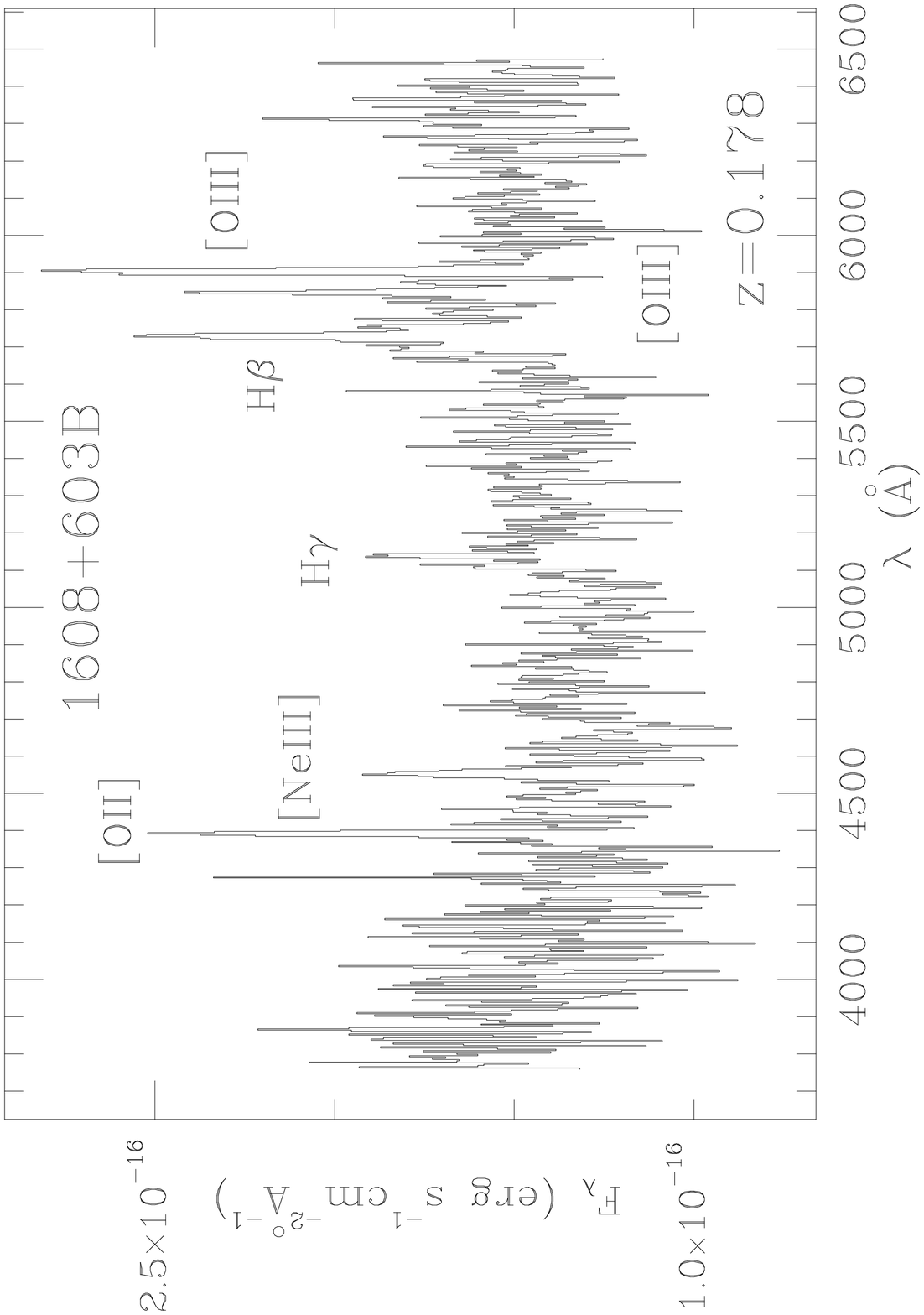,height=7.1cm,width=6.3cm}
\vspace{-7.21in}
\end{minipage}
\hfill
\begin{minipage}[t]{0.3in}
\vfill
\begin{sideways}
Figure 1.103 $-$ 1.108: Spectra of RGB Sources ({\it continued})
\end{sideways}
\vfill
\end{minipage}
\end{figure}

\clearpage
\begin{figure}
\vspace{-0.3in}
\hspace{-0.3in}
\begin{minipage}[t]{6.3in}
\psfig{file=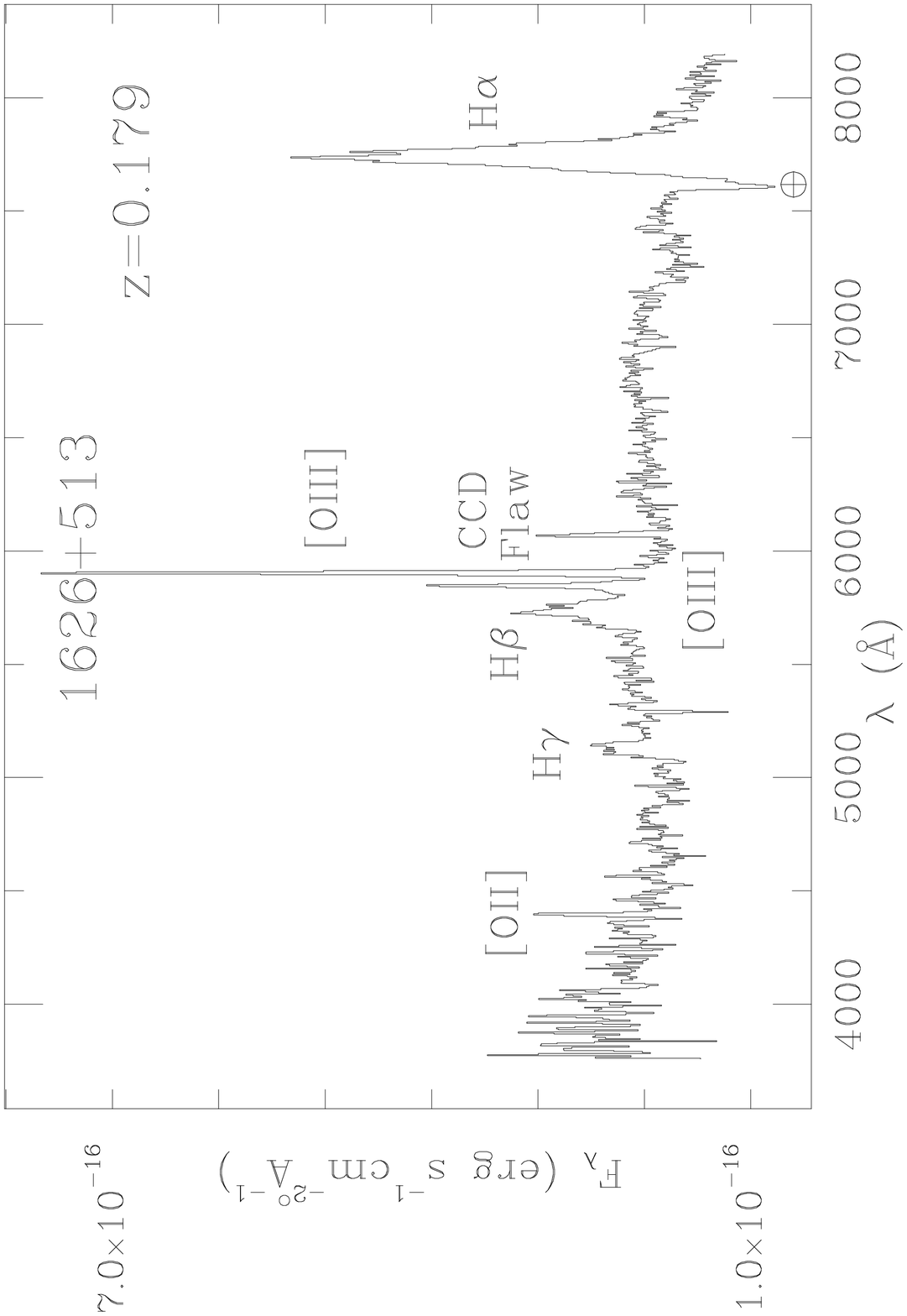,height=7.1cm,width=6.3cm}
\vspace{0.25in}
\psfig{file=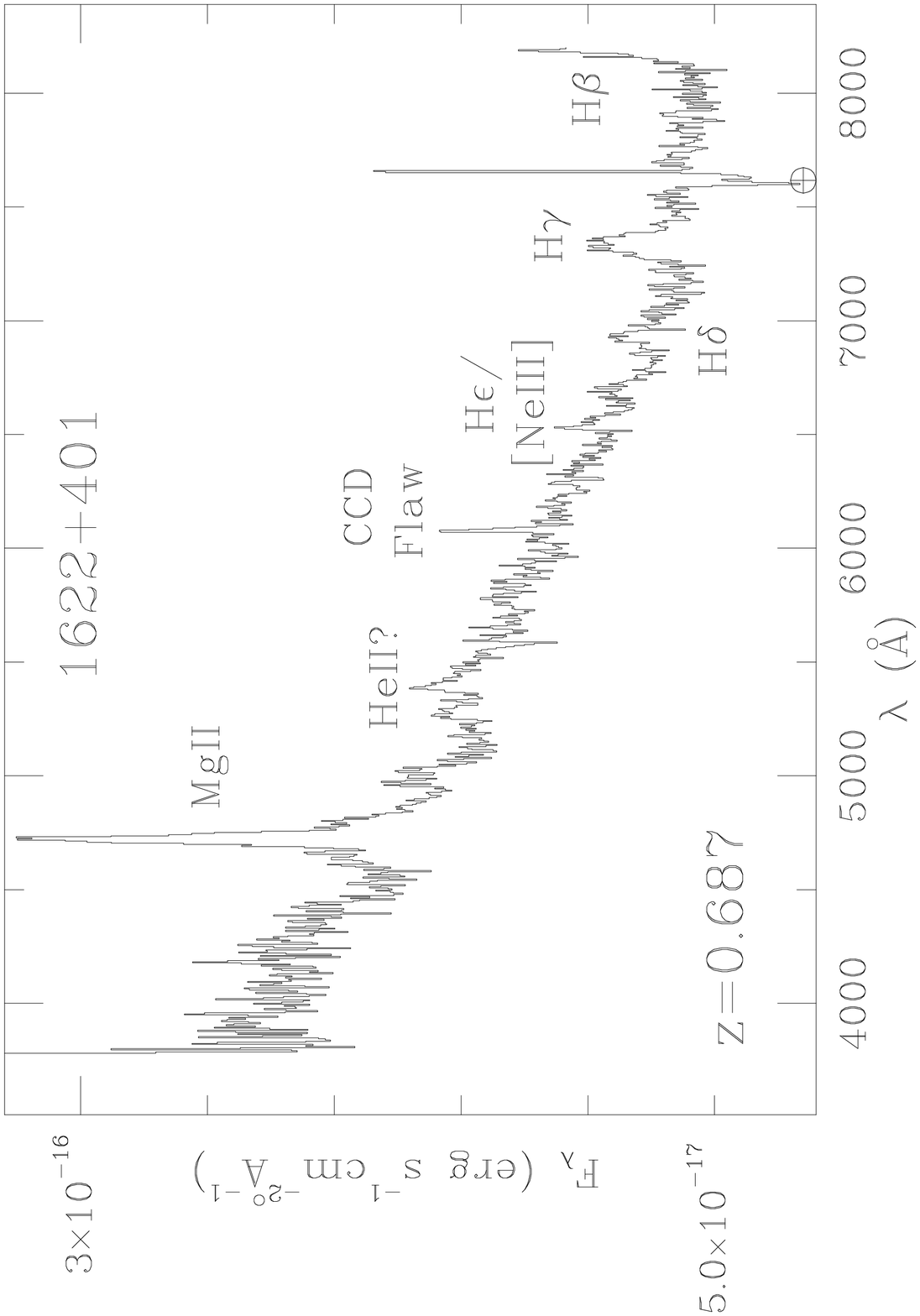,height=7.1cm,width=6.3cm}
\vspace{0.25in}
\psfig{file=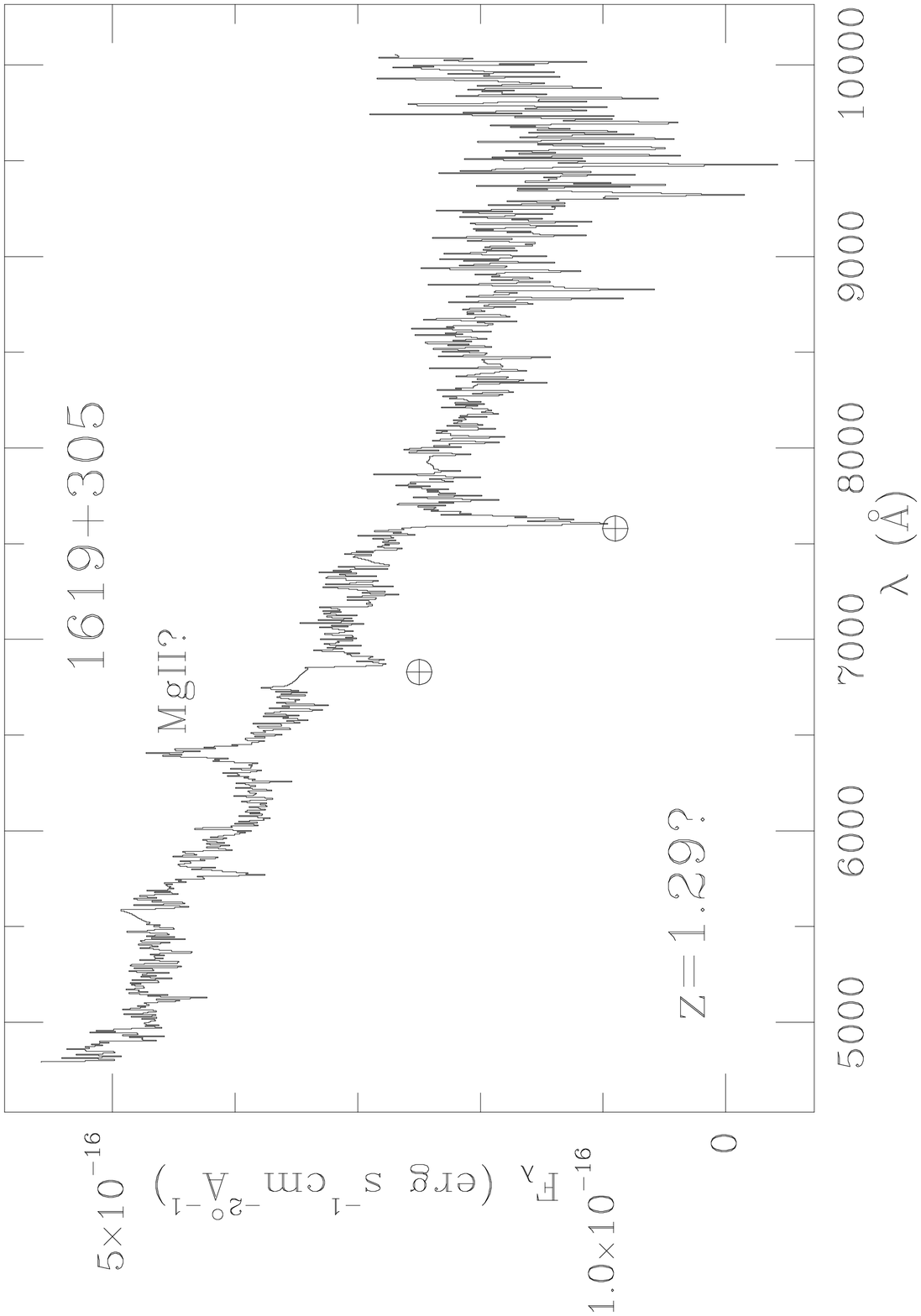,height=7.1cm,width=6.3cm}
\end{minipage}
\hspace{0.3in}
\begin{minipage}[t]{6.3in}
\psfig{file=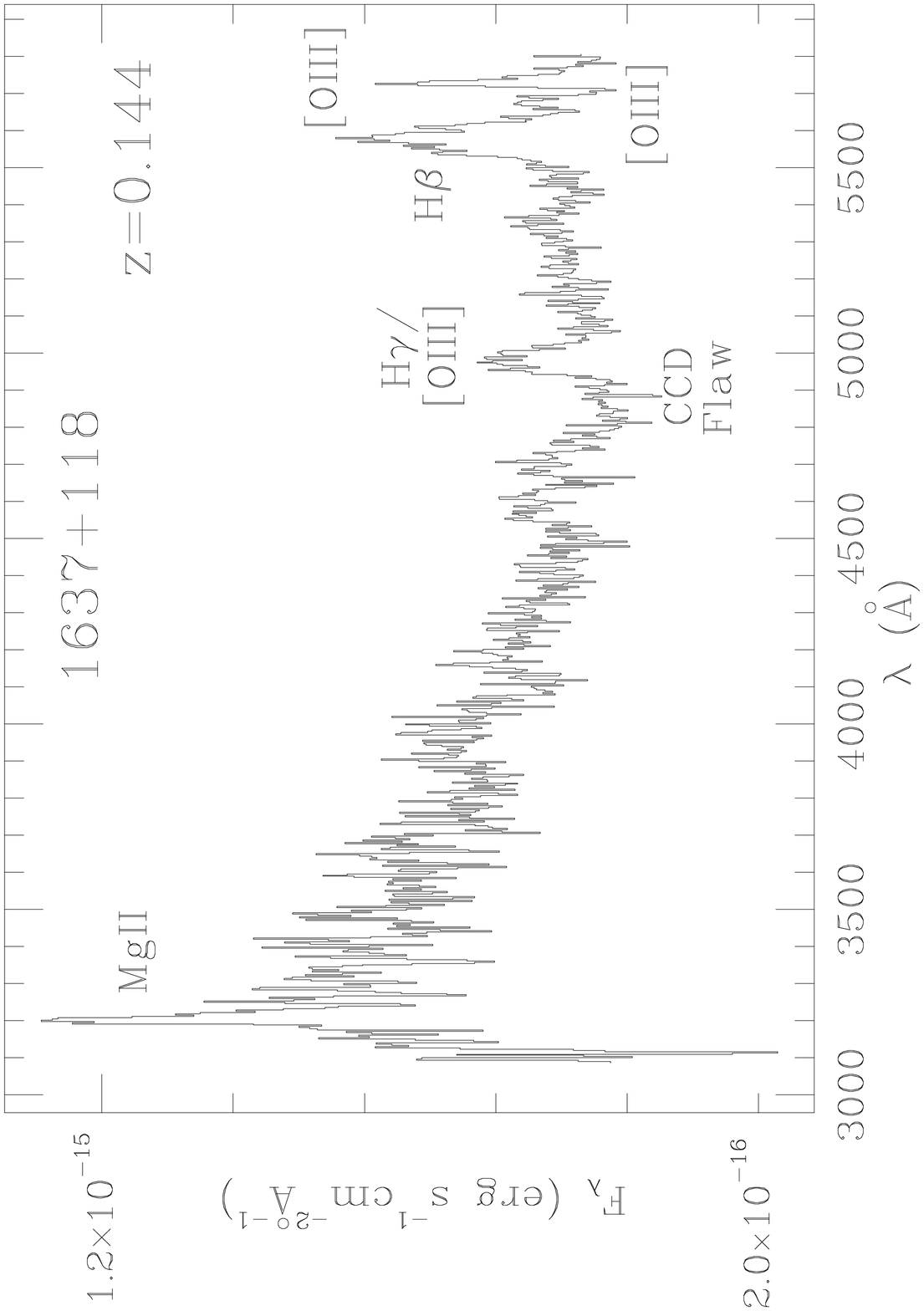,height=7.1cm,width=6.3cm}
\vspace{0.25in}
\psfig{file=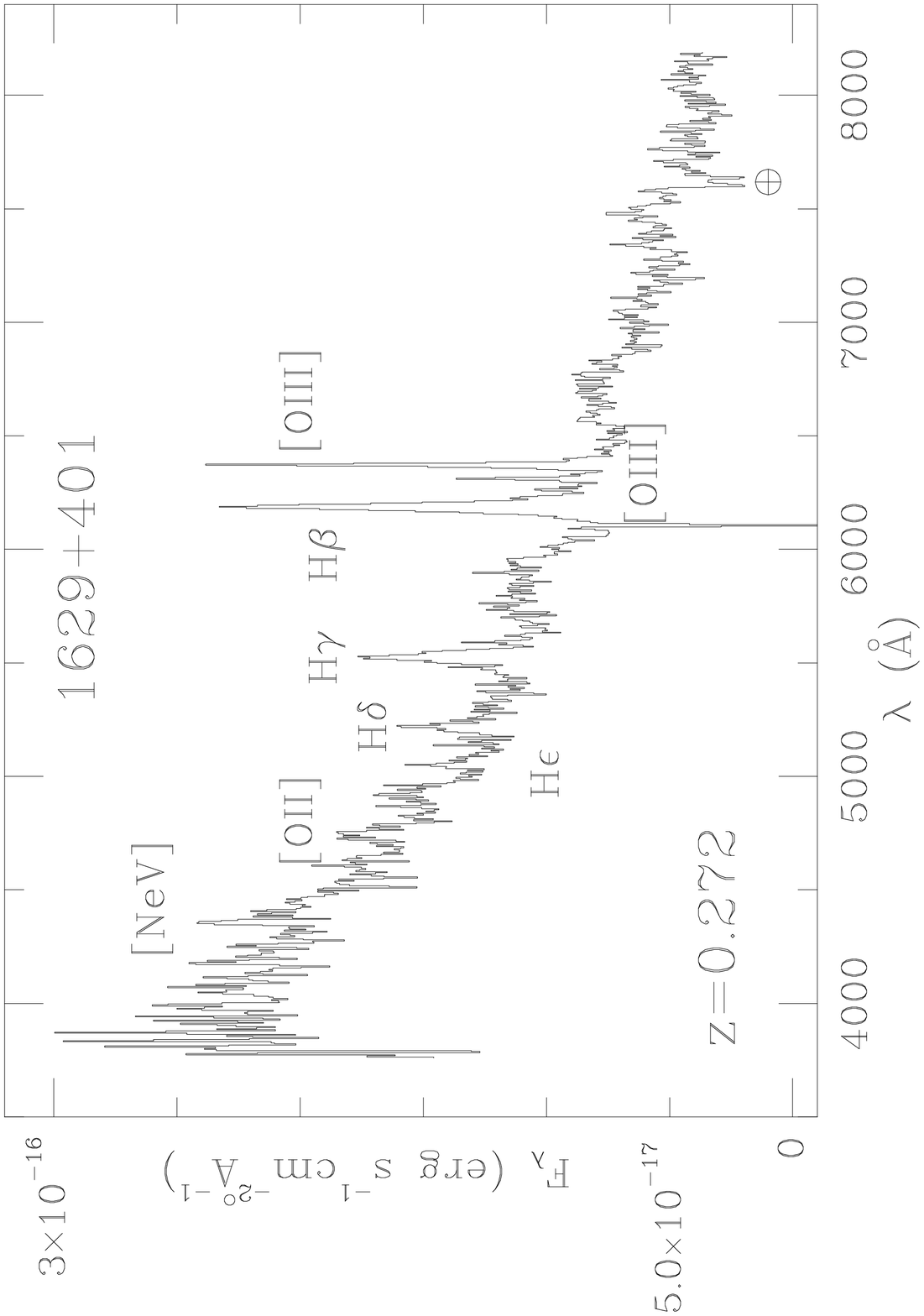,height=7.1cm,width=6.3cm}
\vspace{0.25in}
\psfig{file=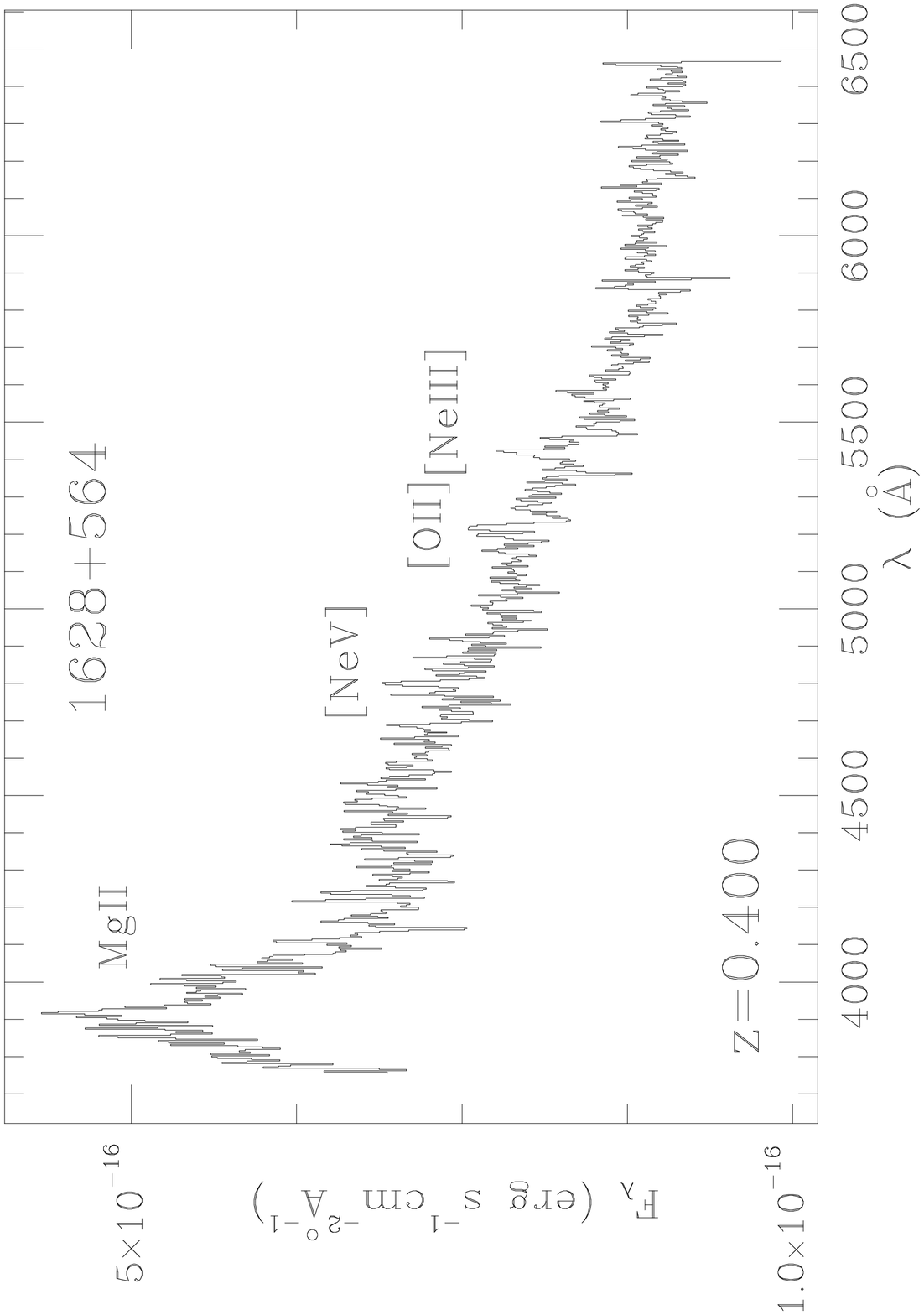,height=7.1cm,width=6.3cm}
\vspace{-7.21in}
\end{minipage}
\hfill
\begin{minipage}[t]{0.3in}
\vfill
\begin{sideways}
Figure 1.109 $-$ 1.114: Spectra of RGB Sources ({\it continued})
\end{sideways}
\vfill
\end{minipage}
\end{figure}

\clearpage
\begin{figure}
\vspace{-0.3in}
\hspace{-0.3in}
\begin{minipage}[t]{6.3in}
\psfig{file=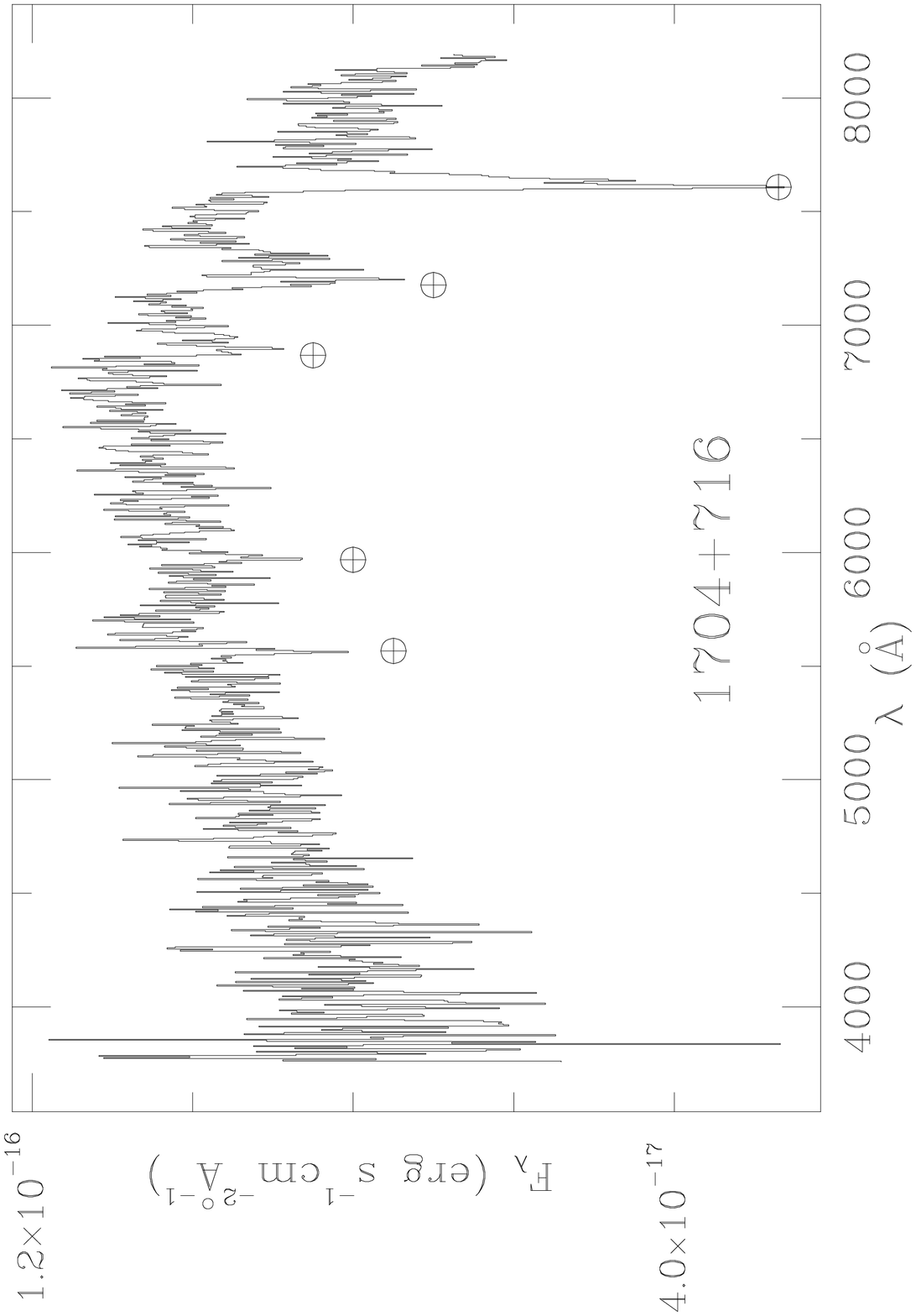,height=7.1cm,width=6.3cm}
\vspace{0.25in}
\psfig{file=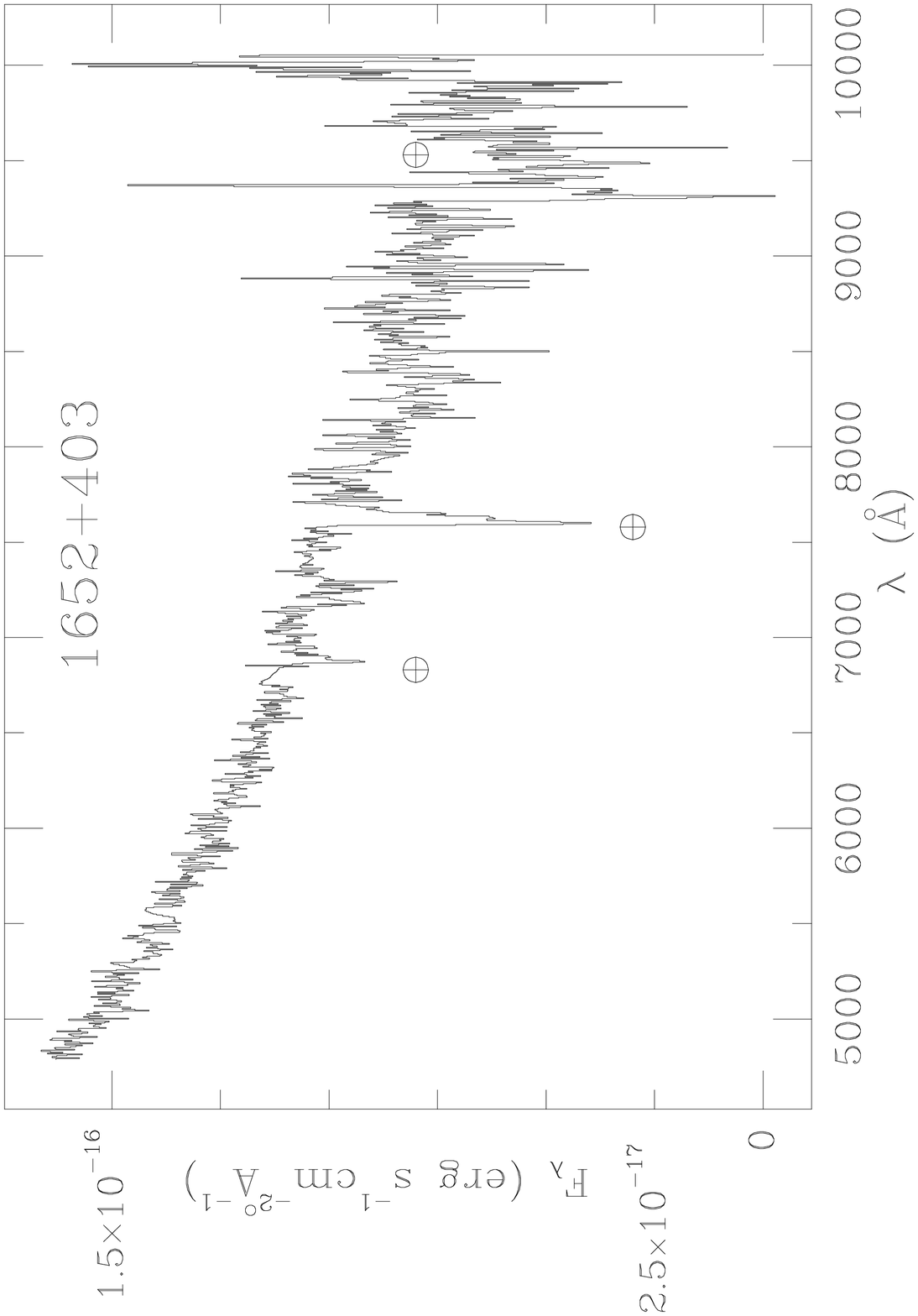,height=7.1cm,width=6.3cm}
\vspace{0.25in}
\psfig{file=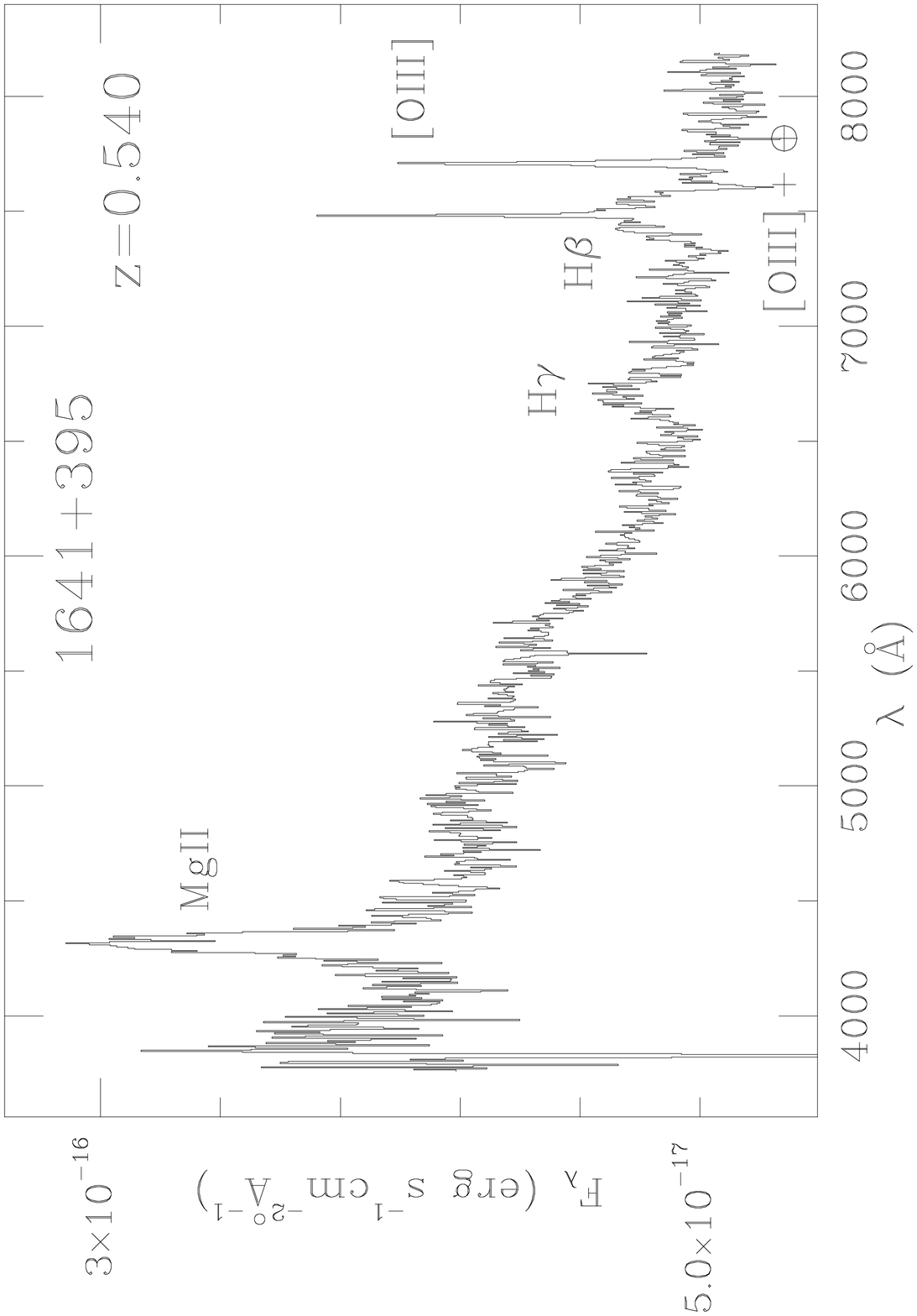,height=7.1cm,width=6.3cm}
\end{minipage}
\hspace{0.3in}
\begin{minipage}[t]{6.3in}
\psfig{file=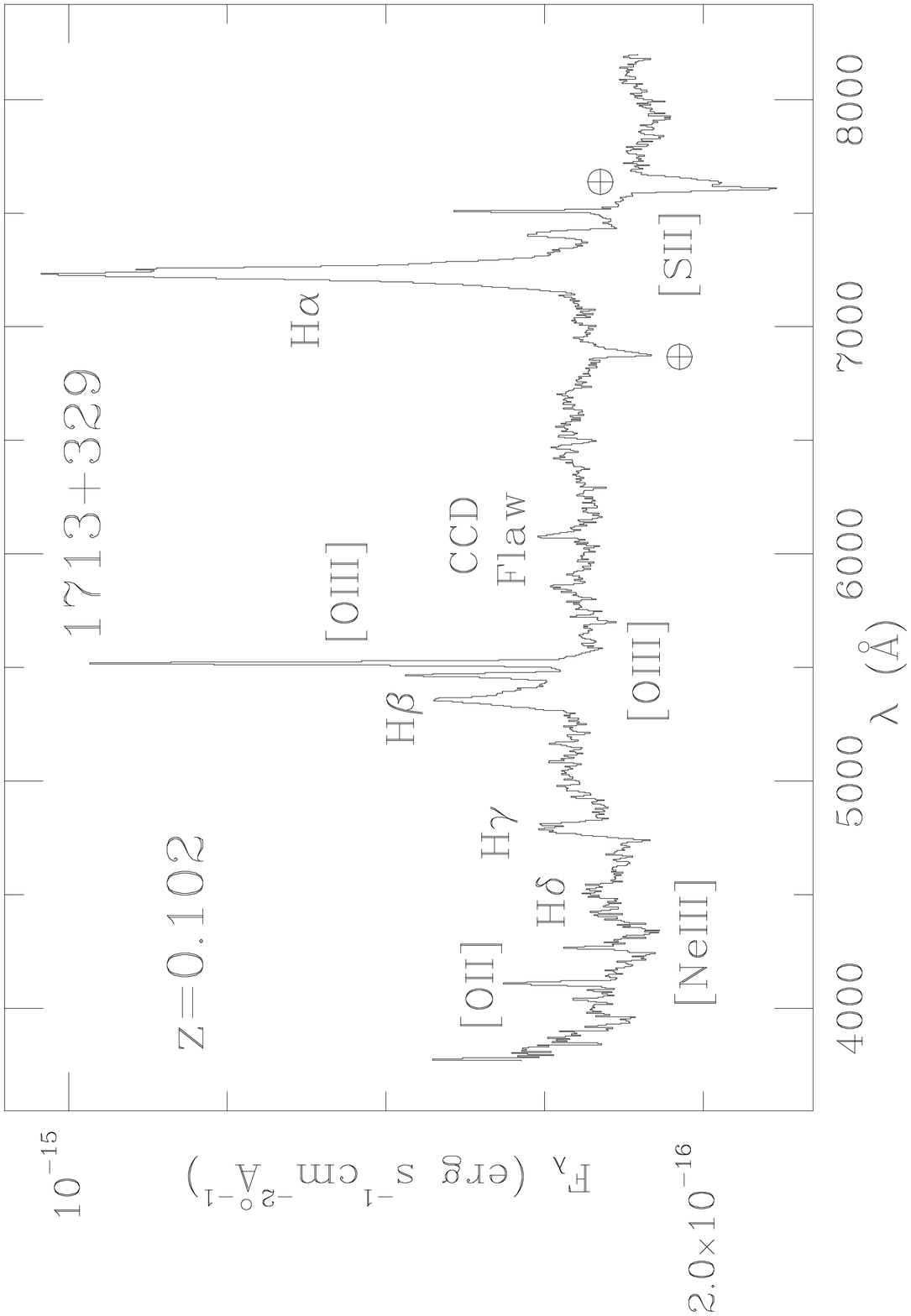,height=7.1cm,width=6.3cm}
\vspace{0.25in}
\psfig{file=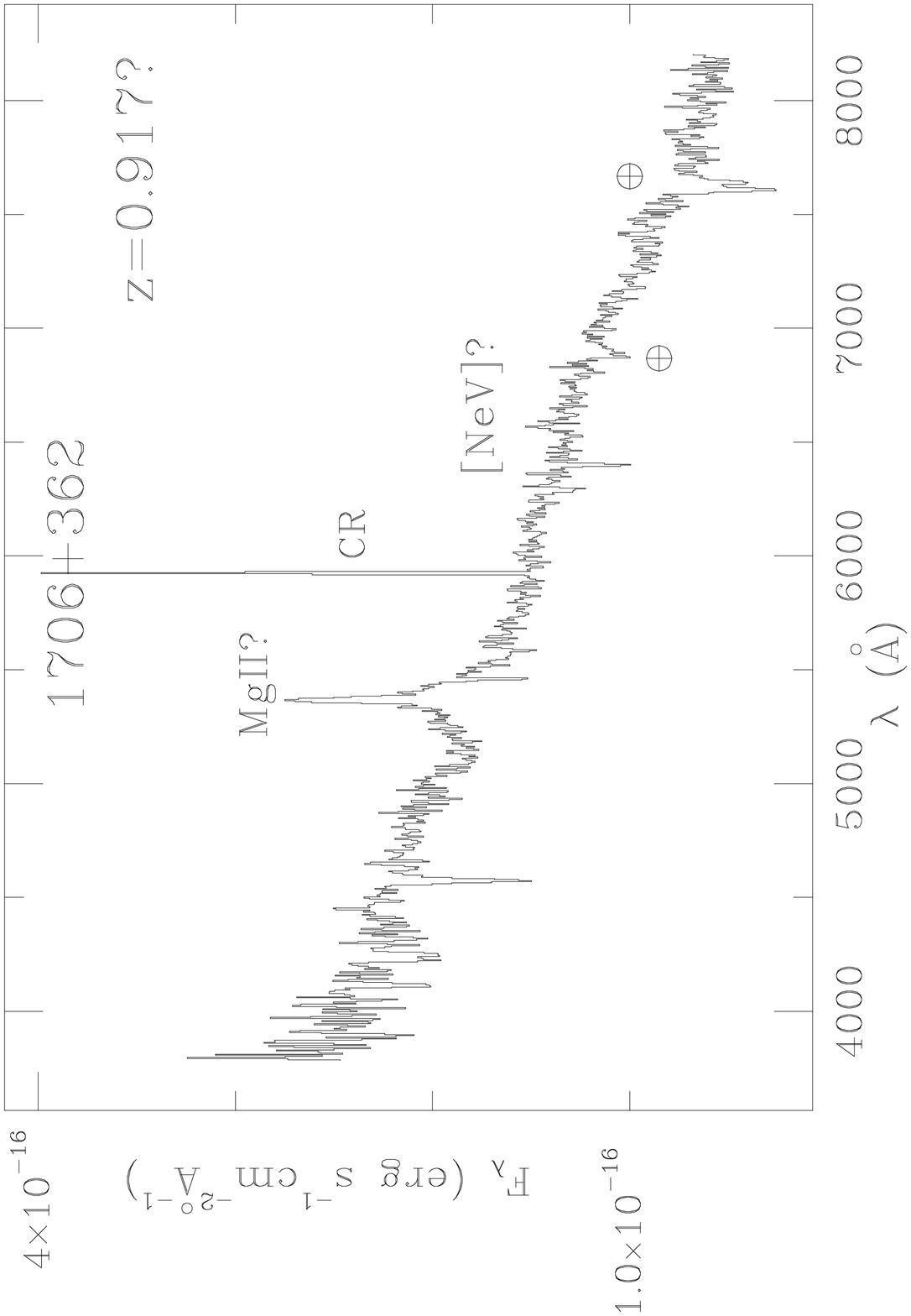,height=7.1cm,width=6.3cm}
\vspace{0.25in}
\psfig{file=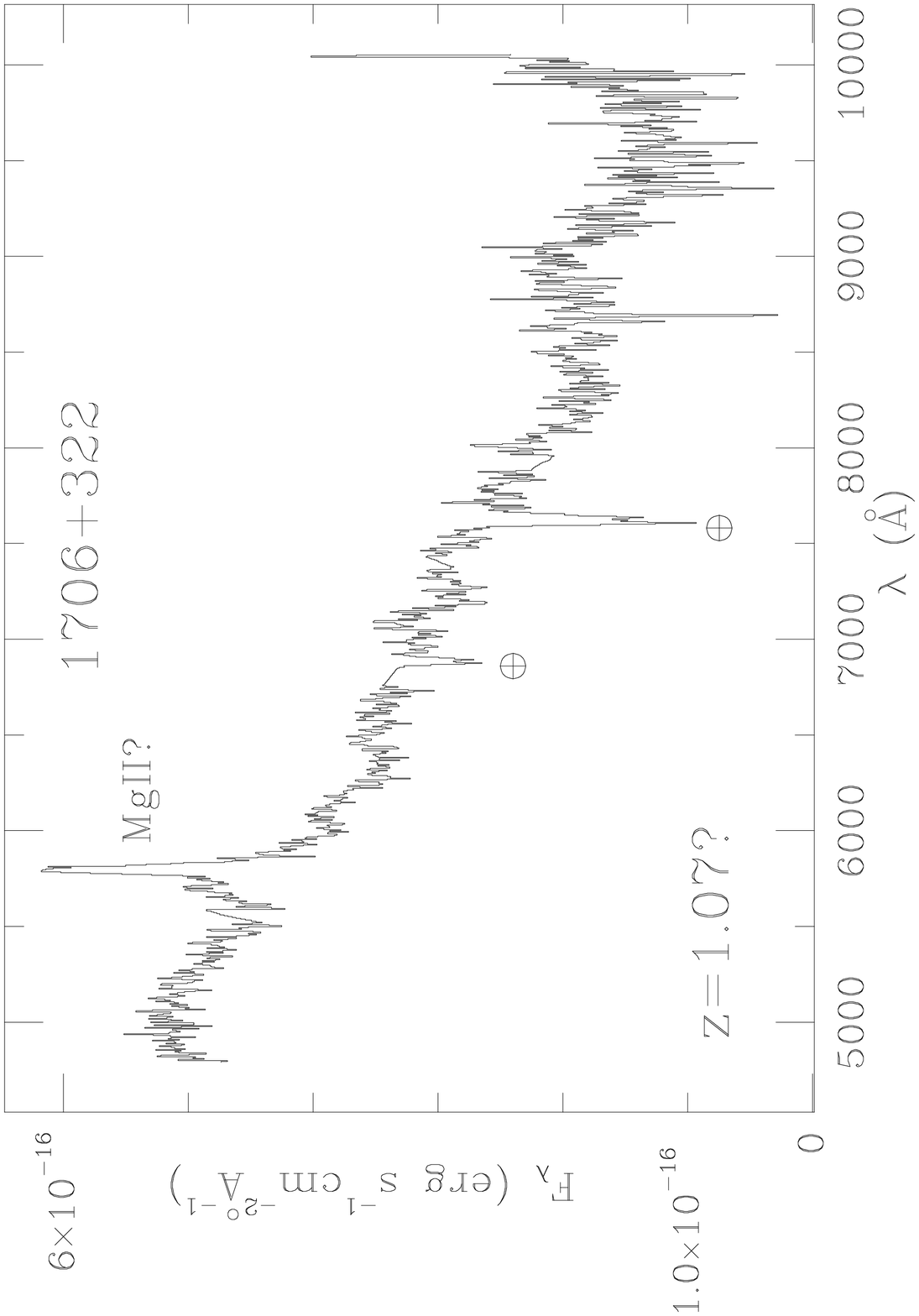,height=7.1cm,width=6.3cm}
\end{minipage}
\hfill
\begin{minipage}[t]{0.3in}
\vfill
\begin{sideways}
Figure 1.115 $-$ 1.120: Spectra of RGB Sources ({\it continued})
\end{sideways}
\vfill
\end{minipage}
\end{figure}

\clearpage
\begin{figure}
\vspace{-0.3in}
\hspace{-0.3in}
\begin{minipage}[t]{6.3in}
\psfig{file=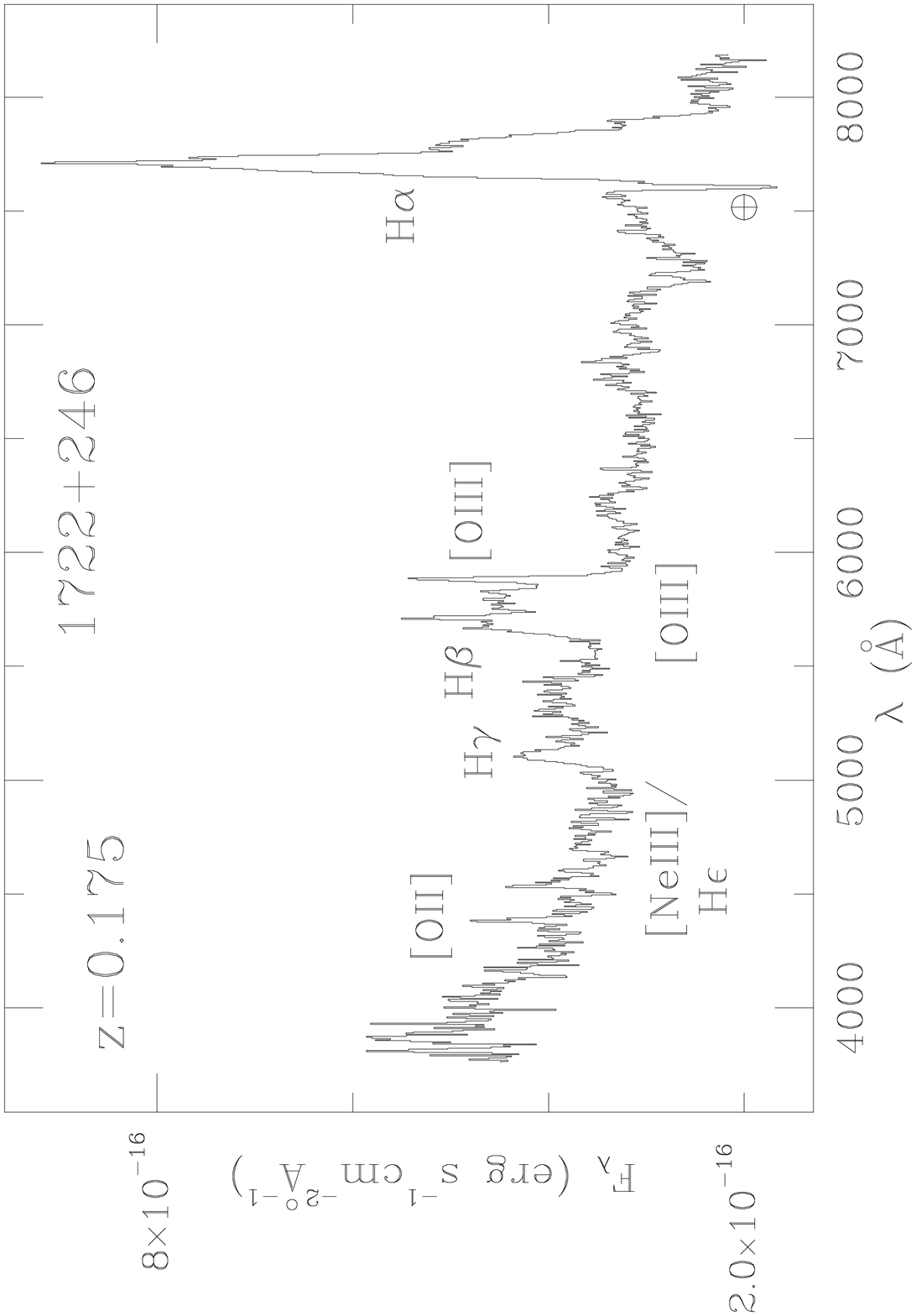,height=7.1cm,width=6.3cm}
\vspace{0.25in}
\psfig{file=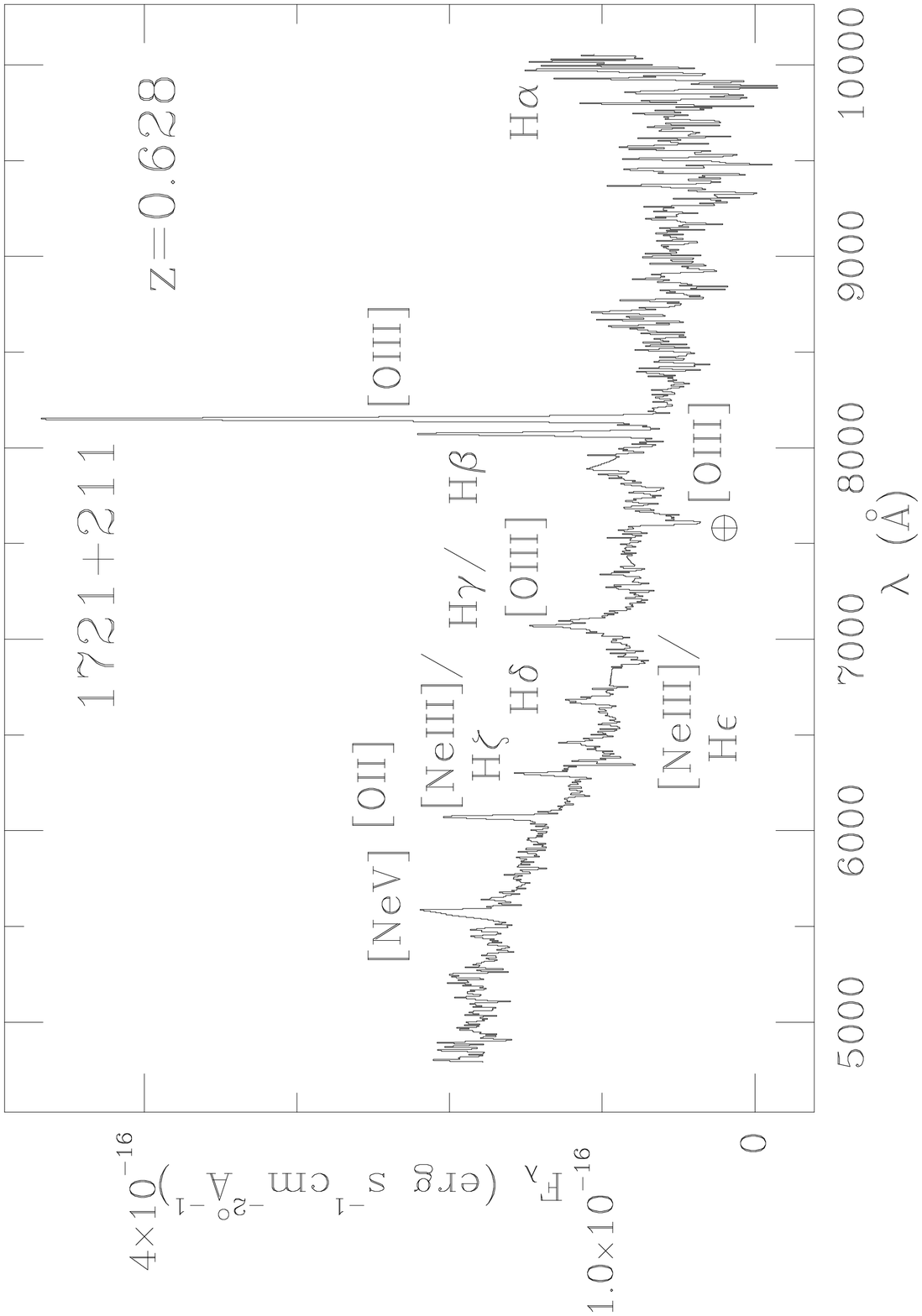,height=7.1cm,width=6.3cm}
\vspace{0.25in}
\psfig{file=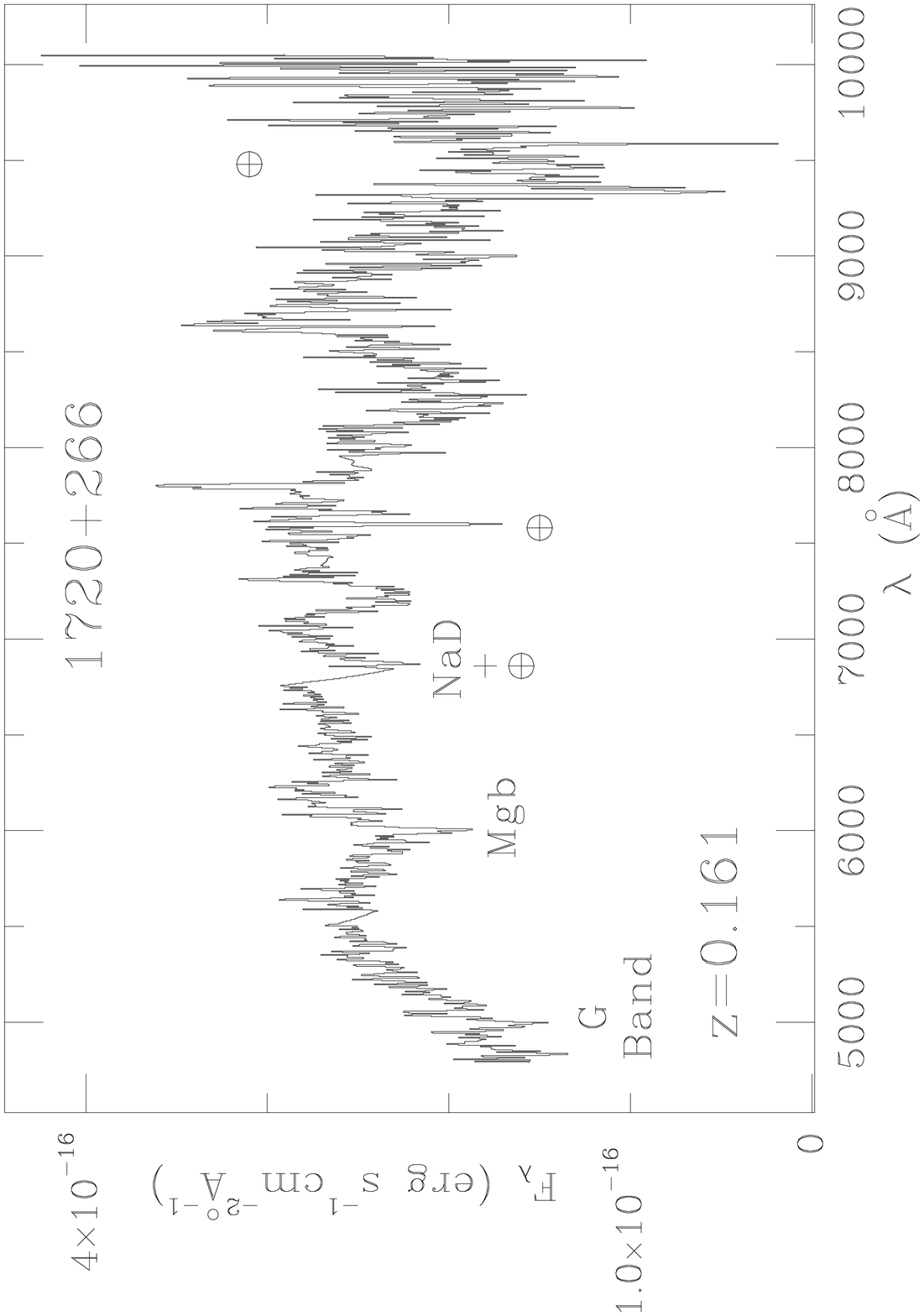,height=7.1cm,width=6.3cm}
\end{minipage}
\hspace{0.3in}
\begin{minipage}[t]{6.3in}
\psfig{file=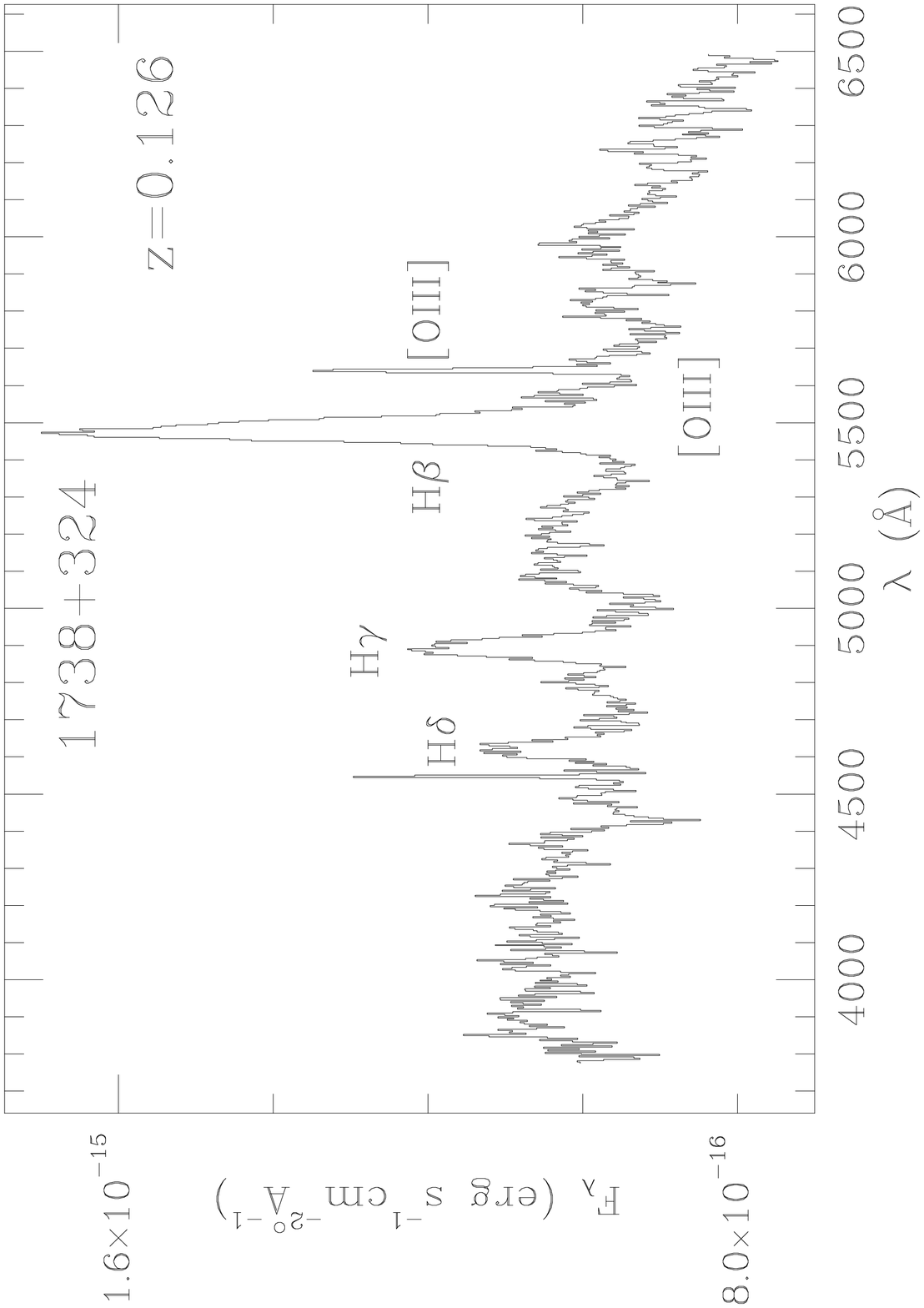,height=7.1cm,width=6.3cm}
\vspace{0.25in}
\psfig{file=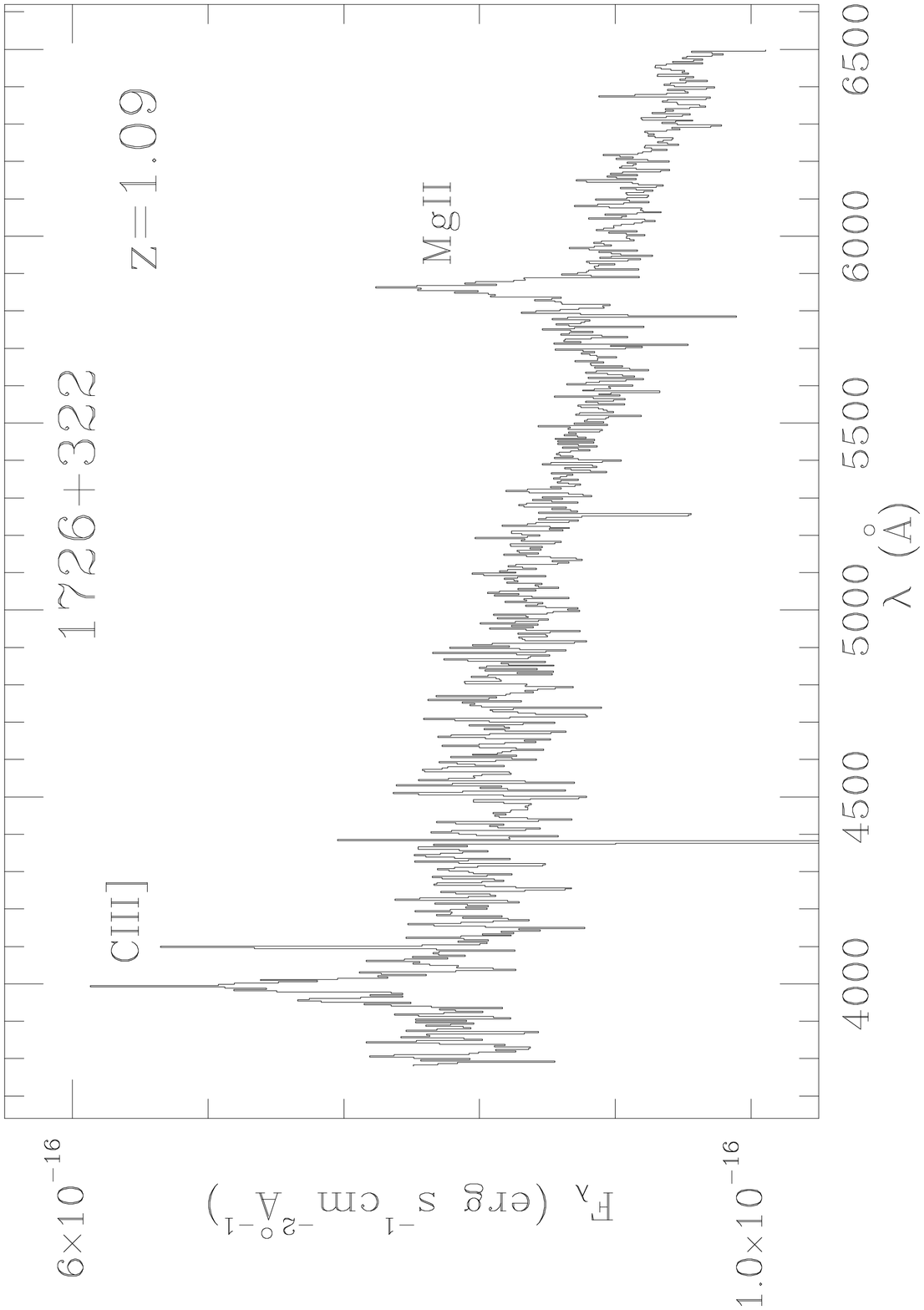,height=7.1cm,width=6.3cm}
\vspace{0.25in}
\psfig{file=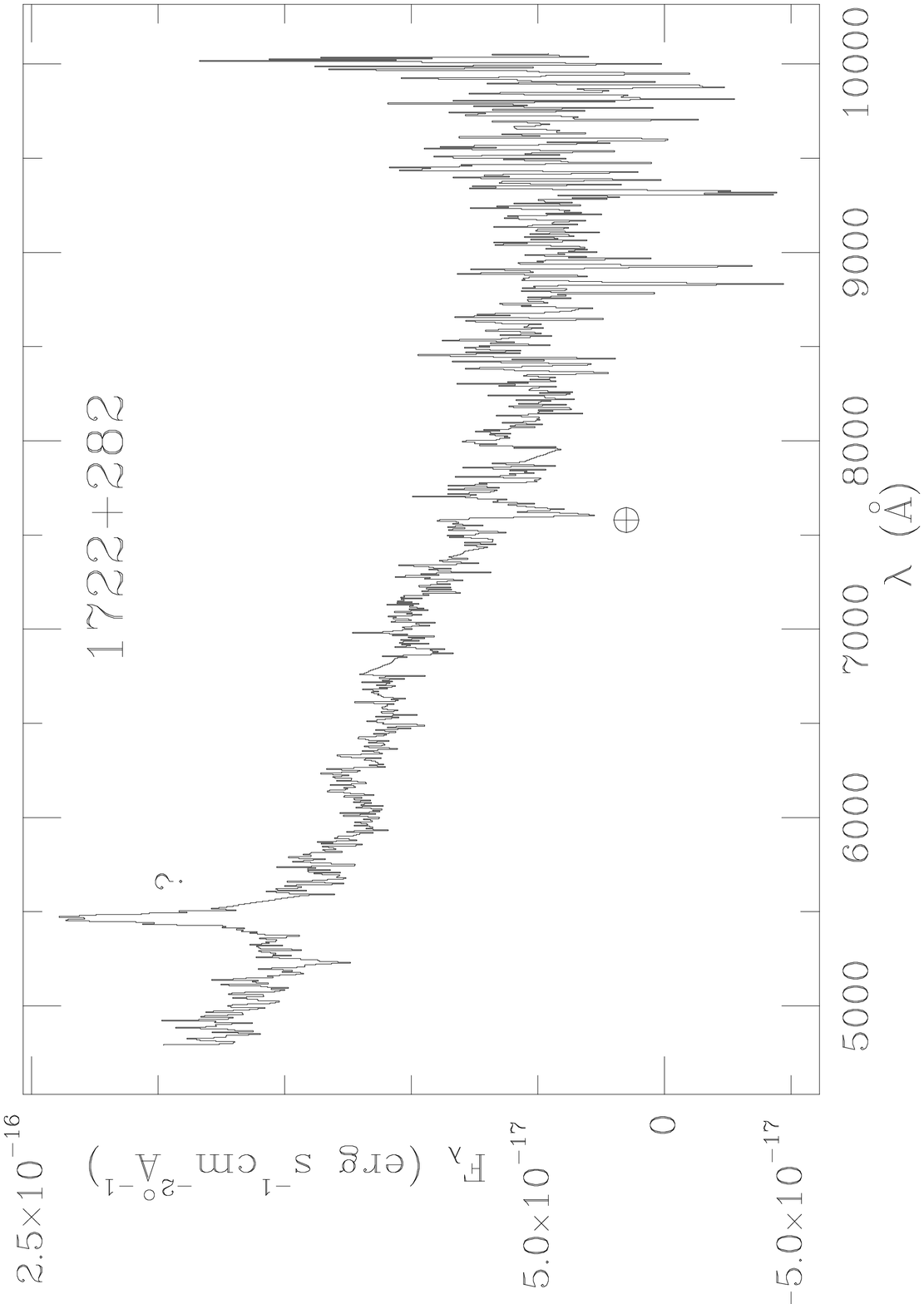,height=7.1cm,width=6.3cm}
\end{minipage}
\hfill
\begin{minipage}[t]{0.3in}
\vfill
\begin{sideways}
Figure 1.121 $-$ 1.126: Spectra of RGB Sources ({\it continued})
\end{sideways}
\vfill
\end{minipage}
\end{figure}

\clearpage
\begin{figure}
\vspace{-0.3in}
\hspace{-0.3in}
\begin{minipage}[t]{6.3in}
\psfig{file=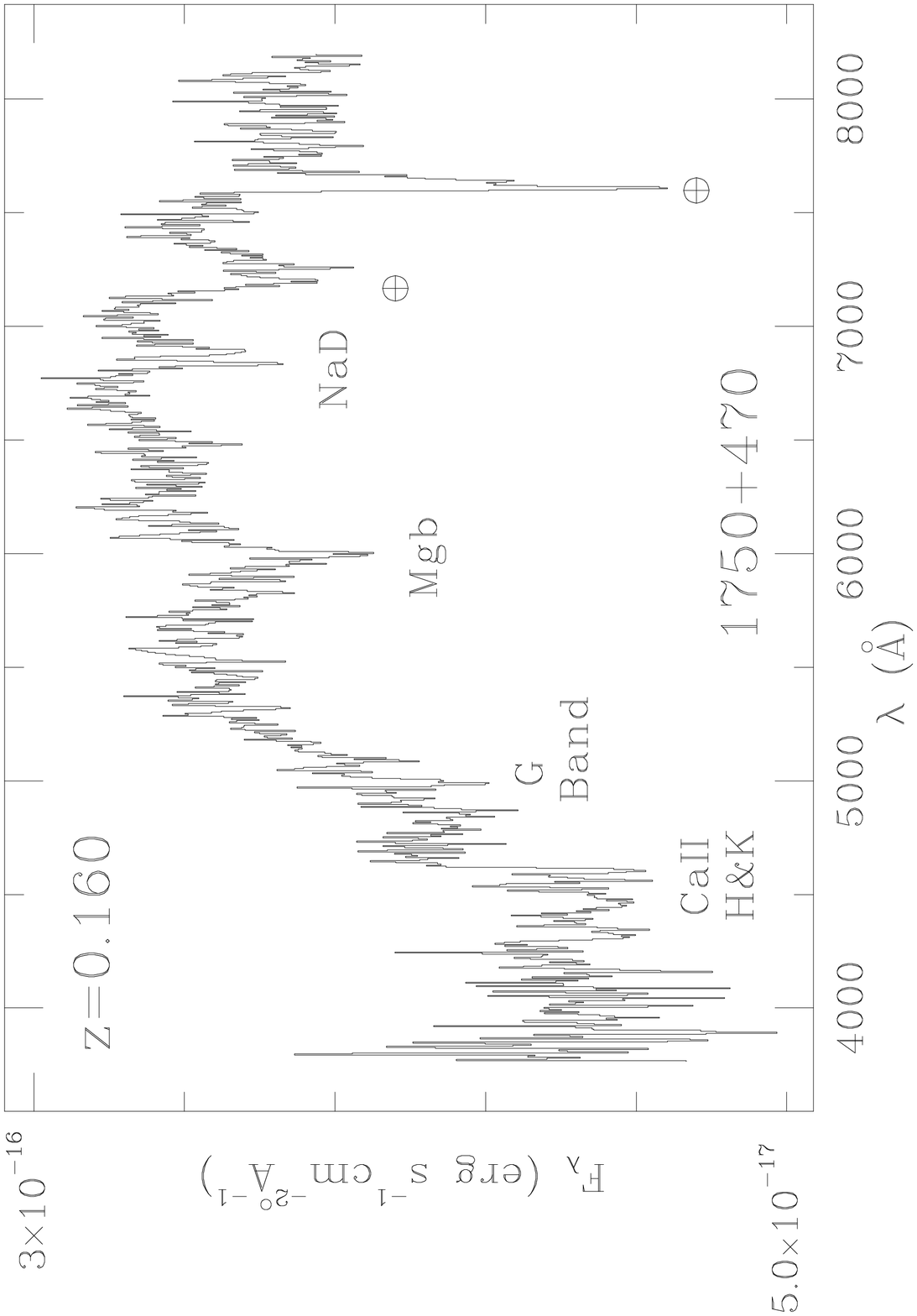,height=7.1cm,width=6.3cm}
\vspace{0.25in}
\psfig{file=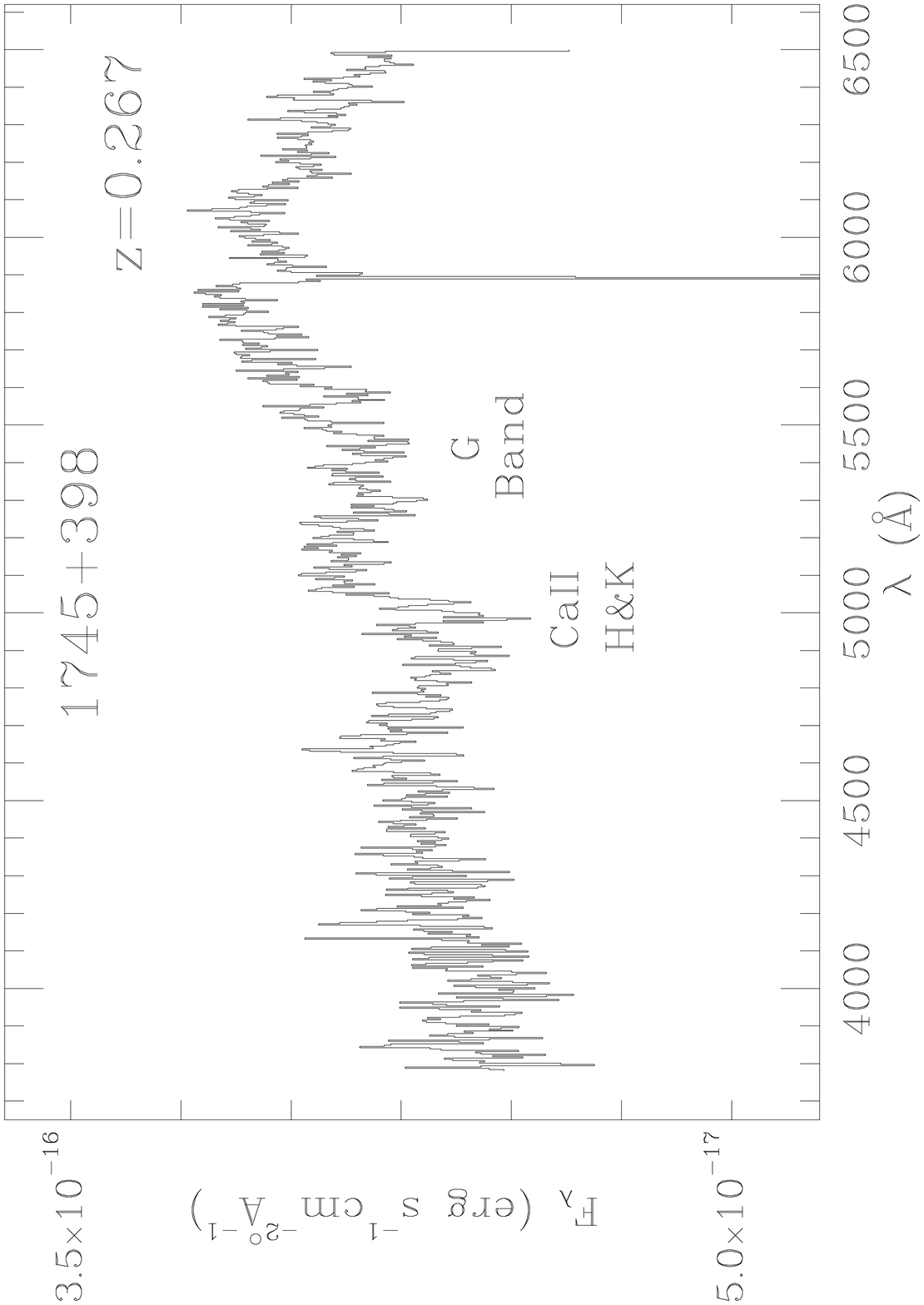,height=7.1cm,width=6.3cm}
\vspace{0.25in}
\psfig{file=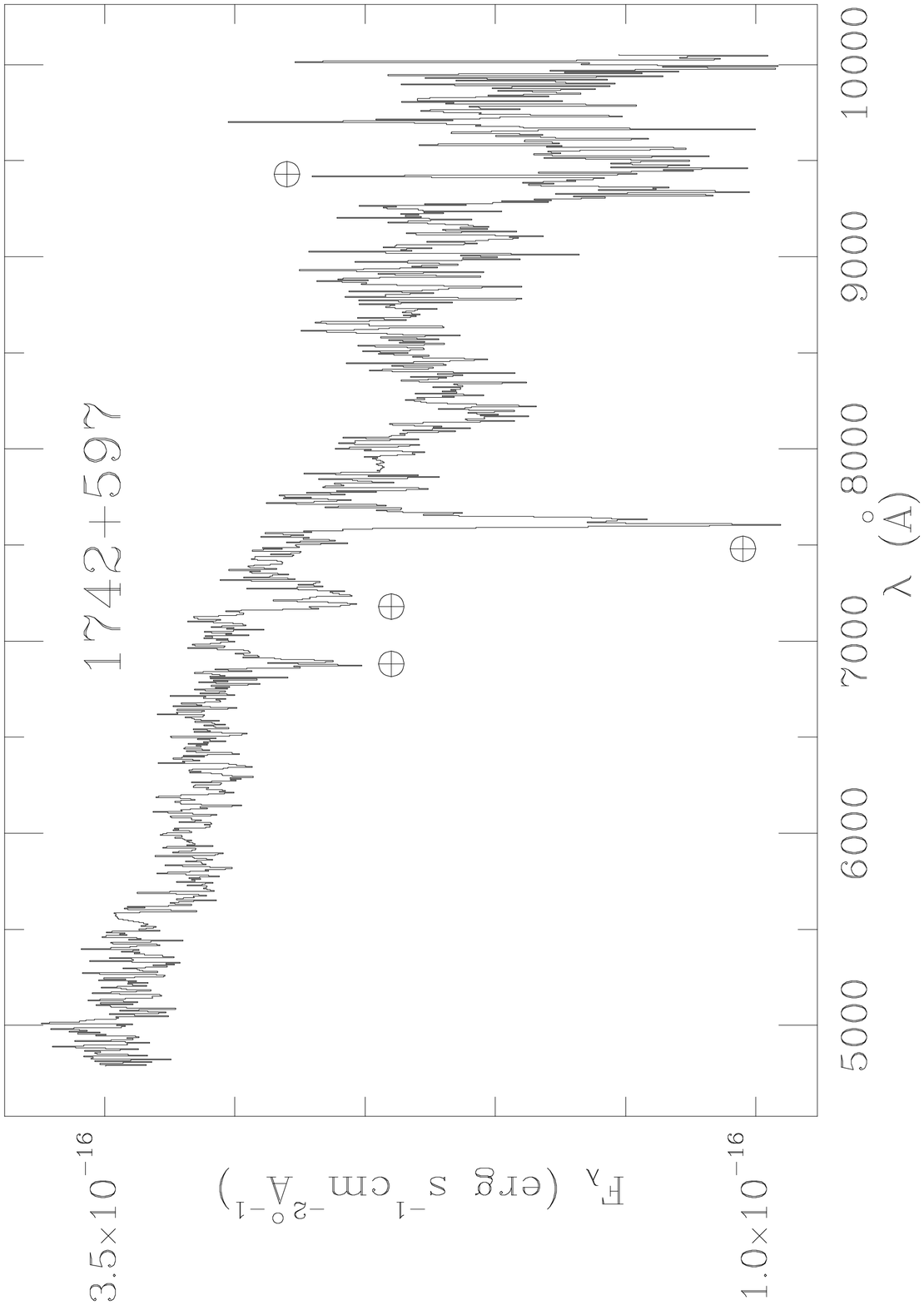,height=7.1cm,width=6.3cm}
\end{minipage}
\hspace{0.3in}
\begin{minipage}[t]{6.3in}
\psfig{file=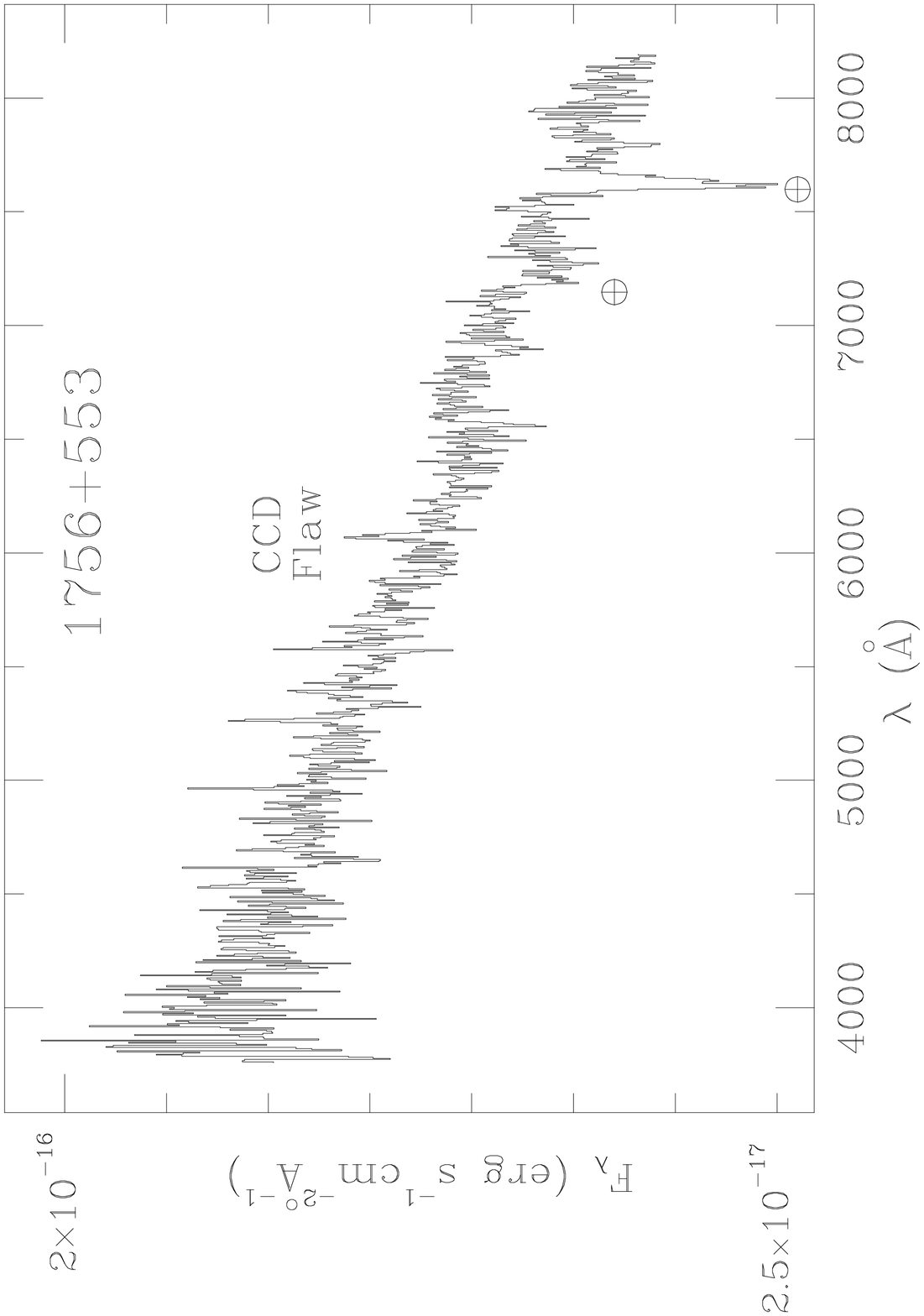,height=7.1cm,width=6.3cm}
\vspace{0.25in}
\psfig{file=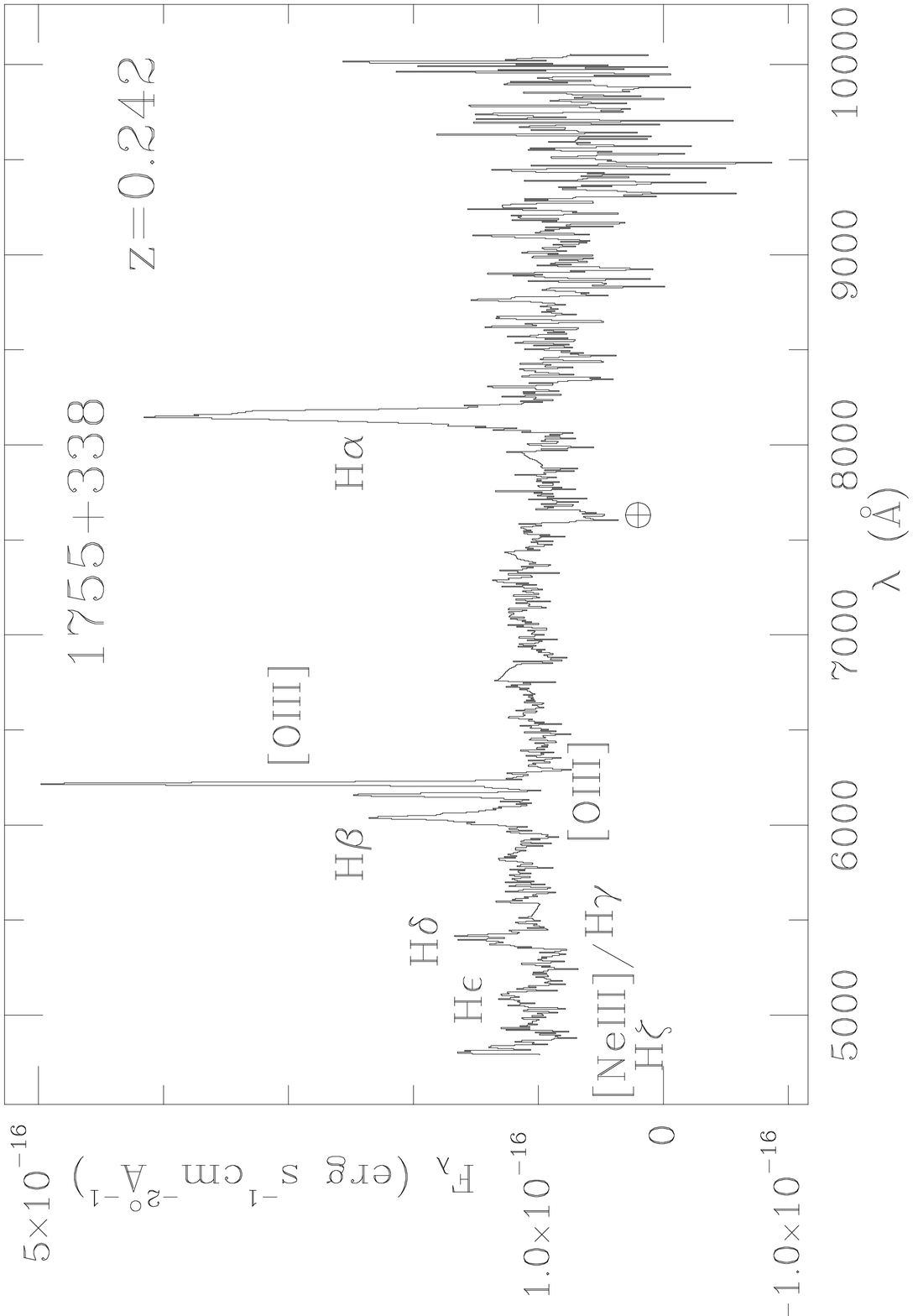,height=7.1cm,width=6.3cm}
\vspace{0.25in}
\psfig{file=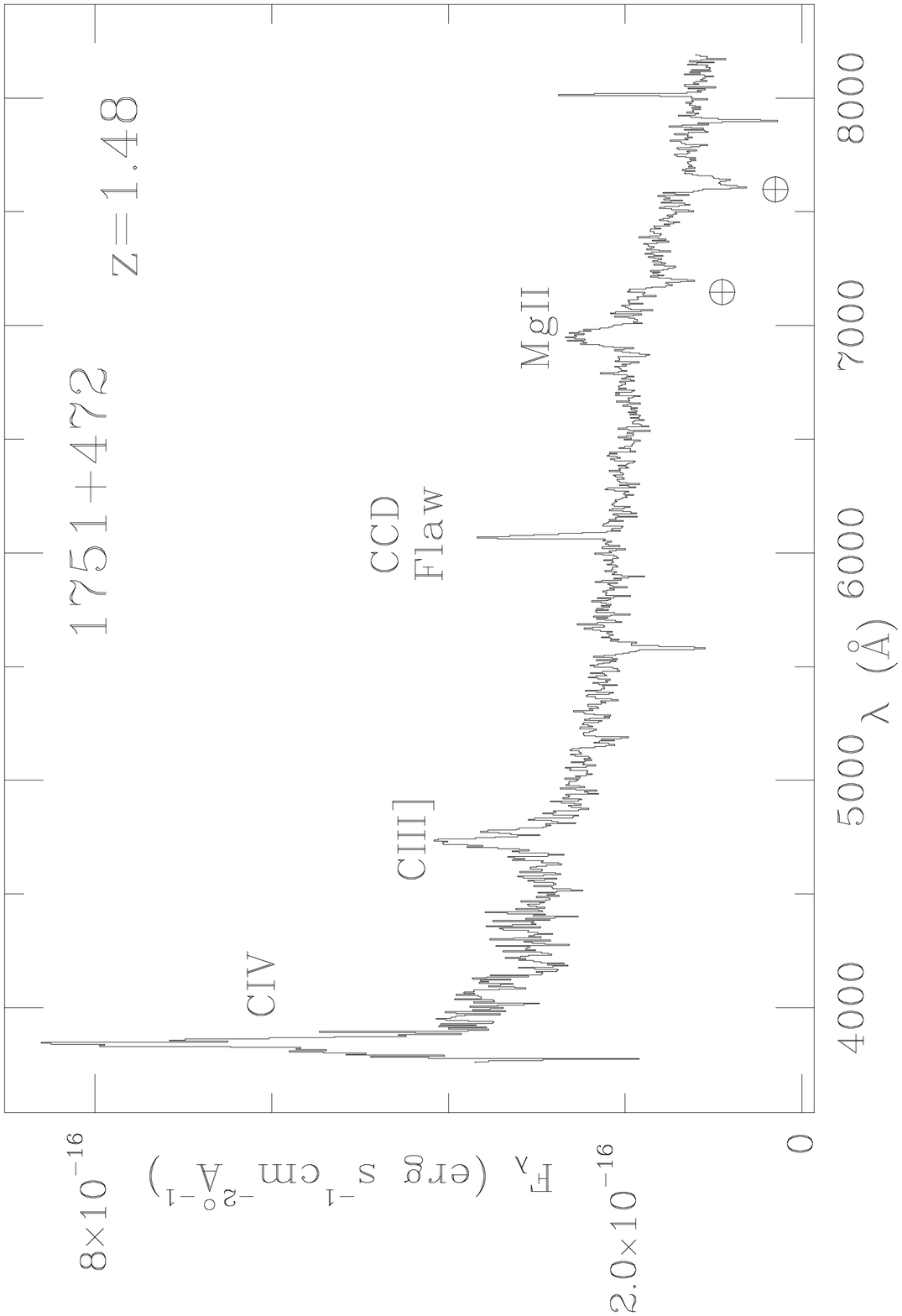,height=7.1cm,width=6.3cm}
\end{minipage}
\hfill
\begin{minipage}[t]{0.3in}
\vfill
\begin{sideways}
Figure 1.127 $-$ 1.132: Spectra of RGB Sources ({\it continued})
\end{sideways}
\vfill
\end{minipage}
\end{figure}

\clearpage
\begin{figure}
\vspace{-0.3in}
\hspace{-0.3in}
\begin{minipage}[t]{6.3in}
\psfig{file=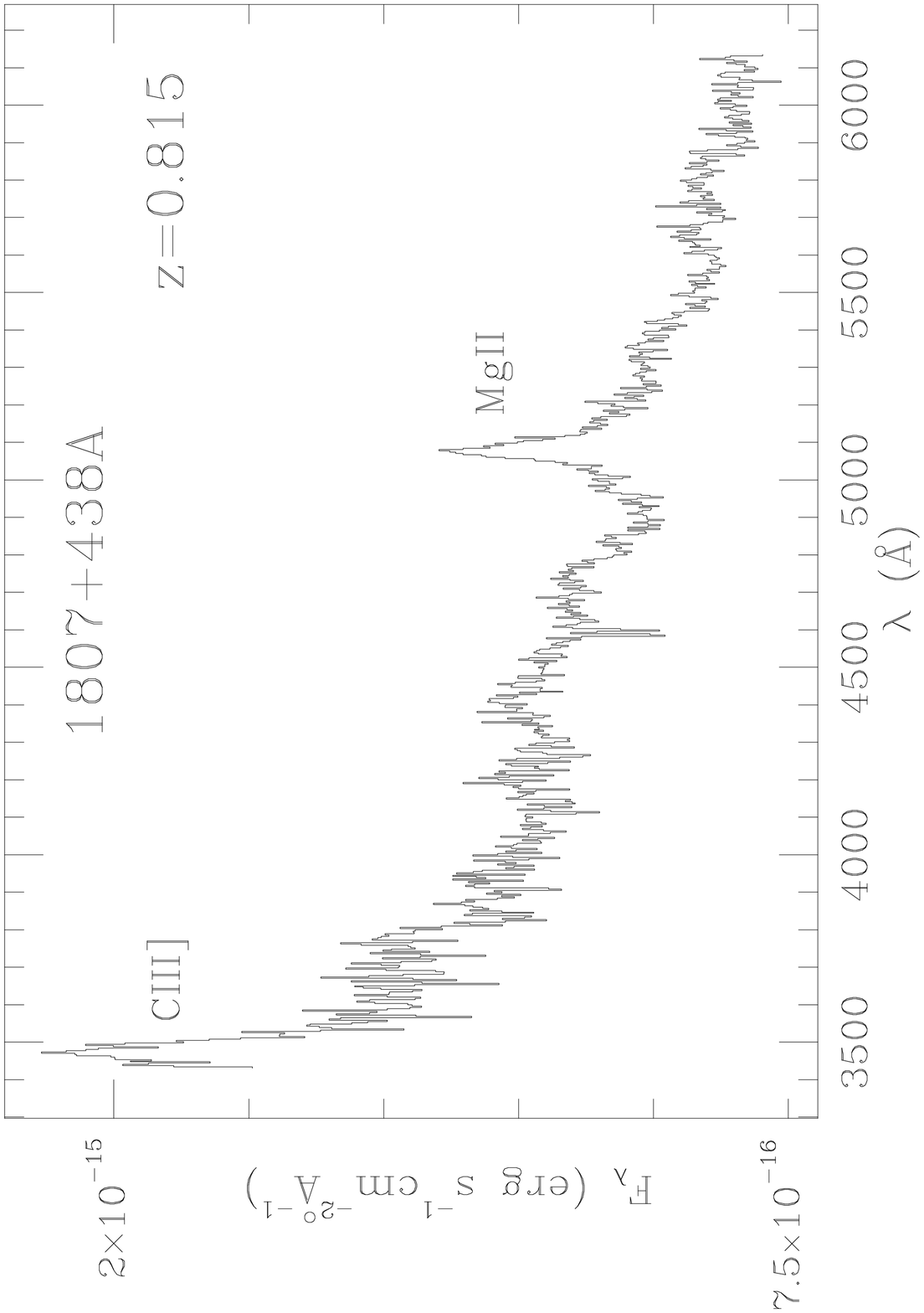,height=7.1cm,width=6.3cm}
\vspace{0.25in}
\psfig{file=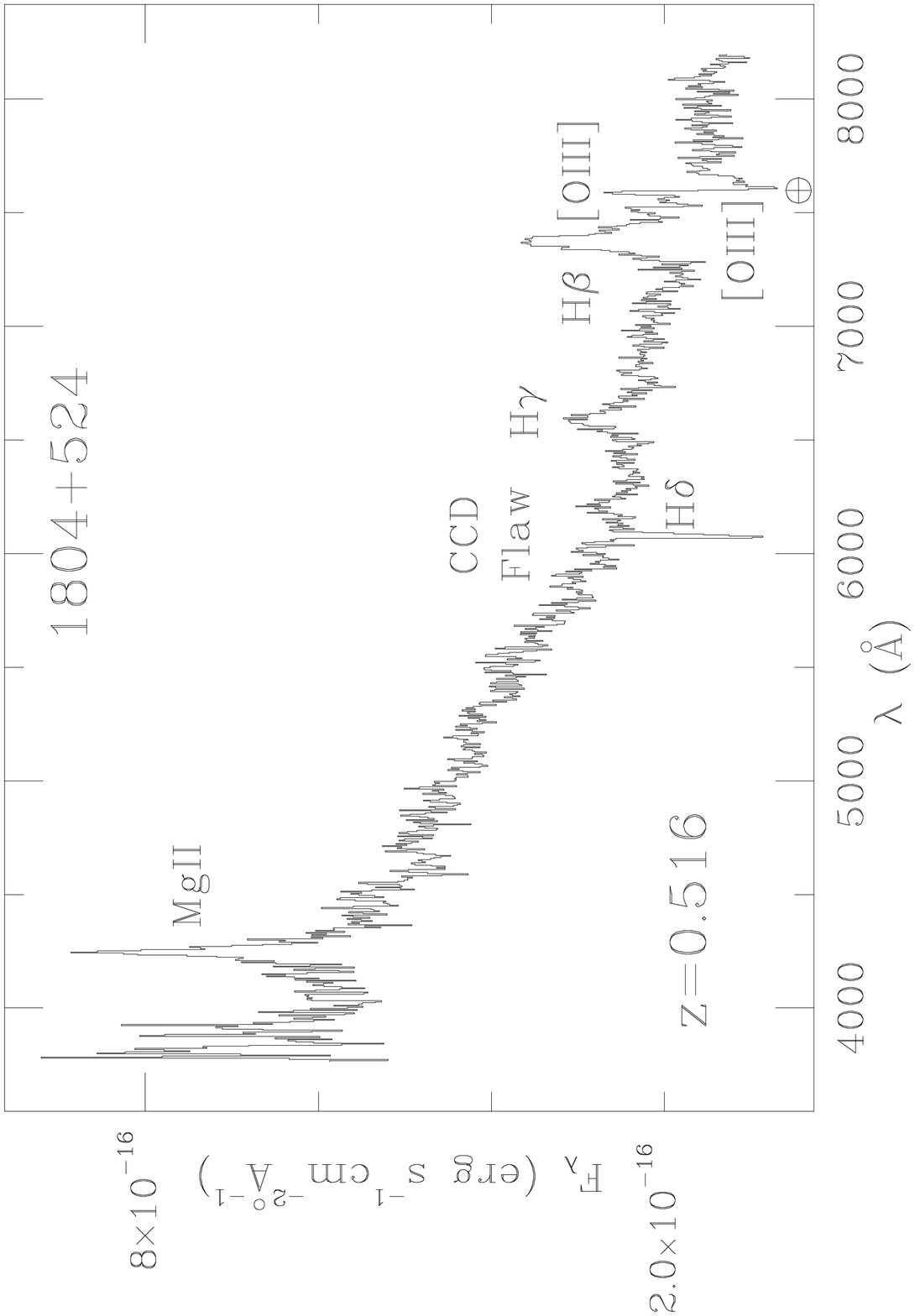,height=7.1cm,width=6.3cm}
\vspace{0.25in}
\psfig{file=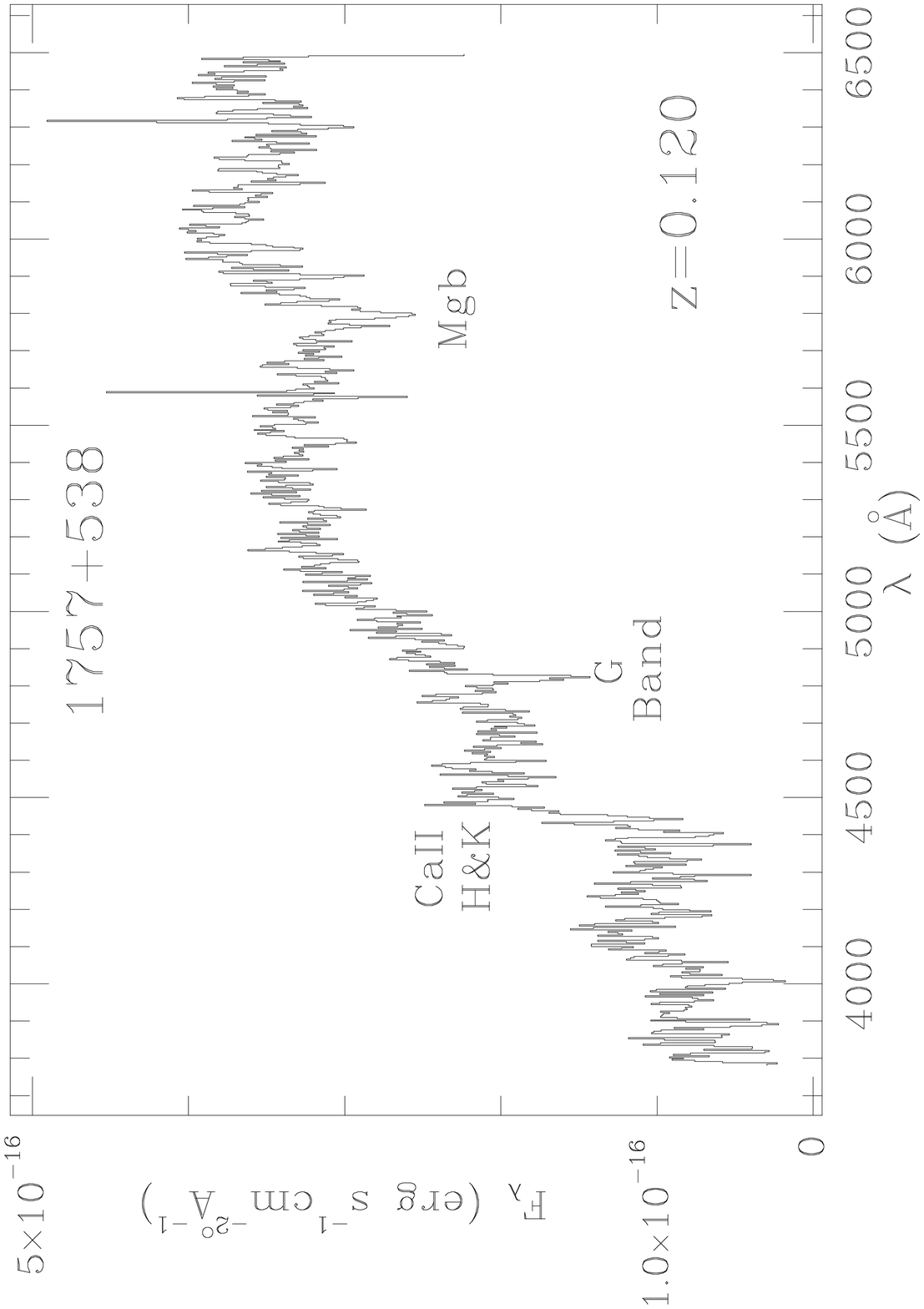,height=7.1cm,width=6.3cm}
\end{minipage}
\hspace{0.3in}
\begin{minipage}[t]{6.3in}
\psfig{file=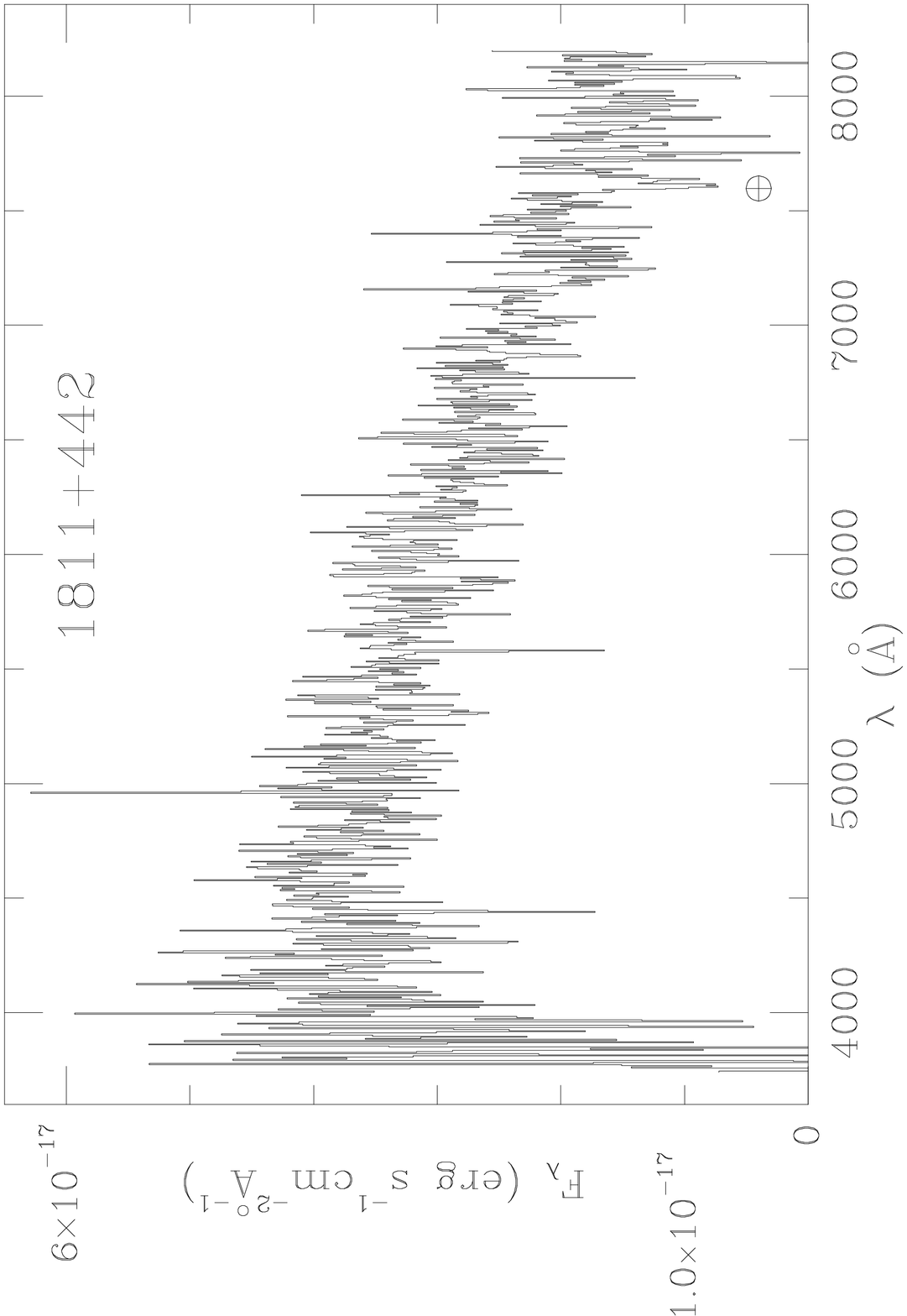,height=7.1cm,width=6.3cm}
\vspace{0.25in}
\psfig{file=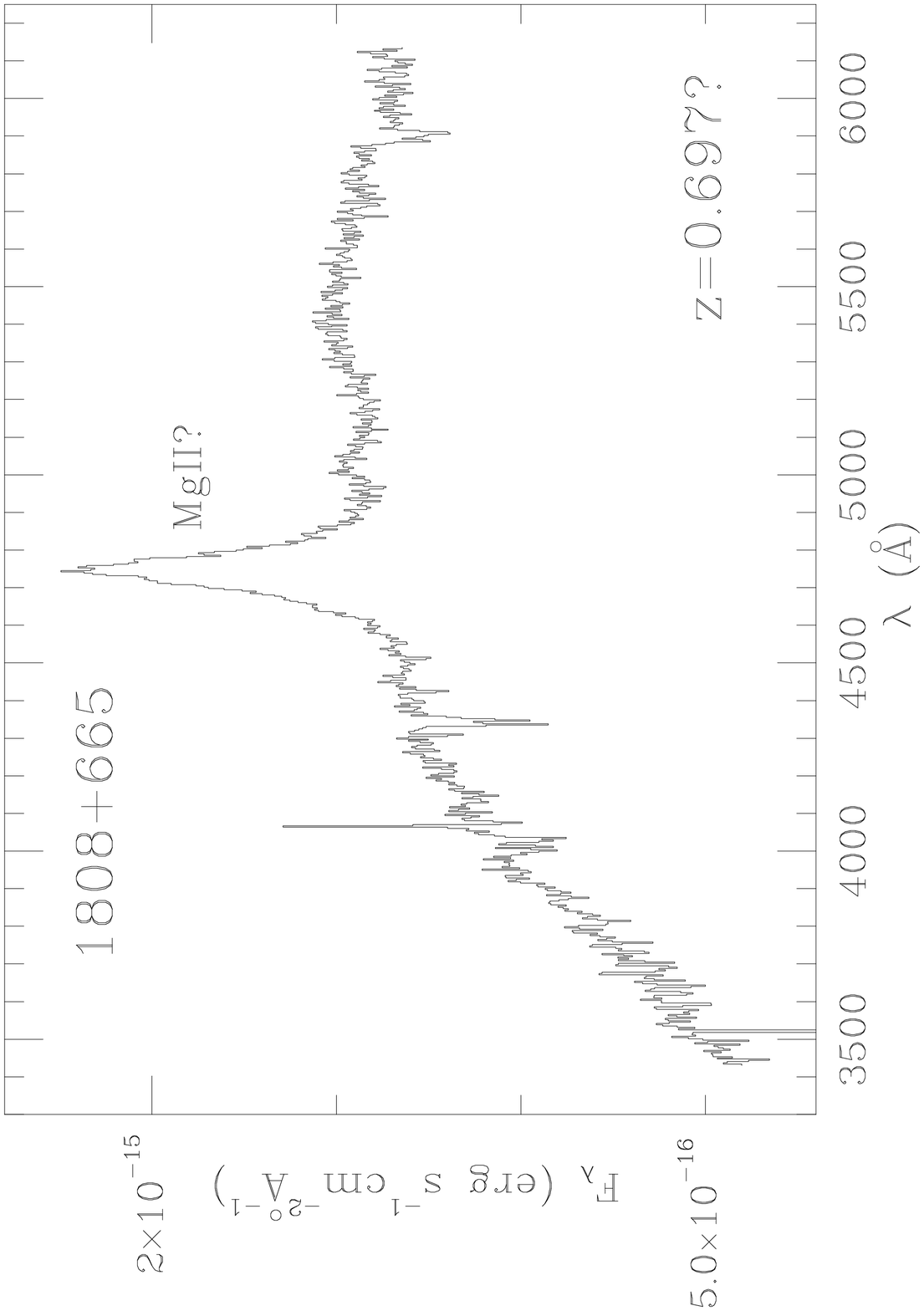,height=7.1cm,width=6.3cm}
\vspace{0.25in}
\psfig{file=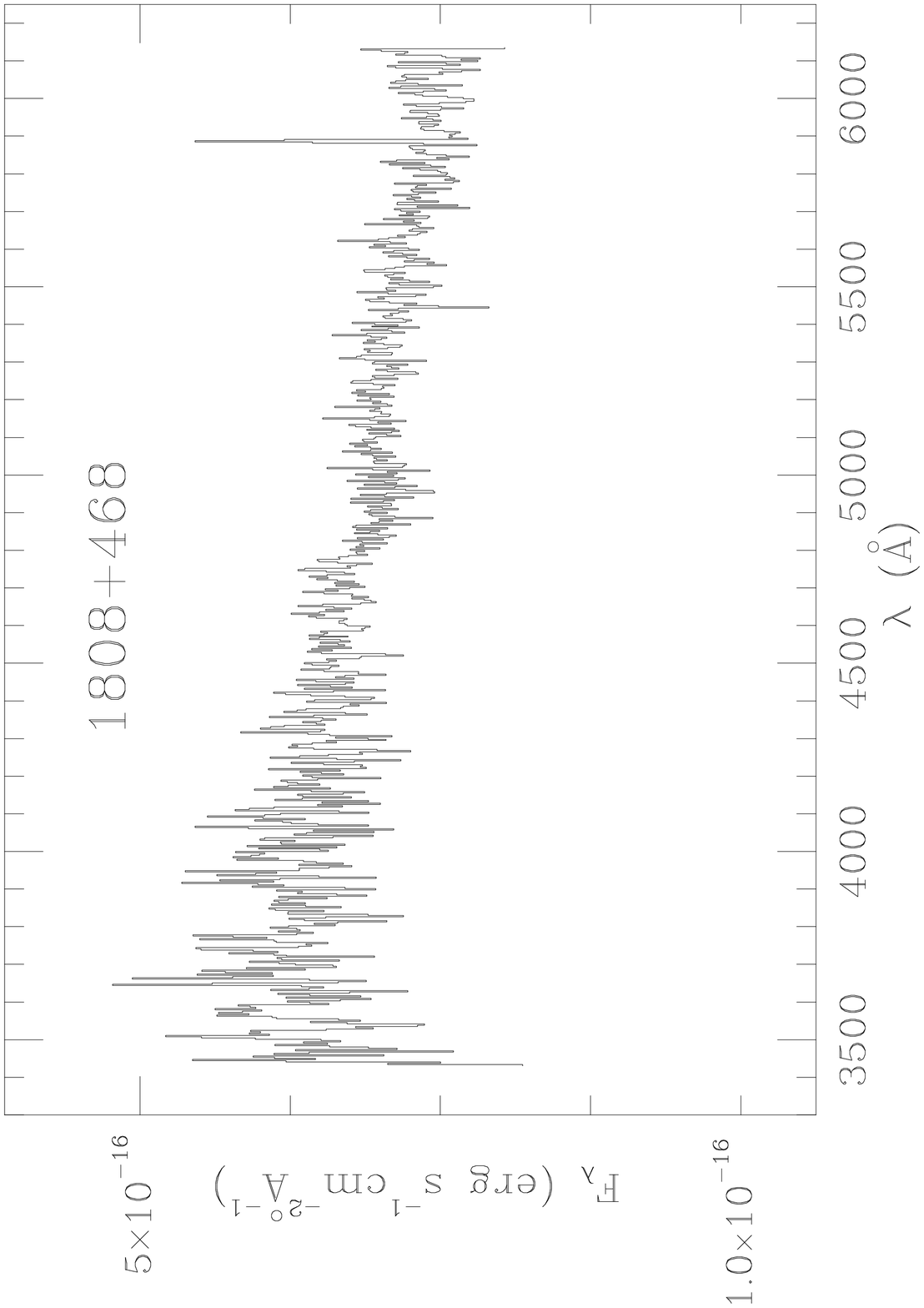,height=7.1cm,width=6.3cm}
\end{minipage}
\hfill
\begin{minipage}[t]{0.3in}
\vfill
\begin{sideways}
Figure 1.133 $-$ 1.138: Spectra of RGB Sources ({\it continued})
\end{sideways}
\vfill
\end{minipage}
\end{figure}

\clearpage
\begin{figure}
\vspace{-0.3in}
\hspace{-0.3in}
\begin{minipage}[t]{6.3in}
\psfig{file=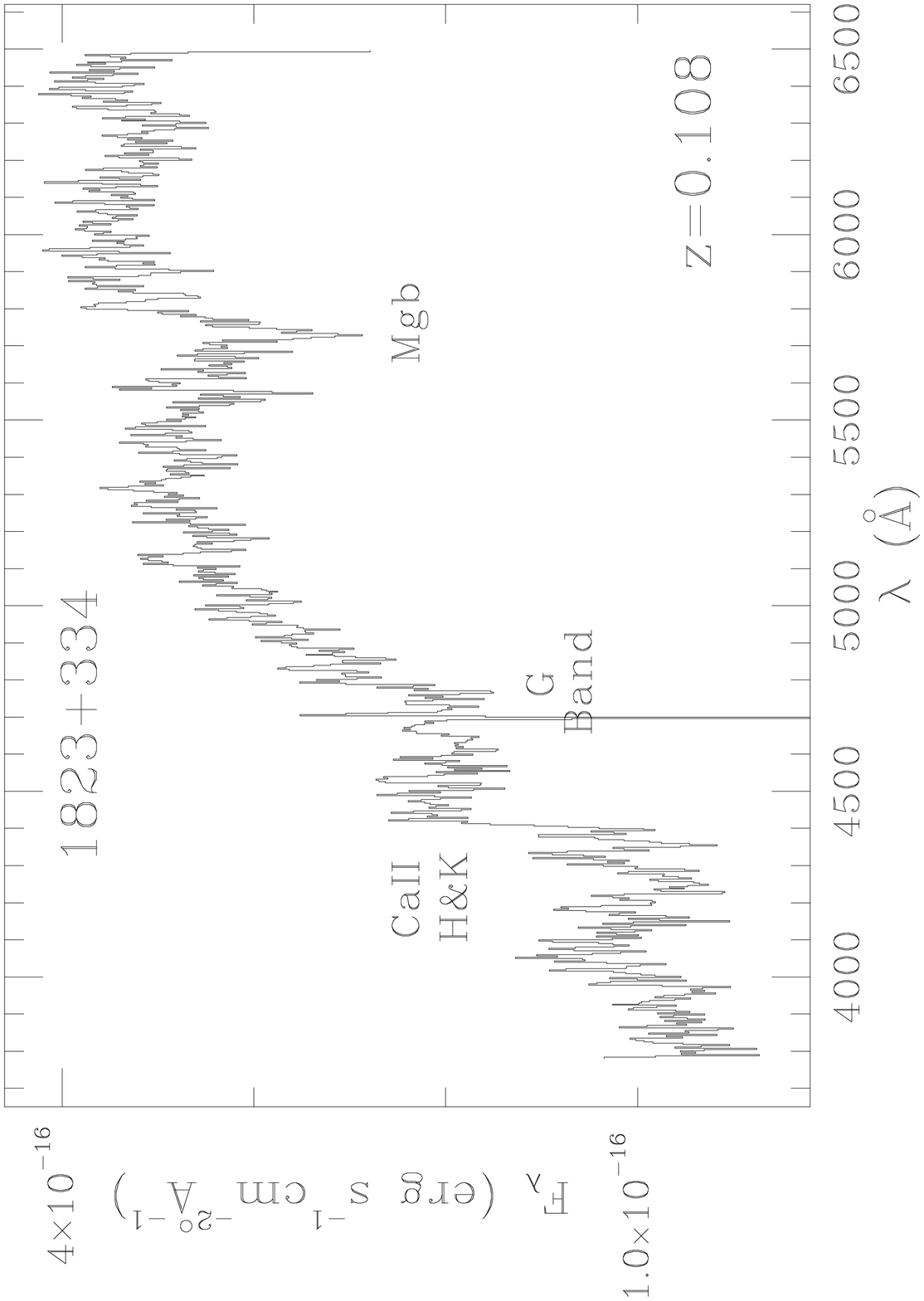,height=7.1cm,width=6.3cm}
\vspace{0.25in}
\psfig{file=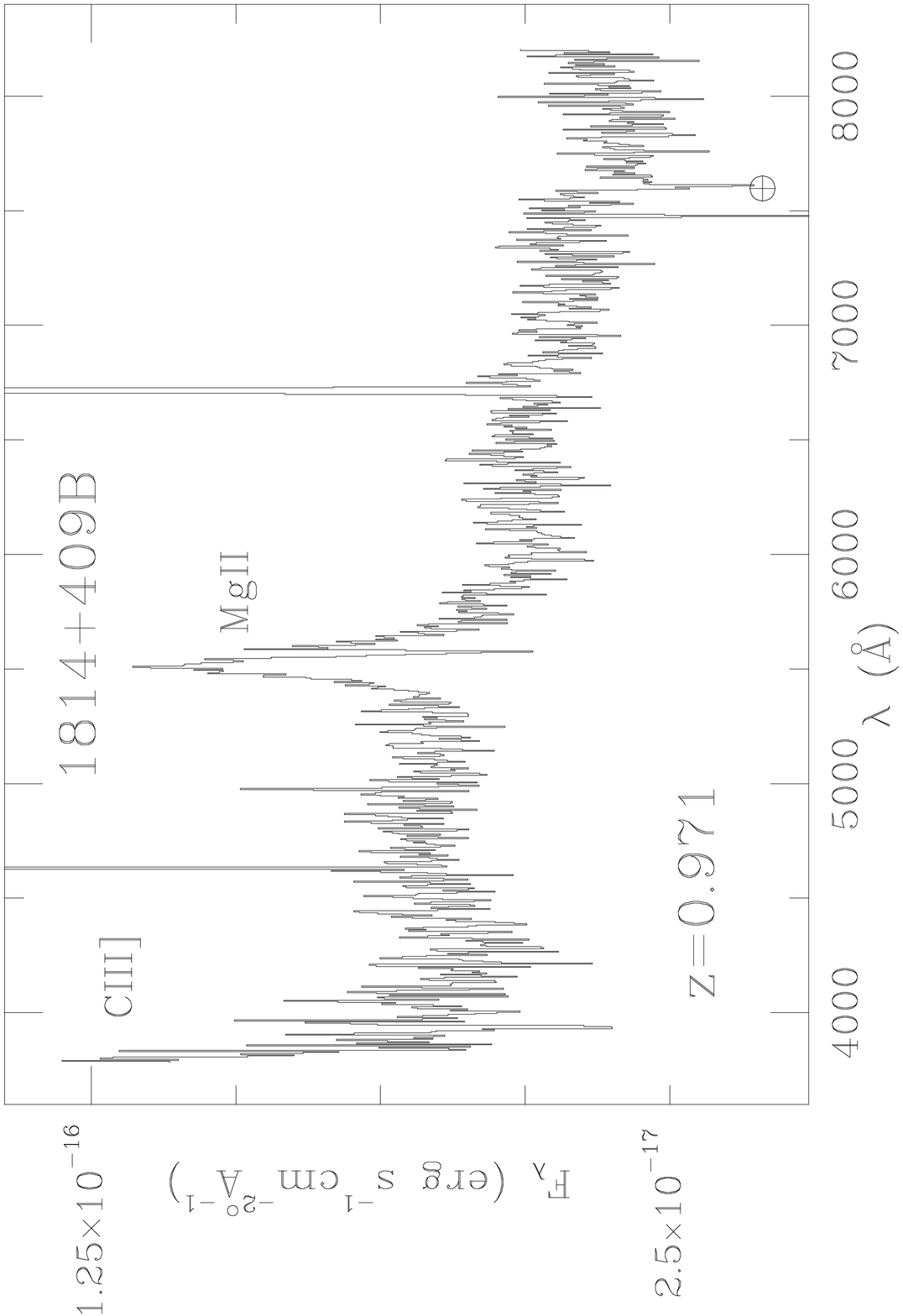,height=7.1cm,width=6.3cm}
\vspace{0.25in}
\psfig{file=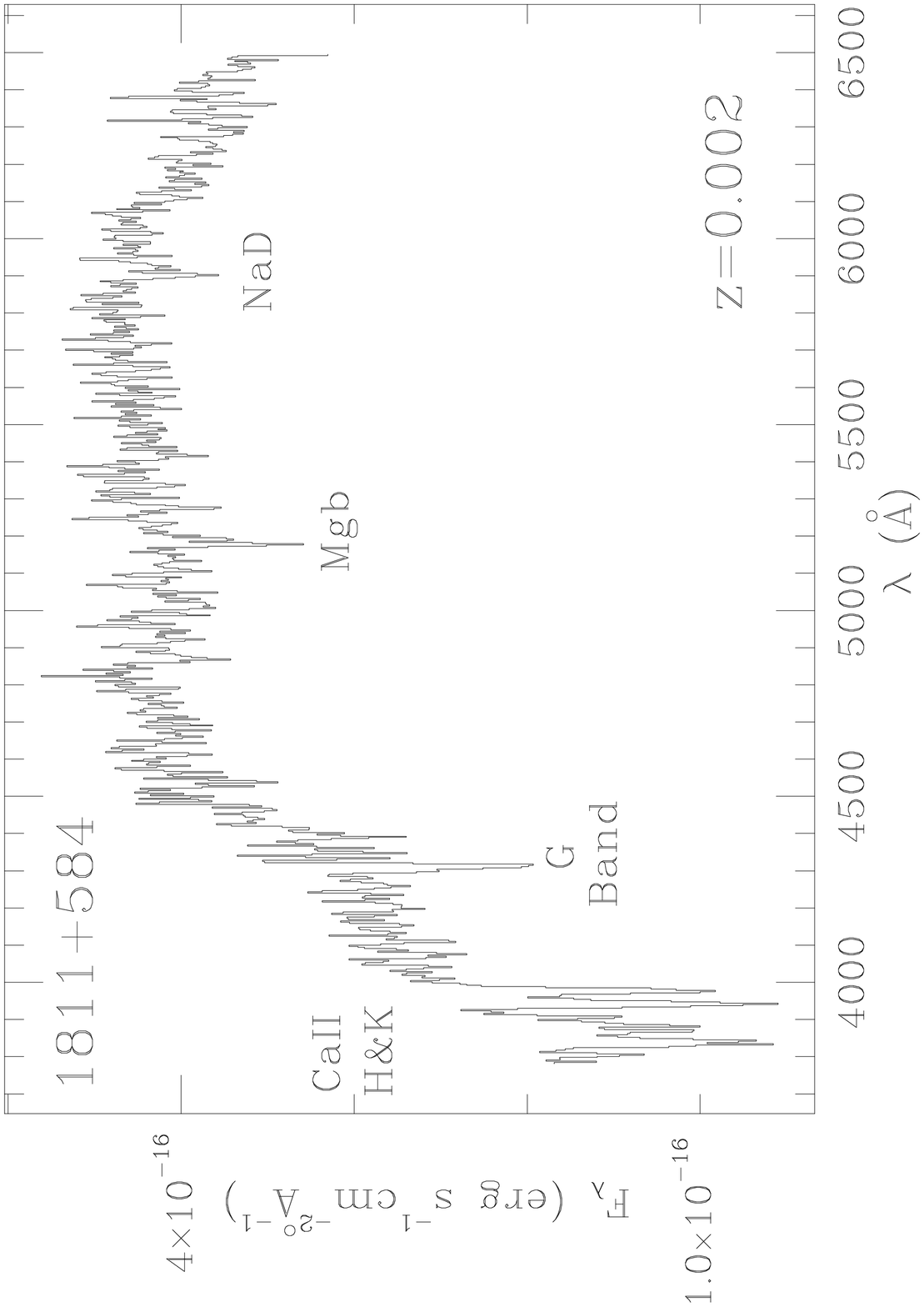,height=7.1cm,width=6.3cm}
\end{minipage}
\hspace{0.3in}
\begin{minipage}[t]{6.3in}
\psfig{file=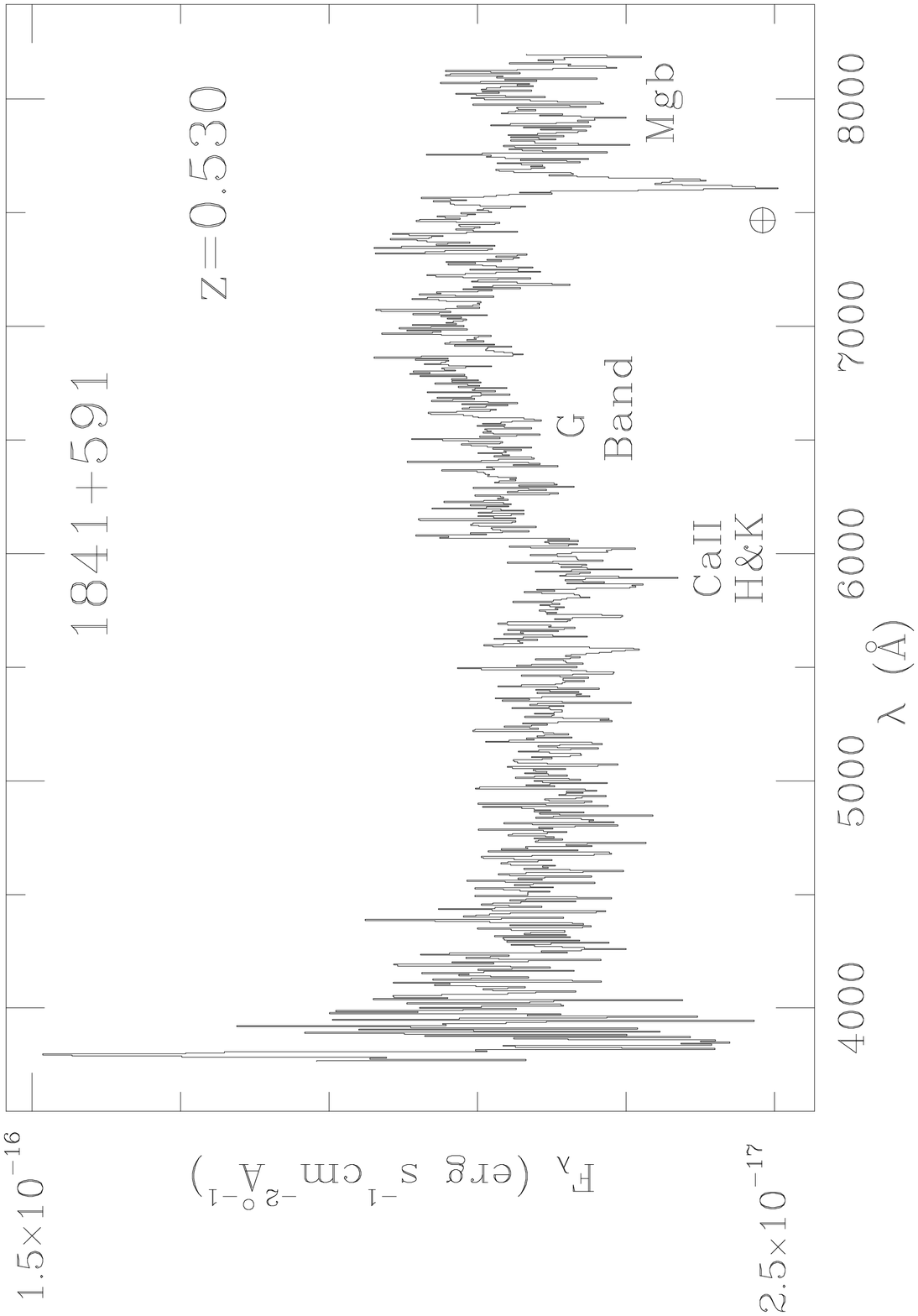,height=7.1cm,width=6.3cm}
\vspace{0.25in}
\psfig{file=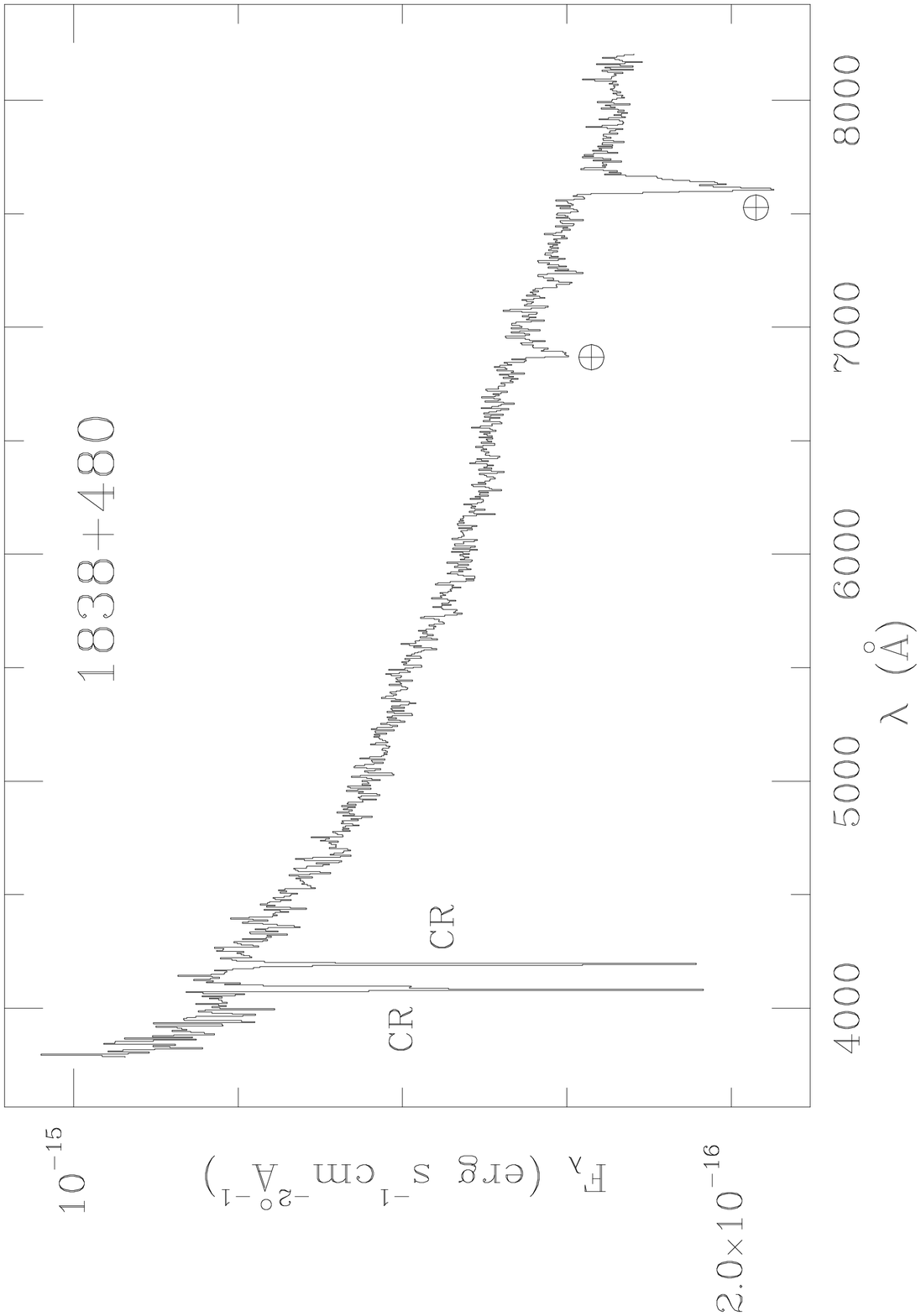,height=7.1cm,width=6.3cm}
\vspace{0.25in}
\psfig{file=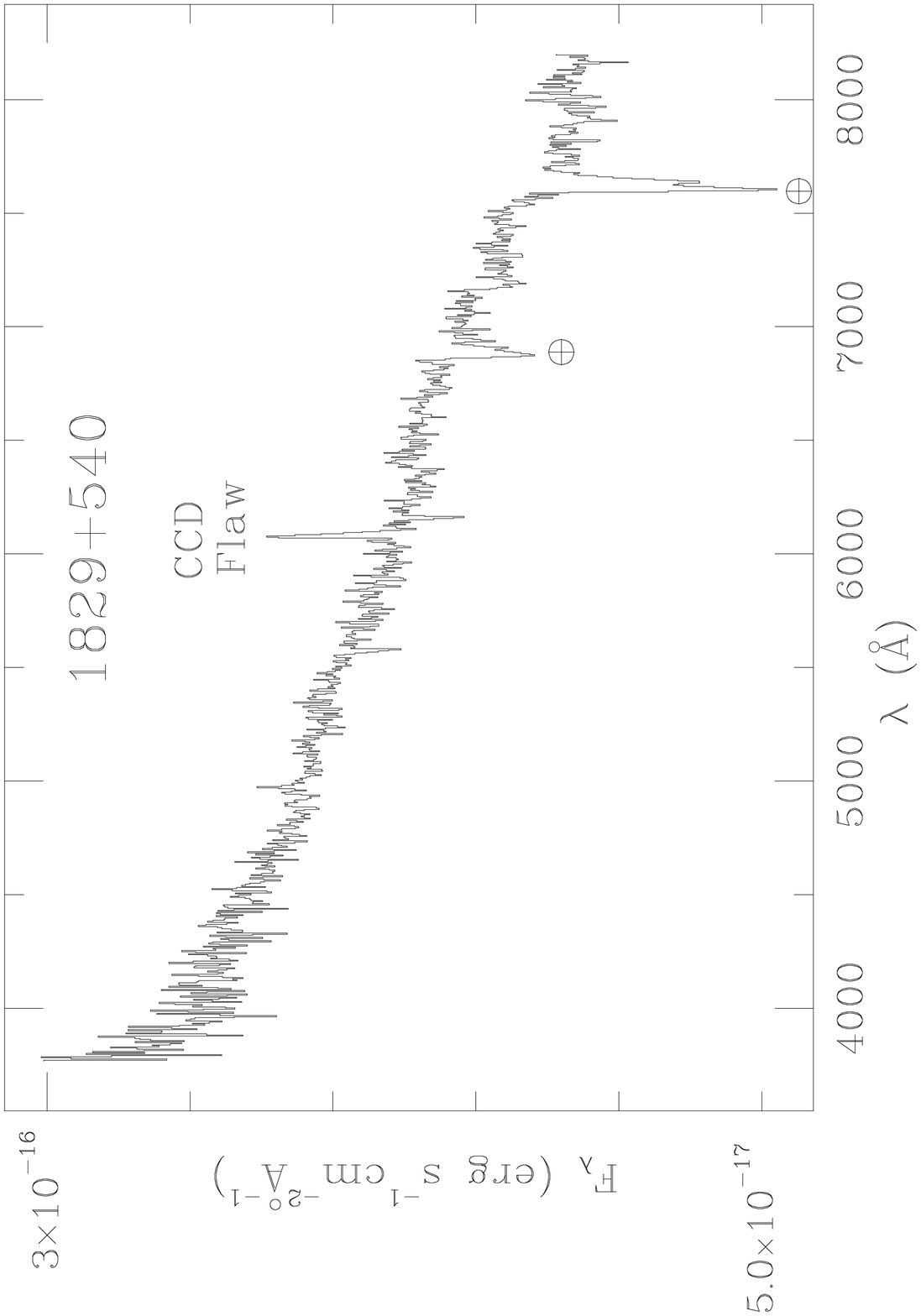,height=7.1cm,width=6.3cm}
\end{minipage}
\hfill
\begin{minipage}[t]{0.3in}
\vfill
\begin{sideways}
Figure 1.139 $-$ 1.144: Spectra of RGB Sources ({\it continued})
\end{sideways}
\vfill
\end{minipage}
\end{figure}

\clearpage
\begin{figure}
\vspace{-0.3in}
\hspace{-0.3in}
\begin{minipage}[t]{6.3in}
\psfig{file=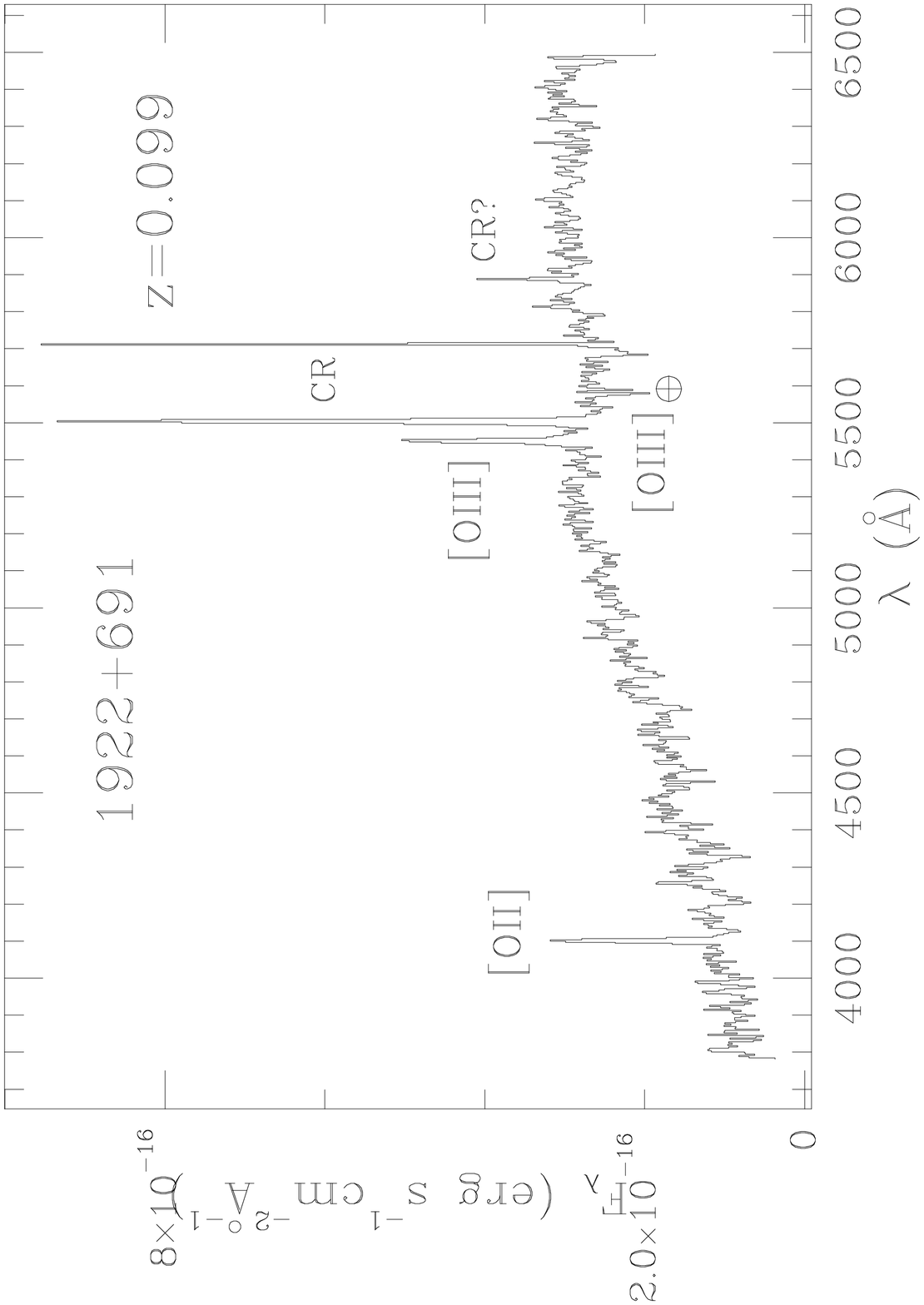,height=7.1cm,width=6.3cm}
\vspace{0.25in}
\psfig{file=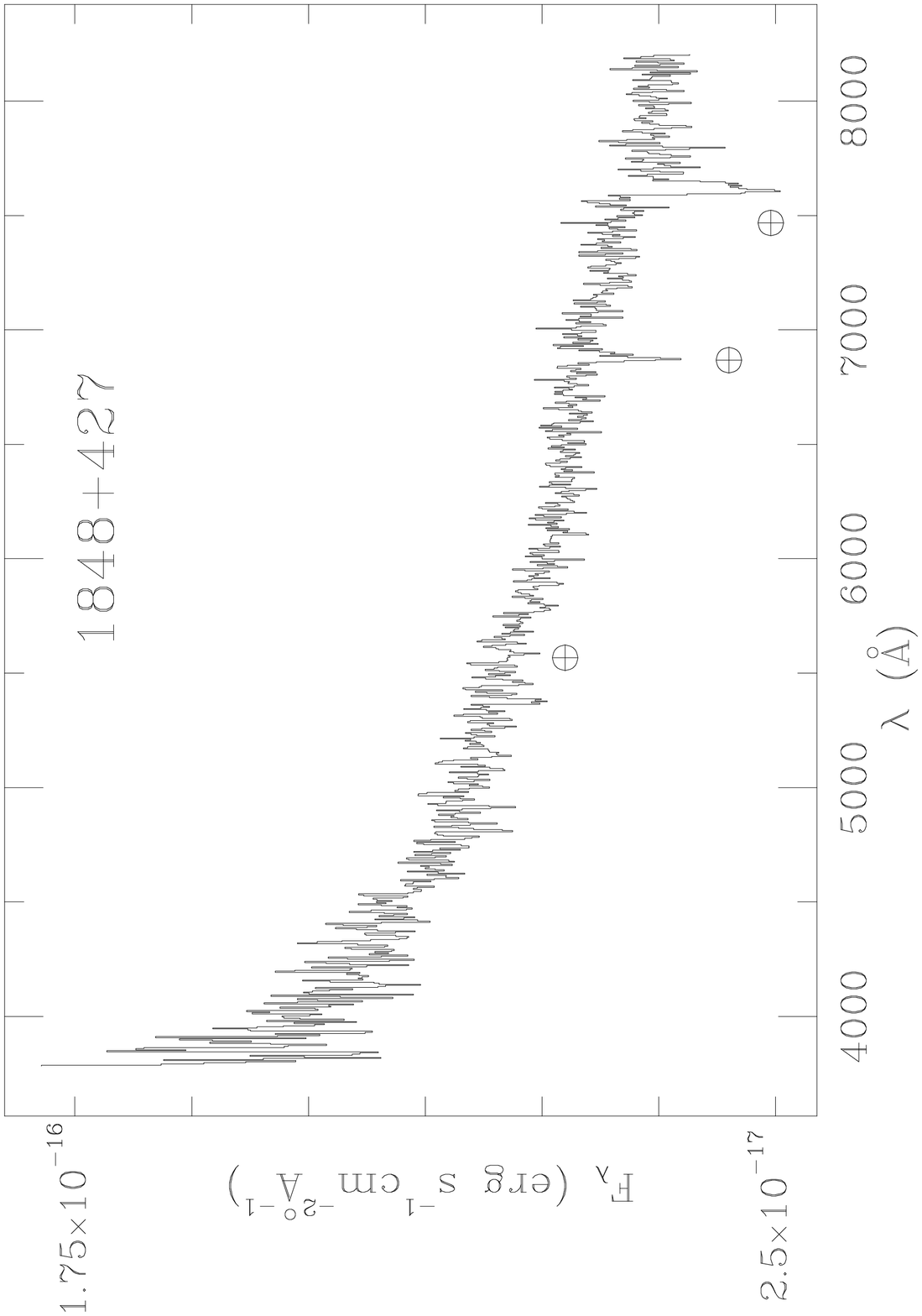,height=7.1cm,width=6.3cm}
\vspace{0.25in}
\psfig{file=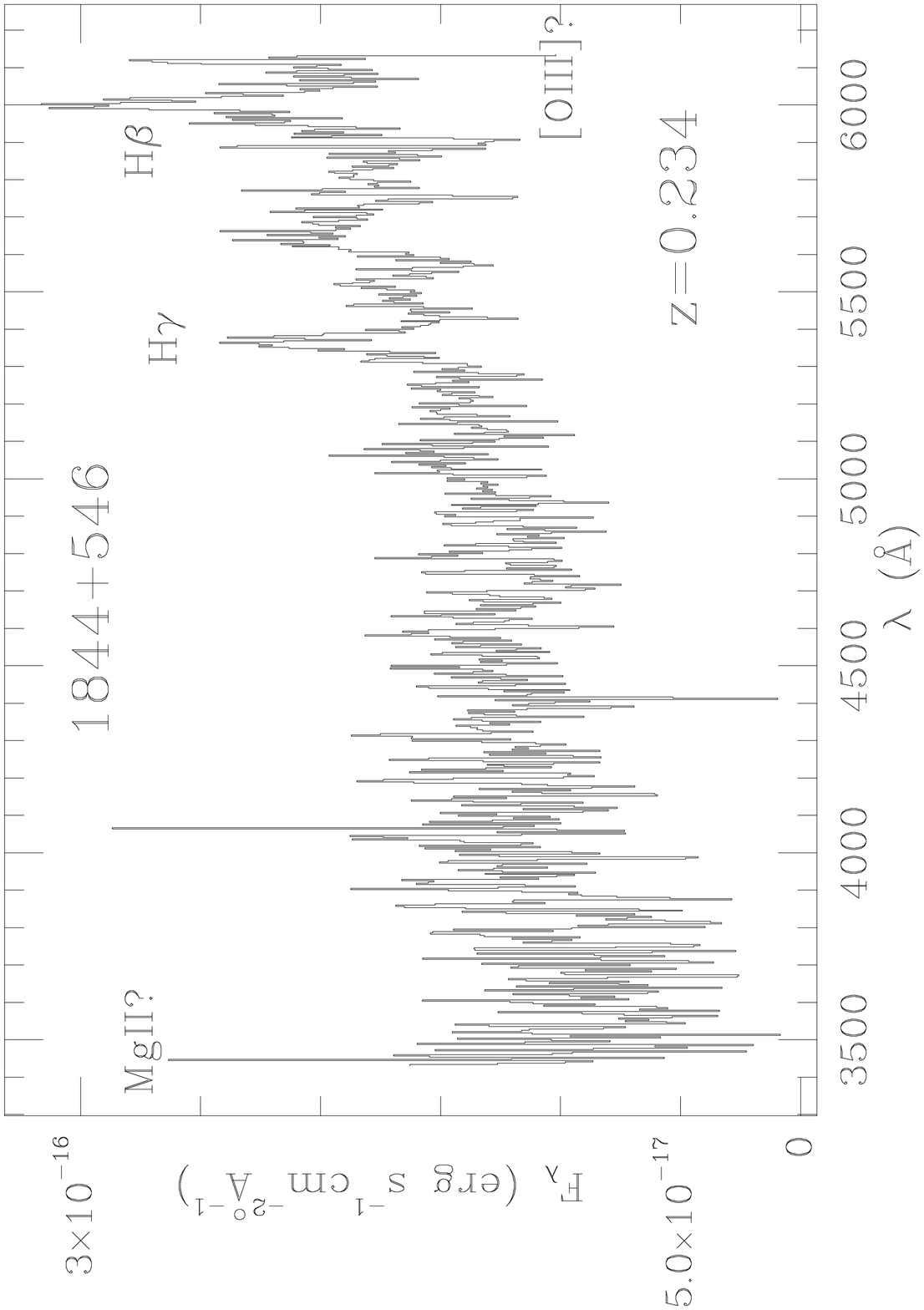,height=7.1cm,width=6.3cm}
\end{minipage}
\hspace{0.3in}
\begin{minipage}[t]{6.3in}
\psfig{file=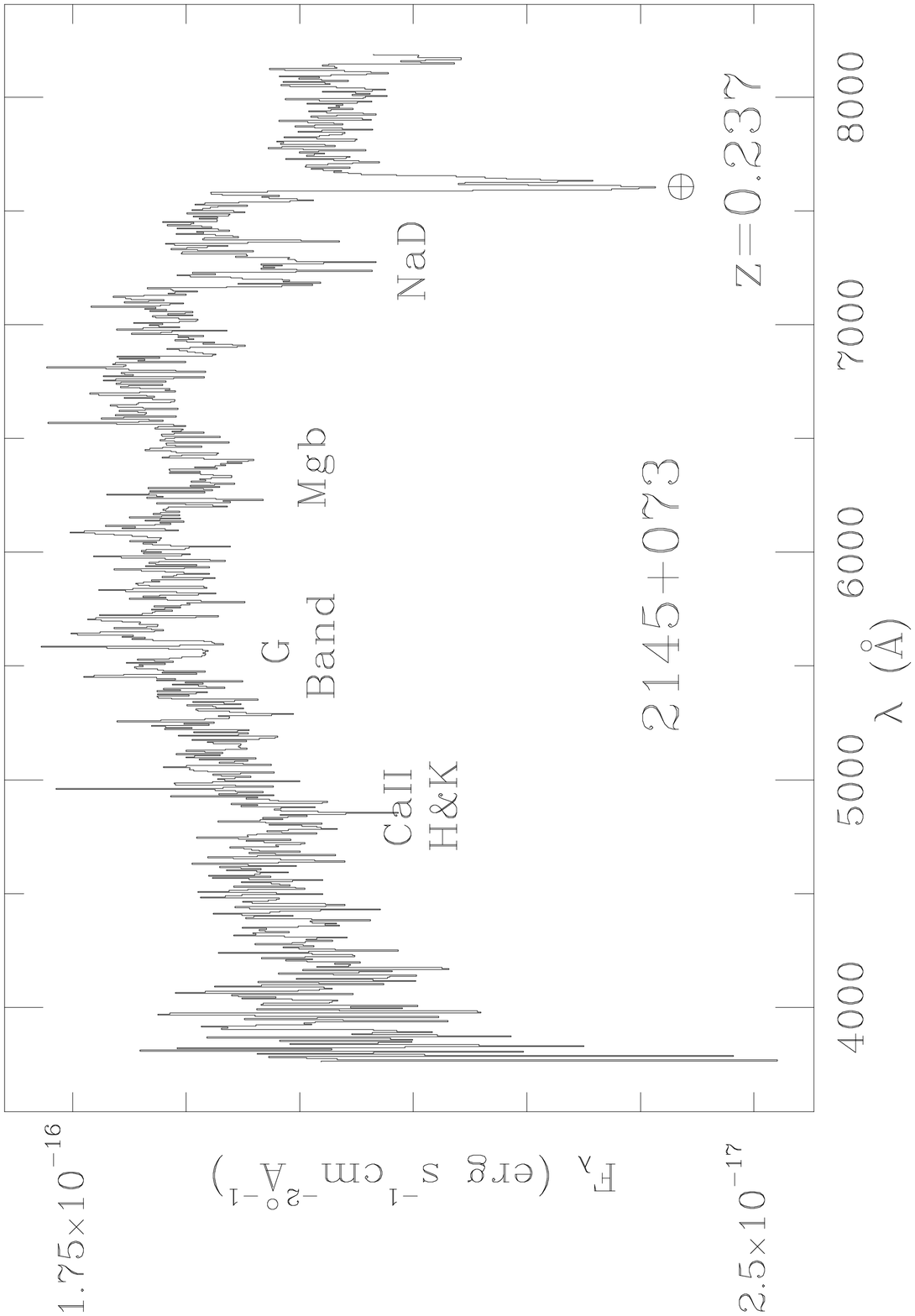,height=7.1cm,width=6.3cm}
\vspace{0.25in}
\psfig{file=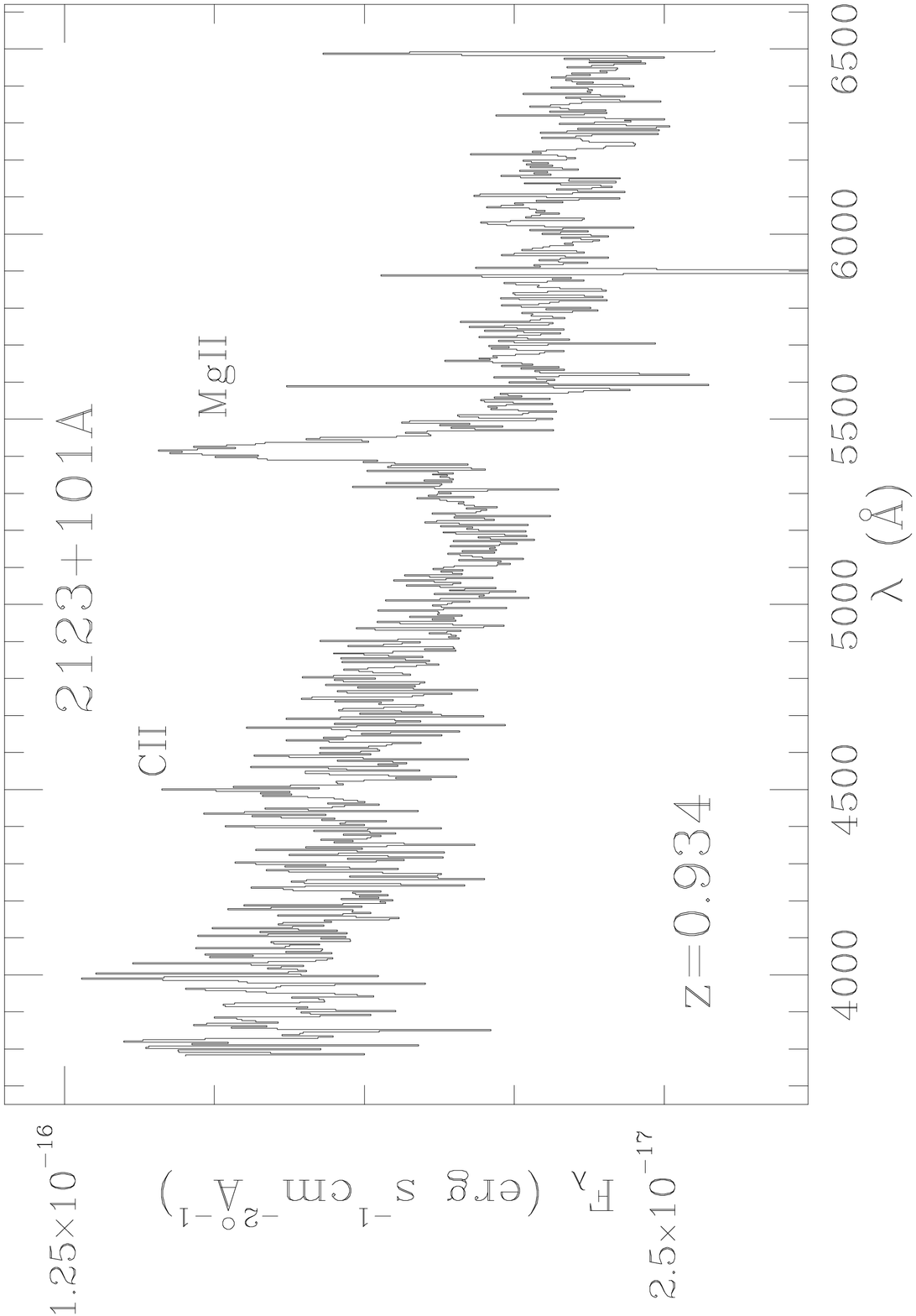,height=7.1cm,width=6.3cm}
\vspace{0.25in}
\psfig{file=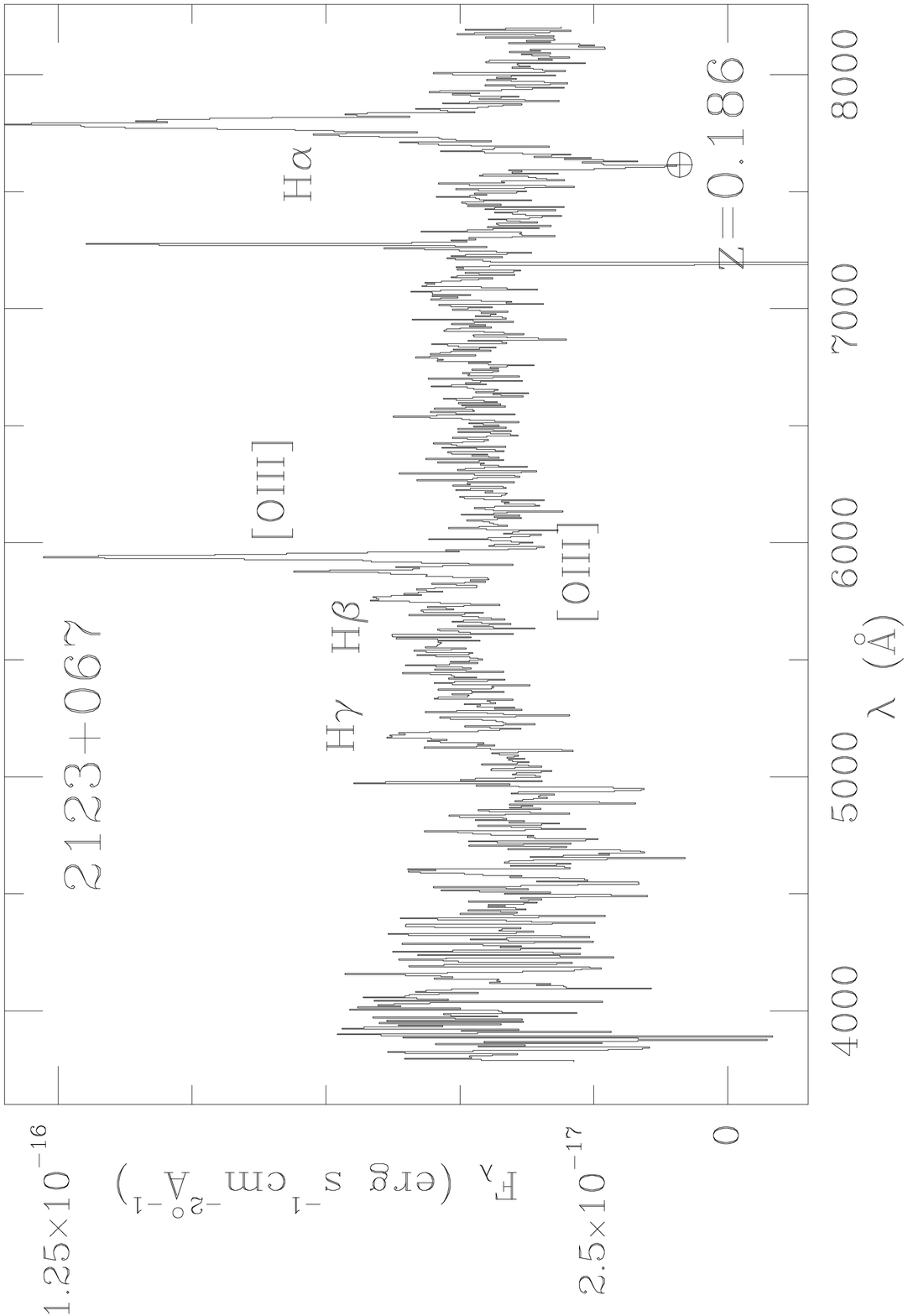,height=7.1cm,width=6.3cm}
\end{minipage}
\hfill
\begin{minipage}[t]{0.3in}
\vfill
\begin{sideways}
Figure 1.145 $-$ 1.150: Spectra of RGB Sources ({\it continued})
\vspace{0.25in}
\end{sideways}
\vfill
\end{minipage}
\end{figure}

\clearpage
\begin{figure}
\vspace{-0.3in}
\hspace{-0.3in}
\begin{minipage}[t]{6.3in}
\psfig{file=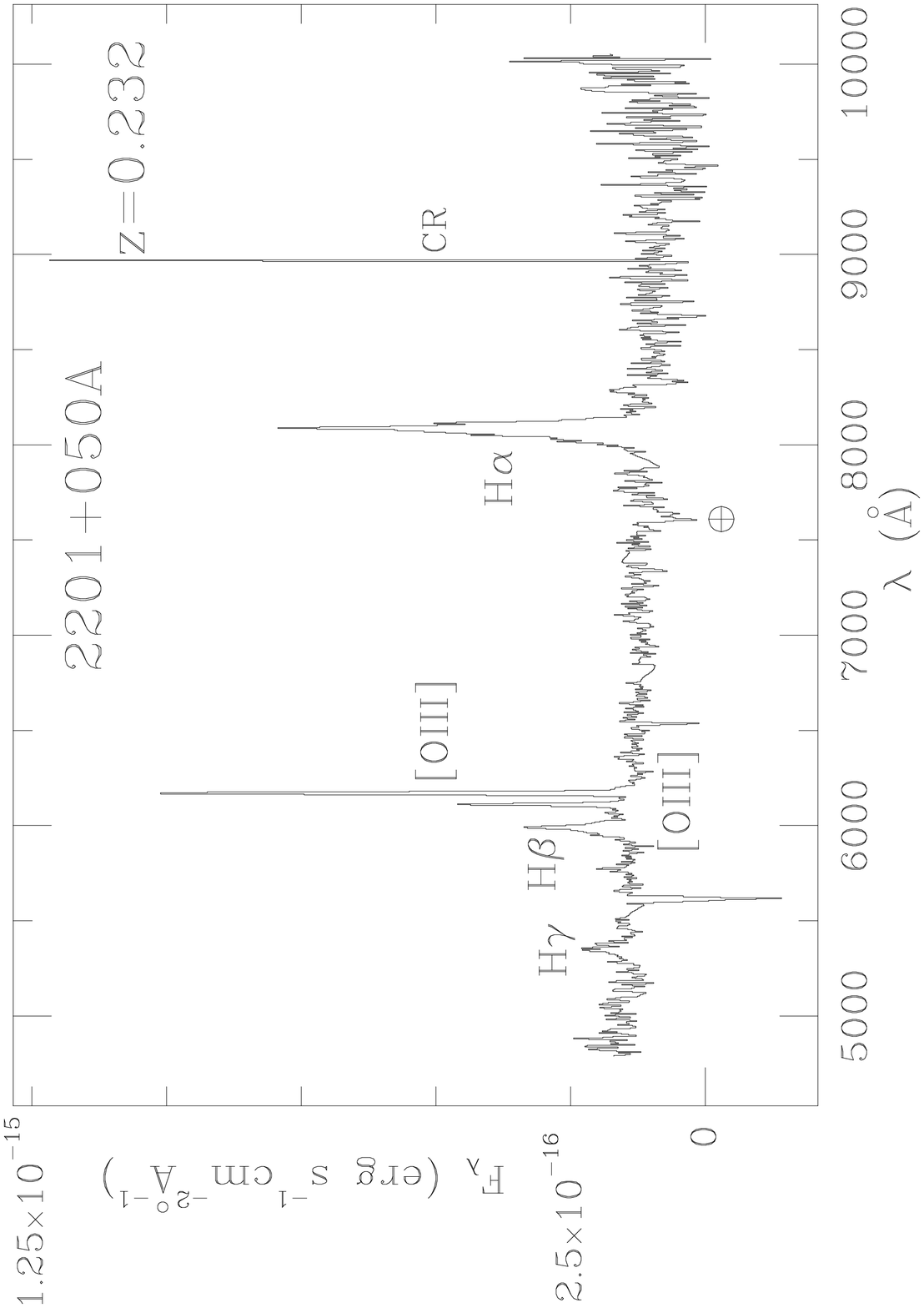,height=7.1cm,width=6.3cm}
\vspace{0.25in}
\psfig{file=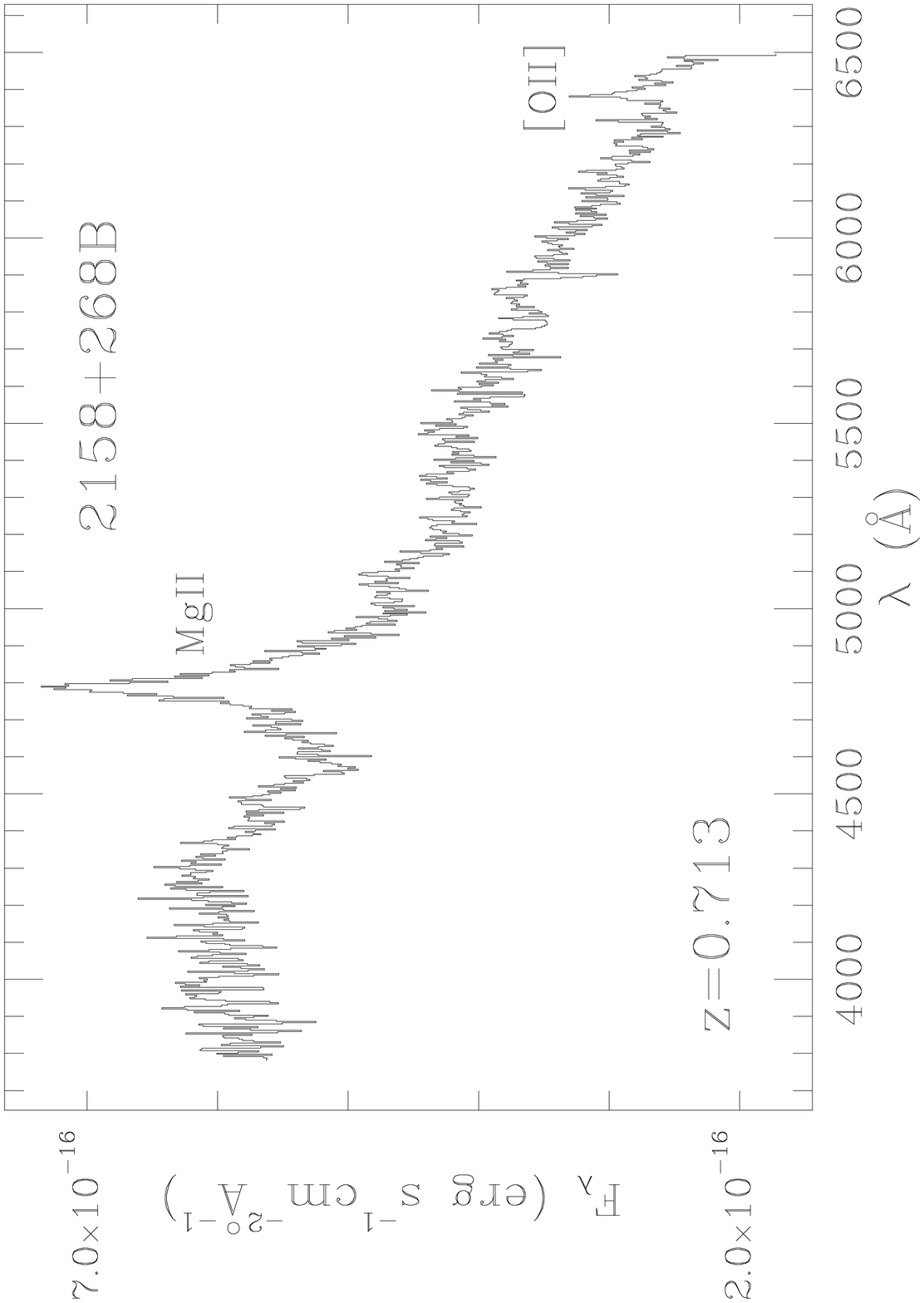,height=7.1cm,width=6.3cm}
\vspace{0.25in}
\psfig{file=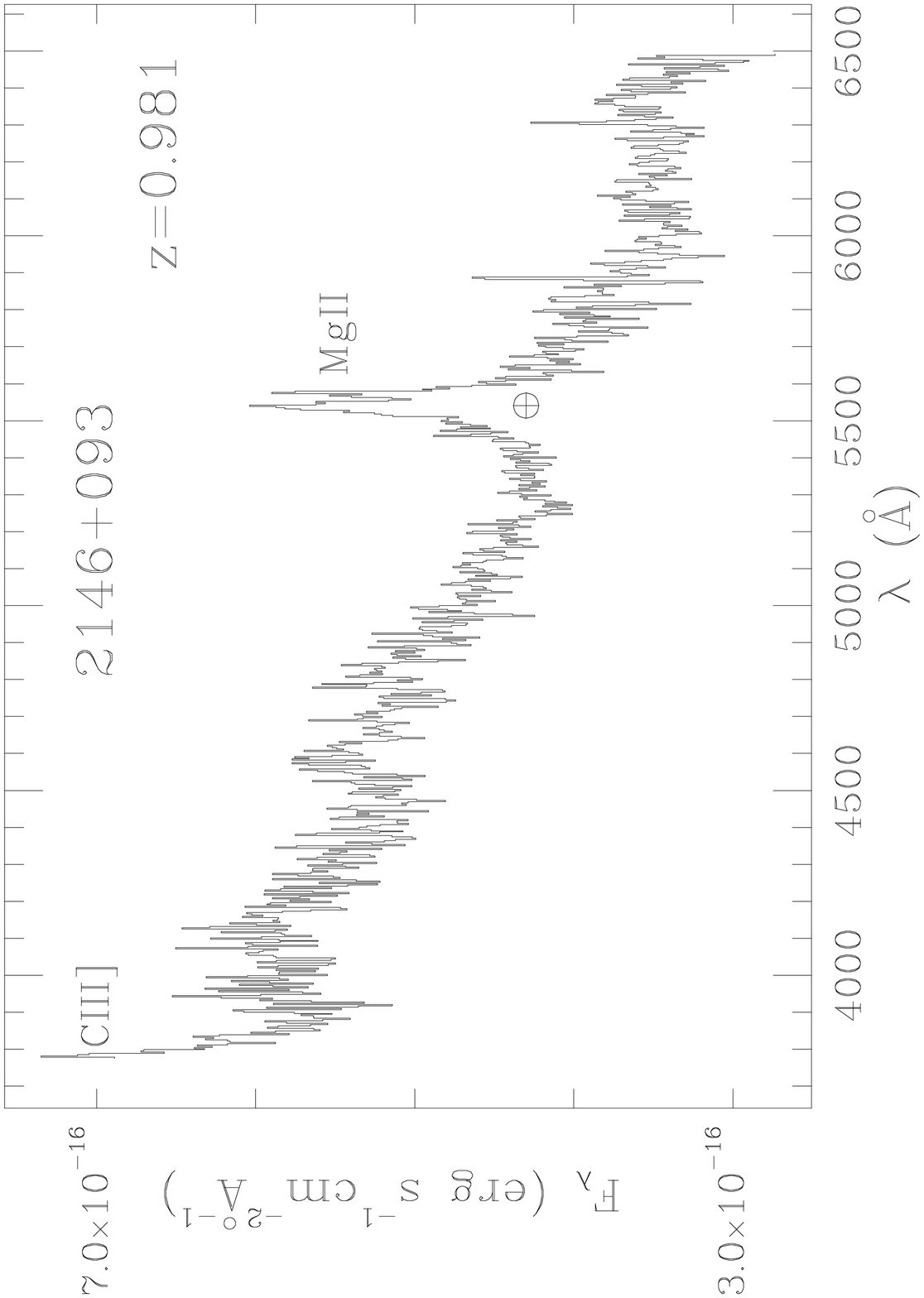,height=7.1cm,width=6.3cm}
\end{minipage}
\hspace{0.3in}
\begin{minipage}[t]{6.3in}
\psfig{file=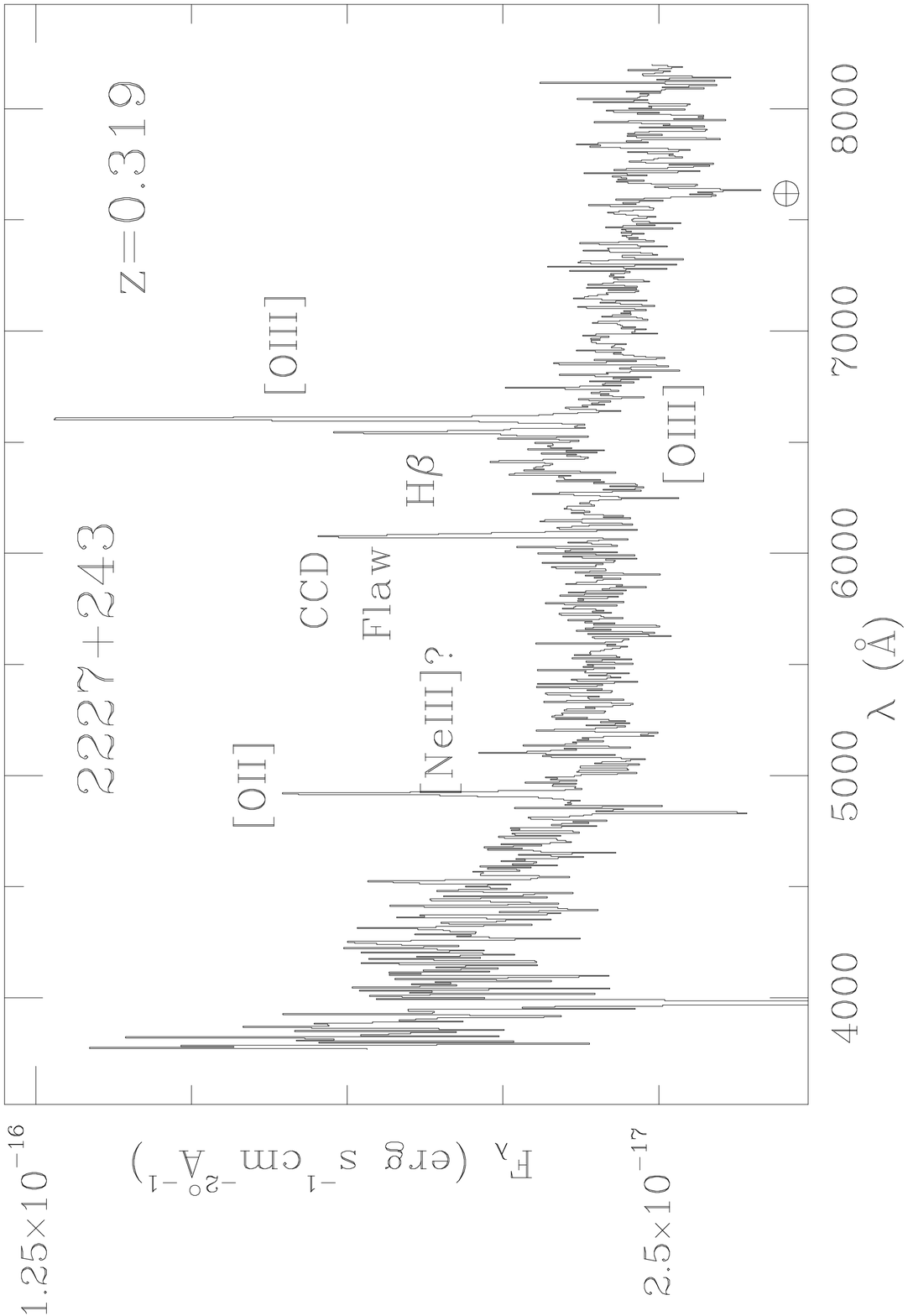,height=7.1cm,width=6.3cm}
\vspace{0.25in}
\psfig{file=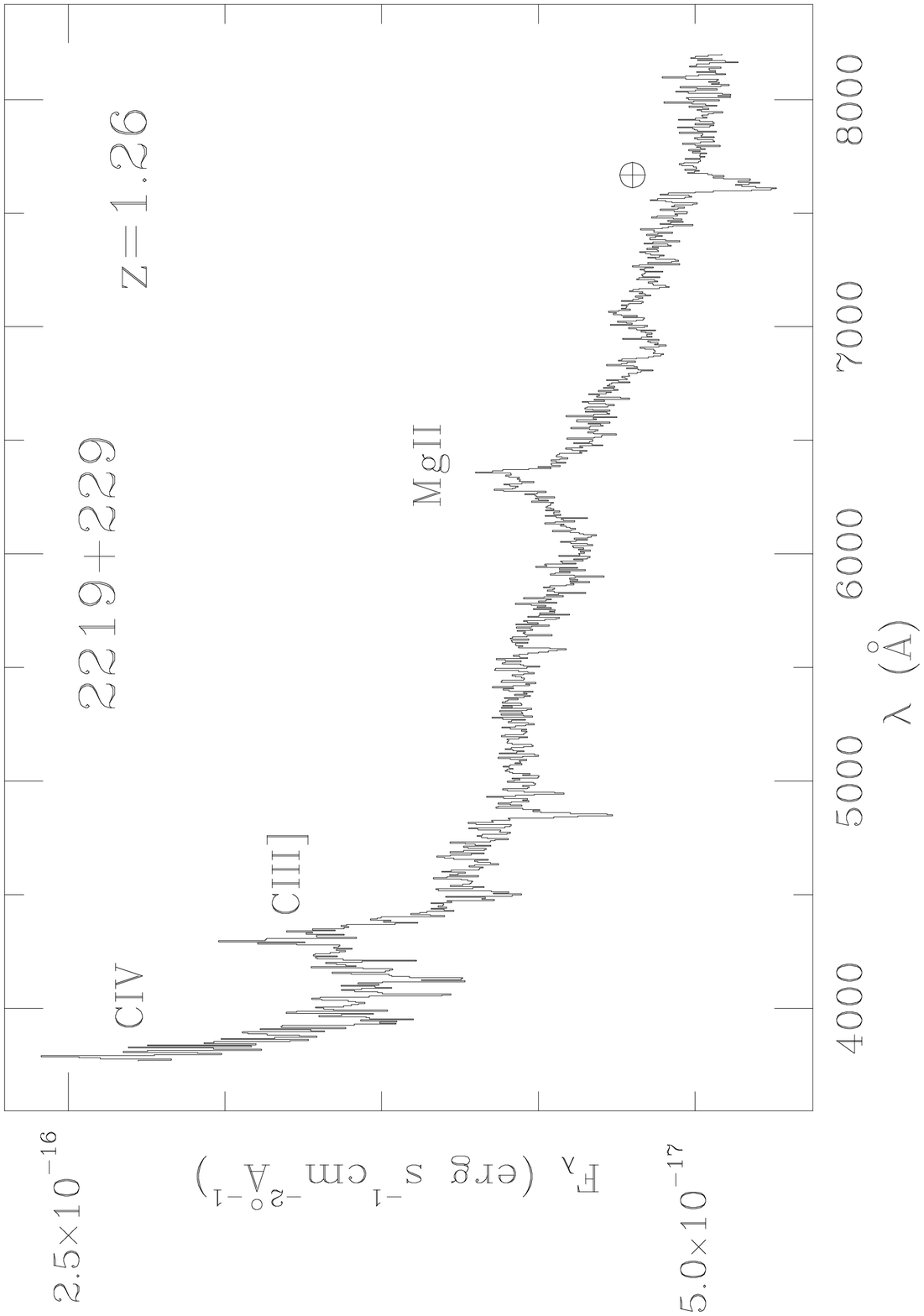,height=7.1cm,width=6.3cm}
\vspace{0.25in}
\psfig{file=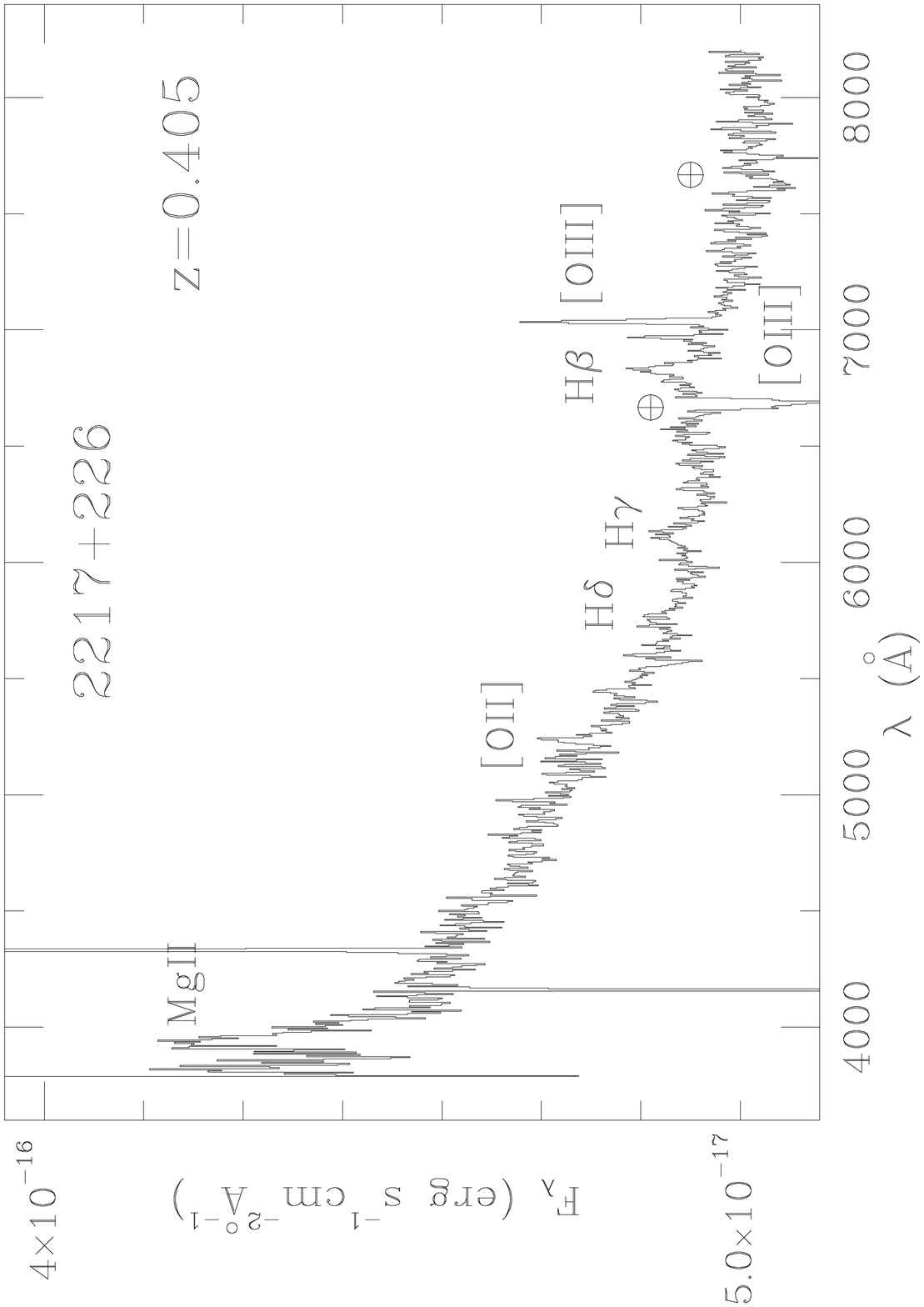,height=7.1cm,width=6.3cm}
\end{minipage}
\hfill
\begin{minipage}[t]{0.3in}
\vfill
\begin{sideways}
Figure 1.151 $-$ 1.156: Spectra of RGB Sources ({\it continued})
\vspace{0.25in}
\end{sideways}
\vfill
\end{minipage}
\end{figure}

\clearpage
\begin{figure}
\vspace{-0.3in}
\hspace{-0.3in}
\begin{minipage}[t]{6.3in}
\psfig{file=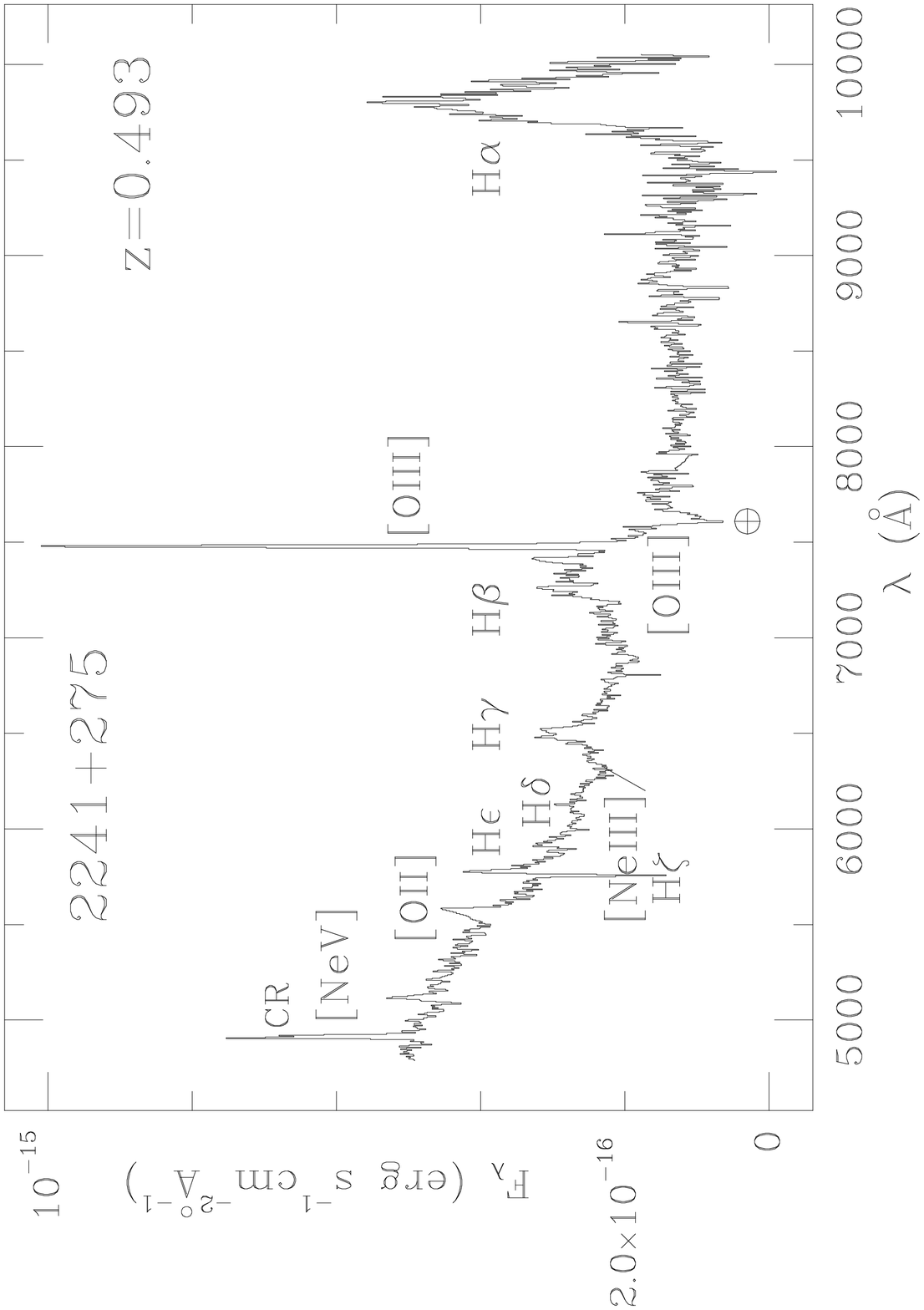,height=7.1cm,width=6.3cm}
\vspace{0.25in}
\psfig{file=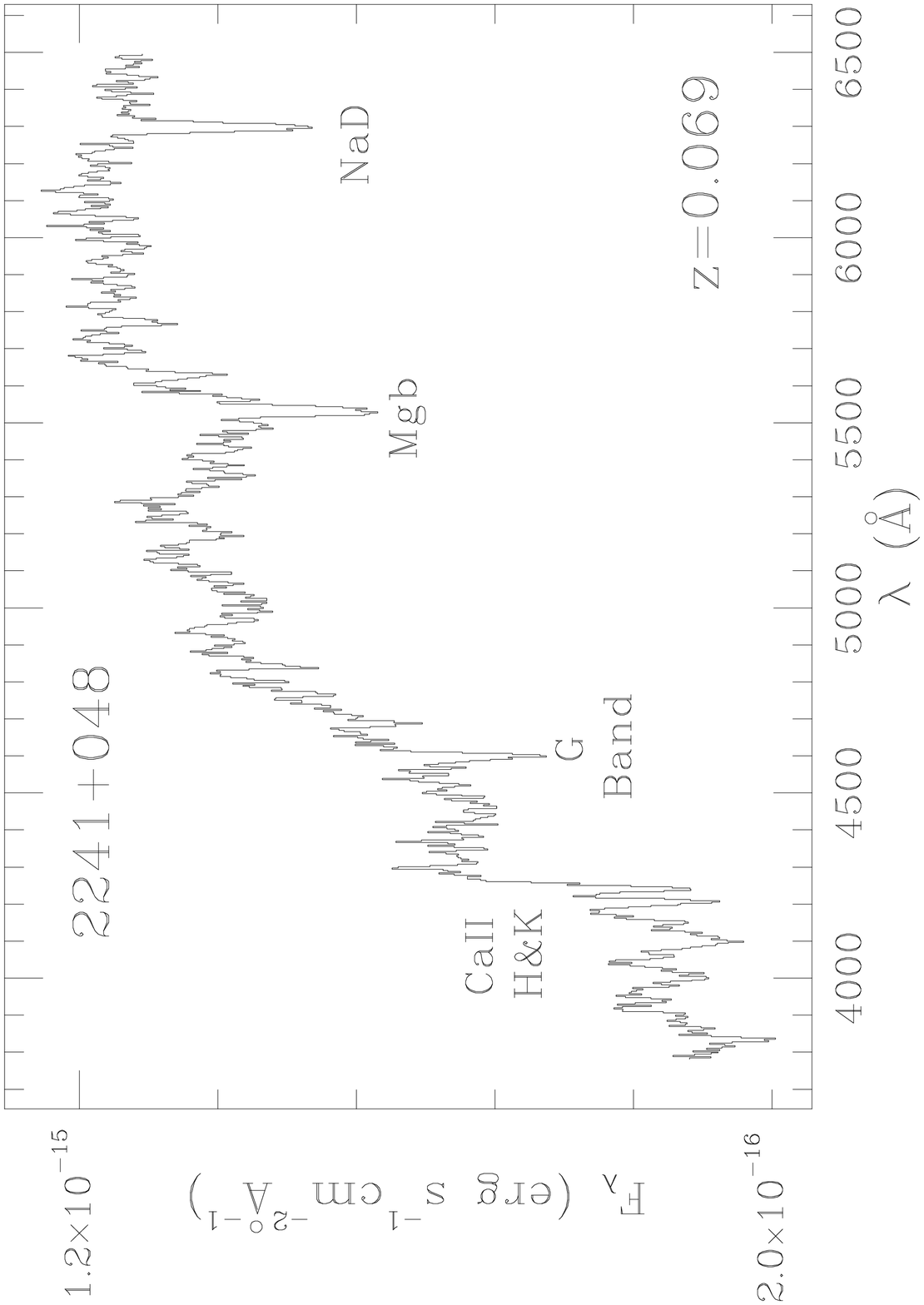,height=7.1cm,width=6.3cm}
\vspace{0.25in}
\psfig{file=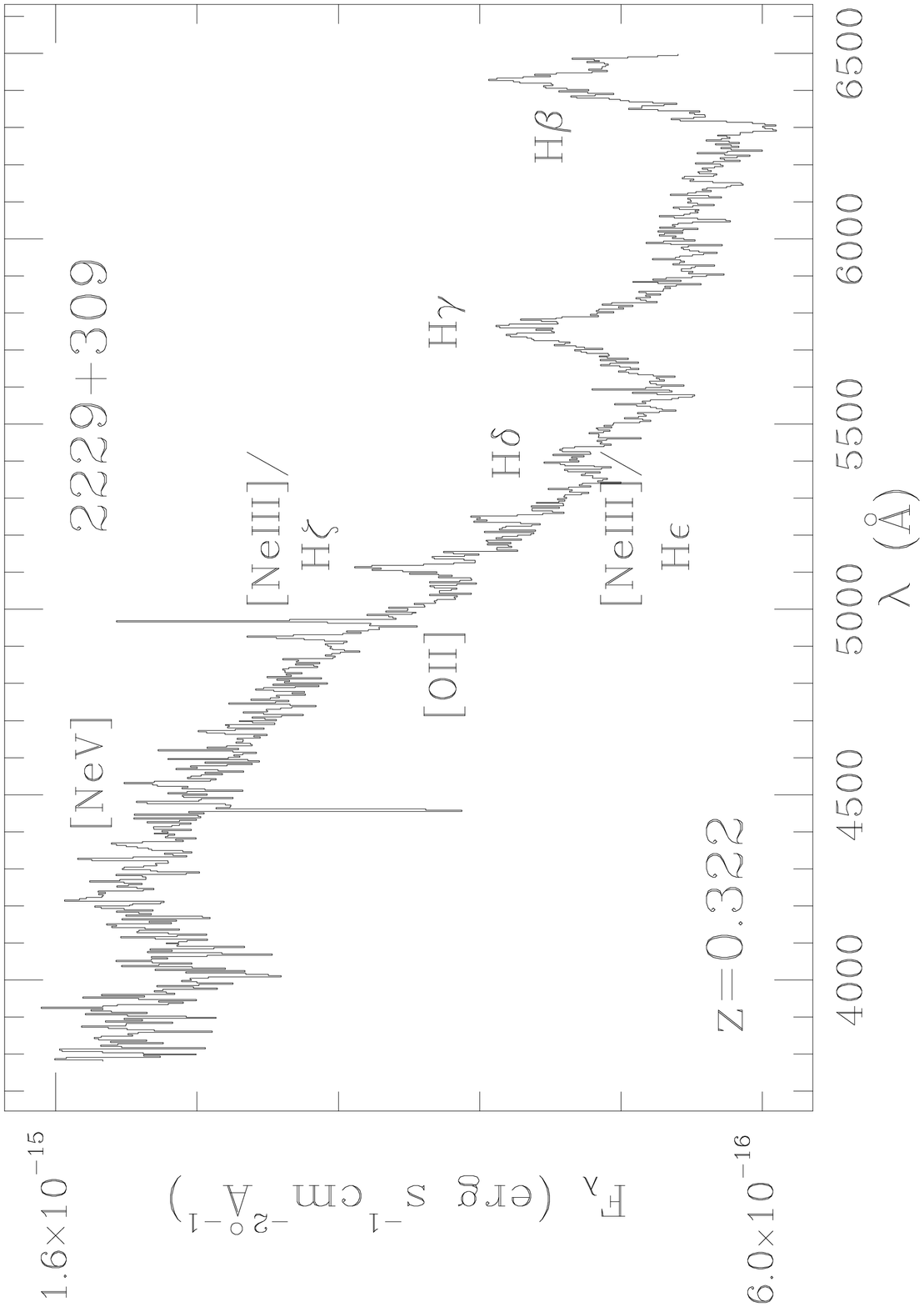,height=7.1cm,width=6.3cm}
\end{minipage}
\hspace{0.3in}
\begin{minipage}[t]{6.3in}
\psfig{file=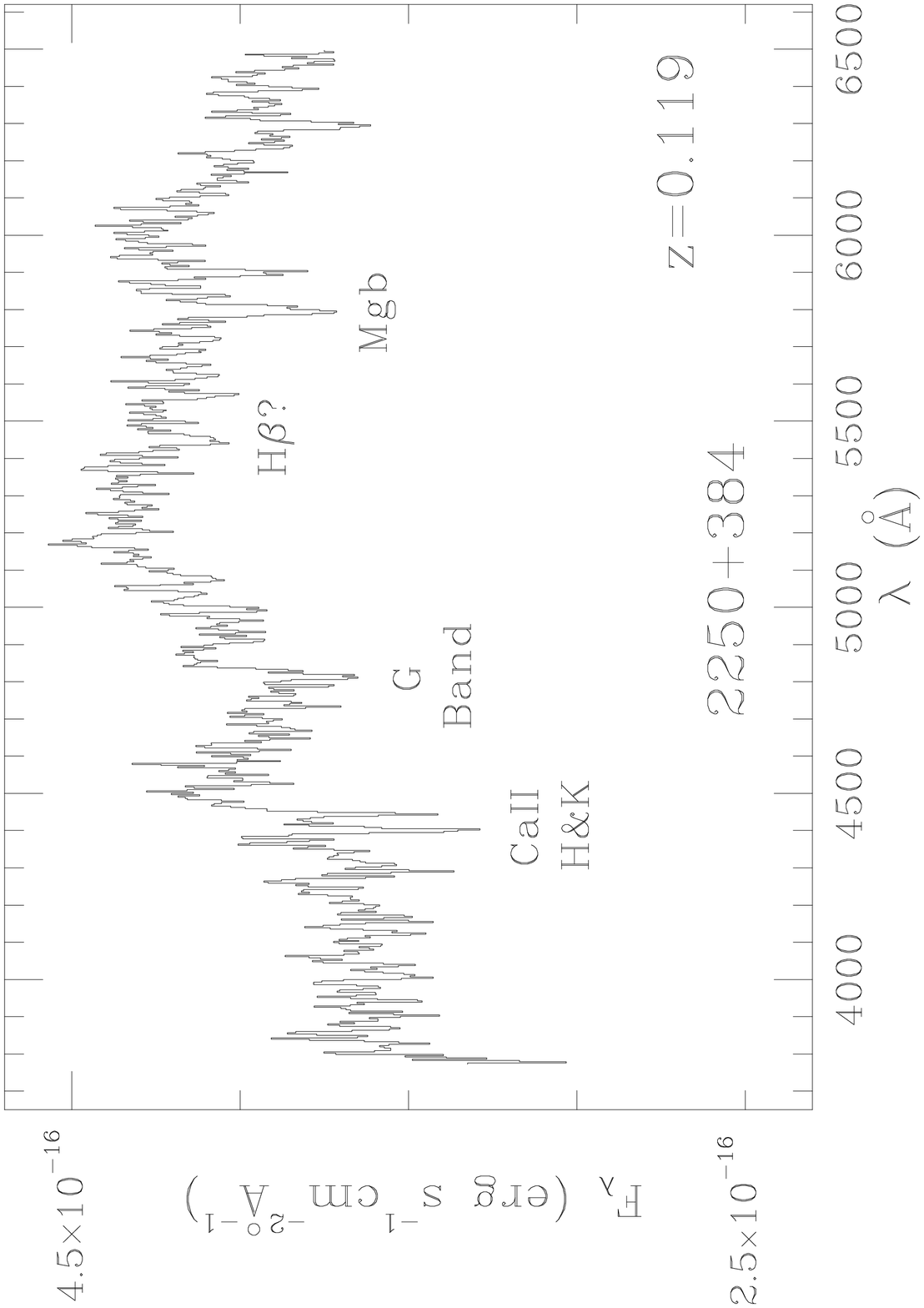,height=7.1cm,width=6.3cm}
\vspace{0.25in}
\psfig{file=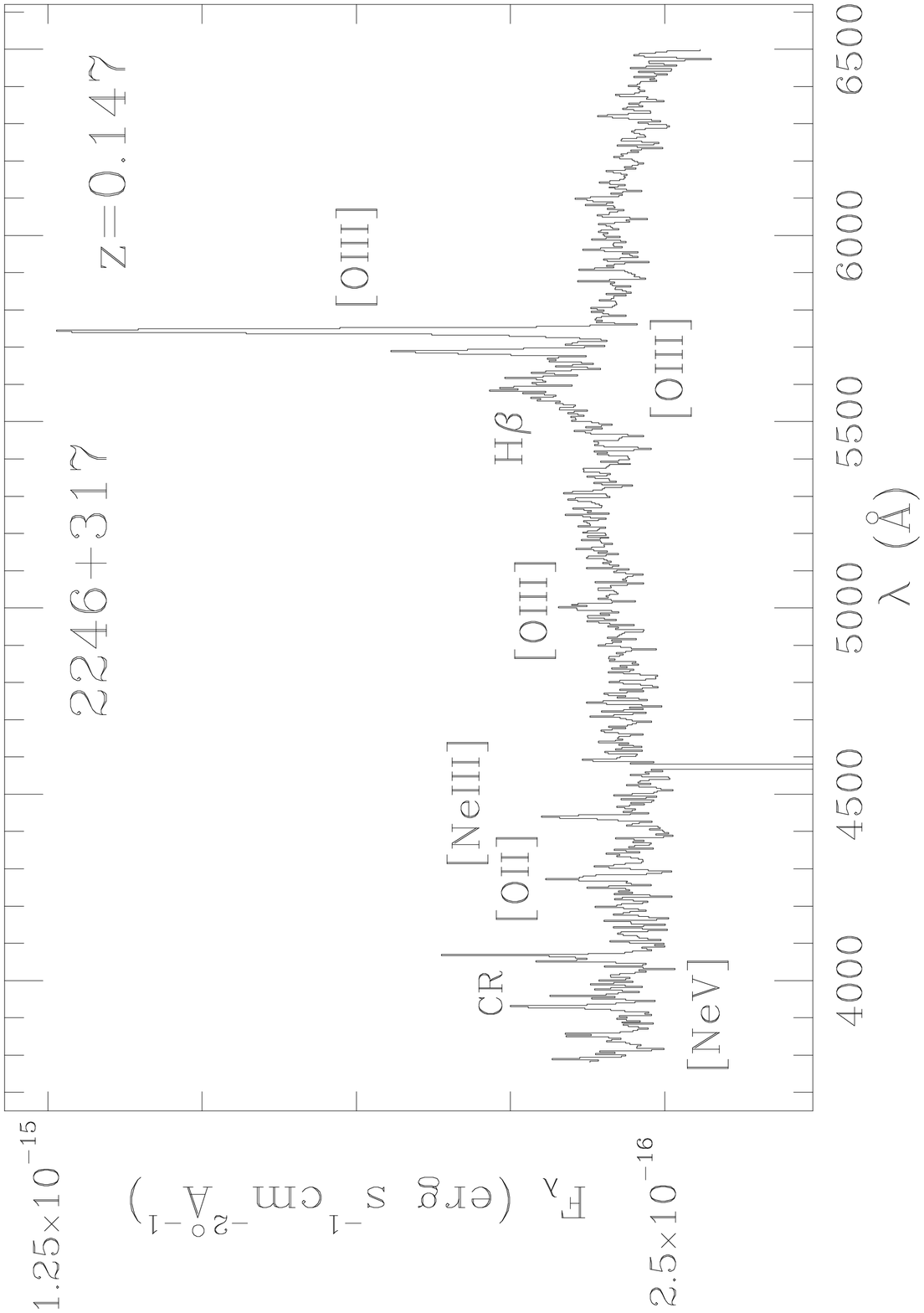,height=7.1cm,width=6.3cm}
\vspace{0.25in}
\psfig{file=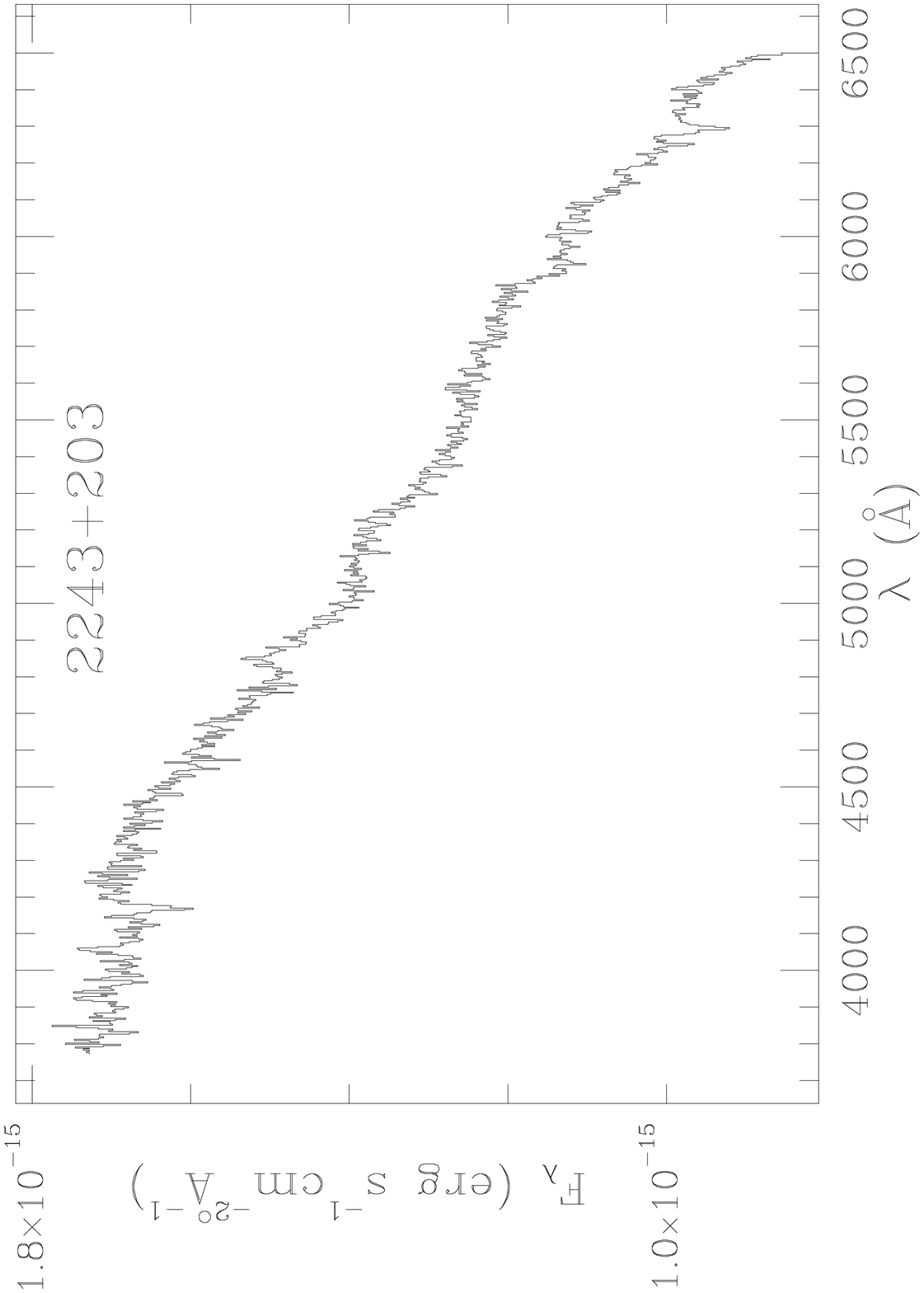,height=7.1cm,width=6.3cm}
\end{minipage}
\hfill
\begin{minipage}[t]{0.3in}
\vfill
\begin{sideways}
Figure 1.157 $-$ 1.162: Spectra of RGB Sources ({\it continued})
\vspace{0.25in}
\end{sideways}
\vfill
\end{minipage}
\end{figure}

\clearpage
\begin{figure}
\vspace{-0.3in}
\hspace{-0.3in}
\begin{minipage}[t]{6.3in}
\psfig{file=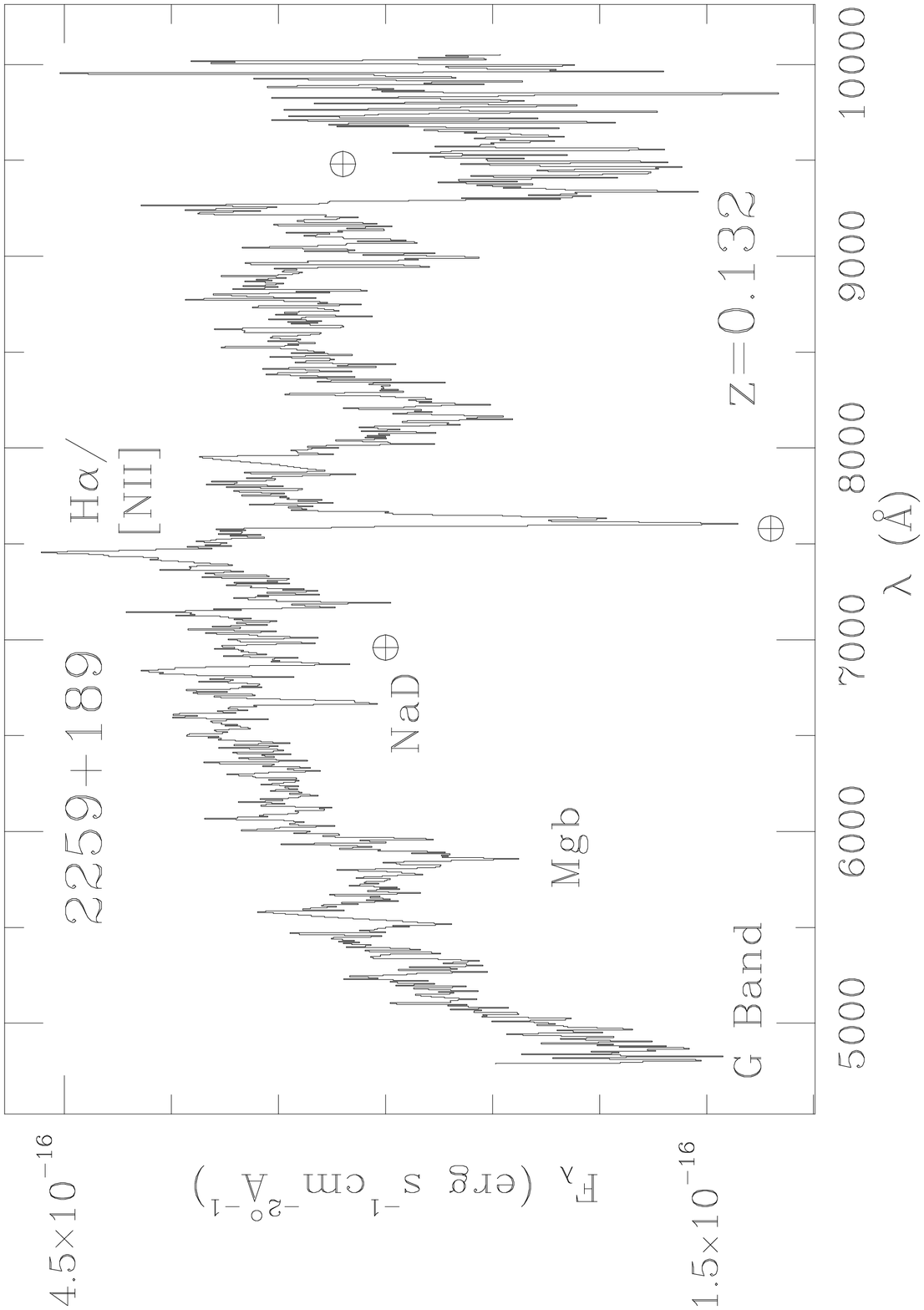,height=7.1cm,width=6.3cm}
\vspace{0.25in}
\psfig{file=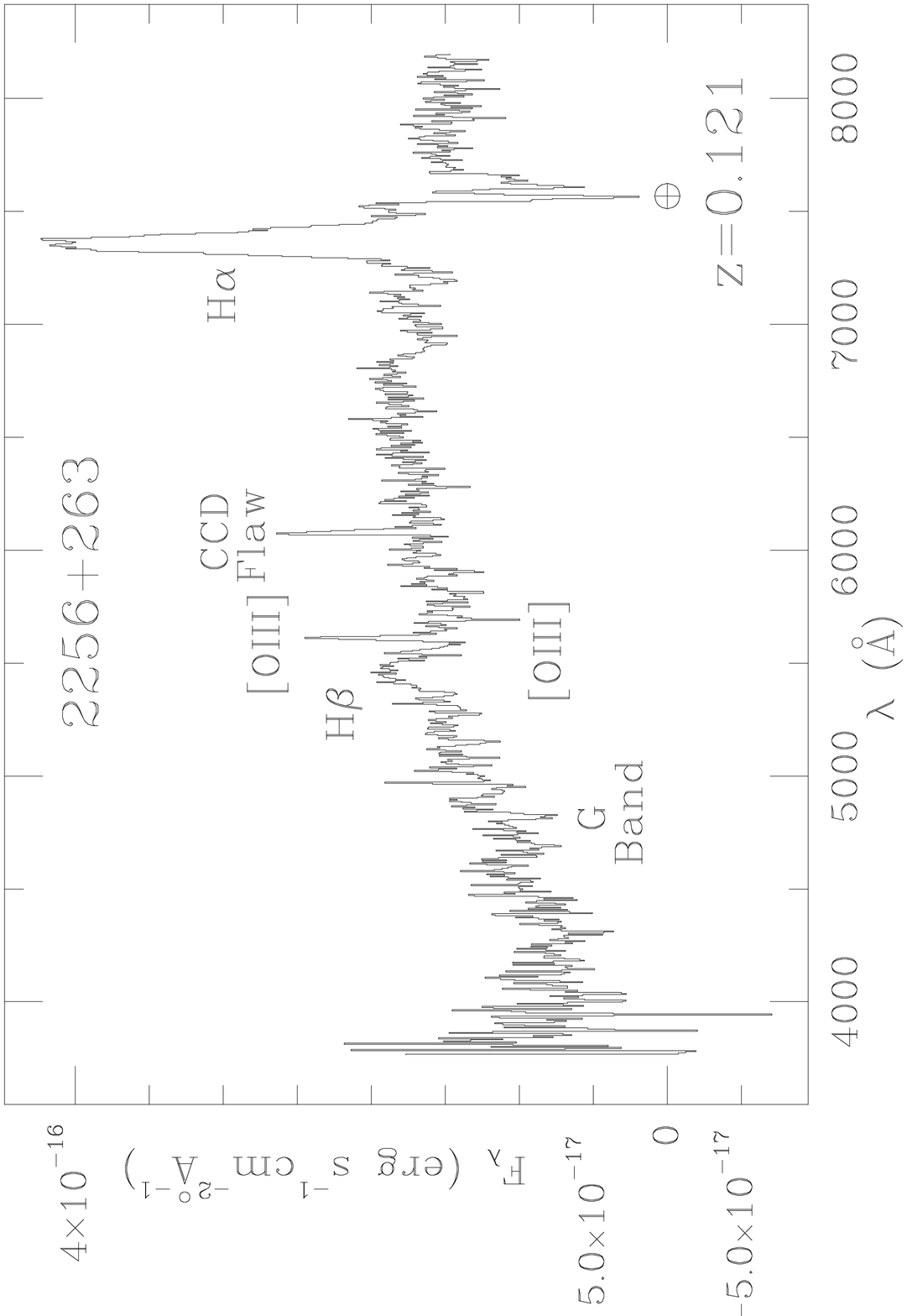,height=7.1cm,width=6.3cm}
\vspace{0.25in}
\psfig{file=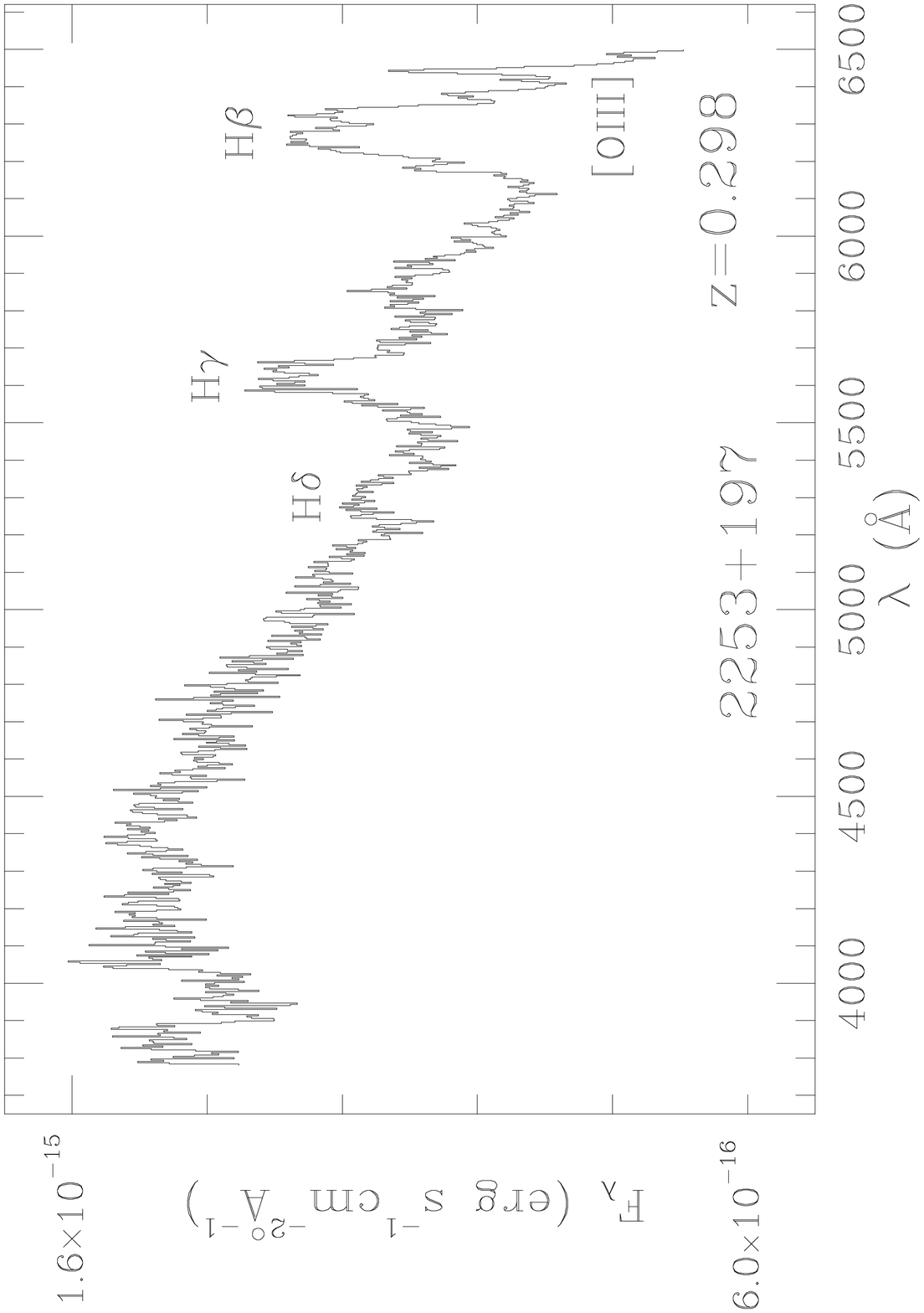,height=7.1cm,width=6.3cm}
\end{minipage}
\hspace{0.3in}
\begin{minipage}[t]{6.3in}
\psfig{file=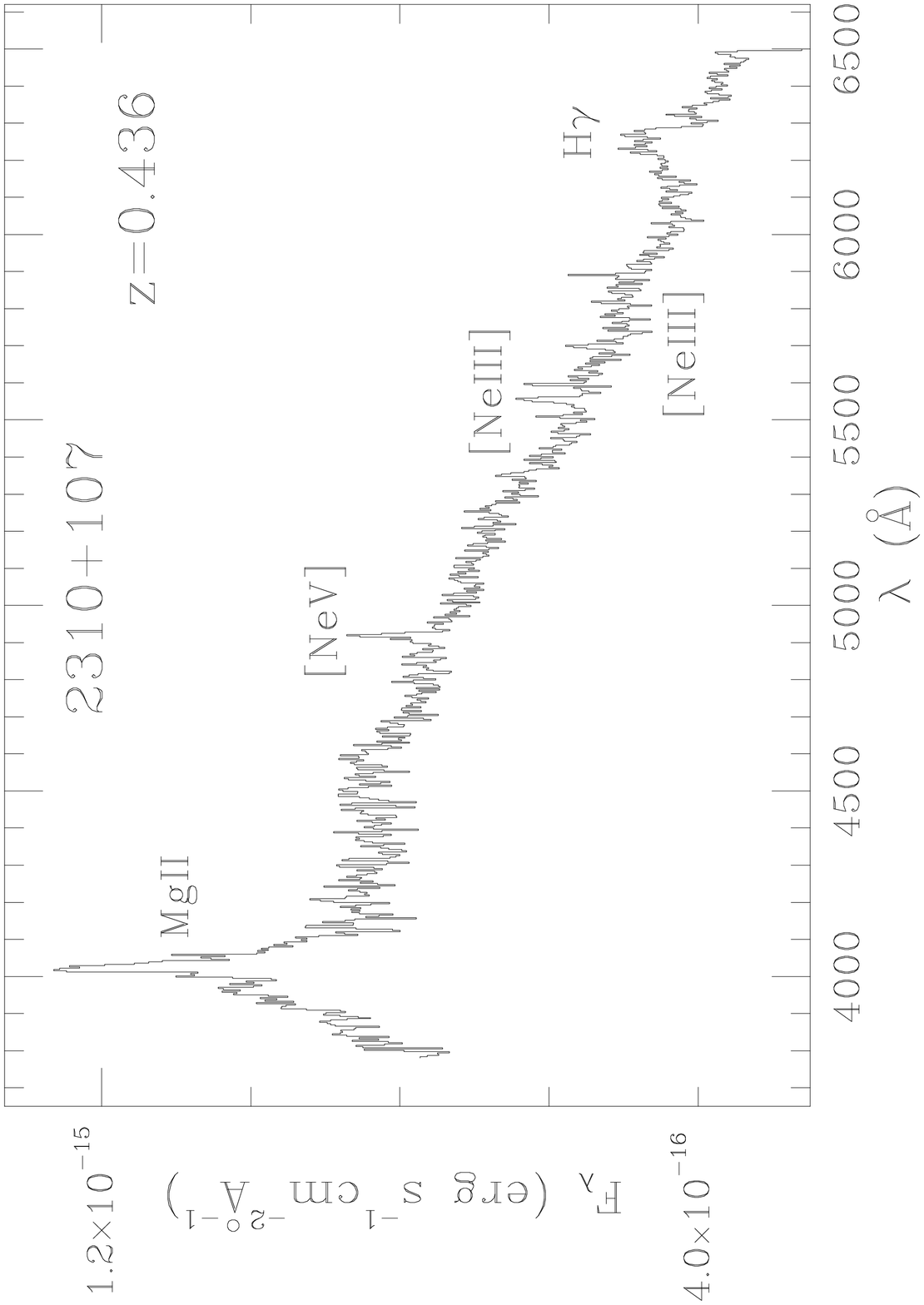,height=7.1cm,width=6.3cm}
\vspace{0.25in}
\psfig{file=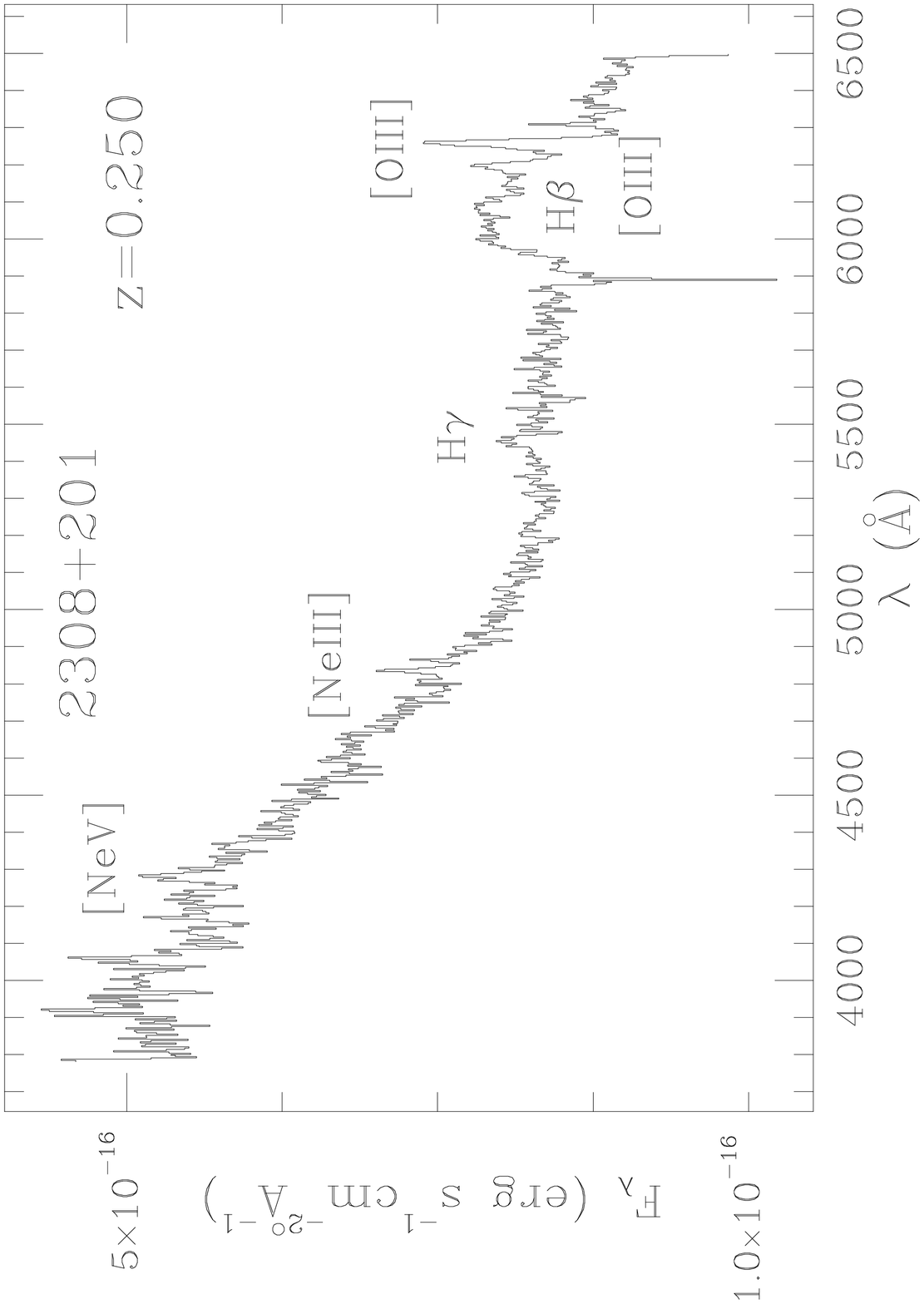,height=7.1cm,width=6.3cm}
\vspace{0.25in}
\psfig{file=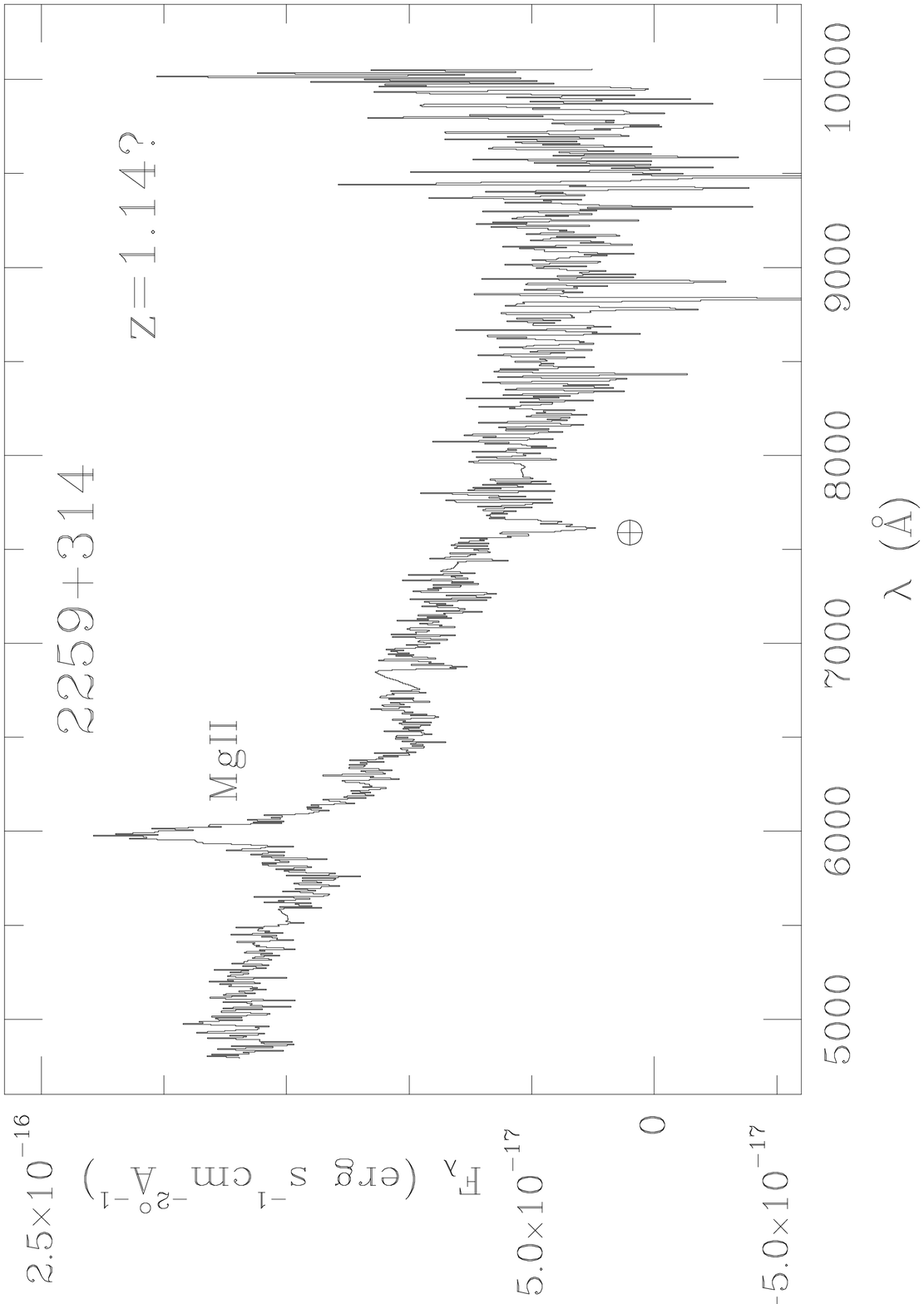,height=7.1cm,width=6.3cm}
\end{minipage}
\hfill
\begin{minipage}[t]{0.3in}
\vfill
\begin{sideways}
Figure 1.163 $-$ 1.168: Spectra of RGB Sources ({\it continued})
\end{sideways}
\vfill
\end{minipage}
\end{figure}

\clearpage
\begin{figure}
\vspace{-0.3in}
\hspace{-0.3in}
\begin{minipage}[t]{6.3in}
\psfig{file=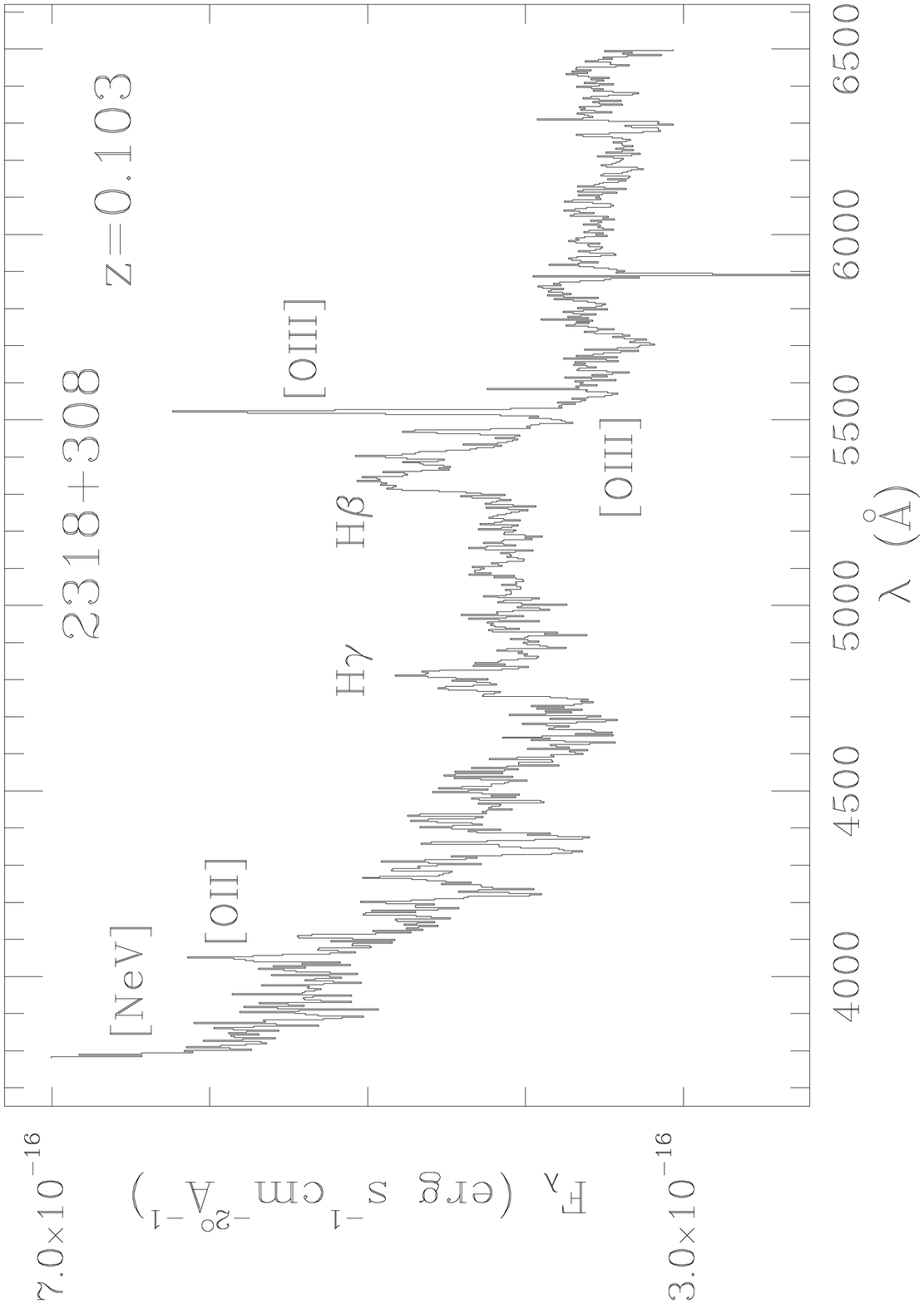,height=7.1cm,width=6.3cm}
\vspace{0.25in}
\psfig{file=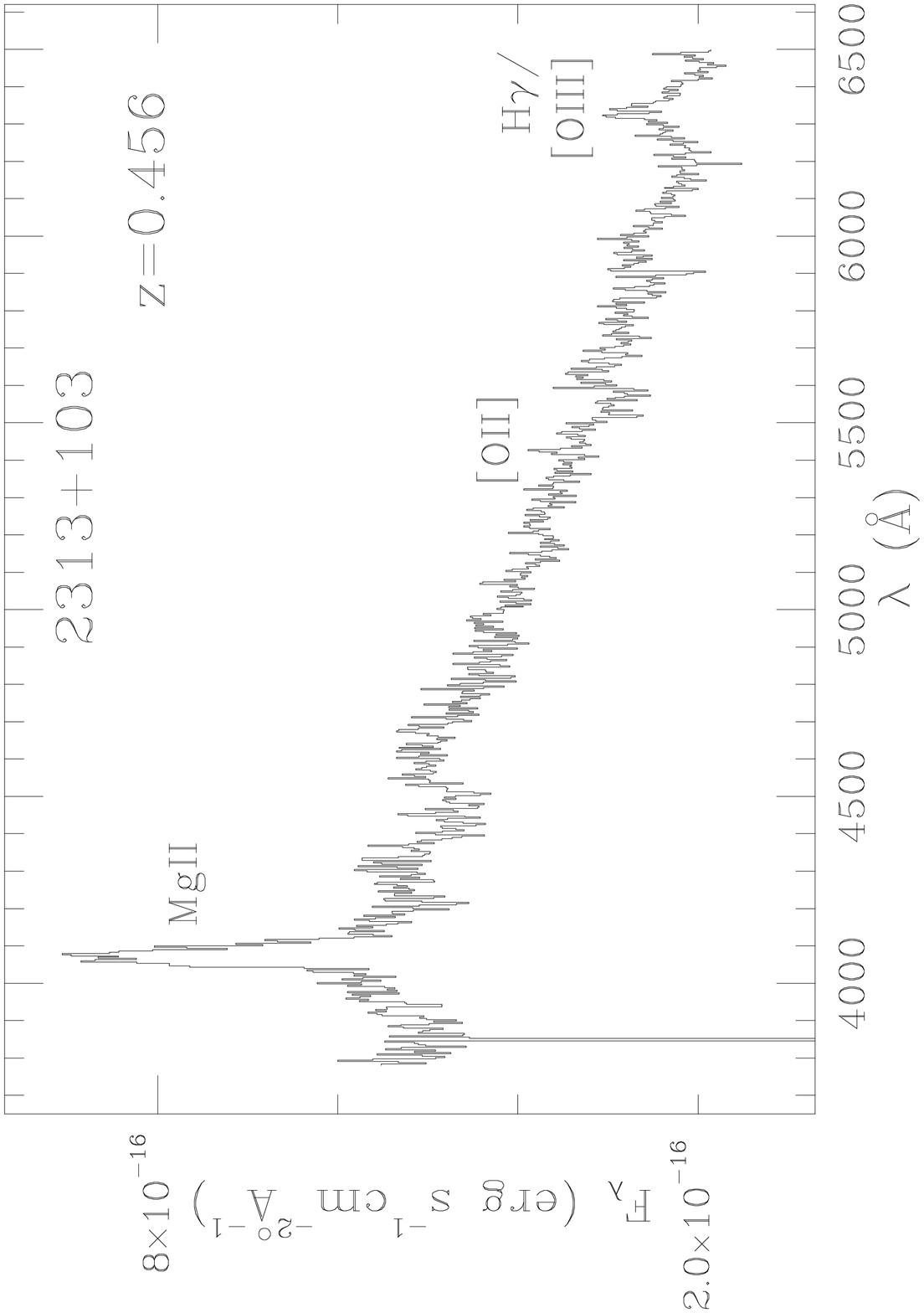,height=7.1cm,width=6.3cm}
\vspace{0.25in}
\psfig{file=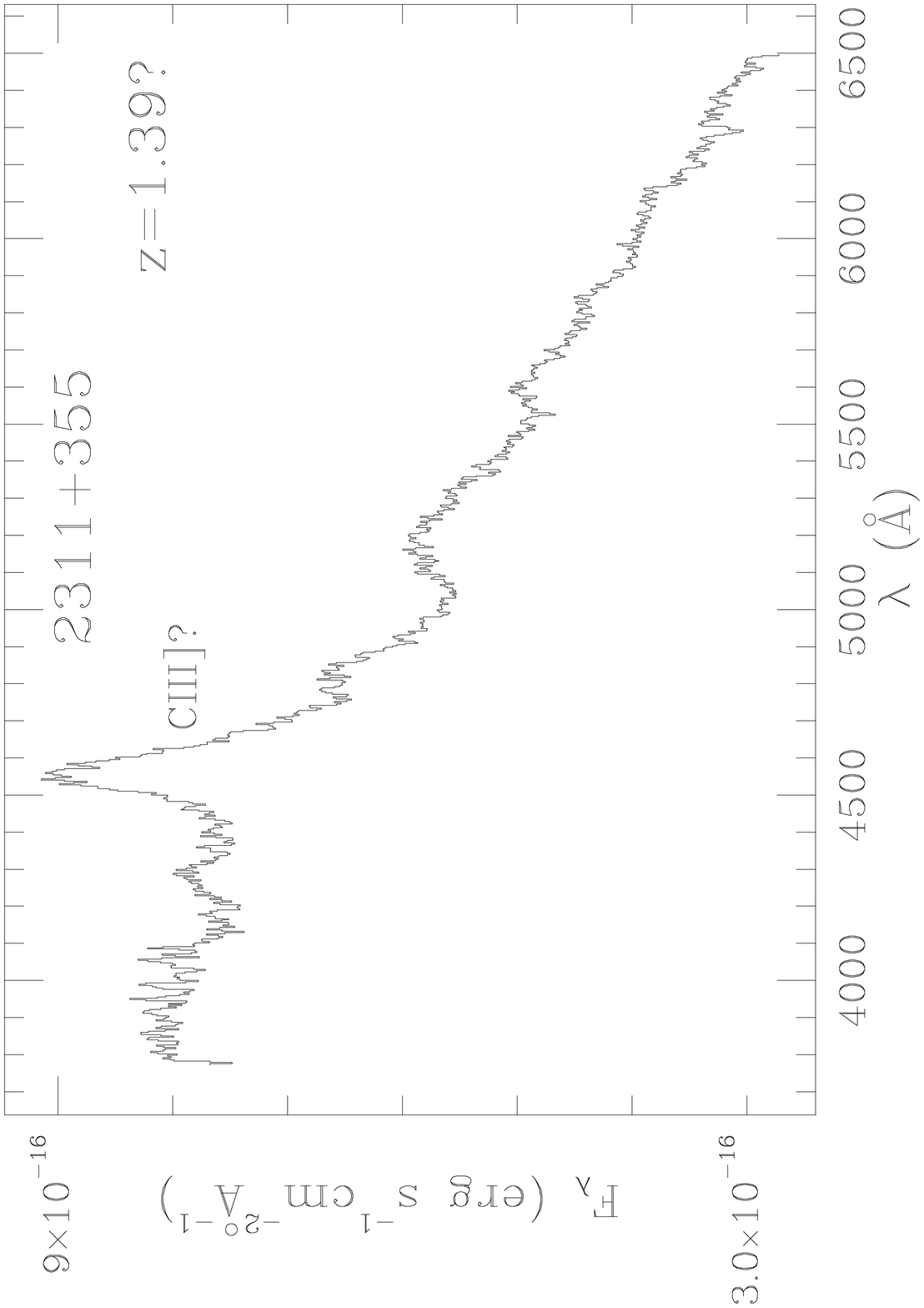,height=7.1cm,width=6.3cm}
\end{minipage}
\hspace{0.3in}
\begin{minipage}[t]{6.3in}
\psfig{file=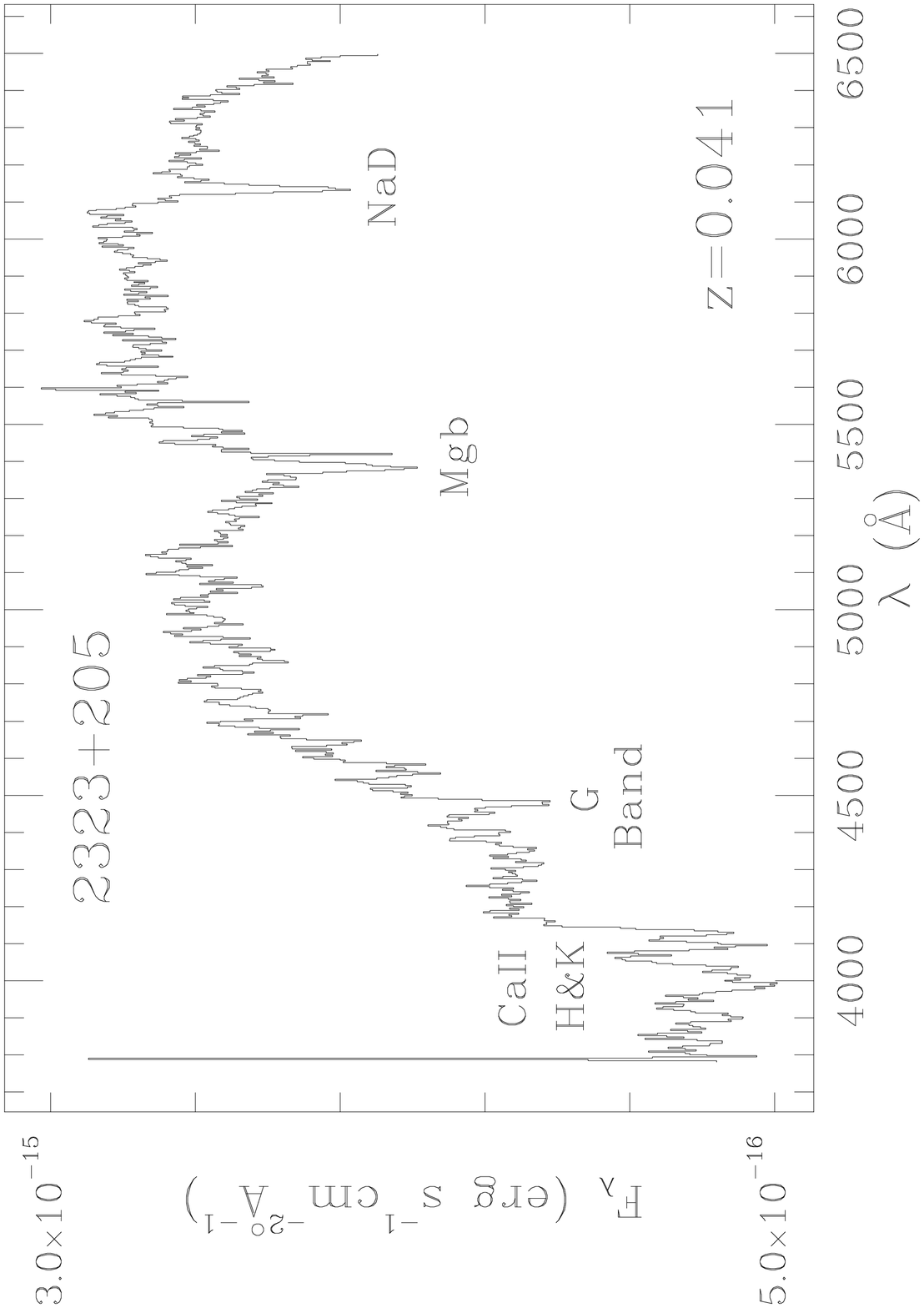,height=7.1cm,width=6.3cm}
\vspace{0.25in}
\psfig{file=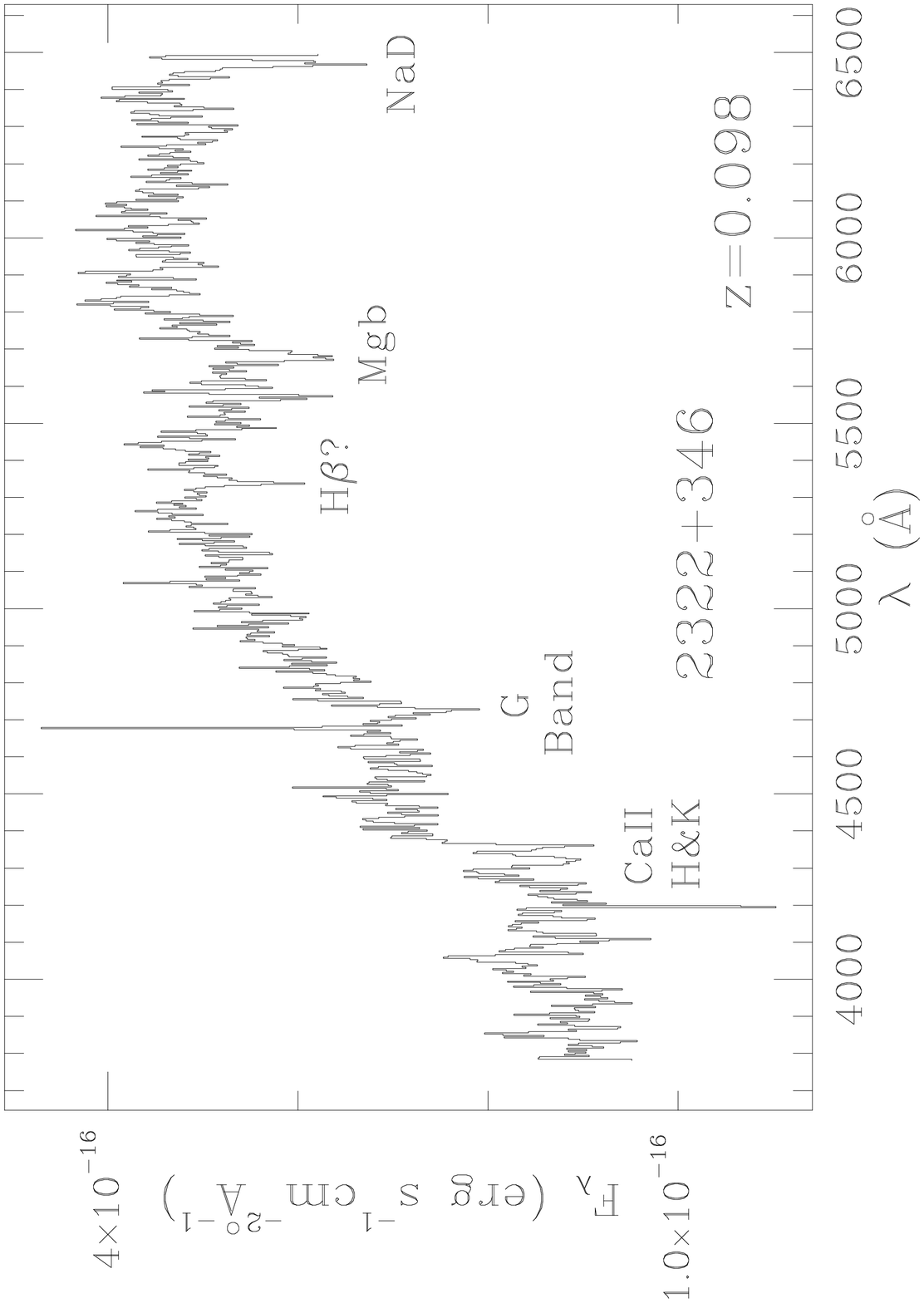,height=7.1cm,width=6.3cm}
\vspace{0.25in}
\psfig{file=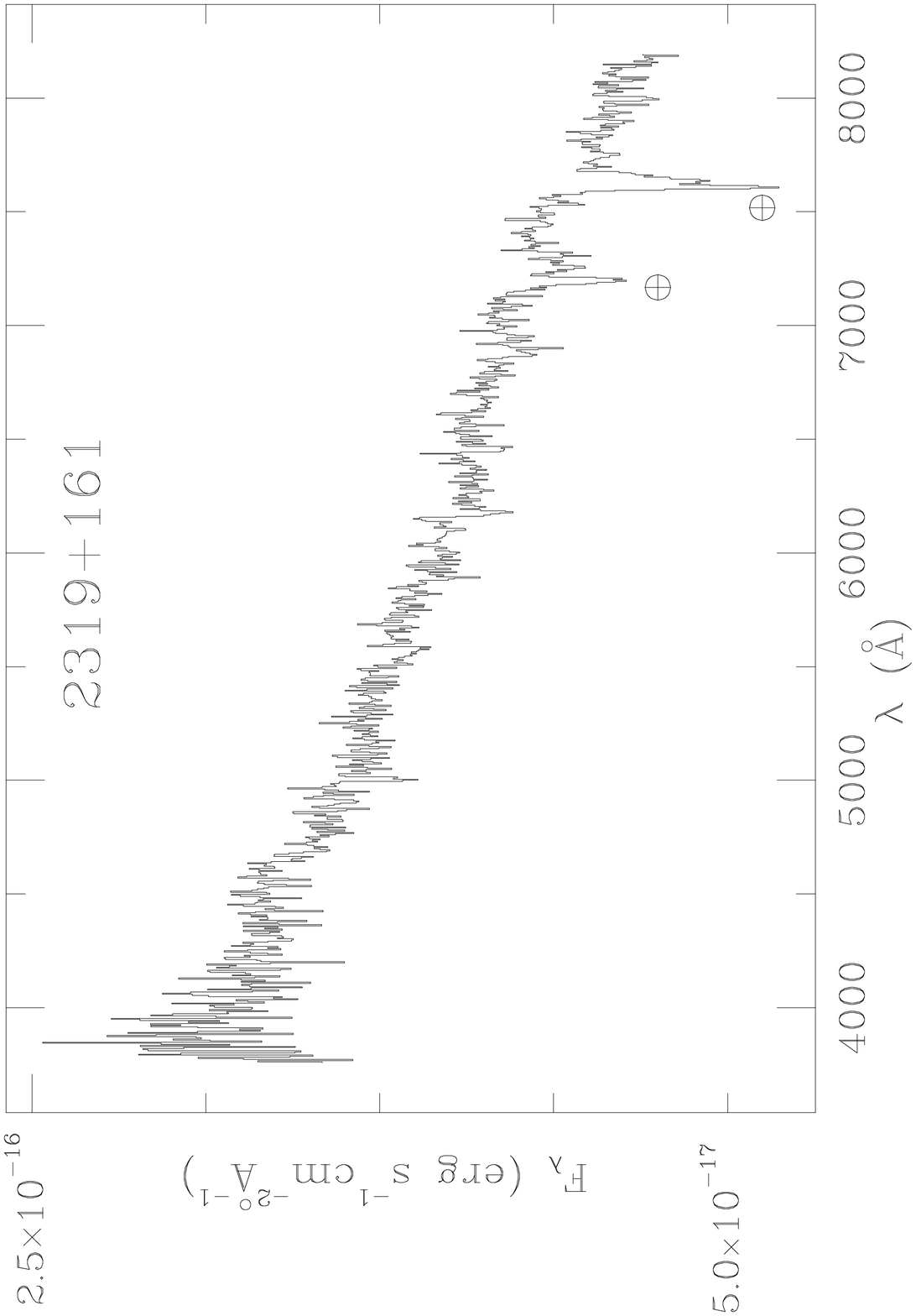,height=7.1cm,width=6.3cm}
\end{minipage}
\hfill
\begin{minipage}[t]{0.3in}
\vfill
\begin{sideways}
Figure 1.169 $-$ 1.174: Spectra of RGB Sources ({\it continued})
\end{sideways}
\vfill
\end{minipage}
\end{figure}

\begin{figure}
\psfig{file=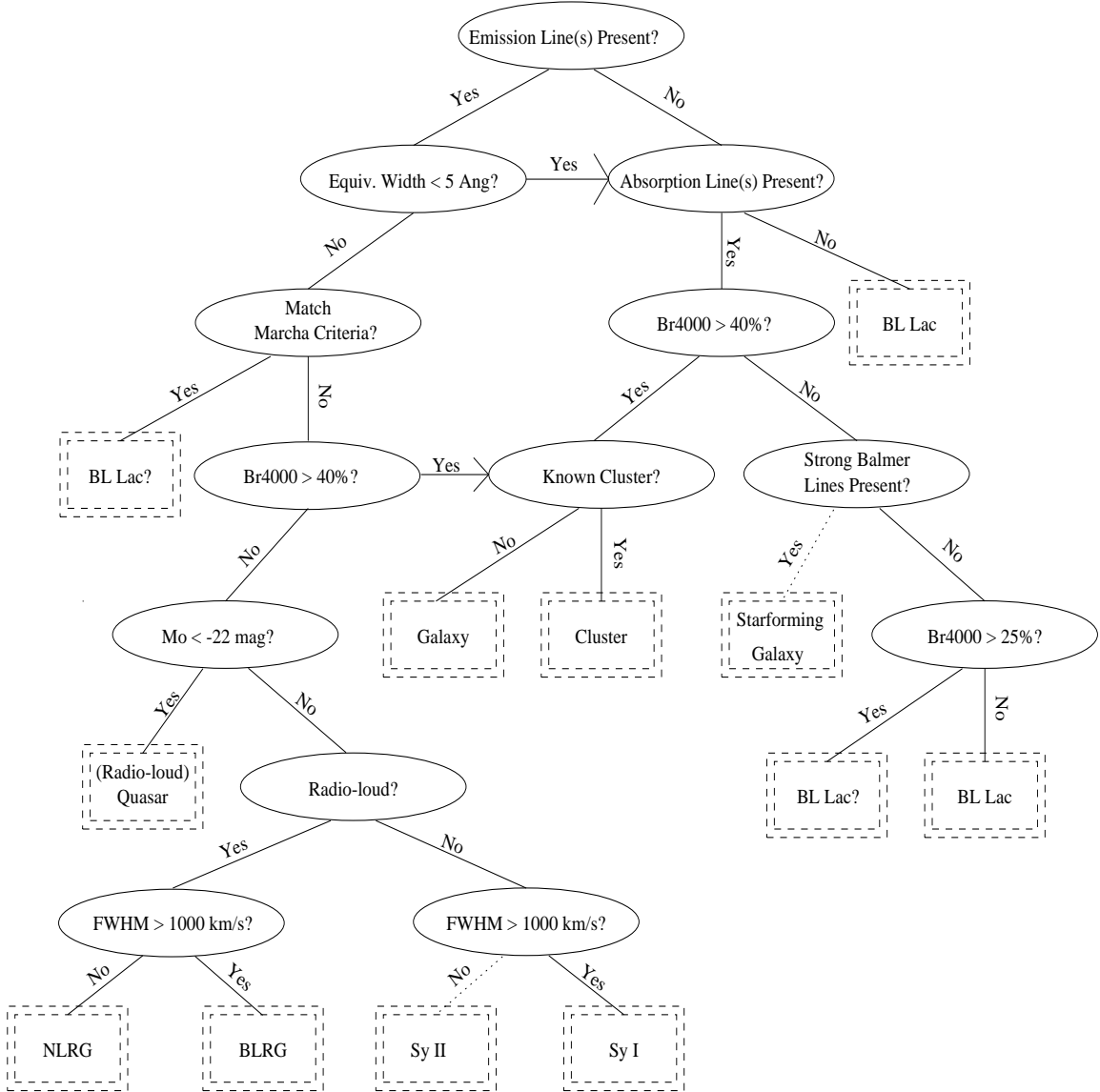,height=6.0in,width=6.0in,angle=-90}
\caption{The RGB classification criteria used in this study.  Dotted boxes
represent the end point of a decision tree.  Only radio-loud quasars are
listed as no radio-quiet quasars were found, although we did allow for that
possibility.  Seyfert~II and starforming galaxies are both listed, although we
found none of these objects in this study.  We therefore use dotted lines
leading to the respective boxes.\label{fig:flow_chart}}
\end{figure}

\begin{figure}
\psfig{file=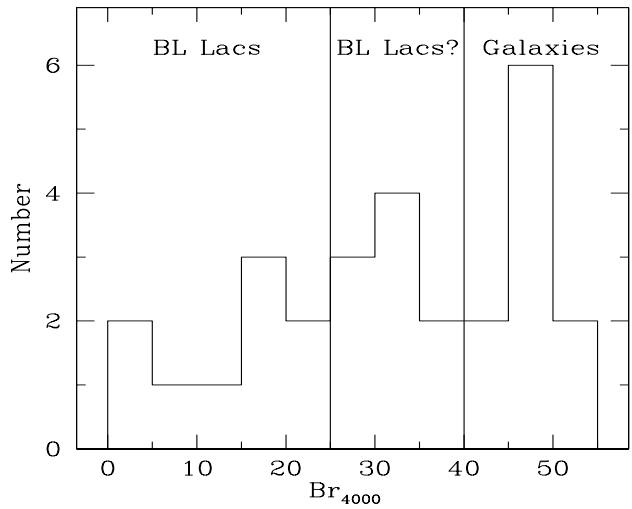,height=7.0cm,width=8.75cm}
\caption{The distribution of \ion{Ca}{2} break strengths (Br$_{4000}$) for the
28 objects for which it was measured.  Objects with Br$_{4000}$$>$40\% have
been classified as galaxies while objects with Br$_{4000}$$<$25\% have been
classified as BL~Lacs.  Objects with intermediate break strengths have been
classified as probable BL~Lacs (``BL~Lacs?'').\label{fig:ca}}
\end{figure}

\begin{figure}
\vspace*{-1.0cm}
\psfig{file=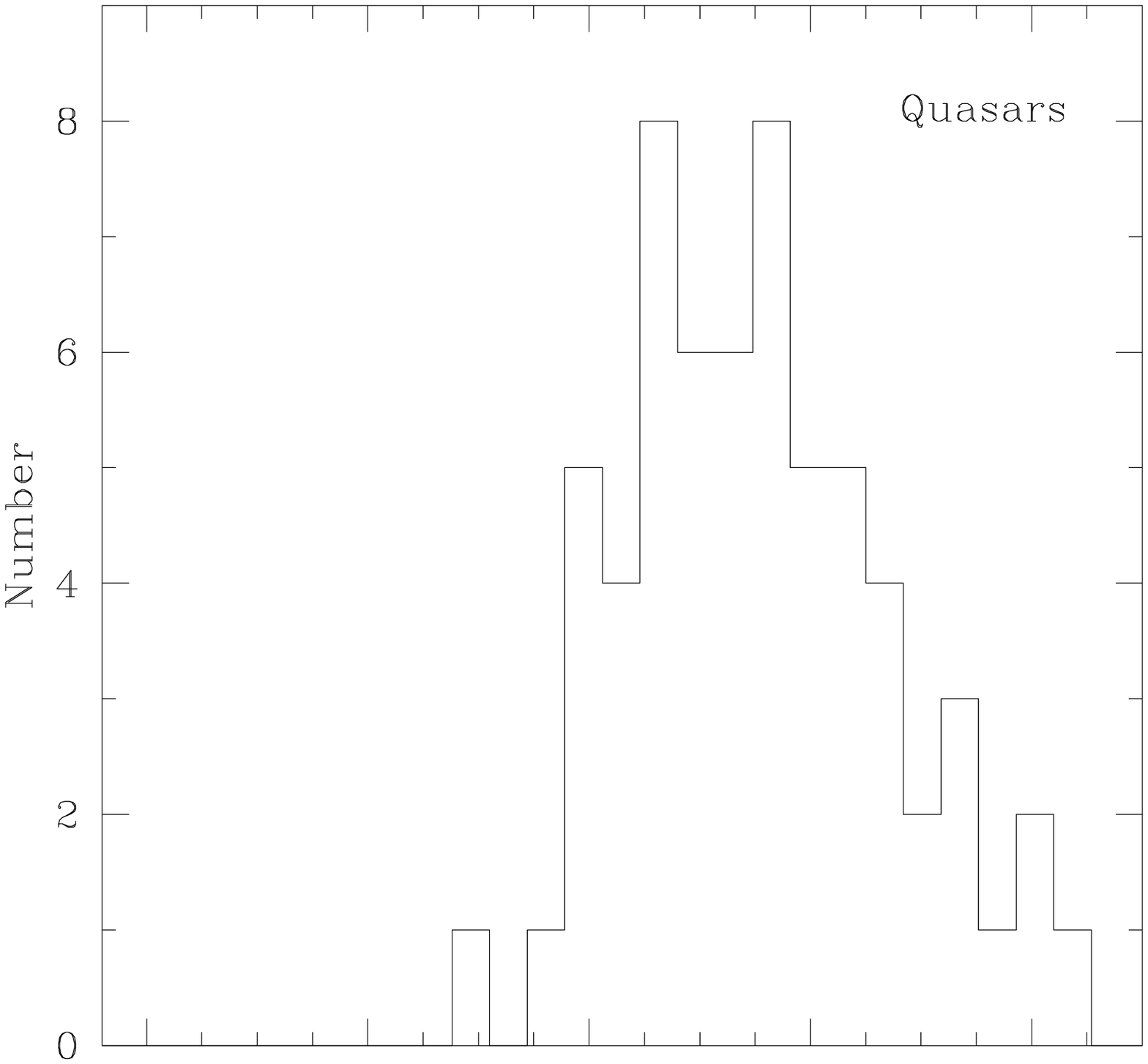,height=6.1cm,width=7.6cm,angle=0}
\vspace*{-1.02cm}
\psfig{file=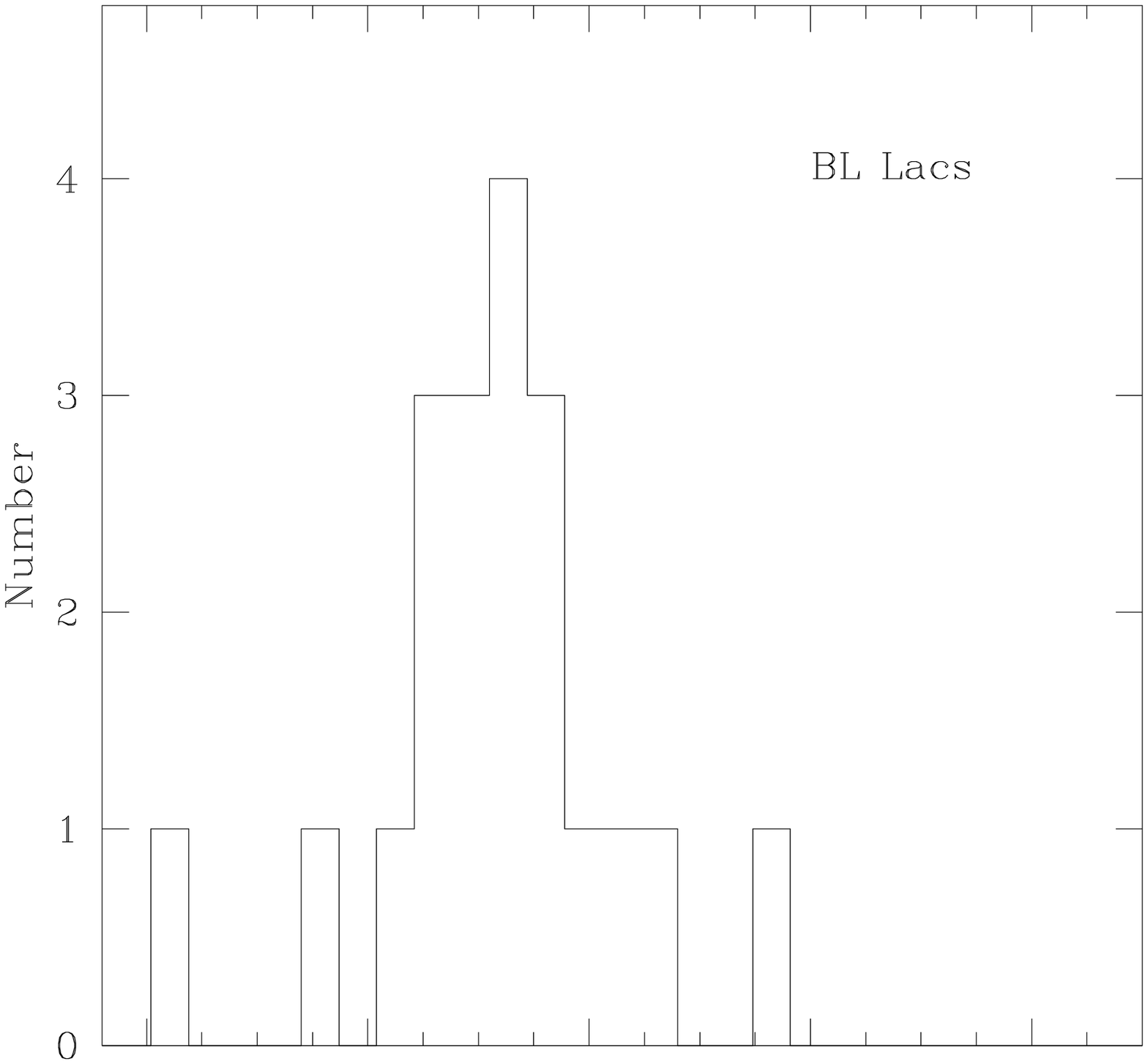,height=6.1cm,width=7.6cm,angle=0}
\vspace*{-1.02cm}
\psfig{file=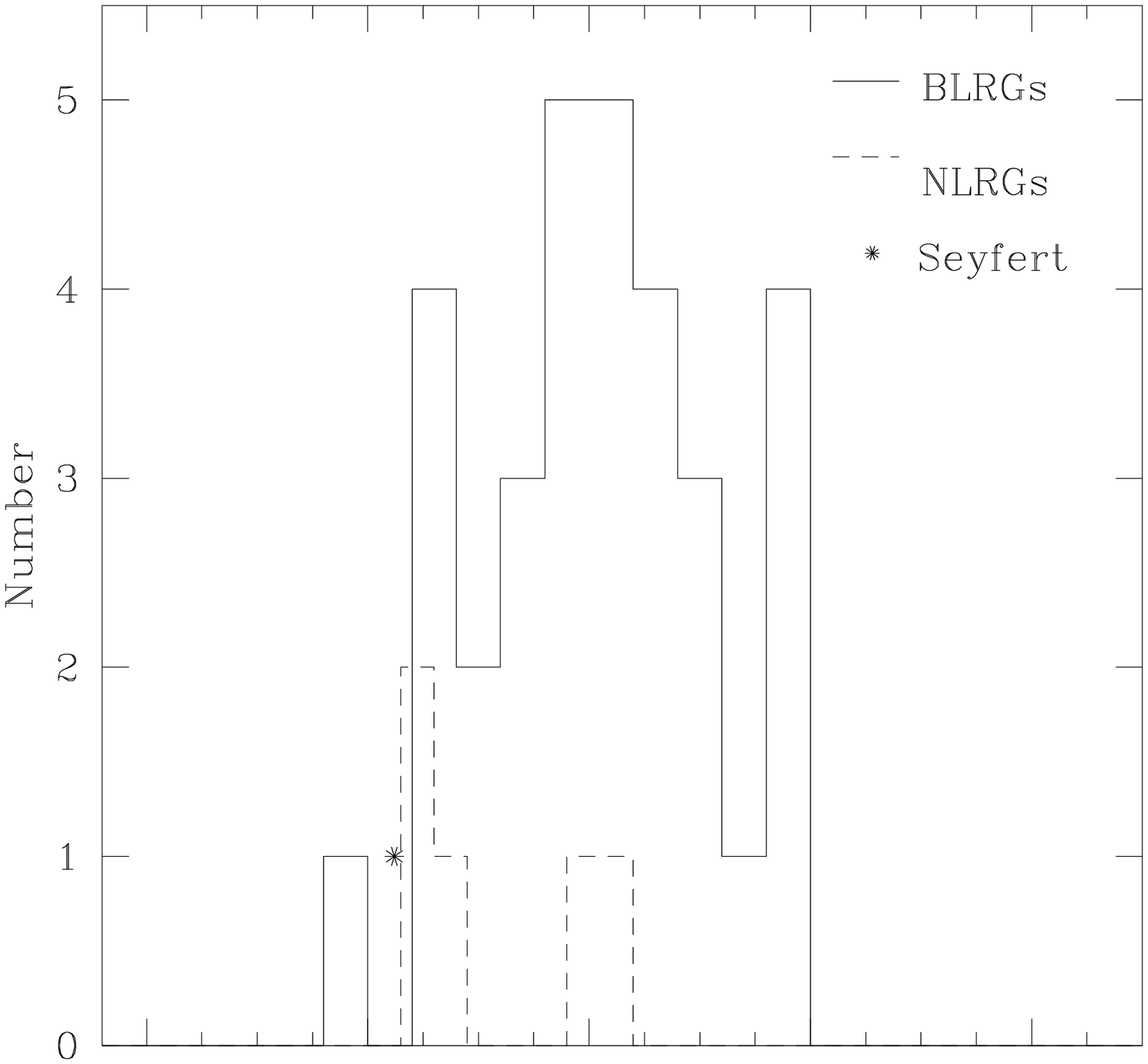,height=6.1cm,width=7.6cm,angle=0}
\vspace*{-1.01cm}
\psfig{file=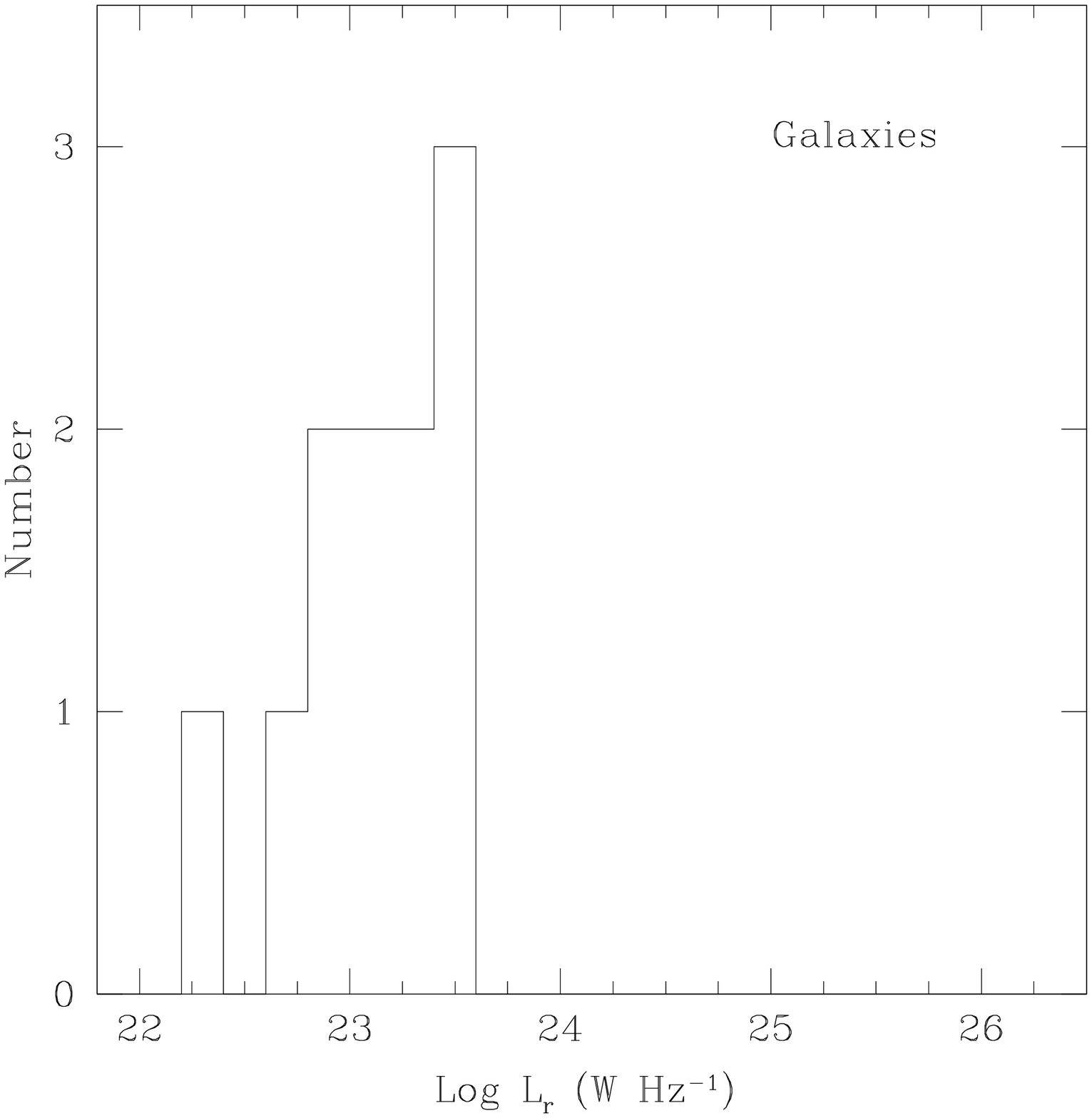,height=6.1cm,width=7.6cm,angle=0}
\caption{The distribution of the logarithm of the radio luminosities for the
newly identified RGB sources.  The panels show the distribution of the various
spectroscopic classes: (1) radio-loud quasars, (2) BL~Lacs, (3) Broad and
Narrow line radio galaxies and the one object we classify as a Seyfert and (4)
galaxies and galaxies in known clusters.\label{fig:lr}}
\end{figure}

\begin{figure}
\vspace*{-1.0cm}
\psfig{file=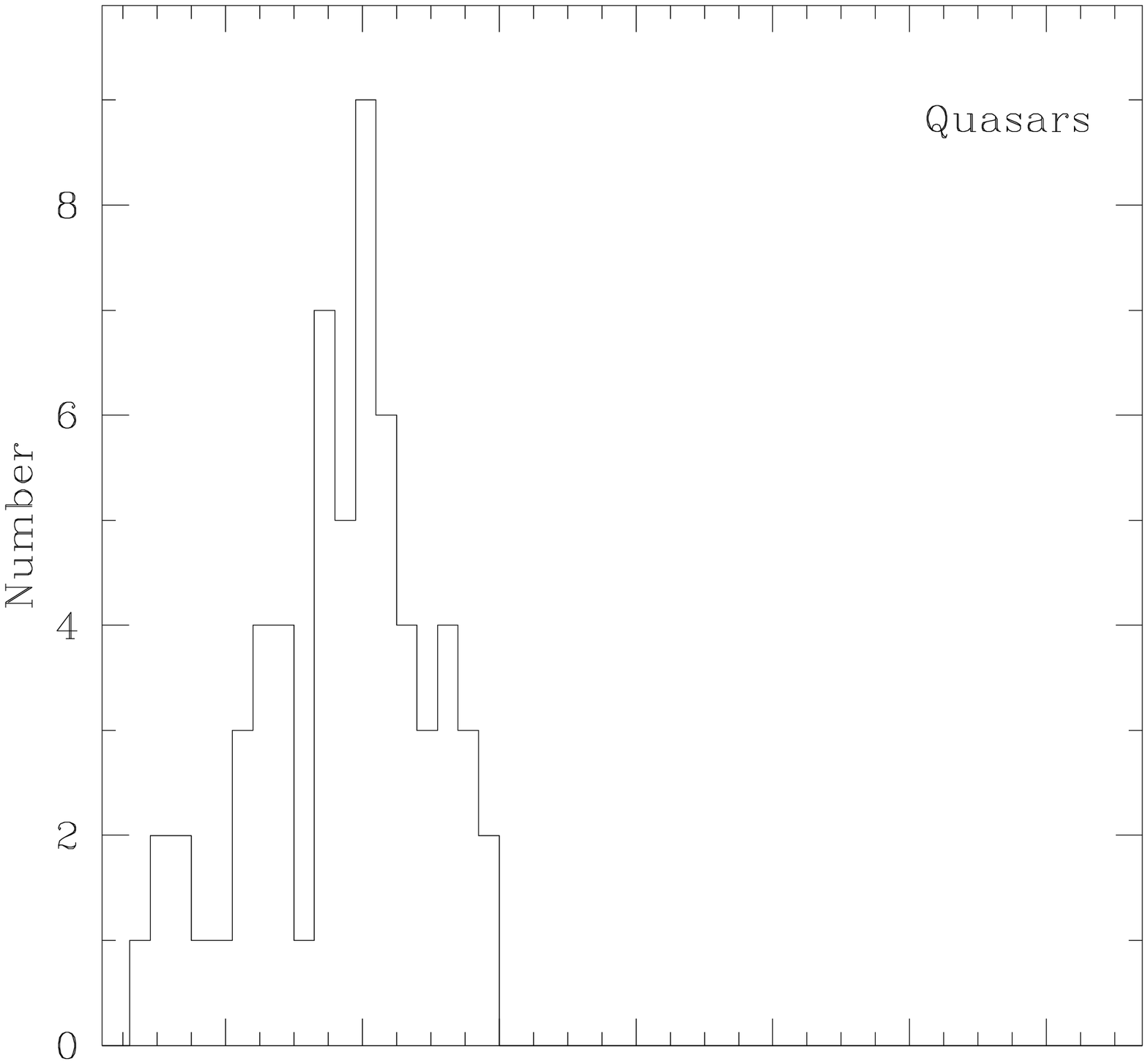,height=6.1cm,width=7.6cm}
\vspace*{-1.02cm}
\psfig{file=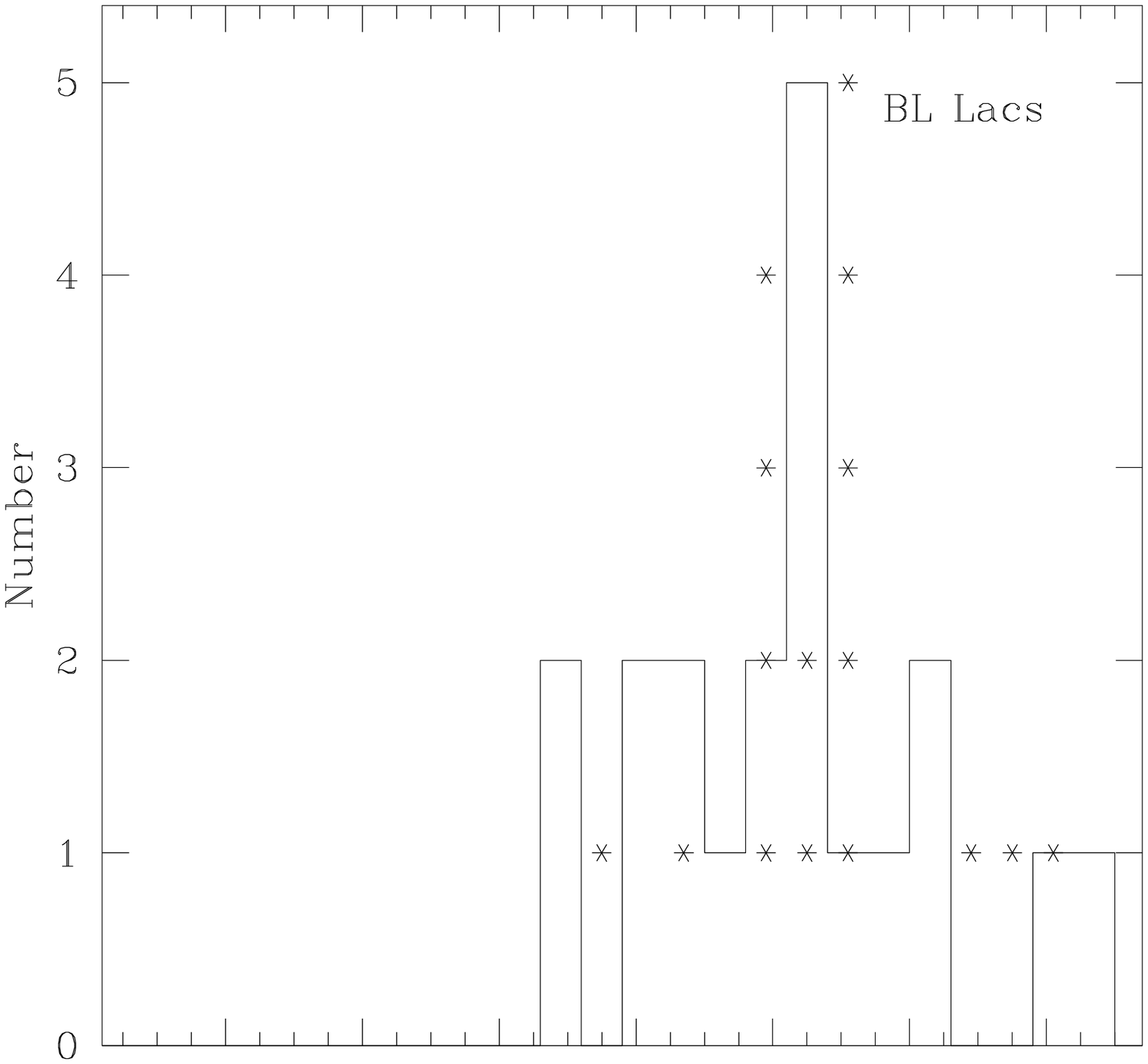,height=6.1cm,width=7.6cm}
\vspace*{-1.02cm}
\psfig{file=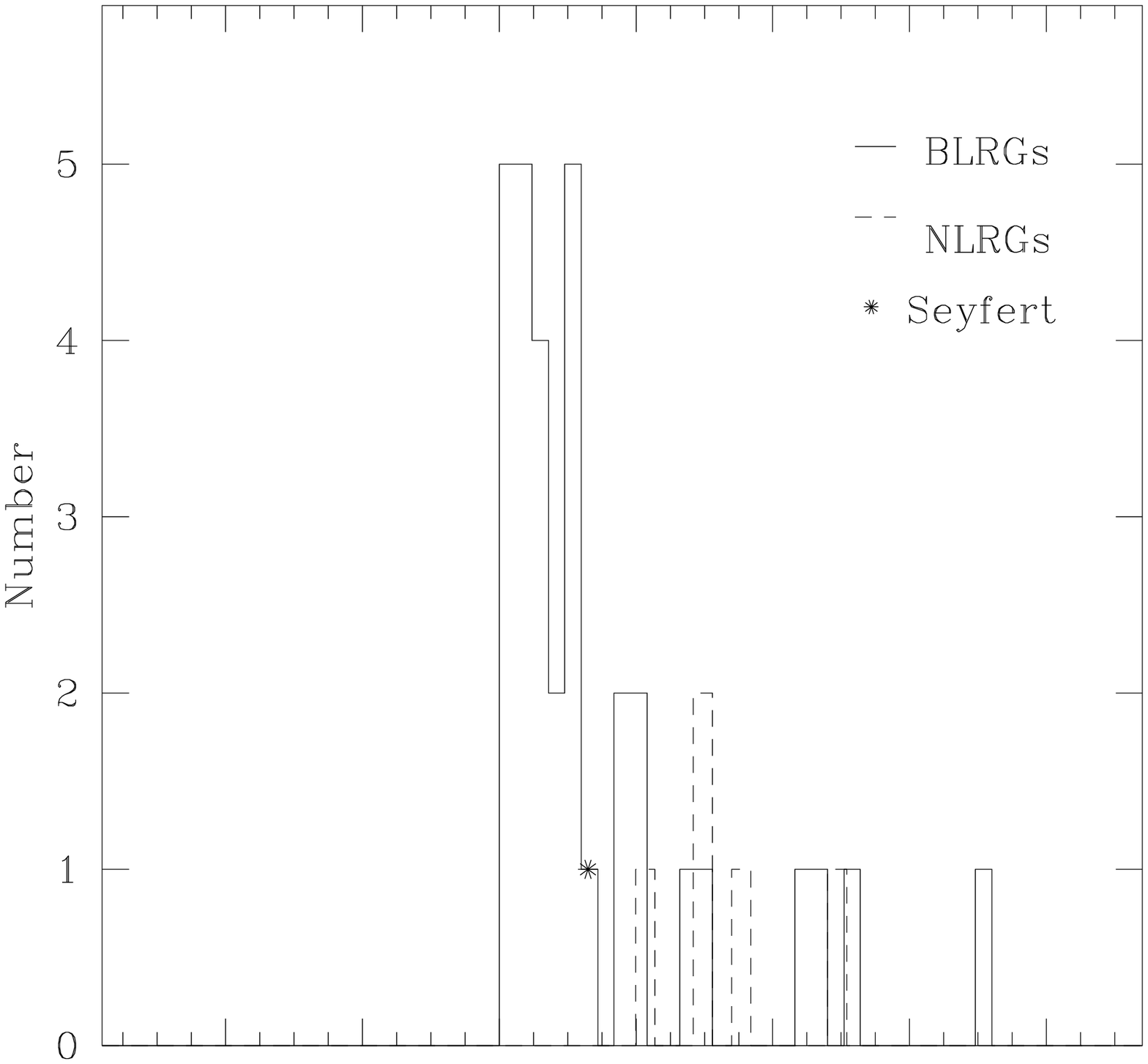,height=6.1cm,width=7.6cm}
\vspace*{-1.01cm}
\psfig{file=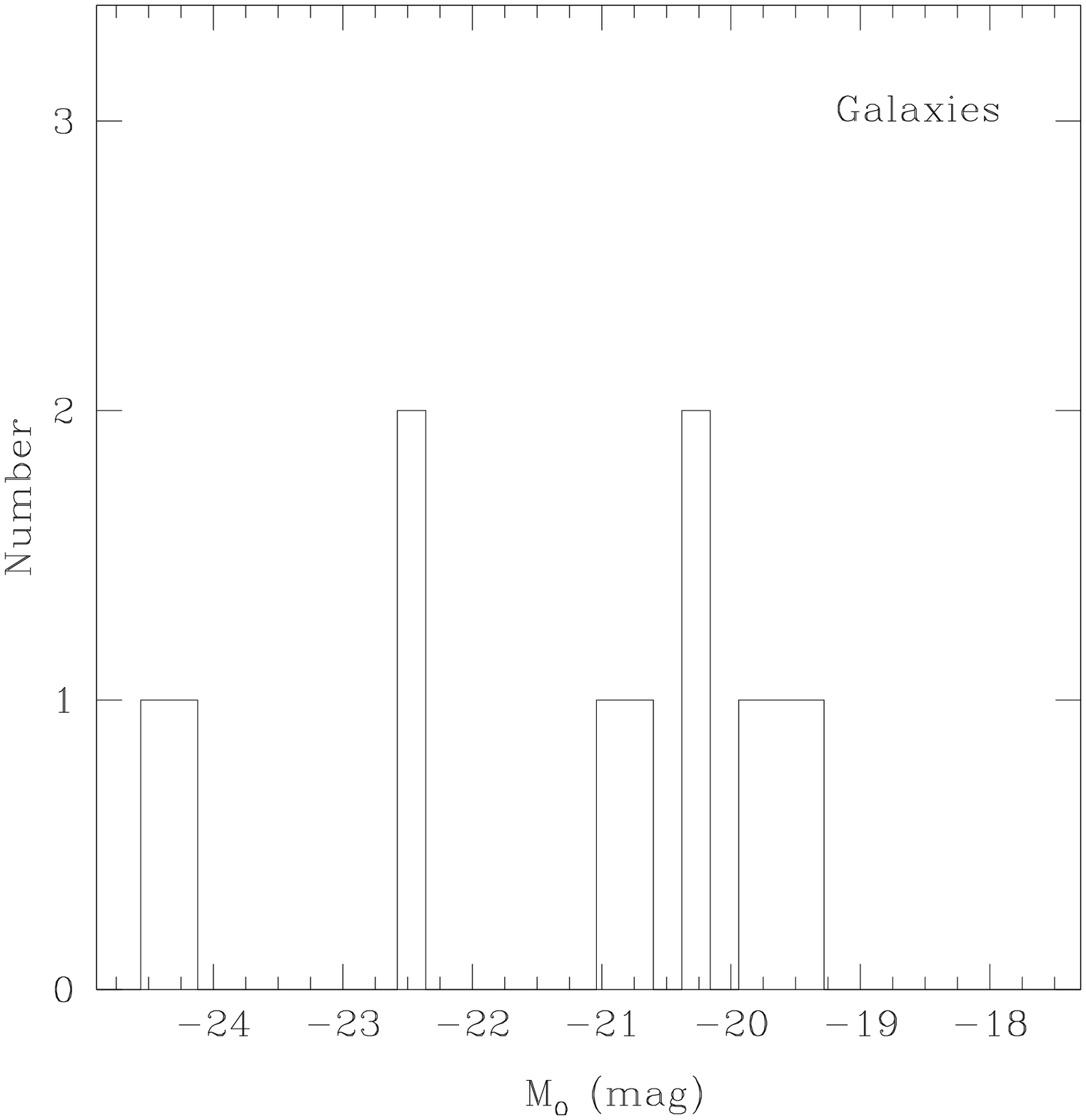,height=6.1cm,width=7.6cm}
\caption{The distribution of absolute O magnitudes for the newly identified
RGB sources, separated by spectroscopic class. The panels are in the same
order and represent the same objects as described in Figure \ref{fig:lr}.  The
asterisks in second panel show the distribution of BL~Lac host galaxy optical
luminosities derived using our method described in \S6.\label{fig:lo}}
\end{figure}

\begin{figure}
\vspace*{-1.0cm}
\psfig{file=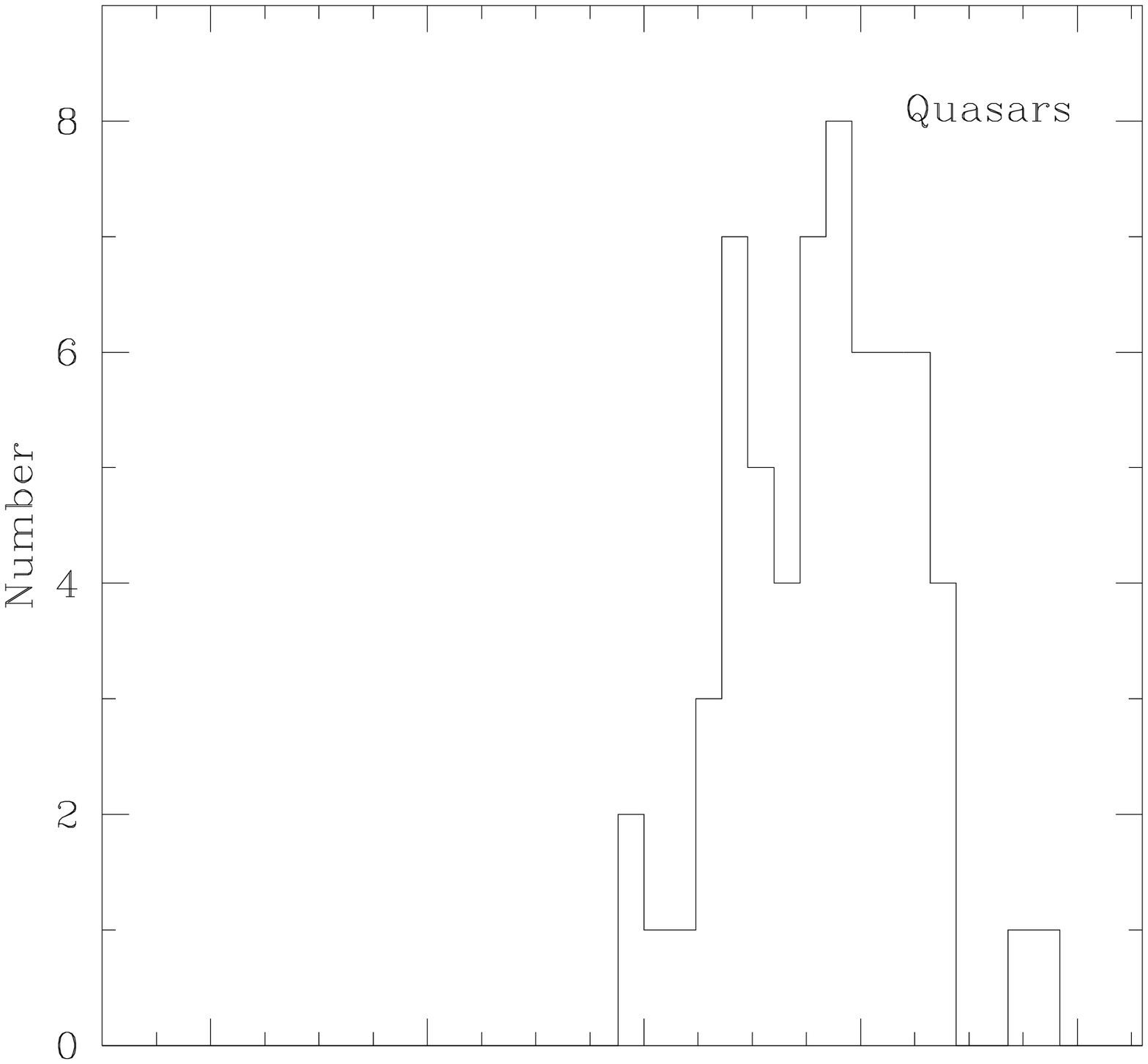,height=6.1cm,width=7.6cm}
\vspace*{-1.02cm}
\psfig{file=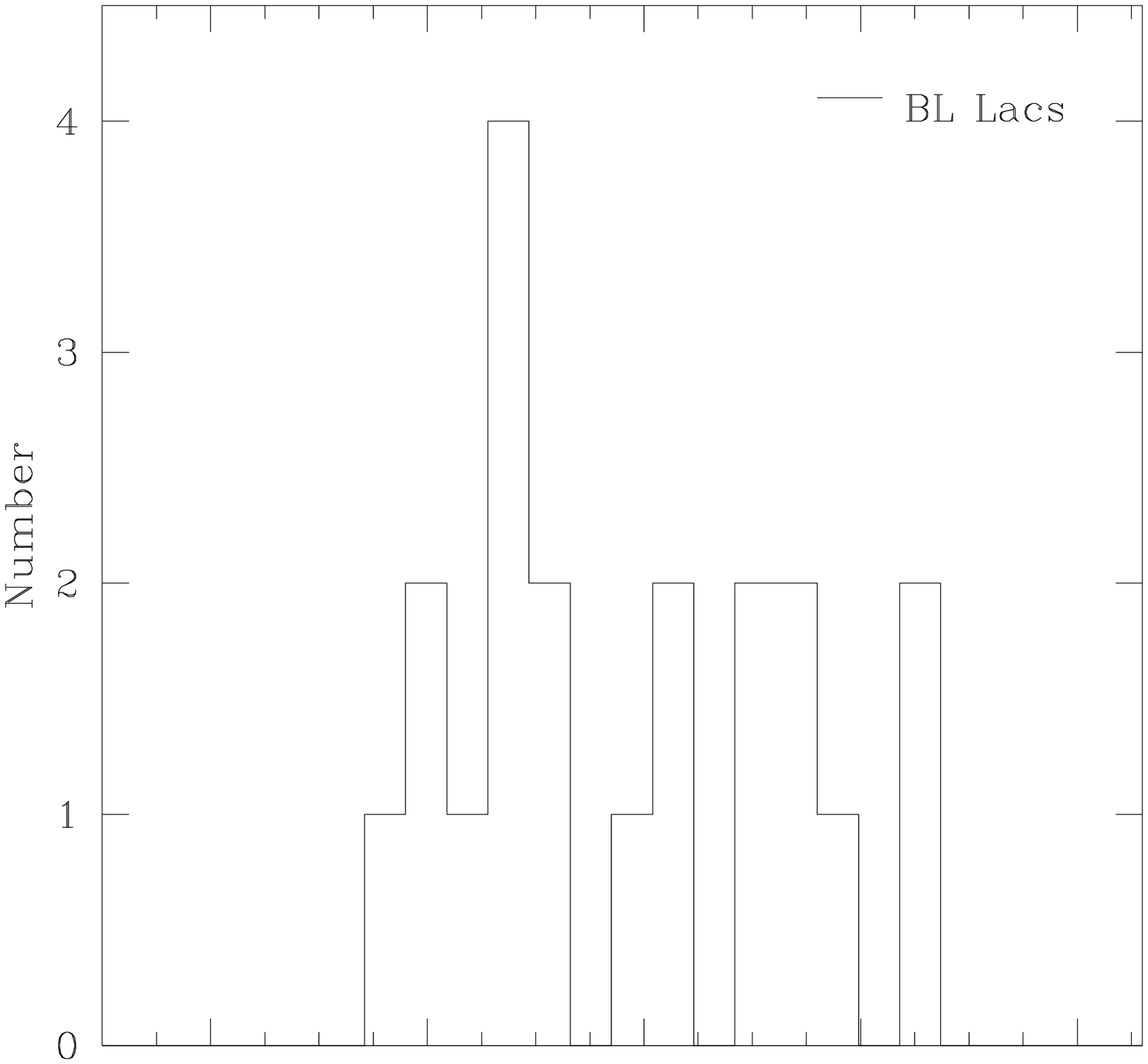,height=6.1cm,width=7.6cm}
\vspace*{-1.02cm}
\psfig{file=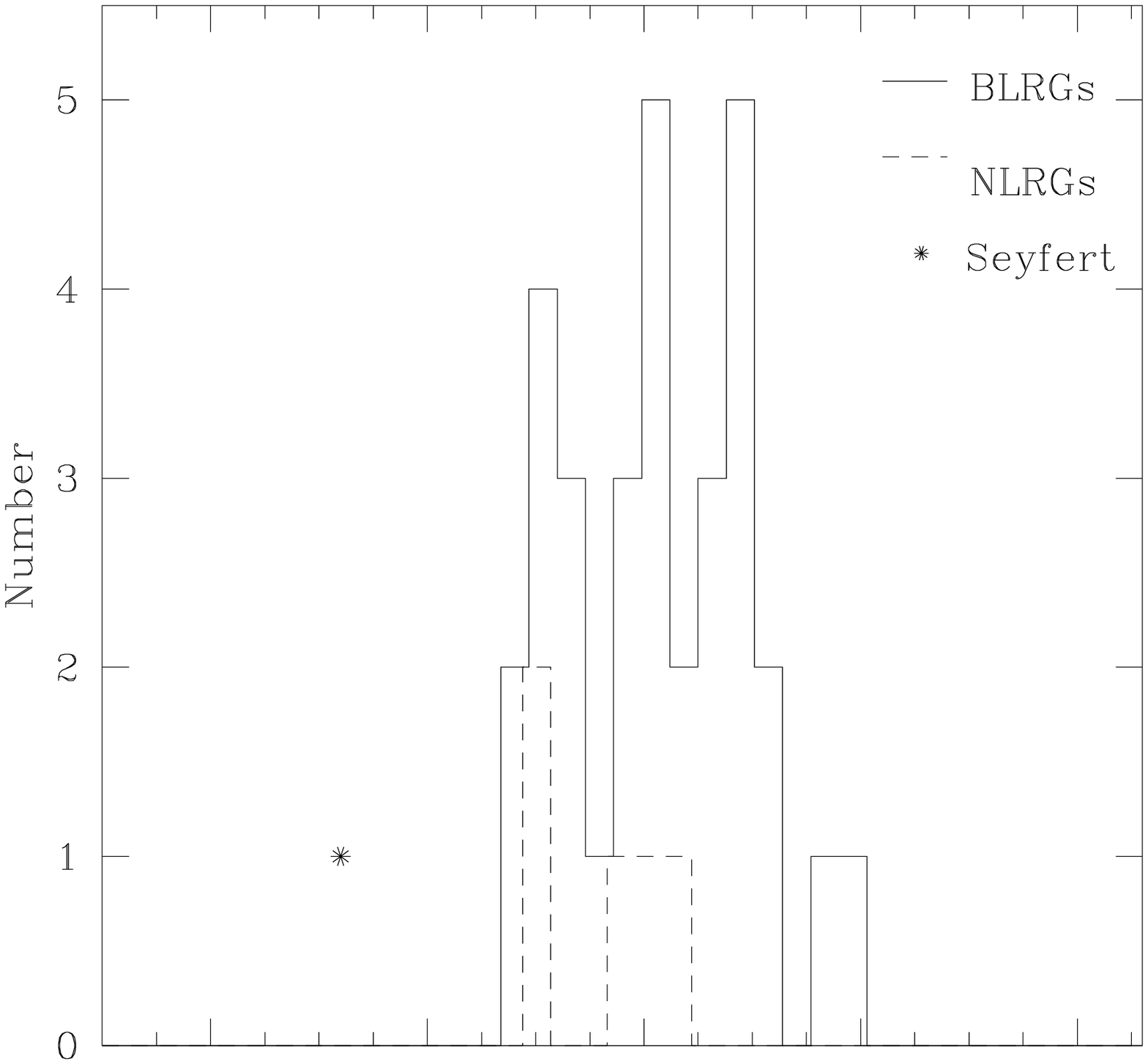,height=6.1cm,width=7.6cm}
\vspace*{-1.01cm}
\psfig{file=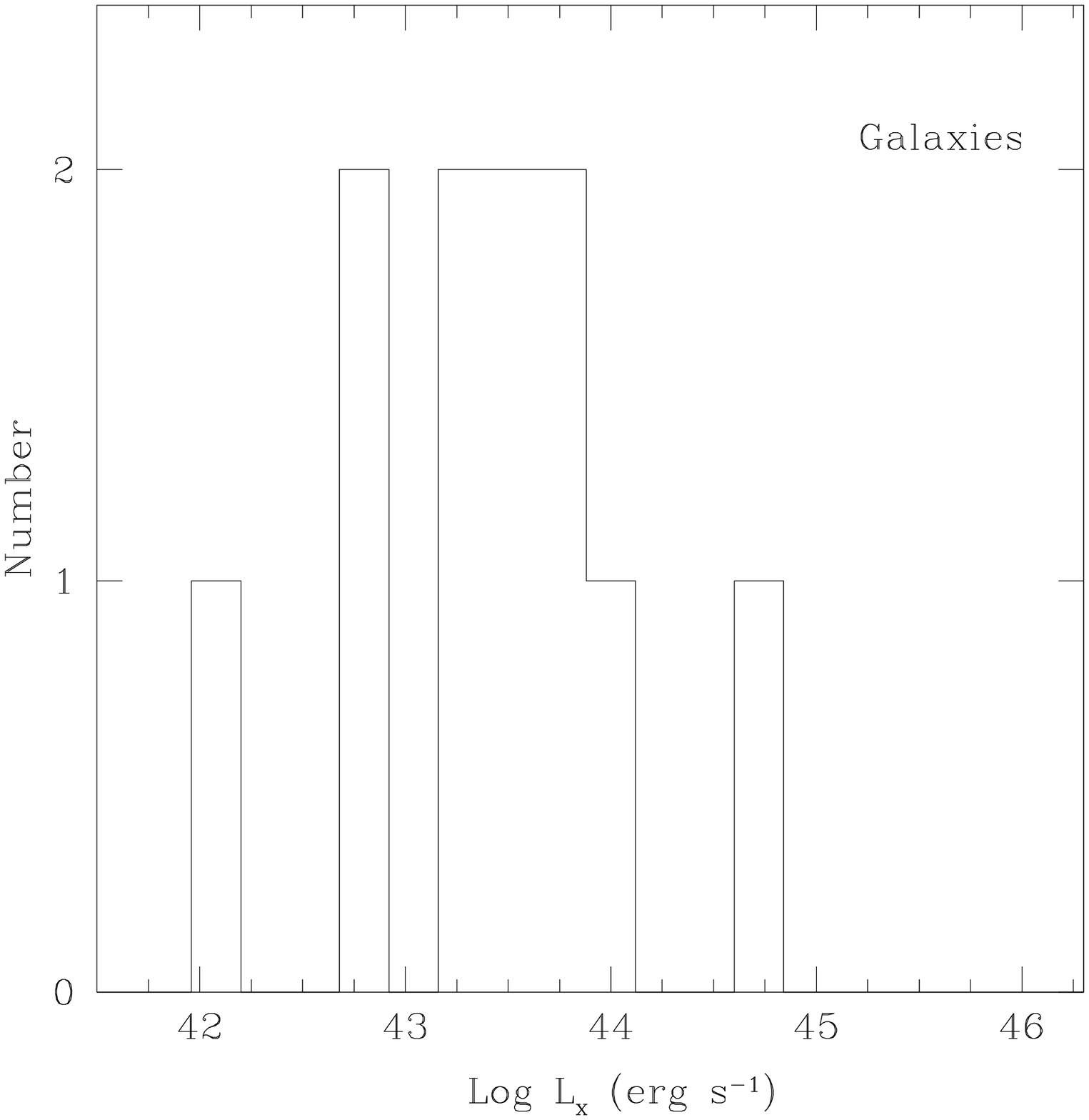,height=6.1cm,width=7.6cm}
\caption{The distribution of the logarithm of the X-ray luminosity for the
newly identified RGB sources.  The panels show the distribution of the various
spectroscopic classes (see Figure \ref{fig:lr}).\label{fig:lx}}
\end{figure}

\begin{figure}
\psfig{file=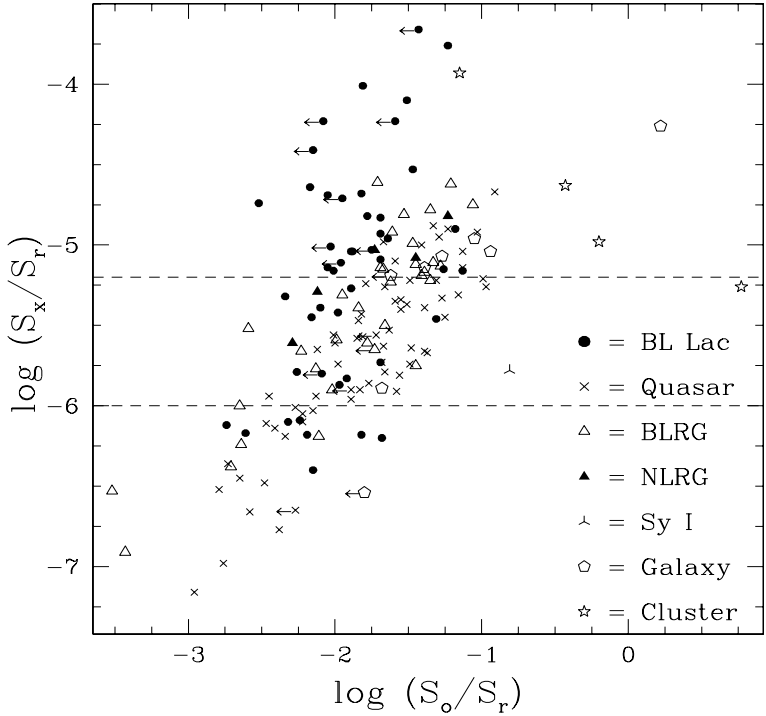,height=10.0cm,width=10.625cm}
\caption{Log~(S$_{\rm x}$/S$_{\rm r}$) as a function of log~(S$_{\rm
o}$/S$_{\rm r}$).  Estimated optical fluxes of only the AGN component are used
for those objects with measured break contrasts (\S6).  The ratio S$_{\rm
o}$/S$_{\rm r}$ is therefore formally an upper limit and left arrows in the
figure denote this.  While there is no clear separation of classes, the
objects with the highest X-ray to radio flux density ratios are preferentially
BL~Lacs (see also Figure 11 in Brinkmann et al.\ 1997).  However, BL~Lacs are
found throughout the diagram, including the region between log~(S$_{\rm
x}$/S$_{\rm r}$)$=$$-6.0$ and log~(S$_{\rm x}$/S$_{\rm r}$)$=$$-5.2$ which
represents a zone of avoidance in which few previously known BL~Lacs
resided.\label{fig:ratio}}
\end{figure}

\end{document}


\pagestyle{empty}

\begin{deluxetable} {lrlrrll}
\tablenum{2}
\tablecolumns{7}
\tablewidth{0pt}
\tablecaption{Observing log and Source Properties \label{tab:obs_times}}
\tablehead{
\colhead{} &
\colhead{} &
\multicolumn{2}{c}{Exposure} &
\colhead{} &
\colhead{} &
\colhead{}\\
\colhead{Name} &
\colhead{Run\tablenotemark{a}} &
\colhead{T$_{\rm tot}$} &
\colhead{\#} &
\colhead{Class} &
\colhead{z} &
\colhead{Notes}\\
\colhead{} &
\colhead{} &
\colhead{(hr)} &
\colhead{} &
\colhead{} &
\colhead{} &
\colhead{} }
\startdata
0006$+$125 &mcd4&0.33&1&      Quasar\phm{?}& 0.980 &                       \nl
0009$+$180 &mcd4&0.33&1&      Quasar\phm{?}& 0.310 &                       \nl
0027$+$452 & kp1&0.67&2&      Quasar\phm{?}& 0.971?&     Uncertain line ID \nl
0043$+$316 &mcd4&0.33&1&      Quasar\phm{?}& 0.631 &                       \nl
0044$+$193 &mcd4&0.33&1&      Quasar?      & 0.181 & Possible Narrow-lined
                                                            Sy~I; See text \nl
0046$+$228 &mcd4&0.33&1&      Quasar\phm{?}& 0.433 &                       \nl
0109$+$318 &mcd4&0.33&1&      Quasar\phm{?}& 1.71  &                       \nl
0110$+$139 &mcd4&0.33&1&     Cluster\phm{?}& 0.061 &    Br$_{4000}$$=$50\% \nl
0110$+$418 &mcd2&0.5 &1&     BL~Lac?       & 0.096 & Br$_{4000}$$=$32\%
                   (W$_{\lambda}^{\rm M}$$\le$25\AA); H$\alpha$ useful \nl
0112$+$383 &mcd2&0.5 &1&      Quasar\phm{?}& 0.333 &  Low resolution VLA catalog \nl
0123$+$262 &mcd4&0.5 &1&      Quasar\phm{?}& 0.849?&     Uncertain line ID \nl
0136$+$391 & kp1&0.33&1&      BL~Lac\phm{?}&\nodata& Low resolution VLA catalog \nl
0152$+$017 &mcd2&0.5 &1&     BL~Lac?       & 0.080 & Low resolution VLA catalog;
    Br$_{4000}$$=$29\% (W$_{\lambda}^{\rm M}$$\le$38\AA); H$\alpha$ useful \nl
0157$+$413 &mcd4&0.33&1&     Cluster\phm{?}& 0.080 & Br$_{4000}$$=$47\%;
                                 W$_{\lambda}$([OII]$\lambda$3727)$=$12\AA \nl
0207$+$295A&mcd4&0.33&1&        NLRG\phm{?}& 0.111 &Alternate radio-quiet
                                                    classification possible\nl
0250$+$172 &mcd4&1.17&1&      BL~Lac\phm{?}&\nodata&                       \nl
0254$+$395 &mcd4&0.33&1&      Quasar\phm{?}& 0.293 &                       \nl
0314$+$247 &mcd1&1.5 &3&     BL~Lac?       & 0.054 & Low resolution VLA catalog;
                                                     Br$_{4000}$$=$30\%; \nl
&&&&&&W$_{\lambda}$([OII]$\lambda$3727)$=$14\AA; W$_{\lambda}$([OIII]$\lambda$5007)$=$12\AA \nl
0321$+$236 &mcd4&0.67&2&      BL~Lac\phm{?}&\nodata&                       \nl
0326$+$024 &mcd1&1.0 &2&      BL~Lac\phm{?}&\nodata&  Low resolution VLA catalog; Prev.\ known BL~Lac \nl
0336$+$225B&mcd1&0.5 &1&      Quasar\phm{?}& 0.563 &  Low resolution VLA catalog \nl
0505$+$042 &mcd4&1.0 &3&      BL~Lac\phm{?}&\nodata&                       \nl
0656$+$426 &mcd4&0.25&1&     BL~Lac?       & 0.059 & Br$_{4000}$$=$36\%
              (W$_{\lambda}^{\rm M}$$\le$40\AA); Star-Forming Elliptical?;
                                                          H$\alpha$ useful \nl
0705$+$548 &mcd1&1.0 &2&      Quasar\phm{?}& 0.830 & $\Delta_{\rm
                                                      rx}$$=$52${\arcsec}$ \nl
0710$+$591 &mcd1&1.5 &3&      BL~Lac\phm{?}& 0.122 & Br$_{4000}$$=$22\%
                   (W$_{\lambda}^{\rm M}$$\le$25\AA); H$\alpha$ useful \nl
0714$+$741 &mcd3&0.67&2&      Quasar\phm{?}& 0.371 &                       \nl
0717$+$645 &mcd1&0.5 &1&      Quasar\phm{?}& 0.594?&     Uncertain line ID \nl
0735$+$712 &mcd3&0.5 &1&      Quasar\phm{?}&\nodata& W$_{\lambda}$$=$27\AA \nl
0746$+$527 &mcd1&0.5 &1&        BLRG\phm{?}& 0.542 &                       \nl
0749$+$451 &mcd3&0.25&1&        BLRG\phm{?}& 0.190 &                       \nl
0801$+$476 &mcd3&0.25&1&        BLRG\phm{?}& 0.155 &                       \nl
0806$+$595 &mcd1&1.75&4&      BL~Lac\phm{?}&\nodata&                       \nl
0806$+$728 &mcd3&0.42&1&        NLRG\phm{?}& 0.098 &Br$_{4000}$$=$16\%;
                              Alternate radio-quiet classification possible\nl
0810$+$504 &mcd3&0.5 &1&      Quasar\phm{?}& 1.20  &                       \nl
0814$+$561 &mcd1&1.0 &2&      Quasar\phm{?}& 0.511?&     Uncertain line ID \nl
0820$+$488 &mcd3&0.83&2&     Galaxy\phm{?} & 0.130 & Br$_{4000}$$=$41\%
                   (W$_{\lambda}^{\rm M}$$\le$47\AA); H$\alpha$ useful     \nl
0854$+$441 &mcd3&1.33&4&      BL~Lac\phm{?}&\nodata&Previously known BL~Lac\nl
0909$+$522 &mcd1&1.0 &2&        BLRG\phm{?}& 0.411 &                       \nl
0912$+$685 &mcd3&0.5 &1&      Quasar\phm{?}& 1.08  &                       \nl
0952$+$656 &mcd1&1.67&3&      BL~Lac\phm{?}&\nodata&                       \nl
1012$+$424 &mcd3&2.0 &4&     BL~Lac?       &\nodata& W$_{\lambda}$$=$10\AA\ \&
                                                                      4\AA \nl
1021$+$453 &mcd1&1.83&4&        BLRG\phm{?}& 0.364 &                       \nl
1031$+$508 &mcd1&1.0 &2&      BL~Lac\phm{?}&\nodata&Previously known BL~Lac\nl
1037$+$571 &mcd1&0.5 &1&      BL~Lac\phm{?}&\nodata&Previously known BL~Lac\nl
           &mcd3&0.5 &1&                   &       & W$_{\lambda}$$=$2\AA  \nl
1045$+$528A&mcd3&0.5 &1&      Quasar\phm{?}& 1.05  &                       \nl
1058$+$564 & kp2&0.5 &2&      BL~Lac?      &\nodata& $\Delta_{\rm
                                 rx}$$=$69${\arcsec}$; Prev.\ known BL~Lac \nl
           &mcd1&0.5 &1&                   &       &                       \nl
1110$+$715 &mcd3&1.5 &3&      BL~Lac\phm{?}&\nodata& $\Delta_{\rm
                                                      rx}$$=$55${\arcsec}$ \nl
1120$+$422 &mcd1&0.83&2&      BL~Lac\phm{?}&\nodata&Previously known BL~Lac\nl
1136$+$676 &mcd1&0.5 &2&      BL~Lac\phm{?}& 0.136 & Br$_{4000}$$=$20\%
  (W$_{\lambda}^{\rm M}$$\le$47\AA); H$\alpha$ useful; Prev.\ known BL~Lac \nl
1203$+$451 &mcd1&0.5 &1&      Quasar\phm{?}& 1.07  &                       \nl
1208$+$526 &mcd1&0.67&2&      Quasar\phm{?}& 0.427?&     Uncertain line ID \nl
1223$+$541 &mcd3&0.33&1&        BLRG\phm{?}& 0.157 &                       \nl
1233$+$483A& kp2&0.94&2&      Quasar\phm{?}&\nodata&                       \nl
1241$+$516B&mcd3&0.5 &1&      Quasar\phm{?}& 0.466?&     Uncertain line ID \nl
1241$+$495 &mcd1&1.33&3&      Quasar\phm{?}& 0.818?&     Uncertain line ID \nl
1248$+$514 & kp2&0.33&1&        Star\phm{?}&\nodata&     See text          \nl
1249$+$447B& kp2&0.33&1&      Quasar\phm{?}& 0.803 &                       \nl
1252$+$648 & kp2&0.33&1&        BLRG\phm{?}& 0.313 &                       \nl
1253$+$509 &mcd1&0.5 &1&      Galaxy\phm{?}& 0.121 & $\Delta_{\rm
                                 rx}$$=$55${\arcsec}$; Br$_{4000}$$=$49\%  \nl
1301$+$713 &mcd3&1.0 &2&        BLRG\phm{?}& 0.275 &                       \nl
1302$+$483 & kp2&0.5 &1&      Quasar\phm{?}& 0.875 &                       \nl
1306$+$554 & kp2&0.33&1&      Quasar\phm{?}& 1.60  &                       \nl
1324$+$576 &mcd1&0.67&2&      Galaxy\phm{?}& 0.115 & Br$_{4000}$$=$41\%    \nl
1334$+$565 & kp2&0.5 &1&        BLRG\phm{?}& 0.343 &                       \nl
1336$+$656 &mcd3&0.5 &1&        BLRG\phm{?}& 0.436?&     Uncertain line ID \nl
1346$+$623 & kp2&0.5 &1&        BLRG\phm{?}& 0.117 &Alternate radio-quiet
                                                    classification possible\nl
1353$+$431 &mcd3&0.33&1&      Quasar\phm{?}&\nodata&    Unidentified lines \nl
1356$+$416 &mcd3&0.33&1&      Quasar\phm{?}& 0.697?&     Uncertain line ID \nl
1411$+$426 & kp1&0.33&1&      Quasar\phm{?}& 0.888 &                       \nl
1413$+$436 & kp2&0.67&2&        BLRG\phm{?}& 0.090 &FWHM(H$\alpha$)$\approx$
                                               1900~km\,s$^{-1}$; See text \nl
1415$+$485 & kp2&1.17&3&      BL~Lac\phm{?}&\nodata&  Calibration error at
                                                               $\sim$9970? \nl
1417$+$257 & kp1&0.5 &1&      BL~Lac\phm{?}& 0.237 & Low resolution VLA
                           catalog; Br$_{4000}$$=$0\%; Prev.\ known BL~Lac \nl
1427$+$541 &mcd3&1.0 &2&     BL~Lac?       & 0.105 & Br$_{4000}$$=$39\%
                       (W$_{\lambda}^{\rm M}$$\le$37\AA); H$\alpha$ useful \nl
1428$+$426 & kp1&1.0 &2&      BL~Lac\phm{?}& 0.130 & Br$_{4000}$$=$1\%;
                                                   Previously known BL~Lac \nl
1437$+$507 & kp2&0.33&1&      Quasar\phm{?}& 0.785 &                       \nl
1439$+$395 & kp1&0.75&2&      BL~Lac\phm{?}&\nodata&Previously known BL~Lac\nl
1439$+$584 & kp2&0.5 &2&        BLRG\phm{?}& 0.425 &                       \nl
1442$+$120 & kp1&0.5 &1&      BL~Lac\phm{?}& 0.162 & Low resolution VLA
                                              catalog; Br$_{4000}$$=$18\%; \nl
 & & & & &&H$\alpha$ \& atmospheric A-band coincident; Prev.\ known BL~Lac \nl
1443$+$520B&mcd3&0.33&1&        NLRG\phm{?}& 0.142 &                       \nl
1448$+$361 & kp1&0.83&2&      BL~Lac\phm{?}&\nodata&                       \nl
1451$+$535 & kp1&0.5 &1&        BLRG\phm{?}& 0.433 &                       \nl
1454$+$514 & kp2&1.0 &2&     BL~Lac?       &\nodata& Flat fielding error at
                                                               $\sim$8380? \nl
1459$+$336 & kp1&0.5 &1&      Quasar\phm{?}& 0.644 &                       \nl
1504$+$568 &mcd3&1.5 &3&      Quasar\phm{?}&\nodata& Unidentified lines;
    W$_{\lambda}(\lambda6309)$$=$22\AA; W$_{\lambda}(\lambda3790)$$>$22\AA \nl
1509$+$573 & kp1&0.5 &1&      Quasar\phm{?}& 0.814?&     Uncertain line ID \nl
1515$+$479B& kp2&0.5 &1&      Quasar\phm{?}& 0.914 &                       \nl
1516$+$293 & kp1&0.83&2&     BL~Lac?       & 0.130 & Br$_{4000}$$=$32\%;
  W$_{\lambda}({\rm H}\alpha,{\rm [NII]})$$=$9\AA; Star-forming ellpitical?\nl
1517$+$654B& kp1&0.5 &2&      BL~Lac\phm{?}&\nodata&Previously known BL~Lac\nl
1518$+$485 & kp1&1.0 &2&      Quasar\phm{?}& 0.576 &                       \nl
1518$+$407 & kp2&0.42&2&       Sy\,I\phm{?}& 0.065 & FWHM(H$\alpha$)$\approx$
                                                         1900~km\,s$^{-1}$ \nl
           &mcd3&0.33&1&                   &       &                       \nl
1522$+$667B&mcd3&0.33&1&      Quasar\phm{?}& 0.629 &                       \nl
1523$+$636 &mcd3&0.33&1&        BLRG\phm{?}& 0.204 &                       \nl
1527$+$745C&mcd3&0.33&1&      Quasar\phm{?}& 0.921?&     Uncertain line ID \nl
1531$+$439 & kp2&0.58&2&      Quasar\phm{?}& 0.452 &                       \nl
1532$+$302 &mcd3&0.17&1&      BL~Lac?      & 0.064 &    Br$_{4000}$$=$29\% \nl
1533$+$342 & kp1&1.0 &2&      BL~Lac\phm{?}&\nodata&                       \nl
1534$+$372 & kp2&1.0 &3&     BL~Lac?       & 0.143 &
   W$_{\lambda}$(H$\alpha$,[NII]??)$=$4\AA; Br$_{4000}$ needed; O$-$E$=$1.6\nl
1534$+$586 & kp2&0.5 &1&      Quasar\phm{?}& 1.90? &     Uncertain line ID \nl
1542$+$614 &mcd3&0.67&2&      BL~Lac\phm{?}&\nodata&                       \nl
1545$+$348 & kp2&0.42&2&      Quasar\phm{?}& 0.518 &                       \nl
1601$+$179 & kp1&0.5 &1&      Quasar\phm{?}& 0.659?&     Uncertain line ID \nl
1602$+$308 & kp2&1.5 &3&      BL~Lac\phm{?}&\nodata&Previously known BL~Lac\nl
1608$+$603B&mcd3&1.0 &2&      Quasar\phm{?}& 0.178 &                       \nl
1610$+$671B& kp2&1.0 &2&      BL~Lac?      & 0.067?& Uncertain line ID;
       $\Delta_{\rm rx}$$=$43${\arcsec}$; Br$_{4000}$ needed; O$-$E$=$1.54 \nl
1610$+$728B&mcd3&1.0 &3&      Quasar\phm{?}& 0.592?&     Uncertain line ID \nl
1619$+$305 & kp2&0.33&1&      Quasar\phm{?}& 1.29? &     Uncertain line ID \nl
1622$+$401 & kp1&0.5 &1&      Quasar\phm{?}& 0.687 &                       \nl
1626$+$513 & kp1&0.33&1&        BLRG\phm{?}& 0.179 & $\Delta_{\rm
                                                      rx}$$=$42${\arcsec}$ \nl
1628$+$564 &mcd3&0.75&2&      Quasar\phm{?}& 0.400 &                       \nl
1629$+$401 & kp1&0.5 &1&        BLRG\phm{?}& 0.272 &                       \nl
1637$+$118 &mcd1&0.92&3&        BLRG\phm{?}& 0.144 & $\Delta_{\rm
        rx}$$=$56${\arcsec}$; Alternate radio-quiet classification possible\nl
1641$+$395 & kp1&0.33&1&      Quasar\phm{?}& 0.540 &                       \nl
1652$+$403 & kp2&2.0 &4&      BL~Lac\phm{?}&\nodata&                       \nl
1704$+$716 & kp1&1.0 &2&      BL~Lac\phm{?}&\nodata&Previously known BL~Lac\nl
1706$+$362 & kp1&0.5 &1&      Quasar\phm{?}& 1.07  &                       \nl
1706$+$322 & kp2&0.33&1&      Quasar\phm{?}& 0.917?&     Uncertain line ID \nl
1713$+$329 & kp1&0.5 &1&        BLRG\phm{?}& 0.102 &Alternate radio-quiet
                                                    classification possible\nl
1720$+$266 & kp2&1.0 &2&     Cluster?      & 0.161 & Br$_{4000}$ needed;
                                                               O$-$E$=$2.4 \nl
1721$+$211 & kp2&1.33&3&        BLRG\phm{?}& 0.628 &                       \nl
1722$+$246 & kp1&0.33&1&        BLRG\phm{?}& 0.175 &                       \nl
1722$+$282 & kp2&0.83&2&        BLRG\phm{?}&\nodata&  Unidentified line;
                                                     W$_{\lambda}$$>$60\AA \nl
1726$+$322 &mcd4&0.5 &1&      Quasar\phm{?}& 1.09  &                       \nl
1738$+$324 &mcd4&0.33&1&        BLRG\phm{?}& 0.126 &                       \nl
1742$+$597 & kp2&1.33&4&      BL~Lac\phm{?}&\nodata&                       \nl
1745$+$398 &mcd4&1.0 &2&      BL~Lac\phm{?}& 0.267 & Br$_{4000}$$=$17\%
                       (W$_{\lambda}^{\rm M}$$\le$21\AA); H$\alpha$ useful \nl
1750$+$470 & kp1&0.83&2&      BL~Lac?      & 0.160 & Br$_{4000}$$=$29\%;
                               H$\alpha$ \& atmospheric A-band coincident  \nl
1751$+$472 & kp1&0.33&1&      Quasar\phm{?}& 1.48  &                       \nl
1755$+$338 & kp2&0.33&1&        BLRG\phm{?}& 0.242 &                       \nl
1756$+$553 & kp1&0.83&2&      BL~Lac\phm{?}&\nodata&                       \nl
1757$+$538 &mcd4&0.33&1&     Cluster\phm{?}& 0.120 &    Br$_{4000}$$=$50\% \nl
1804$+$524 & kp1&0.5 &2&      Quasar\phm{?}& 0.516 &                       \nl
1807$+$438A&mcd2&0.5 &1&      Quasar\phm{?}& 0.815 &   $\Delta_{\rm
                                                      rx}$$=$51${\arcsec}$ \nl
1808$+$468 &mcd2&1.0 &2&      BL~Lac\phm{?}&\nodata&   See text            \nl
1808$+$665 &mcd2&0.5 &1&      Quasar\phm{?}& 0.697?&     Uncertain line ID \nl
1811$+$584 &mcd4&0.33&1&        Star\phm{?}&\nodata&              See text \nl
1811$+$442 & kp1&2.94&5&      BL~Lac\phm{?}&\nodata&                       \nl
1814$+$409B& kp1&0.5 &1&      Quasar\phm{?}& 0.971 &                       \nl
1823$+$334 &mcd4&0.33&1&      Galaxy\phm{?}& 0.108 &    Br$_{4000}$$=$49\% \nl
1829$+$540 & kp1&1.0 &2&      BL~Lac\phm{?}&\nodata&                       \nl
1838$+$480 & kp1&0.5 &1&      BL~Lac\phm{?}&\nodata&                       \nl
1841$+$591 & kp1&1.5 &3&      BL~Lac\phm{?}& 0.530 &   Br$_{4000}$$=$17\% 
                       (W$_{\lambda}^{\rm M}$$\le$10\AA); H$\alpha$ useful \nl
1844$+$546 &mcd2&0.5 &1&        BLRG\phm{?}& 0.234 &                       \nl
1848$+$427 & kp1&1.5 &3&      BL~Lac\phm{?}&\nodata&                       \nl
1922$+$691 &mcd4&0.33&1&              NLRG?& 0.099 &Alternate radio-quiet
                class.\ possible;No Balmer lines; See text \nl
2123$+$101A&mcd4&0.5 &1&        BLRG\phm{?}& 0.186 &                       \nl
2123$+$067 & kp1&0.5 &1&      Quasar\phm{?}& 0.934 &                       \nl
2145$+$073 & kp1&0.5 &1&      BL~Lac\phm{?}& 0.237 &   Br$_{4000}$$=$10\%
                    (W$_{\lambda}^{\rm M}$$\le$10\AA); Prev.\ known BL~Lac \nl
2146$+$093 &mcd4&0.33&1&      Quasar\phm{?}& 0.981 &                       \nl
2158$+$268B&mcd4&0.33&1&      Quasar\phm{?}& 0.713 &                       \nl
2201$+$050A& kp2&0.25&1&        BLRG\phm{?}& 0.232 &                       \nl
2217$+$226 & kp1&0.5 &1&      Quasar\phm{?}& 0.405 &                       \nl
2219$+$229 & kp1&1.33&3&      Quasar\phm{?}& 1.26  &                       \nl
2227$+$243 & kp1&0.5 &1&        NLRG\phm{?}& 0.319 &                       \nl
2229$+$309 &mcd4&0.33&1&      Quasar\phm{?}& 0.322 &                       \nl
2241$+$275 & kp2&0.5 &1&      Quasar\phm{?}& 0.493 &                       \nl
2241$+$048 &mcd4&0.33&1&      Galaxy\phm{?}& 0.069 &    Br$_{4000}$$=$45\% \nl
2243$+$203 &mcd4&1.17&3&      BL~Lac\phm{?}&\nodata&                       \nl
2246$+$317 &mcd4&0.33&1&        BLRG\phm{?}& 0.147 &Alternate radio-quiet
                                                    classification possible\nl
2250$+$384 &mcd4&1.17&3&      BL~Lac?      & 0.119 &   Br$_{4000}$$=$7\%
                             (W$_{\lambda}^{\rm M}$$\le$10\AA);
                                Star-forming elliptical?; H$\alpha$ useful \nl
2253$+$197 &mcd4&0.33&1&      Quasar\phm{?}& 0.298 &                       \nl
2256$+$263 & kp1&0.5 &1&        BLRG\phm{?}& 0.121 &Br$_{4000}$$=$25\%     \nl
2259$+$189 & kp2&0.75&2&     Galaxy?       & 0.132 &
                      (W$_{\lambda}$(H$\alpha$,[NII])$=$9\AA); O$-$E$=$2.7 \nl
2259$+$314 & kp2&0.5 &1&      Quasar\phm{?}& 1.14? &     Uncertain line ID \nl
2308$+$201 &mcd4&1.0 &3&        BLRG\phm{?}& 0.250 &                       \nl
2310$+$107 &mcd4&0.67&2&        BLRG\phm{?}& 0.436 &                       \nl
2311$+$355 &mcd4&1.17&3&      Quasar\phm{?}& 1.39? &     Uncertain line ID \nl
2313$+$103 &mcd4&0.5 &1&        BLRG\phm{?}& 0.456 &                       \nl
2318$+$308 &mcd4&0.5 &1&        BLRG\phm{?}& 0.103 &Alternate radio-quiet
                                                    classification possible\nl
2319$+$161 & kp1&1.0 &2&      BL~Lac\phm{?}&\nodata&                       \nl
2322$+$346 &mcd4&0.33&1&     BL~Lac?       & 0.098 & Br$_{4000}$$=$33\%
                             (W$_{\lambda}^{\rm M}$$\le$29\AA);
                                Star-forming elliptical?; H$\alpha$ useful \nl
2323$+$205 &mcd4&0.25&1&      Galaxy\phm{?}& 0.041 &    Br$_{4000}$$=$48\% \nl
\enddata
\tablenotetext{a}{See Table 1}
\end{deluxetable}


\pagestyle{empty}

\begin{deluxetable} {lllrccclr}
\tablenum{3}
\tablecolumns{9}
\tablewidth{0pt}
\tablecaption{Source Fluxes and Spectroscopic Classifications\label{tab:flux}}
\tablehead{
\colhead{RGB} &
\multicolumn{2}{c}{RA~~~(J2000)~~~Dec} &
\colhead{S$_{\rm r}$} &
\colhead{O} &
\colhead{$\log \frac{\rm S_{\rm r}}{S_{\rm opt}}$} &
\colhead{F$_{\rm X}$} &
\colhead{z} &
\colhead{Class}\\
\colhead{Name} &
\colhead{} &
\colhead{} &
\colhead{(mJy)} &
\colhead{(mag)} &
\colhead{} &
\colhead{($10^{12} \frac{\rm erg}{\rm s~cm^2}$)} &
\colhead{} &
\colhead{} }
\startdata
0006$+$125       &00 06 23.06&12 35 53.1&156&17.6                    &\phm{$-$}2.88&\phm{0}0.61&0.980         &Quasar\phm{?}      \nl
0009$+$180       &00 09 34.86&18 03 43.1& 70&16.6                     &\phm{$-$}1.95&\phm{0}1.58&0.310         &Quasar\phm{?}     \nl
0027$+$452       &00 27 42.26&45 14 57.1& 91&17.9                     &\phm{$-$}2.76&\phm{0}1.26&0.971?        &Quasar\phm{?}     \nl
0043$+$316       &00 43 59.83&31 37 20.3& 16&16.8                     &\phm{$-$}1.48&\phm{0}1.33&0.631         &Quasar\phm{?}     \nl
0044$+$193       &00 44 59.12&19 21 40.8&  7&16.8                     &\phm{$-$}1.09&\phm{0}2.63&0.181         &Quasar?           \nl
0046$+$228       &00 46 20.27&22 49 43.3& 58&16.3                     &\phm{$-$}1.79&\phm{0}3.04&0.433         &Quasar\phm{?}     \nl
0109$+$318       &01 09 27.88&31 49 56.0&181&17.8                     &\phm{$-$}3.16&\phm{0}1.41&1.71          &Quasar\phm{?}     \nl
0110$+$139       &01 10 03.16&13 58 41.4& 16&11.7                     &      $-$0.77&\phm{0}1.56&0.061         &Cluster\phm{?}    \nl
0110$+$418       &01 10 04.81&41 49 50.7& 18&18.4\tnm{a}\,/17.8\tnm{b}&\phm{$-$}2.00&\phm{0}2.46&0.096         &BL~Lac?           \nl
0112$+$383\tnm{c}&01 12 18.2 &38 18 59  &113&17.3                     &\phm{$-$}2.44&\phm{0}1.34&0.333         &Quasar\phm{?}     \nl
0123$+$262       &01 23 43.10&26 15 22.5&173&18.0                     &\phm{$-$}3.05&\phm{0}0.92&0.849?        &Quasar\phm{?}     \nl
0136$+$391\tnm{c}&01 36 32.7 & 39 06 00 & 49&15.8                     &\phm{$-$}1.42&20.2\phm{0}&\nodata       &BL~Lac\phm{?}     \nl
0152$+$017\tnm{c}&01 52 39.7 &01 47 18  & 65&17.1\tnm{a}\,/16.7\tnm{b}&\phm{$-$}2.03&\phm{0}4.97&0.080         &BL~Lac?           \nl
0157$+$413       &01 57 05.01&41 20 30.6& 23&14.3\tnm{d}              &\phm{$-$}0.45&\phm{0}9.50&0.081         &Cluster\phm{?}    \nl
0207$+$295A      &02 07 02.22&29 30 46.1& 19&16.5                     &\phm{$-$}1.27&\phm{0}5.05&0.110         &NLRG\phm{?}       \nl
0250$+$172       &02 50 37.96&17 12 08.5& 35&17.5                     &\phm{$-$}1.95&\phm{0}5.71&\nodata       &BL~Lac\phm{?}     \nl
0254$+$395       &02 54 42.63&39 31 34.8&273&16.4                     &\phm{$-$}2.45&\phm{0}3.10&0.293         &Quasar\phm{?}     \nl
0314$+$247\tnm{c}&03 14 02.7 &24 44 31  &  6&18.3\tnm{a}\,/17.9\tnm{b}&\phm{$-$}1.47&\phm{0}2.12&0.054         &BL~Lac?           \nl
0321$+$236       &03 21 59.93&23 36 11.0& 56&16.8                     &\phm{$-$}1.87&\phm{0}0.65&\nodata       &BL~Lac\phm{?}     \nl
0326$+$024\tnm{c}&03 26 14.0 &02 25 16  & 68&17.5                     &\phm{$-$}2.24&24.9\phm{0}&\nodata       &BL~Lac\phm{?}     \nl
0336$+$225B\tnm{c}&03 36 04.9&22 35 35  & 64&17.8                     &\phm{$-$}2.47&\phm{0}1.10&0.563         &Quasar\phm{?}     \nl
0505$+$042       &05 05 34.78&04 15 54.7& 90&17.6                     &\phm{$-$}2.40&\phm{0}7.61&\nodata       &BL~Lac\phm{?}     \nl
0656$+$426       &06 56 10.67&42 37 02.7&138&16.5\tnm{a}\,/15.4\tnm{b}&\phm{$-$}2.11&\phm{0}3.87&0.059         &BL~Lac?           \nl
0705$+$548       &07 05 44.69&54 48 09.9& 13&18.0                     &\phm{$-$}1.91&\phm{0}1.26&0.830         &Quasar\phm{?}     \nl
0710$+$591       &07 10 30.07&59 08 20.5& 34&18.0\tnm{a}\,/18.3\tnm{b}&\phm{$-$}2.13&35.6\phm{0}&0.125         &BL~Lac\phm{?}     \nl
0714$+$741       &07 14 36.19&74 08 10.2& 65&17.1                     &\phm{$-$}2.14&\phm{0}2.85&0.371         &Quasar\phm{?}     \nl
0717$+$645       &07 17 54.04&64 30 48.4&  8&17.7                     &\phm{$-$}1.51&\phm{0}1.89&0.594?        &Quasar\phm{?}     \nl
0735$+$712       &07 35 02.36&71 15 06.1& 72&18.1                     &\phm{$-$}2.69&\phm{0}1.46&\nodata\tnm{e}&Quasar\phm{?}     \nl
0746$+$527       &07 46 57.09&52 46 19.9&  8&18.4                     &\phm{$-$}1.79&\phm{0}1.69&0.542         &BLRG\phm{?}       \nl
0749$+$451       &07 49 06.51&45 10 33.9& 51&16.5                     &\phm{$-$}1.73&\phm{0}2.85&0.190         &BLRG\phm{?}       \nl
0801$+$476       &08 01 31.96&47 36 16.1& 42&16.4                     &\phm{$-$}1.59&11.5\phm{0}&0.155         &BLRG\phm{?}       \nl
0806$+$595       &08 06 25.94&59 31 06.9& 29&17.2                     &\phm{$-$}1.75&\phm{0}6.07&\nodata       &BL~Lac\phm{?}     \nl
0806$+$728       &08 06 38.97&72 48 20.6& 20&17.7\tnm{a}\,/18.6\tnm{b}&\phm{$-$}1.77&\phm{0}3.31&0.098         &NLRG\phm{?}       \nl
0810$+$504       &08 10 02.70&50 25 38.6& 12&17.8                     &\phm{$-$}1.88&\phm{0}0.98&1.20          &Quasar\phm{?}     \nl
0814$+$561       &08 14 32.12&56 09 56.8& 49&17.7                     &\phm{$-$}2.30&\phm{0}1.94&0.511?        &Quasar\phm{?}     \nl
0820$+$488       &08 20 28.10&48 53 47.5&  5&17.5                     &\phm{$-$}1.05&\phm{0}0.98&0.130         &Galaxy\phm{?}     \nl
0854$+$441       &08 54 09.88&44 08 30.3& 31&17.6                     &\phm{$-$}1.93&\phm{0}4.97&\nodata       &BL~Lac\phm{?}     \nl
0909$+$522       &09 09 24.68&52 16 32.7& 33&18.4                     &\phm{$-$}2.37&\phm{0}1.27&0.411         &BLRG\phm{?}       \nl
0912$+$685       &09 12 36.59&68 34 25.1&104&17.8                     &\phm{$-$}2.80&\phm{0}0.61&1.08          &Quasar\phm{?}     \nl
0952$+$656       &09 52 32.19&65 38 01.1& 27&18.0                     &\phm{$-$}2.04&\phm{0}1.82&\nodata       &BL~Lac\phm{?}     \nl
1012$+$424       &10 12 44.30&42 29 57.2& 29&18.1                     &\phm{$-$}2.11&10.4\phm{0}&\nodata       &BL~Lac?           \nl
1021$+$453       &10 21 06.04&45 23 31.9& 23&18.2                     &\phm{$-$}2.12&\phm{0}1.06&0.364         &BLRG\phm{?}       \nl
1031$+$508       &10 31 18.52&50 53 35.8& 23&16.3                     &\phm{$-$}1.28&71.5\phm{0}&\nodata       &BL~Lac\phm{?}     \nl
1037$+$571       &10 37 44.31&57 11 55.8& 89&17.4                     &\phm{$-$}2.32&\phm{0}2.58&\nodata       &BL~Lac\phm{?}     \nl
1045$+$528A      &10 45 42.18&52 51 12.6& 43&18.1                     &\phm{$-$}2.53&\phm{0}0.68&1.05          &Quasar\phm{?}     \nl
1058$+$564       &10 58 37.73&56 28 11.3&178&15.8                     &\phm{$-$}1.98&\phm{0}4.69&\nodata       &BL~Lac?           \nl
1110$+$715       &11 10 37.59&71 33 56.6& 22&18.0                     &\phm{$-$}1.95&\phm{0}2.12&\nodata       &BL~Lac\phm{?}     \nl
1120$+$422       &11 20 48.07&42 12 12.5& 19&17.1                     &\phm{$-$}1.51&10.0\phm{0}&\nodata       &BL~Lac\phm{?}     \nl
1136$+$676       &11 36 30.08&67 37 04.4& 40&18.0\tnm{a}\,/18.3\tnm{b}&\phm{$-$}2.21&27.4\phm{0}&0.136         &BL~Lac\phm{?}     \nl
1203$+$451       &12 03 35.39&45 10 49.6& 53&17.7                     &\phm{$-$}2.47&\phm{0}0.88&1.07          &Quasar\phm{?}     \nl
1208$+$526       &12 08 22.45&52 40 13.5& 19&16.7                     &\phm{$-$}1.46&\phm{0}2.01&0.427?        &Quasar\phm{?}     \nl
1223$+$541       &12 23 13.10&54 09 07.8& 42&16.2                     &\phm{$-$}1.51&\phm{0}1.33&0.157         &BLRG\phm{?}       \nl
1233$+$483A      &12 33 54.45&48 20 50.0&  7&18.0                     &\phm{$-$}1.64&\phm{0}0.51&\nodata\tnm{e}&Quasar\phm{?}     \nl
1241$+$516B      &12 41 16.49&51 41 30.0& 20&17.9                     &\phm{$-$}1.97&\phm{0}0.95&0.466?        &Quasar\phm{?}     \nl
1241$+$495       &12 41 39.73&49 34 05.5&  7&17.6                     &\phm{$-$}1.47&\phm{0}1.57&0.818?        &Quasar\phm{?}     \nl
1248$+$514       &12 48 34.30&51 28 07.9& 61&18.4                     &\nodata      &\phm{0}0.67&\nodata       &Star\tnm{f}\phm{?}\nl
1249$+$447B      &12 49 23.56&44 44 50.5& 12&17.7                     &\phm{$-$}1.75&\phm{0}0.90&0.803         &Quasar\phm{?}     \nl
1252$+$648       &12 52 23.81&64 51 38.0& 26&17.7                     &\phm{$-$}1.96&\phm{0}1.88&0.313         &BLRG\phm{?}       \nl
1253$+$509       &12 53 26.16&50 54 28.2&  7&17.7                     &\phm{$-$}1.27&\phm{0}1.06&0.121         &Galaxy\phm{?}     \nl
1301$+$713       &13 01 30.37&71 20 13.1& 30&17.4                     &\phm{$-$}1.89&\phm{0}1.32&0.275         &BLRG\phm{?}       \nl
1302$+$483       &13 02 17.20&48 19 17.6& 90&17.7                     &\phm{$-$}2.65&\phm{0}0.27&0.875         &Quasar\phm{?}     \nl
1306$+$554       &13 06 03.36&55 29 44.0&282&17.9                     &\phm{$-$}3.37&\phm{0}0.35&1.60          &Quasar\phm{?}     \nl
1324$+$576       &13 24 00.92&57 39 16.4& 25&16.6                     &\phm{$-$}1.42&\phm{0}3.21&0.115         &Galaxy\phm{?}     \nl
1334$+$565       &13 34 37.49&56 31 47.9& 36&17.8                     &\phm{$-$}2.15&\phm{0}0.81&0.343         &BLRG\phm{?}       \nl
1336$+$656       &13 36 55.53&65 41 16.1&  6&18.0                     &\phm{$-$}1.45&\phm{0}0.83&0.436?        &BLRG\phm{?}       \nl
1346$+$623       &13 46 17.59&62 20 45.5&  7&17.9                     &\phm{$-$}1.35&\phm{0}2.07&0.117         &BLRG\phm{?}       \nl
1353$+$431       &13 53 41.74&43 10 52.4& 24&16.9                     &\phm{$-$}1.74&\phm{0}0.77&\nodata\tnm{e}&Quasar\phm{?}     \nl
1356$+$416       &13 56 07.38&41 36 15.2& 47&16.4                     &\phm{$-$}1.81&\phm{0}1.02&0.697?        &Quasar\phm{?}     \nl
1411$+$426       &14 11 59.74&42 39 50.4& 46&17.2                     &\phm{$-$}2.16&\phm{0}0.90&0.888         &Quasar\phm{?}     \nl
1413$+$436       &14 13 43.72&43 39 45.1& 34&16.0                     &\phm{$-$}1.32&\phm{0}4.47&0.090         &BLRG\phm{?}       \nl
1415$+$485       &14 15 36.82&48 30 30.5& 58&17.6                     &\phm{$-$}2.21&\phm{0}0.41&\nodata       &BL~Lac\phm{?}     \nl
1417$+$257\tnm{c}&14 17 56.7 &25 43 25  & 40&17.5                     &\phm{$-$}2.04&25.3\phm{0}&0.237         &BL~Lac\phm{?}     \nl
1427$+$541       &14 27 30.33&54 09 23.5& 24&17.4\tnm{a}\,/16.1\tnm{b}&\phm{$-$}1.73&\phm{0}0.80&0.105         &BL~Lac?           \nl
1428$+$426       &14 28 32.62&42 40 21.1& 21&16.9                     &\phm{$-$}1.56&80.8\phm{0}&0.130         &BL~Lac\phm{?}     \nl
1437$+$507       &14 37 26.17&50 45 55.9&  8&18.1                     &\phm{$-$}1.73&\phm{0}0.85&0.785         &Quasar\phm{?}     \nl
1439$+$395       &14 39 17.48&39 32 42.7& 38&18.1                     &\phm{$-$}2.22&15.3\phm{0}&\nodata       &BL~Lac\phm{?}     \nl
1439$+$584       &14 39 42.86&58 27 59.1& 16&18.5                     &\phm{$-$}2.09&\phm{0}1.39&0.425         &BLRG\phm{?}       \nl
1442$+$120\tnm{c}&14 42 48.2 &12 00 41  & 45&17.9\tnm{a}\,/18.5\tnm{b}&\phm{$-$}2.23&11.0\phm{0}&0.162         &BL~Lac\phm{?}     \nl
1443$+$520B      &14 43 02.76&52 01 37.3& 65&17.4                     &\phm{$-$}2.18&\phm{0}5.92&0.142         &NLRG\phm{?}       \nl
1448$+$361       &14 48 00.59&36 08 31.0& 29&17.2                     &\phm{$-$}1.75&\phm{0}7.54&\nodata       &BL~Lac\phm{?}     \nl
1451$+$535       &14 51 06.44&53 33 54.0&  6&18.3                     &\phm{$-$}1.57&\phm{0}0.81&0.433         &BLRG\phm{?}       \nl
1454$+$514       &14 54 27.12&51 24 33.6& 80&18.4                     &\phm{$-$}2.67&\phm{0}0.95&\nodata       &BL~Lac?           \nl
1459$+$336       &14 59 58.45&33 37 01.6&105&17.6                     &\phm{$-$}2.62&\phm{0}1.34&0.644         &Quasar\phm{?}     \nl
1504$+$568       &15 04 55.57&56 49 20.3&  6&17.1                     &\phm{$-$}1.21&\phm{0}0.58&\nodata\tnm{e}&Quasar\phm{?}     \nl
1509$+$573       &15 09 40.69&57 18 11.9& 10&18.3                     &\phm{$-$}1.92&\phm{0}0.42&0.814?        &Quasar\phm{?}     \nl
1515$+$479B      &15 15 12.02&47 55 09.0&  9&18.4                     &\phm{$-$}1.94&\phm{0}0.26&0.914         &Quasar\phm{?}     \nl
1516$+$293       &15 16 41.59&29 18 09.2& 34&18.2\tnm{a}\,/17.6\tnm{b}&\phm{$-$}2.21&\phm{0}2.15&0.130         &BL~Lac?           \nl
1517$+$654B      &15 17 47.58&65 25 23.3& 19&17.4                     &\phm{$-$}1.63&19.9\phm{0}&\nodata       &BL~Lac\phm{?}     \nl
1518$+$485       &15 18 30.95&48 32 14.4&  5&18.4                     &\phm{$-$}1.55&\phm{0}0.89&0.576         &Quasar\phm{?}     \nl
1518$+$407       &15 18 38.91&40 45 00.1& 29&15.0                     &\phm{$-$}0.82&\phm{0}0.85&0.065         &Sy\,I\phm{?}      \nl
1522$+$667B      &15 22 58.94&66 45 06.7& 61&17.5                     &\phm{$-$}2.34&\phm{0}1.25&0.629         &Quasar\phm{?}     \nl
1523$+$636       &15 23 45.87&63 39 24.3& 14&16.8                     &\phm{$-$}1.27&\phm{0}5.93&0.204         &BLRG\phm{?}       \nl
1527$+$745C      &15 27 12.08&74 31 50.8&  8&17.5                     &\phm{$-$}1.52&\phm{0}0.50&0.921?        &Quasar\phm{?}     \nl
1531$+$439       &15 31 02.49&43 56 37.6& 21&16.8                     &\phm{$-$}1.54&\phm{0}0.81&0.452         &Quasar\phm{?}     \nl
1532$+$302       &15 32 02.23&30 16 28.9& 47&15.6\tnm{a}\,/17.9\tnm{b}&\phm{$-$}1.29&\phm{0}5.87&0.064         &BL~Lac?           \nl
1533$+$342       &15 33 24.27&34 16 40.2& 33&17.9                     &\phm{$-$}2.09&\phm{0}5.73&\nodata       &BL~Lac\phm{?}     \nl
1534$+$372       &15 34 47.21&37 15 54.4& 20&18.3                     &\phm{$-$}2.01&\phm{0}0.48&0.143         &BL~Lac?           \nl
1534$+$586       &15 34 57.24&58 39 23.5&150&18.1                     &\phm{$-$}3.23&\phm{0}0.28&1.90?         &Quasar\phm{?}     \nl
1542$+$614       &15 42 56.94&61 29 55.3&102&17.2                     &\phm{$-$}2.30&\phm{0}1.46&\nodata       &BL~Lac\phm{?}     \nl
1545$+$348       &15 45 10.95&34 52 46.8&  7&17.1\tnm{g}              &\phm{$-$}1.19&\phm{0}1.49&0.518         &Quasar\phm{?}     \nl
1601$+$179       &16 01 51.56&17 54 09.9&  7&18.4                     &\phm{$-$}1.75&\phm{0}0.49&0.659?        &Quasar\phm{?}     \nl
1602$+$308       &16 02 18.04&30 51 09.3& 20&18.5\tnm{g}              &\phm{$-$}2.11&\phm{0}2.59&\nodata       &BL~Lac\phm{?}     \nl
1608$+$603B      &16 08 20.55&60 18 28.2& 26&16.0                     &\phm{$-$}1.22&\phm{0}2.25&0.178         &Quasar\phm{?}     \nl
1610$+$671B      &16 10 04.07&67 10 26.4& 13&18.4                     &\phm{$-$}1.78&\phm{0}4.83&0.067?        &BL~Lac?           \nl
1610$+$728B      &16 10 49.94&72 50 46.4&  5&17.7                     &\phm{$-$}1.27&\phm{0}0.65&0.592?        &Quasar\phm{?}     \nl
1619$+$305       &16 19 02.48&30 30 51.5& 38&17.1                     &\phm{$-$}2.13&\phm{0}0.94&1.29?         &Quasar\phm{?}     \nl
1622$+$401       &16 22 29.32&40 06 43.7& 30&17.9                     &\phm{$-$}2.21&\phm{0}0.97&0.687         &Quasar\phm{?}     \nl
1626$+$513       &16 26 11.62&51 20 38.3& 14&17.3                     &\phm{$-$}1.47&\phm{0}1.61&0.179         &BLRG\phm{?}       \nl
1628$+$564       &16 28 50.41&56 29 29.2& 18&17.4                     &\phm{$-$}1.71&\phm{0}0.50&0.400         &Quasar\phm{?}     \nl
1629$+$401       &16 29 01.32&40 08 00.0& 21&17.6                     &\phm{$-$}1.80&\phm{0}9.05&0.272         &BLRG\phm{?}       \nl
1637$+$118       &16 37 46.50&11 49 49.0& 11&17.4                     &\phm{$-$}1.39&\phm{0}1.19&0.144         &BLRG\phm{?}       \nl
1641$+$395       &16 41 47.51&39 35 03.2& 47&18.0                     &\phm{$-$}2.41&\phm{0}0.67&0.540         &Quasar\phm{?}     \nl
1652$+$403       &16 52 49.94&40 23 10.2& 11&17.3                     &\phm{$-$}1.35&\phm{0}0.68&\nodata       &BL~Lac\phm{?}     \nl
1704$+$716       &17 04 46.98&71 38 17.6& 17&16.5                     &\phm{$-$}1.23&\phm{0}3.79&\nodata       &BL~Lac\phm{?}     \nl
1706$+$362       &17 06 34.11&36 15 08.0& 15&18.3                     &\phm{$-$}2.16&\phm{0}0.91&1.07          &Quasar\phm{?}     \nl
1706$+$322       &17 06 48.05&32 14 22.8& 20&16.8                     &\phm{$-$}1.65&\phm{0}0.76&0.917?        &Quasar\phm{?}     \nl
1713$+$329       &17 13 22.58&32 56 27.9& 25&16.8                     &\phm{$-$}1.50&\phm{0}4.50&0.102         &BLRG\phm{?}       \nl
1720$+$266       &17 20 09.92&26 37 31.1&  7&17.4                     &\phm{$-$}1.21&14.6\phm{0}&0.161         &Cluster?          \nl
1721$+$211       &17 21 33.25&21 11 02.0& 42&19.2\tnm{g}              &\phm{$-$}2.86&\phm{0}0.75&0.628         &BLRG\phm{?}       \nl
1722$+$246       &17 22 41.25&24 36 19.1& 28&16.5                     &\phm{$-$}1.45&\phm{0}3.33&0.175         &BLRG\phm{?}       \nl
1722$+$282       &17 22 42.16&28 15 00.1&150&20.0                     &\phm{$-$}3.61&\phm{0}0.78&\nodata\tnm{e}&BLRG\phm{?}       \nl
1726$+$322       &17 26 35.13&32 13 23.0&183&17.6                     &\phm{$-$}2.97&\phm{0}1.14&1.09          &Quasar\phm{?}     \nl
1738$+$324       &17 38 40.50&32 24 09.0&209&16.1                     &\phm{$-$}2.16&\phm{0}2.40&0.126         &BLRG\phm{?}       \nl
1742$+$597       &17 42 32.00&59 45 06.8& 77&17.4                     &\phm{$-$}2.26&\phm{0}0.91&\nodata       &BL~Lac\phm{?}     \nl
1745$+$398       &17 45 37.76&39 51 30.8&118&18.3\tnm{a}\,/18.9\tnm{b}&\phm{$-$}2.84&\phm{0}1.59&0.267         &BL~Lac\phm{?}     \nl
1750$+$470       &17 50 05.00&47 00 43.7& 10&19.0\tnm{a}\,/18.7\tnm{b}&\phm{$-$}2.01&\phm{0}3.47&0.160         &BL~Lac?           \nl
1751$+$472       &17 51 31.63&47 13 22.6& 34&17.1                     &\phm{$-$}2.11&\phm{0}1.67&1.48          &Quasar\phm{?}     \nl
1755$+$338       &17 55 11.27&33 50 59.7&158&17.9                     &\phm{$-$}2.80&\phm{0}1.16&0.242         &BLRG\phm{?}       \nl
1756$+$553       &17 56 15.89&55 22 18.1& 10&17.9                     &\phm{$-$}1.55&14.1\phm{0}&\nodata       &BL~Lac\phm{?}     \nl
1757$+$538       &17 57 06.70&53 51 37.6& 31&13.4\tnm{g}              &\phm{$-$}0.24&\phm{0}5.77&0.120         &Cluster\phm{?}    \nl
1804$+$524       &18 04 52.74&52 24 29.4&  7&17.0                     &\phm{$-$}1.15&\phm{0}0.76&0.516         &Quasar\phm{?}     \nl
1807$+$438A      &18 07 59.84&43 50 36.9&  5&18.1\tnm{g}              &\phm{$-$}1.14&\phm{0}1.00&0.815         &Quasar\phm{?}     \nl
1808$+$468       &18 08 01.20&46 49 41.0& 41&18.4                     &\phm{$-$}2.38&\phm{0}0.58&\nodata       &BL~Lac\phm{?}     \nl
1808$+$665       &18 08 49.58&66 34 29.7&  6&17.5                     &\phm{$-$}1.35&\phm{0}0.97&0.697?        &Quasar\phm{?}     \nl
1811$+$584       &18 11 33.92&58 25 15.6&  3&17.8                     &\nodata      &\phm{0}0.49&\nodata       &Star\tnm{f}\phm{?}\nl
1811$+$442       &18 11 53.47&44 16 28.5&  6&18.9\tnm{g}              &\phm{$-$}1.69&\phm{0}0.87&\nodata       &BL~Lac\phm{?}     \nl
1814$+$409B      &18 14 34.48&40 57 46.4&  7&18.7                     &\phm{$-$}1.94&\phm{0}1.30&0.971         &Quasar\phm{?}     \nl
1823$+$334       &18 23 09.84&33 24 39.1&  3&14.9                     &      $-$0.22&\phm{0}2.94&0.108         &Galaxy\phm{?}     \nl
1829$+$540       &18 29 24.29&54 02 59.8& 18&17.6                     &\phm{$-$}1.69&\phm{0}3.51&\nodata       &BL~Lac\phm{?}     \nl
1838$+$480       &18 38 49.17&48 02 34.4& 23&17.6                     &\phm{$-$}1.80&\phm{0}3.82&\nodata       &BL~Lac\phm{?}     \nl
1841$+$591       &18 41 20.31&59 06 08.2&  6&19.7\tnm{a}\,/20.3\tnm{b}&\phm{$-$}2.17&\phm{0}0.74&0.530         &BL~Lac\phm{?}     \nl
1844$+$546       &18 44 30.88&54 41 44.8& 32&17.2                     &\phm{$-$}1.82&\phm{0}1.27&0.234         &BLRG\phm{?}       \nl
1848$+$427       &18 48 47.14&42 45 39.4&  8&18.9                     &\phm{$-$}1.86&13.9\phm{0}&\nodata       &BL~Lac\phm{?}     \nl
1922$+$691       &19 22 14.65&69 11 12.4& 20&17.0\tnm{d}              &\phm{$-$}1.47&\phm{0}2.95&0.099         &NLRG?             \nl
2123$+$101A      &21 23 13.34&10 07 54.9&301&19.0                     &\phm{$-$}3.50&\phm{0}0.65&0.186         &BLRG\phm{?}       \nl
2123$+$067       &21 23 19.33&06 46 22.9& 11&18.0                     &\phm{$-$}1.87&\phm{0}1.55&0.934         &Quasar\phm{?}     \nl
2145$+$073       &21 45 52.30&07 19 27.2& 37&19.0\tnm{a}\,/20.7\tnm{b}&\phm{$-$}2.61&11.9\phm{0}&0.237         &BL~Lac\phm{?}     \nl
2146$+$093       &21 46 54.08&09 20 48.9& 50&17.1                     &\phm{$-$}2.18&\phm{0}1.11&0.981         &Quasar\phm{?}     \nl
2158$+$268B      &21 58 25.60&26 52 41.4& 24&17.8                     &\phm{$-$}2.08&\phm{0}1.11&0.713         &Quasar\phm{?}     \nl
2201$+$050A      &22 01 07.16&05 04 36.3& 13&17.9                     &\phm{$-$}1.71&\phm{0}1.37&0.232         &BLRG\phm{?}       \nl
2217$+$226       &22 17 10.96&22 39 46.8& 19&17.4                     &\phm{$-$}1.72&\phm{0}1.51&0.405         &Quasar\phm{?}     \nl
2219$+$229       &22 19 21.02&22 59 22.7& 10&18.6                     &\phm{$-$}2.13&\phm{0}1.01&1.26          &Quasar\phm{?}     \nl
2227$+$243       &22 27 47.95&24 21 12.4& 22&19.0                     &\phm{$-$}2.41&\phm{0}0.95&0.319         &NLRG\phm{?}       \nl
2229$+$309       &22 29 34.15&30 57 12.2& 59&16.4                     &\phm{$-$}1.79&\phm{0}1.97&0.322         &Quasar\phm{?}     \nl
2241$+$275       &22 41 02.03&27 32 59.5& 35&17.8                     &\phm{$-$}2.18&\phm{0}1.72&0.493         &Quasar\phm{?}     \nl
2241$+$048       &22 41 34.25&04 53 10.6& 41&16.8\tnm{d}              &\phm{$-$}1.70&\phm{0}0.94&0.069         &Galaxy\phm{?}     \nl
2243$+$203       &22 43 54.73&20 21 03.9& 86&16.0                     &\phm{$-$}1.74&\phm{0}0.97&\nodata       &BL~Lac\phm{?}     \nl
2246$+$317       &22 46 21.68&31 42 08.0& 13&16.5                     &\phm{$-$}1.09&\phm{0}4.13&0.147         &BLRG\phm{?}       \nl
2250$+$384       &22 50 05.77&38 24 37.3& 60&16.0\tnm{g,a}~~~/18.0\tnm{b}&\phm{$-$}1.17&\phm{0}7.36&0.119      &BL~Lac?           \nl
2253$+$197       &22 53 07.37&19 42 34.8&254&16.3                     &\phm{$-$}2.38&\phm{0}1.01&0.298         &Quasar\phm{?}     \nl
2256$+$263       &22 56 39.17&26 18 43.8& 40&19.1\tnm{a}\,/19.0\tnm{b}&\phm{$-$}2.64&\phm{0}2.14&0.121         &BLRG\phm{?}       \nl
2259$+$189       &22 59 19.16&18 57 30.3& 17&17.6                     &\phm{$-$}1.67&\phm{0}1.96&0.132         &Galaxy?           \nl
2259$+$314       &22 59 50.77&31 24 52.1& 13&18.4                     &\phm{$-$}2.14&\phm{0}0.85&1.14?         &Quasar\phm{?}     \nl
2308$+$201       &23 08 11.63&20 08 42.3&177&17.6                     &\phm{$-$}2.73&\phm{0}1.80&0.250         &BLRG\phm{?}       \nl
2310$+$107       &23 10 09.74&10 47 24.3& 12&18.1                     &\phm{$-$}1.82&\phm{0}1.50&0.436         &BLRG\phm{?}       \nl
2311$+$355       &23 11 48.98&35 35 41.4& 18&17.2                     &\phm{$-$}1.86&\phm{0}0.74&1.39?         &Quasar\phm{?}     \nl
2313$+$103       &23 13 55.67&10 19 09.8& 35&18.1                     &\phm{$-$}2.29&\phm{0}1.06&0.456         &BLRG\phm{?}       \nl
2318$+$308       &23 18 36.90&30 48 37.0& 22&17.5                     &\phm{$-$}1.73&\phm{0}2.85&0.103         &BLRG\phm{?}       \nl
2319$+$161       &23 19 43.44&16 11 50.1& 17&18.0\tnm{g}              &\phm{$-$}1.83&\phm{0}4.61&\nodata       &BL~Lac\phm{?}     \nl
2322$+$346       &23 22 44.01&34 36 14.0& 30&18.2\tnm{a}\,/17.4\tnm{b}&\phm{$-$}2.14&\phm{0}2.16&0.098         &BL~Lac?           \nl
2323$+$205       &23 23 20.35&20 35 23.6&114&16.0\tnm{d}              &\phm{$-$}1.82&\phm{0}0.59&0.041         &Galaxy\phm{?}     \nl
\enddata
\tablenotetext{a}{Estimate of AGN magnitude}
\tablenotetext{b}{Estimate of Galaxy magnitude}
\tablenotetext{c}{Object from low radio resolution sample}
\tablenotetext{d}{Converted from V to B mag assuming $\alpha_{\rm opt}$$=$$1.0$.}
\tablenotetext{e}{Line identificaton unknown; Redshift undetermined}
\tablenotetext{f}{Optical counterpart is likely a chance coincidence}
\tablenotetext{g}{Converted from E to B mag assuming $\alpha_{\rm opt}$$=$$1.0$.}
\end{deluxetable}